\documentclass[11pt,a4paper]{uolthesis}
\usepackage[T1]{fontenc}
\usepackage{lmodern}
\usepackage{url}                 
\usepackage{lscape}                 
\usepackage{subfigure}
\usepackage{mathrsfs}
\usepackage{graphicx,wrapfig,lipsum}
\usepackage{caption2}
\usepackage{epstopdf}

\usepackage{amssymb}
\usepackage{amsmath}
\usepackage{tensor}
\usepackage{amsthm}
\usepackage{latexsym}
\usepackage[dvips]{epsfig}
\usepackage{eufrak}
\usepackage{bm}
\usepackage{tikz}
\usepackage{authblk}
\usepackage{enumerate}
\usepackage{mathtools}

\begin{document}

\title{Numerical Cauchy evolution of asymptotically AdS spacetimes with no symmetries}

\author{\textbf{Lorenzo Rossi}}
\department{
School of Mathematical Sciences} 
\college{Queen Mary, University of London}
\degree{Doctor of Philosophy} \degreemonth{December} \degreeyear{2021}

\copyrightnoticetext{\copyright ~Queen Mary University of London, 2021}

\dedication{To my family, for showing me how to always be brave in the face of adversities.}


\setcounter{page}{1}
\pagenumbering{roman} 

\maketitle


\thispagestyle{plain}
\begin{declaration}
I, Lorenzo Rossi, confirm that the research included within this thesis is my own work or that where it has been carried out in collaboration with, or supported by others, that this is duly acknowledged below and my contribution indicated. Previously published material is also acknowledged below.

I attest that I have exercised reasonable care to ensure that the work is original, and does not to the best of my knowledge break any UK law, infringe any third party's copyright or other Intellectual Property Right, or contain any confidential material.

I accept that the College has the right to use plagiarism detection software to check the electronic version of the thesis.

I confirm that this thesis has not been previously submitted for the award of a degree by this or any other university.

The copyright of this thesis rests with the author and no quotation from it or information derived from it may be published without the prior written consent of the author.
\newline
\newline
\newline
Signature: \includegraphics[width=2.0in]{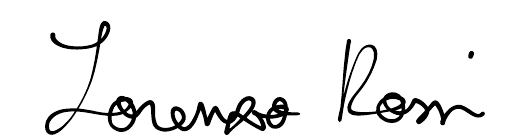}

Date: 25$^{\text{th}}$ December 2021
\end{declaration}

\thispagestyle{plain}
\begin{listpubs}
Many of the ideas of Chapter~\ref{Chapter:NoSym} are based on the following published manuscript.

\begin{itemize}
\item H.~Bantilan, P.~Figueras, L.~Rossi, \emph{Cauchy Evolution of Asymptotically Global AdS Spacetimes with No Symmetries}, {\em Physical Review D} 103 (2021) 086006 \url{https://arxiv.org/abs/2011.12970}.
 
 \end{itemize}
 
Many of the ideas of Chapter~\ref{Chapter:KAdS} are based on preliminary, currently unpublished work.

\end{listpubs}

%

\thispagestyle{plain}
\begin{abstract}
General relativity, the currently accepted classical theory of gravity, in asymptotically anti-de Sitter (AdS) spacetimes has been conjectured to be dual to a strongly interacting conformal field theory (CFT), similar in some respects to real world theories of quantum particles.
This is the paradigm of the AdS/CFT correspondence.
The field of numerical relativity aims to solve the Einstein equations of general relativity, to high accuracy, in computers, in order to investigate problems for which the exact solution is not known and other approximation methods do not apply.
In asymptotically AdS spacetimes, this program can be successful, and thus employed in AdS/CFT-related studies, if certain conditions are imposed on the solution at the AdS boundary.

In this thesis, I present the first numerical scheme able to perform Cauchy evolutions of asymptotically AdS spacetimes with reflective boundary conditions under no symmetry requirements on the solution.
The scheme is based on the generalised harmonic formulation of the Einstein equations.
The main difficulty in removing all symmetry assumptions can be phrased in terms of finding a set of generalised harmonic source functions that are consistent with the AdS boundary conditions.
I detail a prescription to obtain the set of source functions that achieves stable evolution in full generality.
This prescription leads to the first long-time stable 3+1 simulations of four dimensional spacetimes with a negative cosmological constant in Cartesian coordinates.
I show results of gravitational collapse with no symmetry assumptions, and the subsequent ringdown to a static black hole in the bulk, which corresponds to evolution towards a homogeneous state on the boundary.
Furthermore, it is argued that this scheme is well-suited for the study of black hole superradiance -- the amplification of waves scattering off a rotating black hole -- in Kerr-AdS spacetime. This statement is supported by the results of a preliminary simulation of perturbed Kerr-AdS.
\end{abstract}

\thispagestyle{plain}
\markboth{Acknowledgments}{Acknowledgments}
\newif\ifpaper
\paperfalse

\ifpaper
\input{../preamble}
\begin{document}
\fi

\begin{acknowledgments}
During the four years of my Ph.D., life has surprised me in various ways. 
When I found myself in turbulent waters, certain people were the anchors that helped me remain ``close to the shore''.
When the water was calm, certain people were able to turn the peacefulness into happiness.
This allowed me to enjoy the process that led to the research presented in this thesis.
Therefore, it is for me both a necessity and a pleasure to dedicate a few words to those who supported me throughout the years.

I was lucky to do research under the guidance of Pau Figueras and Hans Bantilan, my supervisors. 
Their knowledge eased the learning of topics that were new to me, and their advice helped me navigate the academic world.
Nevertheless, the best aspect of working with them is, perhaps, the environment that they contributed to create around me and the rest of the relativity group at Queen Mary University of London (QMUL).
They motivated me to do my best, while considering my opinions, and understanding the difficulties that a Ph.D. student may face (especially during a pandemic).
They were mentors, and are now friends.
By observing them, I realised that it is possible to be successful in the field of theoretical physics, while maintaining a healthy work-life balance.
I will treasure their example in the future.

My desire for knowledge and my work ethic descend from my parents, Carlo and Valeria.
They inspired me and my sister, Letizia, to never give up in the face of adversity, and showed us that great results can only be achieved through hard work and persistence.
My parents and my sister have always been next to me, either in person or in spirit, to remind me that I am not going through life alone, and that every difficulty can be overcome by joining forces with loved ones.
My family's support was priceless in many occasions. In particular, it made lockdowns and isolation in London bearable. 
Even during a pandemic, even when my father suffered a stroke, he and the rest of the family found a way to come together and deal with these new, unexpected challenges, while still dedicating thoughts and energies to me. It is something that I will never take for granted.

Then, I cannot forget the person that, in more recent times, has supported me as much as my family members: my girlfriend, Algina. Since she agreed to share this journey with me, I have been shown what loving someone really means and feels like. Her presence has brightened my days, and her hugs have made me feel at home.

Finally, I am thankful to all the friends that found time for a chat, a call, or even just a message during my Ph.D. years. 
People often say that it is easy to feel lonely in London and, some add, in life. If that is the case, I must be very lucky to have old and new friendships that enrich my spirit, while I study to enrich my knowledge. 
The time spent with friends has the ability to quickly recharge my batteries, which enables me to work hard without feeling overwhelmed.
I could write a long list of my friends' names here, but it would not say anything about the many ways in which each of these people contributed to my well-being, and consequently my research (on the other hand, writing about their contributions could make this section as long as one of the main chapters, which does not seem appropriate).
Therefore, I will keep the names and the memories with me.
I also intend to do as much as I can to keep these people in my life, as well as be a supportive presence in theirs.

My Ph.D. was funded by a Queen Mary University of London (QMUL) Ph.D. scholarship.
I acknowledge the use of Athena at HPC Midlands+ in this research, as part of the HPC Midlands+ consortium.
This research also utilised Queen Mary's Apocrita HPC facility, supported by QMUL Research-IT.
I am grateful to the UK Materials and Molecular Modelling Hub for computational resources, which is partially funded by EPSRC (EP/P020194/1 and EP/T022213/1).
The author gratefully acknowledges the Gauss Centre for Supercomputing e.V. (www.gauss- centre.eu) for providing computing time on the GCS Supercomputer SuperMUC-NG at Leibniz Supercomputing Centre (www.lrz.de).
\end{acknowledgments}

\ifpaper
\end{document}
\fi

\tableofcontents   
\listoffigures     

\newpage

\setcounter{page}{1} %
\pagenumbering{arabic}
\pagestyle{fancy}

\newif\ifpaper
\paperfalse

\ifpaper
\input{../preamble}
\begin{document}
\fi

\chapter{Introduction}
\label{chap:intro}

Einstein's theory of general relativity \cite{Einstein:1915ca} has proven to be an accurate theory of gravity for many purposes. 
The most recent confirmation of its accuracy is provided by the detection of gravitational waves \cite{LIGOScientific:2016aoc}, whose existence was predicted by Einstein soon after the presentation of his model \cite{Einstein:1916cc,Einstein:1918btx}.
General relativity also predicts the existence of black holes\footnote{Ref. \cite{Schwarzschild:1916uq} is a translation of the first work in which black holes were shown to be a prediction of general relativity.}, peculiar objects whose gravitational attraction does not let any kind of matter (including light) escape.
It is widely accepted that these objects do exist in nature. 
The crucial feature of general relativity is that it assumes spacetime, the set of all possible events, to be a smooth geometric object, and describes gravity as a consequence of the curvature of spacetime. The dynamical quantity that determines the shape, and thus the curvature, of a spacetime, called metric, must satisfy a set of $2^{nd}$ order partial differential equations (PDEs), called Einstein equations. 
As in any other theory of physics, one would like to solve the ``equations of motion'' of the theory in order to determine the evolution of a system, whose state is known at a given initial time. We refer to this problem as the \emph{Cauchy problem}.
Given the difficulty of exactly solving the Einstein equations in general settings, frameworks that find approximated, but highly accurate, solutions have been developed and implemented in computers. 
We will refer to these frameworks as numerical schemes.
Numerical relativity is the research field that aims to obtain and improve such schemes.
Numerical studies have led to great advancements in the understanding of gravitational effects, especially in spacetimes that have zero curvature away from energy and matter sources, called asymptotically flat spacetimes.
By now, there exist several numerical schemes able to solve the Cauchy problem in asymptotically flat spacetimes.

The study of the Cauchy problem of general relativity is of great interest also in other types of spacetimes. Away from asymptotic flatness, the most investigated case, and the one that we consider in this thesis, is that of asymptotically anti-de Sitter (AdS) spacetimes, i.e., spacetimes that reduce to AdS spacetime, the Lorentzian version of hyperbolic space, away from energy and matter sources.
In these settings, the Cauchy problem is an initial-boundary value problem: in addition to the initial configuration, it is necessary to specify also conditions at the boundary of spacetime during the entire evolution, in order to determine the solution of the evolution equations of general relativity for arbitrarily long times.
In recent years, AdS with reflective boundary conditions has proven to be a particularly exciting theoretical laboratory for studying the strong-field regime of general relativity.
This can be understood from the fact that AdS with reflective boundary conditions plays the role of a box that naturally keeps propagating waves confined to its interior, where they are forced to perpetually interact and grow.
Thus, even the smallest perturbations in AdS can enter the strong-field regime, where qualitatively new gravitational phenomena emerge.

In an unprecedented way, the simulation of asymptotically AdS spacetimes has also opened up the field of numerical relativity to the study of phenomena in areas beyond the traditional astrophysical setting.
At the heart of this push to understand AdS is a conjectured duality, now known as the AdS/CFT correspondence, between gravity in AdS and certain quantum conformal field theories (CFTs) living at the boundary of AdS~\cite{Maldacena:1997re,Gubser:1998bc,Witten:1998qj}\footnote{Although this conjecture has not yet been proved at the mathematical level, it is supported by many studies and it is by now widely accepted.}.
Through this connection, the study of AdS spacetimes has become immediately relevant to fundamental questions in many areas in physics, such as fluid dynamics~\cite{Baier:2007ix,Bhattacharyya:2010owp,Hubeny:2011hd}, relativistic heavy ion collisions~\cite{Chesler:2010bi,Casalderrey-Solana:2013aba,Casalderrey-Solana:2013sxa,Chesler:2015wra}, and superconductivity~\cite{Gubser:2008px,Hartnoll:2008kx,Hartnoll:2008vx}.
See, for example, \cite{Ammon:2015wua,Nastase:2015wjb} for reviews of the AdS/CFT duality and its applications; see \cite{CasalderreySolana:2011us,Chesler:2015lsa,Zaanen:2015oix,Hartnoll:2016apf} for more extensive and specialised reviews of the applications.
The reason why the study of AdS is crucial for our understanding of these phenomena is that AdS/CFT provides an important -- and in most cases the only -- window into the real-time dynamics of strongly interacting quantum field theories far from equilibrium.
The dynamical far-from-equilibrium strongly interacting regime is precisely the one that is least explored and understood, and the one that has the best chance of making contact with certain experiments.
According to the prescription of AdS/CFT, this regime can be accessed, from first principles and in a controlled manner, by solving the classical dynamics of gravity in AdS and mapping the observables in the gravitational side to the observables that describe the (non-gravitational, i.e., living on a fixed background spacetime) CFT physics.

Our current understanding of gravity in AdS remains limited for several reasons.
First, numerical evolution in AdS is notoriously hard, in part because the systematic study of the initial-boundary value problem is still in its infancy.
Second, the most interesting phenomena involve spacetimes that have very little or no symmetry, making these evolutions beyond the reach of most numerical codes.
Third, for many of these phenomena, there is a variety of physical scales that must be adequately resolved to correctly capture the relevant physics.

The main purpose of this thesis is to present the first numerical scheme that does not have any of these limitations and makes Cauchy evolution in AdS possible in full generality: no symmetry requirements, any number of dimensions, and any coupling with matter fields. We show the output of the first proof-of-principle Cauchy evolution of asymptotically AdS spacetimes that has been achieved with no symmetry assumptions.
This scheme is also well-suited for many other studies of gravitational dynamics in AdS. In particular, we show that it can evolve initial data describing a perturbed AdS rotating black hole, called Kerr-AdS black hole, and thus lead to simulations of superradiance, that is, the amplification of waves scattering off a rotating black hole.

The content of this manuscript is organised as follows.
In order to make the thesis somewhat self-contained, in Chapter~\ref{sec:GR} we review the elements of the theory of general relativity that will be needed for the discussions that follow.
In Chapter~\ref{chap:ovengr}, we review some of the historical numerical evolution schemes that have been successfully implemented in computers. This gives us the opportunity to introduce important notions in numerical relativity.
In Chapter~\ref{Chapter:NoSym}, we review the properties of asymptotically AdS spacetimes, and we present our numerical scheme to evolve the laws of general relativity in these settings.
In Chapter~\ref{Chapter:KAdS}, the topic of superradiance is reviewed, with particular focus on the implications in asymptotically AdS spacetimes. Furthermore, evidence that our scheme can be used to simulate the superradiant instability of Kerr-AdS is provided.
We summarise our conclusions in Chapter~\ref{Chapter:outconc}, and we discuss open questions about the physics in AdS that our scheme can help address.
The appendices contain technical discussions that are relevant for the topics presented in Chapter~\ref{Chapter:NoSym} and Chapter~\ref{Chapter:KAdS}.

\section{Notation and conventions}
\label{sec:not}

In this thesis, we work in $D=4$ spacetime dimensions, except in Section~\ref{subsec:spastruc}, where we review general notions of differential geometry for spaces with any number $D\geq 2$ of dimensions.

Consider a set of indices, e.g., $\alpha,\beta,\gamma,\dots$, each of which takes values in a given set, e.g., $\alpha\in\{0,1,\cdots,D-1\}$.
We will associate such a set of indices with a set of coordinates, e.g., $x^\alpha=(x^0,x^1,\dots,x^{D-1})$.
In a long manuscript, involving calculations with many sets of indices, the reader might lose track of the coordinate set that corresponds to each set of indices.
In order to avoid this type of confusion, we explicitly state which coordinates are being used.  
In all cases in which the specification is not made, then the reader should assume that an arbitrary set of coordinates is being used.
We use the first few indices of the Greek alphabet, $\alpha,\beta,\gamma,\dots$, to denote indices associated with an arbitrary set of coordinates.
However, we will sometimes associate the set of indices, $\alpha,\beta,\gamma,\dots$, with coordinates $x^\alpha$ that satisfy certain properties, and thus are not arbitrary. This is one of those cases in which we explicitly state the requirements that restrict the choice of $x^\alpha$.

We employ Einstein notation: if an index appears twice in an expression, then a sum over all possible values of that index is intended, except where explicitly stated otherwise. For instance, $V^\alpha \omega_\alpha$ must be intended as $\sum\limits_{\alpha} V^\alpha \omega_\alpha$, where the sum is over all possible values of $\alpha$. 

It is also common to denote partial derivatives with respect to one of the coordinates by a comma. For instance, the derivative of a function $f$ with respect to $x^\alpha$, usually denoted by $\partial_\alpha f$, can also be denoted by $f_{,\alpha}$.

The results of this thesis are expressed in natural units, defined so that $G=c=\hbar=k_B=1$, where $G$ is Newton's gravitational constant, $c$ is the speed of light, $\hbar:=h/(2\pi)$ where $h$ is Planck's constant, and $k_B$ is Boltzmann's constant ($G, c ,\hbar, k_B$ can later be reinserted by dimensional analysis). In particular, this implies that times, lenghts, masses and temperatures have the same units.

\ifpaper
\end{document}
\fi
\newif\ifpaper
\paperfalse

\ifpaper
\input{../preamble}
\begin{document}
\fi

\chapter{Overview of general relativity}
\label{sec:GR}

The theory of general relativity has been shown to describe physics at large scales in our universe with remarkable accuracy. 
This model hinges upon the assumption that the effects of acceleration and gravity are indistinguishable, and it explains these effects as a consequence of the curvature of spacetime.
In this chapter, we briefly review the building blocks of the theory and some of its pivotal theoretical predictions.

\section{Geometric description of spacetime}
\label{subsec:spastruc}

Within the framework of general relativity, the geometry of spacetime is what determines physical, gravitational phenomena.
In this section, we review the mathematical tools that allow to characterise curved spacetimes and their geometric features.

\subsection{Manifolds, coordinates, bases and tensors}
\label{sec:mcbt}

Let us start by introducing the set of points, or \emph{events}, that constitute a spacetime: the manifold.
A $D$-dimensional smooth \emph{manifold} $\mathcal{M}$ is a set defined by requiring that, for each subset $\mathcal{U}\subset\mathcal{M}$, it is possible to construct at least one map, called \emph{chart} or \emph{frame}, from $\mathcal{U}$ to a subset of $\mathbb{R}^D$. Each chart must uniquely identify points of $\mathcal{M}$ with a set of $D$ real numbers, called \emph{coordinates} and denoted by $(x^0,\dots,x^{D-1})$. Furthermore, if two subsets of $\mathcal{M}$ overlap over some region, the transition map between the corresponding coordinate sets over the overlapping region must be smooth.
We define \emph{global coordinates} on $\mathcal{M}$ as coordinates covering the entire manifold $\mathcal{M}$ (possibly with the exception of a subset of points with positive co-dimension).
Coordinates that are defined only in a sufficiently small neighbourhood of a point of $\mathcal{M}$ are called \emph{local}.
The collection of all sets of coordinates is called the \emph{atlas} of $\mathcal{M}$. The choice of a set of coordinates over a certain region can be made arbitrarily within an atlas, without changing the physical predictions of the theory.
In physics, different descriptions of a theory that lead to the same physical predictions are called \emph{gauges}.
Therefore, the choice of coordinates corresponds to a choice of gauge in general relativity. 

Given any coordinate set $x^\alpha=(x^0,\dots,x^{D-1})$ in a neighbourhood of a point $p\in \mathcal{M}$, points on a curve in that neighbourhood with parameter $\lambda$ can be denoted by $x^\alpha(\lambda)$.
Let us consider the curve along which only one of the coordinates, say $x^{\bar\alpha}$, varies, and use $x^{\bar\alpha}$ as parameter along this curve. We denote the tangent to this curve at $p$ by $\frac{\partial}{\partial x^{\bar\alpha}}$. Repeating this construction for all $D$ coordinates $x^\alpha$, we define $D$ vectors at $p$, denoted by $\frac{\partial}{\partial x^\alpha}$, which form a basis for the space of tangent vectors to $\mathcal{M}$ at $p$, denoted by $T_p(\mathcal{M})$. The components of a vector $V$ in this basis are denoted by $V^\alpha$.

\emph{Covectors} at $p$ are linear maps that associate a vector at $p$ to a real number $\mathbb{R}$. The $D$ covectors $dx^\alpha$ that satisfy $dx^\alpha\left(\frac{\partial}{\partial x^\beta}\right)=\delta^\alpha_\beta$ (where $\delta^\alpha_\beta$ is the Kronecker delta) form a basis for covectors at $p$. The components of a covector $\omega$ in this basis are denoted by $\omega_\alpha$.
\emph{Dual vectors} at $p$, i.e., linear maps that associate a covector at $p$ with a real number, and vectors at $p$ are isomorphic under a basis-independent isomorphism. For this reason, we often refer to both types of objects as vectors and we use the same notation for both.

The \emph{tensor product} $\otimes$ allows to combine (dual) vectors and covectors to obtain linear maps that associate any number of vectors and covectors at $p$ with a real number, e.g., $\frac{\partial}{\partial x^\alpha}\otimes dx^\beta\otimes dx^\gamma$ is the linear map that acts as $\bigl(\frac{\partial}{\partial x^\alpha}\otimes dx^\beta\otimes dx^\gamma \bigr)(\omega,V,W)=\frac{\partial}{\partial x^\alpha}(\omega)dx^\beta(V)dx^\gamma(W) $ for any covector $\omega$ and any pair of vectors $V,W$ at $p$. \emph{Tensors} at $p$ are linear combinations of such tensor products, e.g., $T=T^\alpha_{\phantom \alpha \beta\gamma} \frac{\partial}{\partial x^\alpha}\otimes dx^\beta\otimes dx^\gamma$ is a tensor at $p$ of rank $\binom{1}{2}$. The real coefficients $T^\alpha_{\phantom \alpha \beta\gamma}$ are the components of $T$ with respect to the basis associated with the coordinates $x^\alpha$, and they can be computed as $T\left(dx^\alpha,\frac{\partial}{\partial x^\beta},\frac{\partial}{\partial x^\gamma}\right)$. When no confusion regarding the basis under consideration is possible, it is customary to refer to the components $T^\alpha_{\phantom \alpha \beta\gamma}$ as ``the tensor $T$''.
If we change coordinates from $x^\alpha$ to a different set $x^{\alpha'}$, the corresponding vector and covector bases at $p$ transform, respectively, as $\frac{\partial}{\partial x^{\alpha'}}=\frac{\partial x^\beta}{\partial x^{\alpha'}}\frac{\partial}{\partial x^\beta}$ and $dx^{\alpha'}=\frac{\partial x^{\alpha'}}{\partial x^{\beta}}dx^\beta$, where the partial derivatives are evaluated at $p$.
Tensor components transform accordingly, e.g.,  
\begin{equation}
\label{eq:tenstran12}
T^{\alpha'}_{\phantom {\alpha'} \beta'\gamma'}=\frac{\partial x^{\alpha'}}{\partial x^{\alpha}}\frac{\partial x^\beta}{\partial x^{\beta'}}\frac{\partial x^\gamma}{\partial x^{\gamma'}}T^\alpha_{\phantom \alpha \beta\gamma}.
\end{equation}
Since the components of all tensors transform in the same way under a change of coordinates, if an equality between the components of two tensors, called a tensorial equation, holds in some set of coordinates, then it must hold in any set of coordinates. In the following, we use this fact repeatedly, i.e., we prove tensorial equations that hold in any set of coordinates by working in a particular frame.
We conclude this section by defining \emph{tensor fields} as smooth maps that associate a point $p\in \mathcal{M}$ with a tensor at $p$. For instance, in a region of $\mathcal{M}$ covered by coordinates $x^\alpha$, a tensor field $T$ of rank $\binom{1}{2}$ can be written as $T=T^\alpha_{\phantom \alpha \beta\gamma}(x) \frac{\partial}{\partial x^\alpha}\otimes dx^\beta\otimes dx^\gamma$. Notice that a tensor field of rank $\binom{0}{0}$ corresponds to the standard notion of function from $\mathcal{M}$ to $\mathbb{R}$, which is sometimes called \emph{scalar field}, and does not change under a change of coordinates.

\subsection{Metric}

The fundamental field of general relativity is the metric, which captures the notion of length on a manifold.
A metric $g$ is a tensor field of rank $\binom{0}{2}$,
\begin{equation}
g=g_{\alpha\beta}(x) dx^\alpha\otimes dx^\beta,
\end{equation}
such that the corresponding tensor at any point $p\in\mathcal{M}$ is symmetric, i.e., $g_{\alpha\beta}=g_{\beta\alpha}$, and non-degenerate, i.e., $g_{\alpha\beta}$ is invertible.
It is customary to write $g$ as $ds^2$ and omit the symbol $\otimes$. 
The symmetry property can be also written as $g_{\alpha\beta}=g_{(\alpha\beta)}$, where we defined the \emph{symmetrisation} of indices as $g_{(\alpha\beta)}:=\frac{1}{2}(g_{\alpha\beta}+g_{\beta\alpha})$.
We denote the inverse of the matrix $g_{\alpha\beta}$ by $g^{\alpha\beta}$, i.e., $g_{\alpha\beta}g^{\beta\gamma}=\delta^\gamma_\alpha$ . The indices of the components of a tensor at $p$ can be lowered or raised by $g_{\alpha\beta}$ at $p$, and the result is still a tensor at $p$. For instance, given a vector $V$ and a covector $\omega$, $V_\alpha:=g_{\alpha\beta}V^\beta$ are the components of a covector and $\omega^\alpha:=g^{\alpha\beta}\omega_\beta$ are the components of a vector.
For any pair of vectors $V,W$, the scalar quantity $g(V,W)=g_{\alpha\beta}V^\alpha W^\beta=g^{\alpha\beta}V_\alpha W_\beta=V_\alpha W^\alpha$ is called \emph{scalar product} of $V$ and $W$.
A coordinate-independent notion of length of a curve $x^\alpha(\lambda)$, called \emph{proper length}, between two points $x^\alpha(\lambda_0)$ and $x^\alpha(\lambda_1)$ is defined in terms of the metric as 
\begin{equation}
\label{eq:distance}
L:=\int_{\lambda_0}^{\lambda_1}{d\lambda \,g(V,V)|_{x^\alpha=x^\alpha(\lambda)}},
\end{equation}
where $V=\frac{\partial}{\partial \lambda}=\frac{dx^\alpha}{d\lambda}\frac{\partial}{\partial x^\alpha}$ is the tangent vector to the curve.

The symmetry of $g$ implies that $g_{\alpha\beta}$ has $D$ eigenvalues. Since $g_{\alpha\beta}$ is invertible, the eigenvalues are non-vanishing. A metric is said to be \emph{Lorentzian} if exactly one eigenvalue is negative, whereas it is said to be \emph{Riemannian} if all eigenvalues are positive.
In a sufficiently small neighbourhood $\mathcal{U}$ of any point $p$ of a manifold $\mathcal{M}$, it is possible to define \emph{locally inertial coordinates} such that $g_{\alpha\beta}$ at $p$ is the diagonal matrix $\operatorname{diag}(-1,1,1,\dots,1)$ in the Lorentzian case, or $\operatorname{diag}(1,1,\dots,1)$ in the Riemannian case, and all first derivatives of $g_{\alpha\beta}$ vanish at $p$. Importantly, for a generic manifold, it is not possible to define coordinates such that also the second derivatives of $g_{\alpha\beta}$ vanish at any $p$.

\subsection{Causal structure of spacetime}

A \emph{spacetime} is a pair $(\mathcal{M},g)$ where $\mathcal{M}$ is a manifold and $g$ is a Lorentzian metric. From now onwards, the symbols $\mathcal{M}$ and $g$ will denote, respectively, the manifold and the metric of a spacetime.
The metric $g$ determines the causal structure of the spacetime through the following definitions.
The square of a vector $V$, with components $V^\alpha$ in an arbitrary basis, is defined as the scalar $V^2:=V^\alpha V^\beta g_{\alpha\beta}$.
A vector $V$, as well as the corresponding covector $\tilde V$ with components $\tilde V_\alpha=g_{\alpha\beta}V^\beta$ in an arbitrary basis, are said to be \emph{timelike}, \emph{spacelike} or \emph{null}, if $V^2$ is, respectively, negative, positive or vanishing. 
A vector that is either timelike or null is said to be \emph{causal}.
A timelike (spacelike) vector $V$ and the covector $\tilde V$ are said to be \emph{unit} if $V^2=-1$ ($V^2=1$).
We say that a coordinate $x^{\bar\alpha}$ is timelike, spacelike or null, if the corresponding vector $\frac{\partial}{\partial x^{\bar\alpha}}$ is, respectively, timelike, spacelike or null.
A curve $x^\alpha(\lambda)$ parameterised by $\lambda$ is said to be timelike, spacelike, null, or causal if its tangent vector $V=\frac{dx^\alpha}{d\lambda}\frac{\partial}{\partial x^\alpha}$ is, respectively, timelike, spacelike, null, or causal everywhere along the curve.
A causal vector field $T$ in a region of spacetime can be used to define a time-orientation in that region.
The cone that contains $T$ is the \emph{future light cone} at $p$, while the other cone is the \emph{past light cone} at $p$.
The exterior of these cones is the \emph{present} of $p$.
The causal vectors contained in the future (past) light cone at $p$ are called \emph{future (past)-directed}.
The \emph{causal future (past)} of a point $p$ is the region of spacetime that is connected to $p$ by future (past)-directed curves, i.e., curves whose tangent vector $V$ is future (past)-directed everywhere along the curve.

It is postulated that a massive (massless) particle travels in $\mathcal{M}$ along a timelike (null) curve, called \emph{worldline}.
Sometimes, we refer to a massive particle as an \emph{observer}.
Given the worldline of a massive particle starting at a point $p$, the length of the worldline between $p$ and an arbitrary point $q$ of the curve, given by \eqref{eq:distance}, is called \emph{proper time} of the particle at $q$. This quantity, typically denoted by $\tau$, corresponds to the time measured by a clock that moves with the particle from $p$ to $q$. The tangent vector $V$ of a timelike worldline parameterised by proper time $\tau$ is called \emph{4-velocity} of the worldline. It is given by $V= \frac{dx^\alpha}{d\tau}\frac{\partial}{\partial x^\alpha}$, and satisfies $V^2=-1$, i.e., it is a unit timelike vector.

\subsection{Hypersurfaces}
\label{sec:hypsuf}

Consider a \emph{hypersurface} $\Sigma$ of $\mathcal{M}$, i.e., a co-dimension 1 surface in $\mathcal{M}$ at some constant value of a function $f$.
Let us show that $n^\alpha:=(df)^\alpha=g^{\alpha\beta}(df)_\beta$ is normal to $\Sigma$. Consider coordinates $x^\alpha$ one of which, say $x^0$, is chosen to be $f$. Then, $n^\alpha=g^{\alpha\beta}(dx^0)_\beta=g^{\alpha0}$. Let $V$ be an arbitrary vector tangent to $\Sigma$. From the definition of basis vectors in Section~\ref{sec:mcbt}, we see that $V$ can be a linear combination of all the basis vectors $\frac{\partial}{\partial x^\alpha}$ except $\frac{\partial}{\partial x^0}$, thus $V^0=0$. Therefore, $g_{\alpha\beta}n^\alpha V^\beta=g_{\alpha\beta}g^{\alpha0}V^\beta=V^0=0$. $g_{\alpha\beta}n^\alpha V^\beta$ is a scalar quantity so it must vanish in any coordinate system, which shows that $(df)^\alpha$ is orthogonal to all vectors tangent to $\Sigma$, i.e., $n^\alpha=(df)^\alpha$ is normal to $\Sigma$.
The hypersurface $\Sigma$ is said to be timelike, spacelike, or null, if a vector $n^\alpha$ everywhere normal to $\Sigma$ (or, equivalently, $\tilde n_{\alpha}:=g_{\alpha\beta}n^\beta$) is, respectively, spacelike, timelike, or null everywhere on $\Sigma$.
If $\Sigma$ is null, $n$ satisfies $n^2=0$, so $n$ is orthogonal to itself.
As a consequence, a normal to a null hypersurface is also tangent to the hypersurface.
If $\Sigma$ is spacelike, and the timelike vector $\frac{\partial}{\partial f}$ is employed to define a time-orientation, then $-(df)^\alpha$ is the future-directed normal to $\Sigma$.
A spacetime metric $g$ can be restricted to a hypersurface $\Sigma$ to obtain a metric $\gamma$ on $\Sigma$. We write $\gamma=g|_{\Sigma}$. In mathematical terms, this operation is the \emph{pull-back} of $g$ onto $\Sigma$ with respect to the mapping of $\Sigma$ into $\mathcal{M}$, called inclusion map (see \cite{wald:1984} for the details). 
If $\Sigma$ is at a fixed value of one of the coordinates $x^\alpha=(x^0,x^1,\cdots,x^{D-1})$, say $x^0$, then $x^i=(x^1,\cdots,x^{D-1})$ must be valid coordinates on $\Sigma$, i.e., they must belong to an atlas of $\Sigma$.
Using such coordinates, $\gamma=g|_{\Sigma}$ has components $\gamma_{ij}=g_{ij}$, where $i,j=1,2,\dots,D-1$ are the indices associated with the coordinates $x^i$ on $\Sigma$. 
Assuming that $g$ is Lorentzian, if $\Sigma$ is spacelike (timelike), then $\gamma$ must be Riemannian (Lorentzian).
A higher co-dimension surface is said to be timelike, spacelike or null if all the vectors tangent to the surface are, respectively, timelike, spacelike or null.

The \emph{projection operator} onto an hypersurface $\Sigma$ is defined, in any set of coordinates, by
\begin{equation}
\gamma^\alpha_\beta:=\delta^\alpha_\beta\pm n^\alpha n_\beta,
\end{equation}
where the upper sign refers to spacelike $\Sigma$ and the lower sign to timelike $\Sigma$.
(Notice that $\gamma^\alpha_\beta$ is idempotent, i.e., $\gamma^\alpha_\gamma \gamma^\gamma_\beta=\gamma^\alpha_\beta$, as appropriate for a projector.)
This operator can be applied to any tensor at a point $p\in\Sigma$ to obtain the component of that tensor tangent to $\Sigma$ at $p$. For instance, given a vector $V$ at a point $p\in\Sigma$, $V_{||}^\alpha=\gamma^\alpha_\beta V^\beta$ is the component of $V$ tangent to $\Sigma$ at $p$, i.e., $V_{||}^\alpha n_\alpha=0$.
Let us now consider a tensor defined on the tangent space of the spacetime manifold $\mathcal{M}$ at a point $p\in\Sigma$. If the tensor is invariant under projection onto $\Sigma$, then it can be identified with a tensor defined on the tangent space of $\Sigma$ at $p$, under a natural (i.e., basis-independent) isomorphism. For example, $\gamma_{\alpha\beta}=g_{\alpha\beta}+n_\alpha n_\beta$ at points on a spacelike $\Sigma$ can be identified with the Riemannian metric on $\Sigma$ defined as the restriction of the spacetime metric $g$ on $\Sigma$, i.e., the pull-back of the spacetime metric $g$ onto $\Sigma$ with respect to the inclusion map of $\Sigma$ into $\mathcal{M}$. 
Indices of tensors invariant under projection onto $\Sigma_t$ can be raised and lowered by $\gamma_{\alpha\beta}$ or $g_{\alpha\beta}$, equivalently. 
Given any set of coordinates $x^i$ on $\Sigma$, indices $i,j,k,\dots$ of tensors on the tangent space of $\Sigma_t$ can be raised and lowered by $\gamma_{ij}$.

\subsection{Tensorial derivatives and geodesics}
\label{subsec:tender}

We now define derivative operators whose action on tensor fields is also a tensor field. We will be interested in two of the many possible choices for such operators. For simplicity of notation, let us illustrate the definitions by considering a $\binom{1}{2}$ tensor field $T$. The generalisation to other ranks is straightforward.

The \emph{Lie derivative} of $T$ with respect to a vector field $V$ is a $\binom{1}{2}$ tensor field $\mathcal{L}_VT$, whose components in any coordinate basis are given by
\begin{equation}
\label{eq:Lieder}
\left(\mathcal{L}_VT\right)^\alpha_{\phantom\alpha\beta\gamma}:=V^\lambda (\partial_\lambda T^\alpha_{\phantom\alpha\beta\gamma})-(\partial_\lambda V^\alpha) T^\lambda_{\phantom\lambda\beta\gamma}+(\partial_\beta V^\lambda) T^\alpha_{\phantom\alpha\lambda\gamma}+(\partial_\gamma V^\lambda) T^\alpha_{ \phantom\alpha\beta\lambda}.
\end{equation}
A different tensorial derivative operator is the \emph{covariant derivative}.
Its action on $T$ is a $\binom{1}{3}$ tensor field $\nabla T$, whose components in any coordinate basis are given by
\begin{equation}
\label{eq:covder}
\nabla_\delta T^\alpha_{\phantom \alpha \beta\gamma}:=\partial_\delta T^\alpha_{\phantom \alpha \beta\gamma}+\Gamma^\alpha_{\phantom\alpha \lambda\delta} T^{\lambda}_{\phantom\lambda\beta\gamma}-\Gamma^\lambda_{\phantom\lambda\beta\delta} T^{\alpha}_{\phantom\alpha\lambda\gamma}-\Gamma^\lambda_{\phantom\lambda\gamma\delta} T^{\alpha}_{\phantom\alpha\beta\lambda},
\end{equation}
for some functions $\Gamma^\alpha_{\phantom\alpha\beta\gamma}$ called \emph{Christoffel symbols} (which are not tensor components despite the notation). 
Notice that $\nabla_\alpha$ reduces to the standard partial derivative $\partial_\alpha$ when acting on functions on $\mathcal{M}$.
In the following, we will always require that the covariant derivative is torsion free, i.e., $\nabla_\alpha\nabla_\beta f=\nabla_\beta\nabla_\alpha f$ for any scalar function $f$ on $\mathcal{M}$, and compatible with the metric, $\nabla g=0$. These conditions define the unique \emph{Levi-Civita covariant derivative} associated with $g$.
The Christoffel symbols of the Levi-Civita covariant derivative are given, in any coordinate basis, by the following expression in terms of the metric and its first derivatives:
\begin{equation}
\label{eq:LCchris}
\Gamma^\alpha_{\phantom\alpha\beta\gamma}=\frac{1}{2} g^{\alpha\lambda}\left( g_{\beta\lambda,\gamma}-g_{\beta\gamma,\lambda}+g_{\lambda\gamma,\beta} \right).
\end{equation}
Notice that $\Gamma^\alpha_{\phantom\alpha\beta\gamma}$ are symmetric with respect to the exchange of $\beta$ and $\gamma$.
In locally inertial coordinates in a neighbourhood of a point $p\in \mathcal{M}$, we have $g_{\alpha\beta,\lambda}=0$ at $p$, therefore $\Gamma^\alpha_{\phantom\alpha\beta\gamma}=0$ at $p$, and the notion of covariant derivative coincides with the usual notion of partial derivative at $p$.
Given an hypersurface $\Sigma$, the covariant derivative on $\Sigma$ of a tensor field invariant under projection onto $\Sigma$ is defined as the projection onto $\Sigma$ of the covariant derivative $\nabla$ of the tensor field, and we denote it by $D$. For instance, $D_\alpha V_{||}^\beta:=\gamma^\gamma_\alpha \gamma^\beta_\delta \nabla_\gamma V_{||}^\delta$.
If $\Sigma$ is at a fixed value of one of the coordinates $x^\alpha=(x^0,x^1,\cdots,x^{D-1})$, say $x^0$, and we use the remaining coordinates $x^i=(x^1,\cdots,x^{D-1})$ as coordinates on $\Sigma$, then $D_\alpha V_{||}^\beta$ is identified with the tensor on the tangent space of $\Sigma$ given by $D_i X_{||}^j=\gamma^\gamma_i \gamma^j_\delta \nabla_\gamma X_{||}^\delta$. Notice that $D$ is the Levi-Civita covariant derivative associated with $\gamma$, i.e., it is torsion-free and $D_i\gamma_{jk}=0$.

A curve $x^\alpha(\lambda)$ is said to be a \emph{geodesic} of the metric $g$ if its tangent vector $V^\alpha=\frac{dx^\alpha}{d\lambda}$ satisfies the \emph{geodesic equation}
\begin{equation}
\label{eq:geod1}
V^\alpha \nabla_\alpha V^\beta=f V^\beta
\end{equation}
along the curve\footnote{The covariant derivative appearing in \eqref{eq:geod1} is defined only if we generalise $V$ to a vector field over an open neighbourhood of the curve. 
This generalisation can be done in an arbitrary (but smooth) way, since it does not affect the values $V^\alpha$ of the solution to \eqref{eq:geod1} on the curve.}, for some real scalar function $f$.
Geodesics are curves that extremise the proper length \eqref{eq:distance} between two points on $\mathcal{M}$.
It is postulated that the worldline of a massive (massless) particle, moving solely under the force of gravity, is a timelike (null) geodesic. The parameter $\lambda$ can always be chosen so that \eqref{eq:geod1} is satisfied with $f=0$, in which case we say that $\lambda$ is an \emph{affine parameter}. If $\lambda$ is an affine parameter, any other affine parameter $\lambda'$ is related to $\lambda$ by $\lambda'=a\lambda+b$ for some constants $a,b$ with $a>0$.
The proper time $\tau$ of a timelike geodesic is an affine parameter.
This can be seen from the fact that the tangent vector of a geodesic parameterised with proper time, $U^\alpha=\frac{dx^\alpha}{d\tau}$, satisfies two equations, namely, \eqref{eq:geod1} and $U^\beta U_\beta=-1$.
Multiplying the first equation by $U_\beta$ and summing over $\beta$, we get $U_\beta U^\alpha \nabla_\alpha U^\beta=-f$. Acting on the second equation with $\nabla_\alpha$, we get $U^\beta \nabla_\alpha U_\beta=0$. Using the latter result in the former, we find $f=0$.

\subsection{Curvature}

The information about the spacetime curvature, which determines gravitational effects, is contained in the \emph{Riemann tensor}.
The components of this tensor in a coordinate basis are given by
\begin{equation}
R^\alpha_{\phantom\alpha\beta\gamma\delta}=\partial_\gamma \Gamma^\alpha_{\phantom\alpha\beta\delta}-\partial_\delta\Gamma^\alpha_{\phantom\alpha\beta\gamma}+\Gamma^\lambda_{\phantom\lambda\beta\delta}\Gamma^\alpha_{\phantom\alpha\lambda\gamma}-\Gamma^\lambda_{\phantom\lambda\beta\gamma}\Gamma^\alpha_{\phantom\alpha\lambda\delta}.
\end{equation}
We see that the Riemann tensor involves second derivatives of the metric components.
As noted above, for a generic spacetime and a generic point $p$, it is not possible to define coordinates in a neighbourhood of $p$ such that both first and second derivatives of the metric vanish at $p$.
Consequently, the Riemann tensor at $p$ is non-vanishing in any set of coordinates. 
We thus see that the information about the point-wise spacetime curvature, encoded in the Riemann tensor, is physical, in the sense that it cannot be made disappear by a suitable, gauge choice of coordinates.

The Riemann tensor has $D^2(D^2-1)/12$ independent components. This can be shown using the fact that $R_{\alpha\beta\gamma\delta}:=g_{\alpha\lambda}R^\lambda_{\phantom\lambda\beta\gamma\delta}$ satisfies the following relations, which we refer to as \emph{symmetries of the Riemann tensor}:
\begin{eqnarray}
\label{eq:riesym}
&R_{\alpha\beta\gamma\delta}=R_{\gamma\delta\alpha\beta}, \quad R_{\alpha\beta\gamma\delta}=-R_{\alpha\beta\delta\gamma}, \quad 
R_{\alpha\beta\gamma\delta}=-R_{\beta\alpha\gamma\delta}, \nonumber\\
&R_{\alpha\beta\gamma\delta}+R_{\alpha\delta\beta\gamma}+R_{\alpha\gamma\delta\beta}=0.
\end{eqnarray}
In addition to these symmetries, the Riemann tensor satisfies the \emph{Bianchi identity}
\begin{equation}
\nabla_\lambda R^\alpha_{\phantom\alpha\beta\gamma\delta}+\nabla_\gamma R^\alpha_{\phantom\alpha\beta\delta\lambda}+\nabla_\delta R^\alpha_{\phantom\alpha\beta\lambda\gamma}=0.
\end{equation}
It is convenient to define the \emph{Ricci tensor}, whose components in an arbitrary coordinate basis are
\begin{equation}
R_{\alpha\beta}:=R^\lambda_{\phantom\lambda \alpha\lambda\beta}=g^{\lambda\gamma}R_{\gamma\alpha\lambda\beta}.
\end{equation}
Using the first of \eqref{eq:riesym} and the symmetry of $g_{\alpha\beta}$, it can be easily proved that $R_{\alpha\beta}=R_{\beta\alpha}$. We also define the \emph{Ricci scalar} as the trace of $R_{\alpha\beta}$ with respect to $g_{\alpha\beta}$:
\begin{equation}
R:=g^{\alpha\beta}R_{\alpha\beta}.
\end{equation}
A consequence of the Bianchi identity is that the Einstein tensor, given by $G_{\alpha\beta}=R_{\alpha\beta}-\frac{1}{2}R g_{\alpha\beta}$, satisfies the \emph{contracted Bianchi identity}
\begin{equation}
\label{eq:conBiaid}
\nabla^\alpha G_{\alpha\beta}=0.
\end{equation}

\subsection{Isometries}
\label{sec:symm}

The Lie derivative is a useful tool to obtain the variation of tensor fields under certain transformations, namely diffeomorphisms, and thus identify their symmetries, as we explain here.

A map $\phi\colon \mathcal{M}\to \mathcal{M}$ is a \emph{diffeomorphism} if it is bijective, smooth and with smooth inverse. The action of a diffeomorphism on a tensor field $T$ gives a tensor field $\phi_\star(T)$, called the \emph{push-forward} of $T$, such that, for any $p\in \mathcal{M}$, $\phi_\star(T)$ at $\phi(p)$ is given by the tensor $T$ at $p$.
It is important to mention that the action of any diffeomorphism on the \emph{components} of tensor fields on $\mathcal{M}$ is the same as that of an appropriate change of coordinates.  
In the theories that we consider, physical predictions must be independent of the coordinate system, hence diffeomorphisms cannot change the physical predictions on $\mathcal{M}$. We thus say that these theories are \emph{diffeomorphism invariant}. 
Notice that, in this thesis, we will take the ``active'' viewpoint that regards diffeomorphisms as maps that move points of $\mathcal{M}$ around, rather than the ``passive'' viewpoint that regards diffeomorphisms as a change of coordinates.  

Given a vector field $\xi$, it is possible to construct a 1-parameter family of diffeomorphisms, $\phi_\lambda\colon \mathcal{M}\to \mathcal{M}$, as follows. Let $\xi$ have components $\xi^\alpha(x)$ in an arbitrary coordinate basis, and consider the curves $x^\alpha(\lambda)$ with tangent vector $\xi^\alpha(x(\lambda))$ at each point (these are called \emph{integral curves} of $\xi$). Given any point $p$, there is a unique integral curve of $\xi$ that passes through $p$, and we denote the value of the parameter of this curve at $p$ by $\lambda_p$. We define $\phi_\lambda\colon \mathcal{M}\to \mathcal{M}$ as the map that sends the point $p$, with coordinates $x^\alpha(\lambda_p)$, to the point $\phi_\lambda(p)$ with coordinates $x^\alpha(\lambda_p+\lambda)$ along the integral curve of $\xi$ through $p$. If this construction can be repeated for all $\lambda\in\mathbb{R}$, in which case we say that $V$ is \emph{complete}, then the family $\phi_\lambda$ is a 1-parameter group of diffeomorphisms, generated by $\xi$. Conversely, given a 1-parameter group of diffeomorphisms $\phi_\lambda,\lambda\in\mathbb{R}$, it is possible to define the generator $\xi$ as the complete vector field given, at each point $p$, by the tangent vector to the curve $\phi_\lambda(p)$ parameterised by $\lambda$. 
We define the \emph{orbit} of a point $p$ under a group of diffeomorphisms as the set of points obtained by acting on $p$ with all the diffeomorphisms in the group.

The variation of a tensor field $T$ under infinitesimal diffeomorphisms $\phi_\lambda$ in the group generated by $\xi$ is given by $\mathcal{L}_\xi T$. In more practical terms, if we use $\lambda$ as one of the coordinates, we have (for simplicity of notation, we consider the case of a $\binom{1}{2}$ tensor field)
\begin{equation}
\frac{\partial}{\partial \lambda} T^\alpha_{\phantom\alpha\beta\gamma}=\left(\mathcal{L}_\xi T\right)^\alpha_{\phantom\alpha\beta\gamma}
\end{equation}
at any point.
Diffeomorphisms on $\mathcal{M}$ that leave the metric $g$ unchanged are called \emph{isometries} of the spacetime $(\mathcal{M},g)$.
A 1-parameter group of diffeomorphisms on $\mathcal{M}$ generated by $\xi$ is a group of isometries, which we refer to as a \emph{symmetry} of the spacetime, if and only if $\xi$ satisfies the \emph{Killing equation} $\mathcal{L}_\xi g=0$.
Since $(\mathcal{L}_\xi g)_{\alpha\beta}=\nabla_\alpha \xi_\beta+\nabla_\beta \xi_\alpha$ in any set of coordinates, the Killing equation can also be written as
\begin{equation}
\label{eq:Killeq}
\nabla_\alpha \xi_\beta+\nabla_\beta \xi_\alpha=0.
\end{equation}
The solutions $\xi$ to \eqref{eq:Killeq} are called \emph{Killing vector fields}.
A $D$-dimensional spacetime can have at most $D(D+1)/2$ linearly independent Killing vector fields, in which case the spacetime is said to be \emph{maximally symmetric}. 
A spacetime is said to be \emph{spherically symmetric} if its set of isometries contains an $SO(3)$ group whose orbits are 2-dimensional spheres $S^2$, i.e., if it possesses the same symmetries as a round $S^2$.
Given a a Killing vector field $\xi$ and an affinely parameterised geodesic with tangent vector $V^\alpha=\frac{dx^\alpha}{d\lambda}$, it is straightforward to prove that the quantity $\xi^\alpha V_\alpha$ is constant along the geodesic, i.e., $\frac{d\left( \xi^\alpha V_\alpha\right)}{d\lambda}=V^\beta\nabla_\beta\left( \xi^\alpha V_\alpha\right)=0$.

\subsection{Conformal isometries}

Diffeomorphisms on $\mathcal{M}$ that act on the metric $g$ as $g\to \omega^2 g$, for some, non-vanishing $\omega$, are said to be \emph{conformal isometries} of the spacetime $(\mathcal{M},g)$. Clearly, isometries of $(\mathcal{M},g)$ are conformal isometries of $(\mathcal{M},g)$ with $\omega=1$. Notice that, if $\phi$ is a conformal isometry of $(\mathcal{M},g)$, and in particular if it is an isometry of $(\mathcal{M},g)$, then it is a conformal isometry of $(\mathcal{M},g')$, where $g'$ is any metric related to $g$ by a conformal transformation. 
A 1-parameter group of diffeomorphisms on $\mathcal{M}$ generated by a vector field $\xi$ is a group of conformal isometries of $(\mathcal{M},g)$, if and only if $\xi$ satisfies the \emph{conformal Killing equation} $\mathcal{L}_\xi g=\frac{2}{D}(\nabla_\gamma \xi^\gamma) g$ or, equivalently,
\begin{equation}
\label{eq:confKilleq}
\nabla_\alpha \xi_\beta+\nabla_\beta \xi_\alpha=\frac{2}{D}(\nabla_\gamma \xi^\gamma) g_{\alpha\beta},
\end{equation}
in any set of coordinates.
The solutions $\xi$ to \eqref{eq:confKilleq} are called \emph{conformal Killing vector fields} of $(\mathcal{M},g)$.
Notice that Killing vectors of $(\mathcal{M},g)$ are conformal Killing vectors of $(\mathcal{M},g)$. In fact, contracting the Killing equation \eqref{eq:Killeq} with $g^{\alpha\beta}$, we obtain $\nabla_\alpha \xi^\alpha=0$, which allows to write the Killing equation in the form \eqref{eq:confKilleq}.
Furthermore, if $\xi$ is a conformal Killing vector of $(\mathcal{M},g)$, and in particular if it is a Killing vector of $(\mathcal{M},g)$, then it is a conformal Killing vector of $(\mathcal{M},g')$, where $g'$ is any metric related to $g$ by a conformal transformation.

\subsection{Conformally compact spacetimes}
\label{subsubsec:confcomspace}

We now review the definition of conformally compact spacetimes, and the properties of vector fields generating 1-parameter groups of diffeomorpshims on this type of spacetimes.
We first identify the \emph{boundary} of a spacetime manifold $\mathcal{M}$, denoted by $\partial \mathcal{M}$, as the set of points at infinity (see \cite{wald:1984} for a more precise definition).
A spacetime $(\mathcal{M},g)$ is said to be \emph{conformally compact} if the metric $g$ has a second order pole at $\partial \mathcal{M}$, and thus it cannot be extended to $\bar{\mathcal{M}}=\mathcal{M}\cup \partial \mathcal{M}$, but it is possible to identify a \emph{defining function}, i.e., a function $\Omega$ on $\bar{\mathcal{M}}$ satisfying $\Omega>0$ in $\mathcal{M}$, $\Omega=0$ on $\partial \mathcal{M}$, and $d\Omega\neq 0$ on $\partial \mathcal{M}$ (these requirements imply that $d\Omega$ is a covector normal to $\partial \mathcal{M}$), such that the metric $\bar{g}:=\Omega^2 g$ can be smoothly extended to $\bar{\mathcal{M}}$ and is non-degenerate. For such spacetimes, $\mathcal{M}$ is also referred to as the \emph{bulk} of the spacetime. This construction provides a practical way to take the near-boundary limit, that is, we take $\Omega\to 0$.
It is important to mention that the conformal metric $\bar{g}$ does not have the same geometric properties as the spacetime metric $g$. However, the two metrics lead to equivalent definitions of timelike, null, and spacelike objects on $\mathcal{M}$.
Moreover, the restriction of $\bar{g}$ on $\partial \mathcal{M}$, regarded as a hypersurface of $\bar{\mathcal{M}}$, gives a metric $\bar{g}_{(0)}:=\bar{g}|_{\Omega=0}$ on $\partial \mathcal{M}$.
Notice that, if $\Omega$ is a suitable defining function, then also $f\Omega$ is a suitable defining function for any smooth, strictly positive $f$. Therefore, the spacetime metric $g$ does not induce a unique metric on $\partial \mathcal{M}$, but rather an equivalence class, called \emph{conformal class of boundary metrics}, containing metrics related to $\bar{g}_{(0)}$ by conformal transformations. Given a representative of this conformal class of boundary metrics, $g_{(0)}$, the class is uniquely specified and we denote it by $[g_{(0)}]$. Clearly, $[g_{(0)}]=[\bar g_{(0)}]$ for any defining function $\Omega$.
When given a conformal class, $\partial \mathcal{M}$ is called \emph{conformal boundary}. A conformal boundary and its conformal class form a \emph{conformal structure}, denoted by $(\partial \mathcal{M},[g_{0)}])$. In the rest of this thesis, we will only consider conformally compact spacetimes.

By definition, any diffeomorphism $\phi$ must map $\partial \mathcal{M}$ to itself, hence it induces a diffeomorphism at the boundary, $\phi_{(0)}\colon\partial \mathcal{M}\to\partial \mathcal{M}$. By combining $\phi$ and $\phi_{(0)}$, we can extend $\phi$ to a diffeomorphism from $\bar{\mathcal{M}}$ to itself, which we still denote by $\phi$. 
If $\phi$ induces a trivial diffeomorphism $\phi_{(0)}$ at the boundary, i.e., $\phi_{(0)}(p)=p$ for any $p\in \mathcal{M}$, we say that $\phi$ is \emph{pure gauge}, since it has no physical effect on $\partial \mathcal{M}$. In particular, a pure gauge diffeomorphism does not affect physical predictions of any potential theory living on the boundary.\footnote{In saying so, we are anticipating applications of this formalism to the AdS/CFT duality, which states, roughly speaking, that gravity in asymptotically locally AdS spacetimes (see Section~\ref{sec:asyAdS} for the definition) is dual to a quantum conformal field theory (CFT) living on the boundary. See, for example, \cite{Ammon:2015wua} for a review.}
A general 1-parameter group of diffeomorphisms $\phi$ on $\mathcal{M}$, generated by a complete vector field $V$, induces a 1-parameter group of diffeomorphisms $\phi_{(0)}$ on $\partial \mathcal{M}$. Let $V_{(0)}$ be the complete vector field generating $\phi_{(0)}$, constructed as explained in Section~\ref{sec:symm}. From its construction, we see that $V_{(0)}$ must be tangent to $\partial \mathcal{M}$.
By combining $V$ and $V_{(0)}$, we can extend $V$ to a vector field on $\bar{\mathcal{M}}$, which we still denote by $V$. 
In the following, we will only consider cases in which this extension can be done smoothly.
The diffeomorphisms of a 1-parameter group generated by a vector field $V$ that vanishes at the boundary, $V_{(0)}=0$, are pure gauge, and the corresponding $V$ is also said to be pure gauge.

\subsection{Asymptotic properties of diffeomorphisms}
\label{subsubsec:asydiff}

In conformally compact spacetimes, we can also define asymptotic properties of diffeomorphisms. Here we list a few of these.
Let $(\mathcal{M},g)$ be a conformally compact spacetime with conformal boundary structure $(\partial \mathcal{M}, [g_{(0)}])$ for some representative $g_{(0)}$ of the conformal class of boundary metrics.
We say that a spacetime $(\mathcal{M},g)$ is \emph{stationary} if there exists a Killing vector field $K$ that is timelike in a neighbourhood of $\partial \mathcal{M}$. Such $K$, which is defined up to a choice of normalisation constant, is said to be \emph{stationary}. 
If, in addition, $K$ is orthogonal to a family of hypersurfaces, then we say that the spacetime is \emph{static}. A spacetime is said to be \emph{axisymmetric} if there exists a Killing vector field $M$ that is spacelike in a neighbourhood of $\partial \mathcal{M}$, and generates a 1-parameter group of isometries whose elements are in bijective correspondence with elements of the group $U(1)$. The last requirement implies that the integral curves of $M$ are closed (although not necessarily with period $2\pi$).
Such $M$, which is defined up to a choice of normalisation constant, is said to be \emph{axial}.
A spacetime is said to be \emph{stationary and axisymmetric} if it is stationary with stationary vector field $K$, it is axisymmetric with axial vector field $M$, and $M$ is invariant under the diffeomorphisms generated by $K$, i.e., $\mathcal{L}_K M=0$ (or, equivalently, $K$ is invariant under the diffeomorphisms generated by $M$, i.e., $\mathcal{L}_M K=0$). 

Conformal isometries of $(\mathcal{M},g)$, and in particular isometries of $(\mathcal{M},g)$, induce conformal isometries of $(\partial \mathcal{M},g_{(0)})$, i.e., diffeomorphisms $\phi_{(0)}$ on $\partial \mathcal{M}$ that act as $g_{(0)}\to \omega^2 g_{(0)}$ for some non-vanishing $\omega$, thus preserving the conformal class $[g_{(0)}]$.
Given a representative $g_{(0)}$of the boundary class of metrics,
any diffeomorphism $\phi$ on $\mathcal{M}$ that induces a conformal isometry of $(\partial \mathcal{M},g_{(0)})$ is called an \emph{asymptotic conformal isometry} of $(\mathcal{M},g)$.
A 1-parameter group of diffeomorphisms generated by a vector field $\xi$ is a group of asymptotic conformal isometries of $(\mathcal{M},g)$, if and only if the boundary limit $\xi_{(0)}$ is a conformal Killing vector field of $(\partial \mathcal{M},g_{(0)})$. In this case we say that $\xi$ is an asymptotic conformal Killing vector field of $(\mathcal{M},g)$. Notice that these definitions allow for asymptotic conformal isometries and generators that are pure gauge.

The most physically interesting notion is that of an asymptotic symmetry, whose definition requires the imposition of \emph{boundary conditions} at $\partial \mathcal{M}$, i.e., the specification of the asymptotics of $g$ and any matter field involved near the boundary, i.e., for small $\Omega$. Such boundary conditions must be chosen so that the equations of motion of the fields can be solved for certain classes of spacetime solutions, large enough to include physically interesting cases. 
In practice, the specification of boundary conditions is expressed by (partially) fixing a gauge, i.e., by choosing some coordinates $x^\alpha$ in which the metric components $g_{\alpha\beta}$ take a certain form, and then specifying the asymptotics of the deviation $h_{\alpha\beta}$ of the resulting $g_{\alpha\beta}$ from certain values $\hat g_{\alpha\beta}$. The boundary conditions on other fields are imposed in a similar way.
We say that a diffeomorphism $\phi$ \emph{preserves the boundary conditions} on $g$, if its action on $g$, $\phi_\star(g)$, still satisfies the boundary conditions on $g$. Similarly, we say that $\phi$ preserves the boundary conditions on any other field, if the action of $\phi$ on the field satisfies the boundary condition on the field.
A 1-parameter group of diffeomorphisms generated by a vector field $\xi$ preserves the boundary conditions on $g$, if and only if $\left(\mathcal{L}_\xi g\right)_{\alpha\beta}$ is of the same order as $h_{\alpha\beta}$ imposed by the boundary conditions. We write this requirement as the asymptotic Killing equation
\begin{equation}
\label{eq:asyKilleq}
\left(\mathcal{L}_\xi g\right)_{\alpha\beta}=\mathcal{O}(h_{\alpha\beta}).
\end{equation}
The solutions $\xi$ to \eqref{eq:asyKilleq} are called \emph{asymptotic Killing vector fields} of the spacetime $(\mathcal{M},g)$. 
The group of diffeomorphisms preserves the boundary conditions on other fields, if and only if $\xi$ satisfies the analog of \eqref{eq:asyKilleq} for the other fields.

According to these definitions, boundary-conditions-preserving diffeomorphisms and their generators $\xi$ can be pure gauge. These do not correspond to interesting physical transformations of a potential theory living at the boundary, so we do not want to include them in any notion of asymptotic symmetry.
A 1-parameter group of diffeomorphisms that preserves the boundary conditions on $g$ and any matter field involved, and that is not pure gauge, is said to be an \emph{asymptotic symmetry} of the spacetime. 
In order to uniquely identify the boundary limit $\xi_{(0)}$ of any generator $\xi$ of an asymptotic symmetry, we assume that some normalisation has been imposed on $\xi_{(0)}$.
If two generators of asymptotic symmetries, $\xi$ and $\xi'$ agree at the boundary, i.e., $\xi'_{(0)}=\xi_{(0)}$, then they can differ at most by a pure-gauge asymptotic Killing vector field. In this case, we say that $\xi$ and $\xi'$ are equivalent. Asymptotic symmetries with equivalent generators are identified and regarded as just one asymptotic symmetry. For instance, a spacetime is said to have an asymptotic time-translation symmetry if there exists a non-pure gauge asymptotic Killing vector field $k$ that is timelike near $\partial \mathcal{M}$. We also require that $k_{(0)}$ is normalised as $k^2_{(0)}=-1$ for a given choice of the representative of the conformal class of boundary metrics.
A spacetime is said to have an asymptotic rotation symmetry if there exists a non-pure gauge asymptotic Killing vector field $m$ that is spacelike near $\partial \mathcal{M}$, whose boundary limit $m_{(0)}$ generates a $U(1)$ group of diffeomorphisms at $\partial \mathcal{M}$. We require that $m_{(0)}$ is normalised so that its integral curves have period $2\pi$.

\subsection{Asymptotic physical notions}
\label{subsubsec:asyphysnot}

The asymptotic symmetries of a conformally compact spacetime can be used to define fundamental physical notions, as we now explain. 
Consider an asymptotic time symmetry generated by $k_{(0)}$ at the boundary. $k_{(0)}$ is completely determined once its normalisation is fixed as in Section~\ref{subsubsec:asydiff}.
We define a notion of ``time'' at the boundary as the parameter of integral curves of $k_{(0)}$.
Given a frame $x^\alpha=(x^0,x^1,x^2,x^3)$, if $x^0$ reduces to our notion of time at the boundary, we say that $x^0$ is a \emph{time coordinate}.
In any such frame, we have $k_{(0)}=\frac{\partial}{\partial x^0}$. At infinity $\partial \mathcal{M}$, we can now define an observer \emph{at rest}, or \emph{stationary}.
This is an observer (i.e., someone on a timelike curve) at $\partial \mathcal{M}$ whose worldline is an integral curve of $k_{(0)}$, i.e., the observer's 4-velocity $u$ is proportional to $k_{(0)}$. The 4-velocity normalisation $u^2=-1$, where $u^2$ is computed with the same boundary metric $g_{(0)}$ that we used to impose $k_{(0)}^2=-1$, implies that $u$ must be exactly equal to $k$. In the frame $x^\alpha$, we thus have $u^\alpha=\frac{dx^\alpha}{d\tau}=(1,0,0,0)$, which tells us that the ``non-time'' coordinates, also called \emph{spatial coordinates}, of the observer are fixed, and that the time $\tau$ measured by the observer varies as our notion of time $x^0$. This explains why such an observer is considered at rest.
If the spacetime admits a Killing vector field $K$ that is timelike in a bulk neighbourhood $\mathcal{U}$ of $\partial \mathcal{M}$, i.e., a stationary vector field, and we fix the normalisation of $K$ so that its boundary limit is $k_{(0)}$, then we can extend the notion of time to the bulk region $\mathcal{U}$, that is, we define time as the parameter of integral curves of $K$.
We can also extend the definition of stationary observer to $\mathcal{U}$ by requiring that the observer's 4-velocity is proportional to $K$. In any frame $x^\alpha$ adapted to $k$, i.e., such that $K=\frac{\partial}{\partial x^0}$, $x^0$ is clearly a time coordinate, and is the only coordinate changing along the worldline of a stationary observer, i.e., $u^1=u^2=u^3=0$ (but $u^0$ is in general not equal to 1 in the bulk).
Let us now assume that the spacetime has also an asymptotic rotation symmetry, generated by $m_{(0)}$, which is completely defined once its normalisation is fixed as as in Section~\ref{subsubsec:asydiff}.
Similarly to the previous discussion, we define a notion of ``azimuthal angle'' at the boundary as the parameter of integral curves of $m_{(0)}$.
Given a frame $x^\alpha$, if $x^3$ reduces to our notion of ``azimuthal angle'' at the boundary, we say that $x^3$ is an \emph{azimuthal coordinate}. In any such frame, we have $m_{(0)}=\frac{\partial}{\partial x^3}$.
If there exists an axial Killing vector field $M$, normalised so that its boundary limit is $m_{(0)}$, we can extend the notion of azimuthal angle to the region where $M$ remains spacelike, that is, we define the azimuthal angle as the parameter along integral curves of $M$.

Equipped with these notions, we can define the \emph{angular velocity of a causal curve}. 
Consider first a frame $x^\alpha$ where $x^0$ is a time coordinate and $x^3$ is an azimuthal coordinate. 
Then, consider an arbitrary causal curve parameterised by $\lambda$, with tangent vector $V^\alpha=\frac{dx^\alpha}{d\lambda}$.
The angular velocity of the points of this curve, measured at $\partial \mathcal{M}$ by employing our notions of time and azimuthal angle, i.e., with respect to stationary observers at $\partial \mathcal{M}$, is given by $\frac{dx^3}{d x^0}=\frac{dx^3}{d\lambda}/\frac{dx^0}{d\lambda}=\frac{V^3}{V^0}$.

We now use the angular velocity of a particular class of curves to define the \emph{angular velocity} $\Omega$ for a stationary and axisymmetric spacetime.\footnote{The angular velocity should not be confused with the defining function, although we use the same symbol for these two quantities. Since these are very different notions, whether we are referring to one or the other should be clear from the context.}
We wish to define this quantity at any point $p$ as a measure, performed at $\partial \mathcal{M}$, of the dragging effect of the rotation of spacetime on an observer at $p$. 
Let $K$ be the stationary vector field and $M$ be the axial vector field, normalised so that their boundary limit is given by $k_{(0)}$ and $m_{(0)}$, respectively. 
Let us consider coordinates $x^\alpha=(x^0,x^1,x^2,x^3)$, where $x^0$ is a time coordinate and $x^3$ is an azimuthal coordinate, thus $K=\frac{\partial}{\partial x^0}$ and $M=\frac{\partial}{\partial x^3}$. 
Now, consider an observer at $p$ whose 4-velocity $u$ is proportional to the normal to hypersurfaces $\Sigma_{x^0}$ at fixed $x^0$, i.e., $u^\alpha=Ng^{\alpha\beta}(dx^0)_\beta$ where the function $N$ can be determined from the 4-velocity normalisation $u^2=-1$.
For such observers, the quantity $L:=g_{\alpha\beta} u^\alpha M^\beta$, which has the interpretation of the measure at $\partial \mathcal{M}$ of the angular momentum per unit mass of the observer, vanishes. 
Because of this fact, which we will prove shortly, these observers can only rotate with the local spacetime geometry, i.e., they experience no rotation. For this reason, we refer to them as \emph{locally non-rotating observers}.
Their angular velocity measured at infinity (which is in general non-vanishing), given by $\frac{dx^3}{d x^0}=\frac{u^3}{u^0}$, can only be due to the dragging effect mentioned above, therefore we use it to define the angular velocity $\Omega$ of the spacetime at the position $p$ of the observer.
Verifying that $L$ vanishes for locally non-rotating observers is simple. In the frame $x^\alpha$, we have $L=N g_{\alpha\beta} g^{\alpha\gamma}(dx^0)_\gamma M^\beta=N (dx^0)_\gamma M^\gamma=N( \partial_\gamma x^0 )\delta^\gamma_3=N\delta^0_3=0$. Since $L$ is a scalar, it must vanish in any other frame.
Under fairly general assumptions, which are satisfied by the spacetimes that we are going to study, it is also possible to choose $x^1$ and $x^2$ so that $g_{01}=g_{02}=g_{13}=g_{23}=0$ (see Chapter 7 of \cite{wald:1984}).
In this frame, the angular velocity of locally non-rotating observers can be computed from a simple formula.
In fact, the condition $L=0$ gives $g_{03}u^0+g_{33}u^3=0$, which can be rearranged as $\frac{u^3}{u^0}=-\frac{g_{03}}{g_{33}}$. Thus, we obtain
\begin{equation}
\label{eq:angveldef}
\Omega=-\frac{g_{03}}{g_{33}}.
\end{equation}
In summary, to compute the angular velocity of a stationary and axisymmetric spacetime, we use the expression \eqref{eq:angveldef} in a frame $x^\alpha$ in which $x^0$ is a time coordinate, $x^3$ is an azimuthal coordinate, and $g_{01}=g_{02}=g_{13}=g_{23}=0$.

\section{Equations of motion}
\label{subsec:EinEq}

The theory of general relativity postulates that the dynamics of physical spacetimes is determined by certain equations of motion for the metric, called Einstein equations.
In order to study gravitational physics with different types of matter, matter fields must be included in the model, and their couplings with the metric field must be specified.
Then, the equations of motion of the resulting theory must be solved simultaneously for the metric and all the matter fields involved.
Here we introduce the equations of motion of general relativity, possibly including matter fields, and we present their simplest solution.
From now onwards, we consider the case of $D=4$ spacetime dimensions.
 
In any set of coordinates $x^\alpha$, the \emph{Einstein equations} with \emph{cosmological constant} $\Lambda$ read
\begin{equation}
\label{eq:EFE}
R_{\alpha\beta}-\frac{1}{2}R g_{\alpha\beta}+\Lambda g_{\alpha\beta}=8\pi T_{\alpha\beta},
\end{equation}
where $T_{\alpha\beta}$ is the \emph{energy-momentum tensor}, which contains the information about energy and momentum of matter fields. 
The Einstein equations \eqref{eq:EFE} can be obtained from the action of general relativity, $S = S_{\text{EH}}+S_{\text{mat}}$, by considering an arbitrary variation of $g_{\alpha\beta}$ that vanishes outside of a compact spacetime region, and requiring that the corresponding variation of the action, $\delta_g S$, vanishes in that region.
The first term of the action, $S_{\text{EH}}$, is the so-called Einstein-Hilbert term. It only depends on the metric field, and describes gravitational physics in the absence of matter. It is given by
\begin{equation}
S_{\text{EH}}=\frac{1}{16\pi}\int_{\mathcal{M}} d^4 x \sqrt{-g} \left( R - 2\Lambda \right),
\end{equation}
where the symbol $g$ in the square root denotes the determinant of the matrix $g_{\alpha\beta}$ (this is negative for Lorentzian metrics).
The term $S_{\text{mat}}$ depends in general on both the metric and the matter fields, and describes how matter couples with gravity. It also determines the energy-momentum tensor through $T_{\alpha\beta}:=-\frac{2}{\sqrt{-g}}\frac{\delta S_{\text{mat}}}{\delta g^{\alpha\beta}}$. This definition shows that, as for the metric $g$, $T$ is symmetric, i.e., $T_{\alpha\beta}=T_{\beta\alpha}$.
In \emph{vacuum}, i.e., in the absence of matter fields, we have $T_{\alpha\beta}=0$.

If we wish to study the interactions between the spacetime geometry and some type of matter, \eqref{eq:EFE} must be solved together with the equations of motion of the matter fields involved.
Similarly to the case of the metric, the equations of motion of any matter field $\varphi$ can be obtained by requiring that $\delta_\varphi S_{\text{mat}}$ vanishes for any compactly supported variation of $\varphi$.
For a solution of the matter equations of motion, it can be shown that the conservation of the energy-momentum tensor, $\nabla^\alpha T_{\alpha\beta}=0$, is a consequence of the diffeomorphism invariance of the theory.
Notice that $\nabla^\alpha T_{\alpha\beta}=0$ is consistent with \eqref{eq:EFE}, since the covariant divergence of the left hand side of \eqref{eq:EFE} also vanishes due to the contracted Bianchi identity \eqref{eq:conBiaid} and the compatibility of $\nabla$ with the metric, $\nabla_\gamma g_{\alpha\beta}=0$. 
We are only interested in matter fields that are thought to potentially describe physical matter in the universe. Since the current of energy and momentum detected by someone travelling along integral curves of $V$ is described by $-T^\alpha_\beta V^\beta$, and we require that physical matter does not move faster than light, then matter is considered ``physical'' if it satisfies the so-called \emph{Dominant Energy Condition}: at any point $p$, given any future-directed timelike vector $V$, $-T^\alpha_\beta V^\beta$ is a future-directed causal vector (or it vanishes).
For instance, this condition is trivially satisfied in vacuum, as $T_{\alpha\beta}=0$. It is also satisfied by matter sourced by a real scalar field $\varphi$ of mass $\mu\geq 0$ minimally coupled with gravity. This is the type of matter on which we focus in this thesis.
Its action is
\begin{equation}
S_{\text{mat}}= \int_{\mathcal{M}} d^4 x \sqrt{-g}\left( - \frac{1}{2}g^{\alpha\beta} \nabla_\alpha \varphi \nabla_\beta \varphi -\frac{1}{2}\mu^2\varphi^2\right).
\end{equation}
The corresponding energy-momentum tensor is given by
\begin{equation}
\label{eq:KHmomtenscov}
T_{\alpha\beta}=\nabla_\alpha \varphi \nabla_\beta \varphi +g_{\alpha\beta}\left(- \frac{1}{2} g^{\gamma\delta} \nabla_{\gamma} \varphi \nabla_{\delta} \varphi -\frac{1}{2}\mu^2\varphi^2\right)\,.
\end{equation}
The equation of motion for $\varphi$ is the 
\emph{Klein-Gordon equation}
\begin{equation}
\label{eq:m0KGeq}
(\Box-\mu^2)\varphi=0,
\end{equation}
where $\Box:=g^{\alpha\beta}\nabla_\alpha\nabla_\beta$.
We see that \eqref{eq:EFE} and \eqref{eq:m0KGeq} form a system of $2^{nd}$ order coupled partial differential equations (PDEs) in the metric components and the matter fields.
Exact solutions to this complicated system of PDEs are known only in very simplified settings, with a large number of symmetries. 

In $\mathcal{M}=\mathbb{R}^4$, the maximally symmetric solution of the Einstein equations with $\Lambda=0$ in vacuum is the Minkowski metric, written in global Cartesian coordinates $(t,x,y,z)\in\mathbb{R}^4$ as
\begin{equation}
\label{eq:flat}
\eta=-dt^2+dx^2+dy^2+dz^2.
\end{equation}
In (global) spherical coordinates $(t,r,\theta,\phi)\in(-\infty,+\infty)\times(0,+\infty)\times(0,\pi)\times(0,2\pi)$, defined by
$x=r \cos\theta,
y=r\sin\theta\cos\phi,
z=r\sin\theta\sin\phi$,
\eqref{eq:flat} reads
\begin{equation}
\label{eq:Minsph}
\eta=-dt^2+dr^2+r^2 d\Omega_2^2,
\end{equation}
where $d{\Omega_2}^2 = d\theta^2 + \sin^2\theta d\phi^2$ is the unit round metric on the 2-dimensional sphere $S^2$.
$\mathbb{M}_4=(\mathbb{R}^4,\eta)$, called \emph{Minkowski spacetime}, is flat, i.e., the Riemann tensor vanishes everywhere. The isometries of Minkowski spacetime form the 10-dimensional Poincar\'e group, consisting of the 6-dimensional group of Lorentz transformations, $O(3,1)$, and the 4-dimensional group of spacetime translations. 
It can be shown that $\mathbb{M}_4$ is conformally compact (see, for example, \cite{wald:1984}). 

For the purposes of this thesis, we can define \emph{asymptotically flat spacetimes} as conformally compact spacetimes that solve \eqref{eq:EFE} with $\Lambda=0$, together with the equations of motion of any matter field involved, and approach Minkowski spacetime near the conformal boundary, i.e., at small values of some defining function $\Omega$. 
In asymptotically flat spacetimes, a theorem by Choquet-Bruhat (see \cite{wald:1984} for a review) tells us that, once initial data for the metric, the matter fields, and their ``time derivatives''\footnote{We define what we mean by ``time derivative'' of the metric more precisely in Chapter~\ref{chap:ovengr}.}, are specified on a spacelike hypersurface in a way that is consistent with the equations of motion of general relativity, the evolution of the metric and the matter fields governed by the equations of motion can be determined uniquely, up to a gauge choice of coordinates. In other words, the Cauchy problem in asymptotically flat spacetimes is an initial value problem. The spacetime region whose metric is obtained as a result of the evolution is called \emph{future Cauchy development} of the prescribed initial data. 
In Chapter~\ref{Chapter:NoSym}, we will see that the Cauchy problem is fundamentally different if we consider conformally compact solutions of \eqref{eq:EFE} with $\Lambda<0$. In fact, in this case we have asymptotically AdS spacetimes (see Chapter~\ref{Chapter:NoSym} for the precise definition) in which the Cauchy problem is an initial-boundary value problem, i.e., in addition to initial data, we also need to prescribe boundary conditions at $\partial \mathcal{M}$ on all evolved fields throughout the entire evolution.
Numerical relativity has allowed to use computers to find approximate solutions to the field equations for rather general scenarios, with few or no symmetries, in asymptotically flat spacetimes. The most fundamental aspects of these methods, as well as a few examples of their implementation, are reviewed in Chapter~\ref{chap:ovengr}. In Chapter~\ref{Chapter:NoSym}, we explain how we applied some of these tools to obtain the numerical solution of the Cauchy problem in asymptotically AdS spacetimes in full generality.

\section{Black holes and horizons}
\label{sec:AH}

One of the most important predictions of general relativity is the existence of solutions, $g$, of the Einstein equations describing spacetimes that contain one or more \emph{black holes}. These objects are defined classically (i.e. non-quantum mechanically) as regions of spacetimes in which gravitational attraction is so strong that nothing can escape, not even light. Remarkably, black hole solutions exist even in vacuum (see Section~\ref{subsubsec:asyblaspa} for two examples of asymptotically AdS black hole spacetimes in vacuum). From a geometric point of view, black holes are regions of spacetime causally disconnected from the rest of the spacetime. Therefore, an observer outside a black hole cannot receive any signal from the black hole interior, although signals can enter the black hole. The boundary of the black hole region is called \emph{(future) event horizon}, and is denoted by $\mathcal{H}^+$. We refer to a spacelike slice of the black hole region at a fixed value of a time coordinate $t$ as the \emph{black hole at $t$}. The boundary of this region, which is a 2-dimensional spacelike slice of $\mathcal{H}^+$, is referred to as the \emph{cross-section of the event horizon at $t$}, or simply as the \emph{event horizon at $t$}.
In the rest of this section, we will review certain properties of black holes that are central to the discussions that follow.
These properties have been investigated and proved mainly for black holes in asymptotically flat spacetimes.
However, black holes in asymptotically flat spacetimes approximate black holes in asymptotically AdS spacetimes, if the latter are sufficiently far from the conformal boundary $\partial \mathcal{M}$, i.e., if they are sufficiently small, or equivalently if the value of the negative cosmological constant $\Lambda$ is sufficiently close to zero. 
Therefore, we can presume that some version of the results outlined in the following holds also for black holes in asymptotically AdS spacetimes.

\subsection{Black hole mechanics}
\label{subsec:blaholmech}

Ref.~\cite{Bardeen:1973gs} proved that black holes satisfy four laws, which show a clear resemblance with the laws of thermodynamics. 
These laws are briefly reviewed in this section. The following discussion will omit several technical details, and is simply intended to give a flavour of these historical results.
A thorough explanation of some historical versions of the laws in asymptotically flat black hole spacetimes can be found, for example, in \cite{wald:1984,Wald94}, although the literature on black hole mechanics in general relativity and other theories of gravity is vast.

Let us first introduce the crucial notion of Killing horizon. A null hypersurface $\mathcal{N}$ is said to be a \emph{Killing horizon} if there exists a Killing vector field $\xi$ that is normal to $\mathcal{N}$.
There exist several results, called rigidity theorems, that explore the relation between a Killing horizon and the event horizon of a stationary black hole, under different assumptions; see, for example, \cite{Hawking:1971vc,Hawking:1973uf,Friedrich:1998wq,Hollands:2006rj,Moncrief:2008mr}. 
In particular, the results of \cite{Hollands:2006rj} are valid also for asymptotically AdS spacetimes. 
These results suggest that the event horizon of a stationary black hole is a Killing horizon, which is what we assume in the following.
Let us normalise the stationary vector field $K$ by requiring that its boundary limit $k_{(0)}$ satisfies $k^2_{(0)}=-1$ with respect to the representative of the conformal class of boundary metrics that we use for the normalisation of the asymptotic time symmetry generators. In this way, $K$ is completely determined, and an asymptotic time symmetry generator. 
If $\xi$ is not proportional to $K$ at $\mathcal{H}^+$, then the spacetime must have an additional Killing vector field $\xi-K$, which is orthogonal to $\xi$ at $\mathcal{H}^+$. The rigidity theorems show that $\xi- K$ is an axial vector field, and the spacetime is also axisymmetric. 

Given any black hole spacetime for which the rigidity theorems hold, we can pick the normalisation of $\xi$ so that $\xi=K+c M$, where $M$ is an axial vector field and $c$ is a constant. 
If we require that $M$ is normalised so that its boundary limit has integral curves of period $2\pi$, then the constant $c$ is the angular velocity of the horizon, $\Omega_H$, as we will prove shortly. This tells us that the horizon rotates rigidly, which explains the terminology ``rigidity theorems''.
The redefinition $x^3\to-x^3$ changes the sign of the constant $\Omega_H$. In the following, we pick the azimuthal coordinate in such a way that $\Omega_H\geq 0$.
For the proof, let us consider a frame $x^\alpha$ adapted to $K$ and $M$, i.e., $K=\frac{\partial}{\partial x^0}$ and $M=\frac{\partial}{\partial x^3}$, and such that $g_{01}=g_{02}=g_{13}=g_{23}=0$. Notice that this is a frame in which the angular velocity of the spacetime, defined in Section~\ref{subsubsec:asyphysnot}, is given by the expression $\Omega=-g_{03}/g_{33}$. 
In this frame, $\xi^\alpha=(1,0,0,c)$ and $(\xi-K)^\alpha=(0,0,0,c)$, and the orthogonality condition between $\xi$ and $\xi-K$ at $\mathcal{H}^+$ gives $g_{03}+cg_{33}=0$. Therefore, the angular velocity at the horizon is $\Omega_H=-g_{03}/g_{33}=c$, as anticipated.

The Killing vector field
\begin{equation}
\label{eq:corotfie}
\xi=K+\Omega_H M,
\end{equation}
normal to $\mathcal{H}^+$, is said to be the vector field co-rotating with the horizon, in the sense that its integral curves (everywhere in the spacetime) rotate with the angular velocity of the horizon, $\Omega_H$. To see this, we use again the coordinate system $x^\alpha$ adapted to $K$ and $M$. Each integral curve of $\xi$ satisfies $\frac{dx^\alpha}{d\lambda}=\xi^\alpha$, where $\xi^\alpha=(1,0,0,\Omega_H)$ in this frame. Therefore, the angular velocity of points on each curve is $\frac{dx^3}{dx^0}=\frac{dx^3}{d\lambda}/\frac{dx^0}{d\lambda}=\xi^3/\xi^0=\Omega_H$.
Since $\xi$ is tangent to $\mathcal{H}^+$, integral curves of $\xi$ through points of $\mathcal{H}^+$ are null curves that lie within $\mathcal{H}^+$.
We now prove that these curves are geodesics.
Since $\xi^2$ has a (vanishing) fixed value at the Killing horizon $\mathcal{H}^+$, then $\left(d(\xi^2)\right)_\alpha=\partial_\alpha (\xi^2)=\nabla_\alpha (\xi^2)$ must be a covector normal to $\mathcal{H}^+$, and thus proportional to $\xi_\alpha$, i.e., $\nabla_\alpha (\xi^\beta\xi_\beta)=-2\kappa \xi_\alpha$ for some function $\kappa$ on $\mathcal{H}^+$, called \emph{surface gravity}. Assuming that the time-orientation is chosen so that $\xi$ is future-directed on $\mathcal{H}^+$, $\kappa$ is non-negative. The left hand side of the last result can be written as $2\xi^\beta\nabla_\alpha \xi_\beta$.
Using also the Killing equation $\nabla_\alpha \xi_\beta=-\nabla_\beta \xi_\alpha$, we obtain
\begin{equation}
\label{eq:surgrav}
\xi^\beta\nabla_\beta \xi^\alpha=\kappa \xi^\alpha
\end{equation} 
at the horizon.
This shows that $\xi$ is the tangent vector of geodesics at $\mathcal{H}^+$. We regard \eqref{eq:surgrav} as the definition of surface gravity of a Killing horizon.

We are now in a position to present the \emph{four laws of black hole mechanics}.
The \emph{zeroth law} states that $\kappa$ is constant on the event horizon of a stationary black hole spacetime. Notice that the fact that the metric solves the Einstein equations is crucial to prove this result.
The version of the \emph{first law} that will be relevant for our discussion is the ``physical process'' version, originally presented in \cite{Hawking:1972hy}. 
Let us define the area of the event horizon at a given time by the area of the cross-section of $\mathcal{H}^+$ at that time.
Consider a stationary vacuum black hole spacetime with energy $E$ and angular momentum $J$. Suppose that $\mathcal{H}^+$ has area $A$, surface gravity $\kappa$, and rotates with angular velocity $\Omega_H$.
Let us now imagine that the black hole is perturbed via a scattering process with a small amount of matter with infinitesimal energy $\delta E$ and infinitesimal angular momentum $\delta J$. By energy and angular momentum conservation, $E+\delta E,J+\delta J$ must be, respectively, the energy and angular momentum of the spacetime once this physical process has occured. The first law states that these variations are related to the variation $\delta A$ of the area of $\mathcal{H}^+$ by
\begin{equation}
\label{eq:firstlaw}
\delta E=\frac{\kappa}{8\pi}\delta A+\Omega_H \delta J.
\end{equation}
It should be mentioned that this version of the first law differs from the one presented in \cite{Bardeen:1973gs}, which is typically referred to as the ``equilibrium state'' version. In fact, the latter shows that \eqref{eq:firstlaw} holds if we perturb a black hole solution $g$ in such a way that the perturbed metric still satisfies the Einstein equations.
This was proved in \cite{Bardeen:1973gs} under rather restrictive assumptions, and later generalised to rather general cases by \cite{Sudarsky:1992ty,Wald:1993ki} and \cite{Wald:1993nt,Iyer:1994ys}.\footnote{Ref. \cite{Rossi:2020cko} presents a detailed review of these results.}
The first law has also been shown to hold for asymptotically locally AdS spacetimes (see Section~\ref{subsec:asylocAdS} for the definition) in \cite{Papadimitriou:2005ii}.
The \emph{second law}, also called area theorem, states that, given two cross-sections of $\mathcal{H}^+$, $C_1$ and $C_2$, such that $C_2$ is in the casual future of $C_1$, the area of $C_2$ is not smaller than the area of $C_1$. In other words, the area of the event horizon does not decrease in time, $\delta A\geq 0$. 
Finally, the \emph{third law} states that it is not possible to reduce $\kappa$ to 0 by a finite sequence of physical operations.

The resemblance of these laws with the four laws of thermodynamics suggests that black holes are thermodynamical objects and, as such, have a certain temperature and emit thermal radiation. This seems to be in contrast with the definition of black holes as regions of spacetimes from which nothing can escape. However, this definition does not take quantum mechanical effects into consideration. In \cite{Hawking:1974sw}, Hawking showed that the thermal properties of black holes arise from a quantum treatment of matter on a fixed black hole background spacetime. In particular, he found that the temperature of a black hole is given by $T_H=\frac{\kappa}{2\pi}$. We refer to this quantity as the \emph{Hawking temperature}.
This result also indicates that black holes have also an entropy, as originally suggested by Bekenstein \cite{Bekenstein:1972tm,Bekenstein:1973ur}. The entropy is given by the Bekenstein-Hawking formula $S=A/4$.

\subsection{Apparent horizons}
\label{subsec:apphor}

Identifying the boundary of causally disconnected regions, i.e., event horizons, requires knowledge of the entire spacetime. However, in numerical simulations, one typically only knows a portion of spacetime, namely the part given by the evolution of initial data up to a certain time. For this reason, it is useful to define regions of spacetime that can be determined simply by the knowledge of the metric solution at a given time, and that typically approximate slices of event horizons at that time. These are called \emph{apparent horizons} and are defined as follows.

Consider a spacetime region $\mathcal{U}$. Then, consider a set of spherical coordinates $x^\alpha=(t,\rho,\theta,\phi)$ on $\mathcal{U}$ defined so that hypersurfaces of constant $t$ are spacelike slices of $\mathcal{U}$, and surfaces of constant $t$ and constant $\rho$ are 2-dimensional spheres with coordinates $(\theta,\phi)$. We also require that $\frac{\partial}{\partial t}$ is future-directed. 
Now, let $\Sigma_t$ be a spacelike slice of $\mathcal{U}$ at fixed $t$. The future-directed unit covector normal to each point of $\Sigma_t$ is given by
$n_\alpha=-\frac{\partial_\alpha t}{\sqrt{-g^{\beta\gamma} \partial_\beta t \partial_\gamma t}}$.
The tensor $\gamma_{\alpha\beta}=g_{\alpha\beta}+n_\alpha n_\beta$ gives the metric induced on $\Sigma_t$ by $g$ in 4-dimensional form.
Next, consider a 2-dimensional spacelike surface $S$ in $\Sigma_t$.
Let $R(\theta,\phi)$ be the function such that $F(\rho,\theta,\phi):=\rho-R(\theta,\phi)=0$ on $S$.
Regarding $S$ as an hypersurface of $\Sigma_t$ with metric $\gamma$, we identify the outward-pointing (i.e., pointing away from $S$) unit covector normal to $S$ and tangent to $\Sigma_t$ as
$s_\alpha=\frac{\gamma_\alpha^\beta \partial_\beta F}{\sqrt{\gamma^{\gamma\sigma}\partial_\gamma F \partial_\sigma F}}$.
Given any 2-dimensional spacelike $S$, at each point $p\in S$, there are two future-directed null directions orthogonal to $S$, corresponding to normal light rays that move towards the interior of $S$ and normal light rays that move towards the exterior of $S$. The ``inward'' direction is identified by the vector $u=n-s$, while the ``outward'' direction is identified by the vector $l=n+s$.
The expansion of the ingoing family of light rays leaving $S$, also called \emph{inward null expansion} of $S$, is defined as
\begin{equation}
\label{eq:inexp}
\Theta_{in}:=\gamma^{\alpha\beta}\nabla_\alpha u_\beta.
\end{equation}
Similarly, the expansion of the outgoing family of light rays leaving $S$, also called \emph{outward null expansion} of $S$, is defined as
\begin{equation}
\label{eq:outexp}
\Theta_{out}:=\gamma^{\alpha\beta}\nabla_\alpha l_\beta.
\end{equation}
We define a \emph{trapped surface} as any 2-dimensional spacelike surface $S$ such that both $\Theta_{in}<0$ and $\Theta_{out}<0$ everywhere on $S$.
The boundary of the region containing trapped surfaces is called \emph{apparent horizon}. If the apparent horizon is made of disconnected pieces, it is customary to refer to each piece as a different apparent horizon, and say that the spacetime has multiple apparent horizons.

Intuitively, light rays moving towards the interior of $S$ are expected to converge, which implies $\Theta_{in}<0$ on $S$, while light rays moving towards the exterior of $S$ are expected to diverge, which implies $\Theta_{out}>0$ on $S$.
These intuitive results are no longer valid in the black hole region, where we expect both families to converge, due to the fact that light cannot escape a black hole. In other words, we expect that all 2-dimensional spacelike surfaces inside the black hole region are trapped, and thus that the boundary of the region containing trapped surfaces, i.e., the apparent horizon, is a good approximation for the slice of the event horizon at a given time $t$ in the evolution.
Furthermore, if we accept this result, by continuity between the interior of the black hole, where $\Theta_{in}<0, \Theta_{out}<0$, and the exterior, where $\Theta_{in}<0, \Theta_{out}>0$, we can also expect that the apparent horizon is the set of points with $\Theta_{out}=0$.
There are theorems confirming these results under certain, physically reasonable assumptions, which are typically satisfied by generic initial data evolved in simulations (see, e.g., \cite{wald:1984}).
This discussion explains the reason why, in simulations, it is customary to look for apparent horizons as 2-dimensional spacelike surfaces at fixed evolution time $t$ satisfying $\Theta_{out}=0$, and use apparent horizons as approximations for the position of a slice of the event horizon at time $t$.

\subsection{Weak cosmic censorship conjecture}
\label{subsec:WCCC}

The well-established existence of black holes in Nature addresses an important question about \emph{curvature singularities}, i.e., spacetime points at which the curvature diverges and the spacetime metric is no longer regular.
The singularity theorem by Penrose \cite{Penrose:1964wq} (see, for example, \cite{wald:1984} for a review), which has recently earned him the Nobel prize, demonstrates that singularities can be generically expected in general relativity dynamics whenever the gravitational field is strong enough to form trapped surfaces. The \emph{weak cosmic censorship conjecture (WCCC)} assumes that singularities are always hidden behind an horizon, so they cannot influence distant observers, which explains why we do not experience diverging curvature effects.
Although this conjecture is widely accepted in astrophysical scenarios, it has not yet been proved under general conditions and, in fact, it is possible to construct (non-astrophysical) counterexamples. The study of these violations of WCCC is crucial for a complete understanding of any gravity theory: the formation of singularities not hidden behind an event horizon (also called \emph{naked singularities}) signal limits to the predictability power of the theory, since predictions in the future of a curvature singularity cannot be made.
A quantum theory of gravity is needed to determine the physics of a system in which a classical (i.e., non-quantum mechanical) theory would detect a naked singularity.
For this reason, it is particularly interesting to study dynamical WCCC violations in AdS/CFT, since the dual CFT provides a fully quantum viewpoint of the formation of the naked singularity occurring in the bulk.

\ifpaper
\end{document}
\fi
\newif\ifpaper
\paperfalse

\ifpaper
\input{../preamble}
\begin{document}
\fi

\chapter{Overview of numerical general relativity}
\label{chap:ovengr}

Numerical methods have been successfully applied to obtain solutions of the Einstein equations, possibly coupled with matter, in computers. 
In particular, in this chapter we discuss numerical schemes able to determine initial data on a spacelike hypersurface and the evolution of such data according to the Einstein equations coupled with matter, i.e., solve the Cauchy problem in general relativity. Numerous techniques have been developed over the years to tackle this problem, especially in the case of asymptotically flat spacetimes. 
A comprehensive review of these tools is beyond the scope of this work. Instead, we limit the discussion to the most fundamental aspects, with particular focus on those that we applied to obtain the fully general, long-lived simulations of asymptotically AdS spacetimes, presented in Chapter~\ref{Chapter:NoSym} and Chapter~\ref{Chapter:KAdS}.

This chapter is structured as follows.
In Section~\ref{sec:3p1splitt}, we review one of the hystoric formulations of the Cauchy problem, the so-called Arnowitt-Deser-Misner (ADM) formulation \cite{PhysRev.116.1322}.
In Section~\ref{sec:BSSNOK}, we review the BSSNOK (Baumgarte, Shapiro, Shibata, Nakamura, Oohara and Kojima) formalism \cite{10.1143/PTPS.90.1,PhysRevD.52.5428,Baumgarte:1998te}, which improves the ADM formalism and has become a common tool in numerical relativity.
In Section~\ref{sec:genharform}, we introduce the generalised harmonic formalism, due to \cite{Foures-Bruhat:1952grw}, which has been shown to lead to stable simulations in asymptotically flat spacetimes \cite{Pretorius:2004jg} and later applied to simulate asymptotically AdS spacetimes \cite{Bantilan:2012vu,Bantilan:2017kok,Bantilan:2020pay,Bantilan:2020xas}.
In Section~\ref{sec:findiffmet}, we describe numerical methods that have been used to implement the above-mentioned formalisms in computers, in order to determine highly accurate approximated solutions of the Einstein equations (possibly coupled with matter).
When discussing practical implementations of these methods, we focus on the ones that are relevant for the scheme that we employ.
In the rest of this thesis, we will always consider the case of $D=4$ spacetime dimensions, except where explicitly stated otherwise.

\section{ADM splitting}
\label{sec:3p1splitt}

In this section we present the main aspects of the \emph{Arnowitt-Deser-Misner (ADM) formulation} of the Cauchy problem in general relativity, also called 3+1 formulation \cite{PhysRev.116.1322} since it involves treating the three spatial directions and the time direction on different footage. A more detailed review can be found in \cite{Alcubierre:1138167}.

Consider a slicing of spacetime $(\mathcal{M},g)$ into 3-dimensional spacelike hypersurfaces, $\Sigma_t$, each of which is at a fixed value of a suitably chosen time coordinate $t$.
We use a set of coordinates $x^\alpha$ given by $x^0=t$ and three coordinates $x^i$ describing points on each slice. 
We employ $\frac{\partial}{\partial t}$ to define the time-orientation of the spacetime.
The future-directed timelike unit normal to each hypersurface is
\begin{equation}
\label{eq:uninormal}
n^\alpha=-N g^{\alpha\beta} (dt)_\beta,
\end{equation}
where $N$ must be chosen as $1/\sqrt{-g^{\alpha\beta}\partial_\alpha t \partial_\beta t}=1/\sqrt{-g^{tt}}$ to ensure that $n$ has unit norm, i.e., $g_{\alpha\beta}n^\alpha n^\beta=-1$. This choice of $N$ is called \emph{lapse} function.

The projection operator onto $\Sigma_t$ was defined in Section~\ref{sec:hypsuf}, and is given by
\begin{equation}
\gamma^\alpha_\beta=\delta^\alpha_\beta+n^\alpha n_\beta.
\end{equation}
As mentioned in Section~\ref{sec:hypsuf}, this can be identified with the Riemannian metric of $\Sigma_t$ defined as the restriction of the spacetime metric $g$ on $\Sigma_t$, i.e., the pull-back of $g$ onto $\Sigma_t$ with respect to the inclusion map that embeds $\Sigma_t$ in $\mathcal{M}$. The result of the pull-back is given by $\gamma_{ij}$ in the spatial coordinates $x^i$.

We then define the \emph{shift} vector as $ N^\alpha:=\gamma^\alpha_\beta \bigl(\frac{\partial}{\partial t}\bigr)^\beta$. This is clearly invariant under projection onto $\Sigma_t$, so it can be identified with a vector in the tangent space of $\Sigma_t$, given by the components $ N^i$. Knowing $N,N^i,\gamma_{ij}$, the spacetime metric can be reconstructed as
\begin{equation}
ds^2=-N^2dt^2+\gamma_{ij}(dx^i + N^i dt)(dx^j+ N^j dt).
\end{equation}
We also have
\begin{equation}
\label{eq:tvec}
\biggl(\frac{\partial}{\partial t}\biggr)^\alpha=N n^\alpha+ N^\alpha.
\end{equation}

As a final ingredient, the projection of $\nabla_\alpha n_\beta$ defines the \emph{extrinsic curvature} of $\Sigma_t$\footnote{We can make sense of covariant derivatives of $n_\alpha$ by extending its definition on $\Sigma_t$ \eqref{eq:uninormal} to a 1-form field over a neighbourhood of $\Sigma_t$, which can be done in an arbitrary way without changing the value of $K_{\alpha\beta}$ on $\Sigma_t$ given by \eqref{eq:extrcurv}.}:
\begin{equation}
\label{eq:extrcurv}
K_{\alpha\beta}:=-\gamma^\gamma_\alpha \gamma^\delta_\beta \nabla_\gamma n_\delta=-\frac{1}{2}\mathcal{L}_n\gamma_{\alpha\beta}.
\end{equation}
$K_{\alpha\beta}$ is identified with the tensor on the tangent space of $\Sigma_t$, given by $K_{ij}$.
The Lie derivative along the normal direction in the second equality suggests that a choice of $K_{\alpha\beta}$ on $\Sigma_{t=0}$ is ``morally'' equivalent to a choice for the time-derivative of the metric components at $t=0$. In fact, this equation can be written in terms of the first order time derivative of $\gamma_{ij}$ as
\begin{equation}
\label{eq:evogamma}
\partial_t\gamma_{ij}=-2N K_{ij}+D_iN_j+D_j  N_i,
\end{equation} 
where $D$ is the Levi-Civita covariant derivative associated with the metric $\gamma_{ij}$ on $\Sigma_t$ (see Section~\ref{subsec:tender}).

The quantities defined so far can be shown to satisfy evolution and constraint equations, equivalent to the Einstein equations \eqref{eq:EFE}, which provide a formulation of the Cauchy problem of general relativity, as we shall now illustrate.
We first separate the Einstein equations \eqref{eq:EFE} into their normal and tangential components to each hypersurface $\Sigma_t$. The ``normal-normal'' projection (i.e., contraction with $n^\alpha n^\beta$) of the Einstein equations gives the Hamiltonian constraint
\begin{equation}
\label{eq:hamconstr}
^{(3)}R-K^{ij}K_{ij}+K^2-2\Lambda=16\pi\rho,
\end{equation}
where $^{(3)}R$ is the Ricci scalar associated with the covariant derivative $D$, $K:=\gamma^{ij}K_{ij}$ and $\rho:=T_{\alpha\beta}n^\alpha n^\beta$ is the matter energy density measured by an observer with 4-velocity $n^\alpha$.
The ``tangent-normal'' projection (i.e., contraction with $\gamma^{\alpha\beta} n^\gamma$) of the Einstein equations gives the momentum constraint
\begin{equation}
\label{eq:momconstr}
D_j {K^j}_i-D_i K=8\pi j_i\,,
\end{equation}
where $j^i:=-T_{\alpha\beta}n^\alpha\gamma^{\beta i}$ is the matter momentum density measured by an observer with 4-velocity $n^\alpha$.
\eqref{eq:hamconstr} and \eqref{eq:momconstr} only involve quantities in the tangent space of each $\Sigma_t$, so they can be regarded as constraint equations to be satisfied on each slice. It can be shown that the contracted Bianchi identity, $\nabla^\alpha \bigl(R_{\alpha\beta}-\frac{1}{2}Rg_{\alpha\beta}\bigr)=0$, ensures that a solution of the Einstein equations satisfies the constraints on all $\Sigma_t$, if it satisfies the constraints on $\Sigma_{t=0}$.

The ``tangential-tangential'' component of \eqref{eq:EFE} is given 
by
\begin{equation}
\label{eq:evoK}
\begin{split}
\partial_t K_{ij}&- N^k\partial_k K_{ij}+\left(\partial_i N^k\right) K_{kj}+\left(\partial_j N^k\right)K_{ik}=\\
&-D_iD_jN +N \left[ \,^{(3)}R_{ij}+K K_{ij}-2 K_{ik}K^k_j\right]+4\pi N \left[\gamma_{ij}(S-\rho)-2 S_{ij}\right],
\end{split}
\end{equation}
where $S_{ij}:=\gamma_i^\alpha \gamma_j^\beta T_{\alpha\beta}$ is the spatial energy-momentum tensor measured by an observer with 4-velocity $n^\alpha$, and $S:=\gamma^{ij}S_{ij}$ is its trace.
This expression involves the first order time derivative of $K_{ij}$, so it is an evolution equation for $K_{ij}$. Moreover, equation \eqref{eq:evogamma} provides an evolution equation for $\gamma_{ij}$. We see that the constraints do not involve $ N$ and $ N^i$ and there are no evolution equations for these two quantities, therefore lapse and shift can be chosen arbitrarily for all $t$. This choice accounts for the gauge freedom in the choice of coordinates $\{t,x^i\}$.
In the ADM formulation, given an initial spacelike slice $\Sigma_0$, we refer to initial data for the metric as a specification of $\gamma_{ij}$ and $K_{ij}$ on $\Sigma_0$. We refer to initial data on for the matter fields as a specification of $\rho$ and $j^i$ on $\Sigma_0$.

In summary, the ADM formulation of the Cauchy problem consists of finding a solution to the evolution equations \eqref{eq:evogamma} and \eqref{eq:evoK}, given a choice of initial data for the metric and the matter fields, satisfying the constraints \eqref{eq:hamconstr}, \eqref{eq:momconstr}. This formulation can be modified in several ways. For instance, an alternative formulation\footnote{This is actually the original ADM formulation.} is obtained by adding a term $-\frac{1}{2} N \gamma_{ij}\mathcal{H}$ to \eqref{eq:evoK}, where $\mathcal{H}:=\frac{1}{2}\left(\,^{(3)}R-K^{ij}K_{ij}+K^2-2\Lambda-16\pi\rho\right)$. Since $\mathcal{H}$ clearly vanishes for solutions of the Hamiltonian constraint, these two formulations provide the same physical solutions. However, it turns out that the second one is ill-posed and thus not suitable for numerical implementation: small changes in the initial data may result in large differences at later times, so small numerical error may grow without bounds during the evolution. The fact that a formulation is well-posed is not a guarantee that its implementation in a numerical scheme will provide stable evolution, i.e., simulations that last, in principle, for arbitrarily long times.
The ADM formulation is an example of a well-posed formulation that has weak stability properties\footnote{This is a consequence of the fact that the ADM system of PDEs is only weakly hyperbolic; see \cite{Alcubierre:1138167} for more details.}, which make it unsuitable for practical implementation. Nevertheless, there are several possible ways in which one can modify the structure of the ADM evolution equations in order to obtain a well-posed formulation of the Cauchy problem of general relativity with strong stability properties. In the next two sections, we present two such modifications of the ADM formulation that have been widely employed.

\section{BSSNOK formalism}
\label{sec:BSSNOK}

In this section, we review how the ADM evolution equations, presented in Section~\ref{sec:3p1splitt}, can be manipulated to give the well-posed \emph{BSSNOK (Baumgarte, Shapiro, Shibata, Nakamura, Oohara and Kojima) formulation} \cite{10.1143/PTPS.90.1,PhysRevD.52.5428,Baumgarte:1998te}, which is used in one form or another by most large three-dimensional codes in numerical relativity. For additional details, see \cite{Alcubierre:1138167}.

We start by defining a conformally rescaled spatial metric as
\begin{equation}
\label{eq:confmet}
\tilde{\gamma}_{ij}\colon =e^{-4\phi}\gamma_{ij}.
\end{equation}
The requirement that $\tilde{\gamma}_{ij}$ has unit determinant implies $\phi=\frac{1}{12}\ln\gamma$, where $\gamma$ is the determinant of $\gamma_{ij}$.
We then separate the extrinsic curvature into its trace $K$ and its tracefree part
\begin{equation}
\label{eq:Ktrfree}
A_{ij}:=K_{ij}-\frac{1}{3}\gamma_{ij}K,
\end{equation}
and define the conformal rescaling of $A_{ij}$ as
\begin{equation}
\label{eq:Atilde}
\tilde A_{ij}:=e^{-4\phi}A_{ij}.
\end{equation}
The BSSNOK method introduces three auxiliary variables, called \emph{conformal connection functions}, given by
\begin{equation}
\label{eq:Gammai}
\tilde\Gamma^i :=\tilde\gamma^{jk}\tilde\Gamma^i_{\phantom i jk}=-\partial_j \tilde\gamma^{ij},
\end{equation}
where $\tilde\Gamma^i_{\phantom i jk}$ are the Christoffel symbols of the conformal metric $\tilde\gamma_{ij}$.

We can now write the ADM evolution equations in terms of the newly defined quantities. We obtain, after a few manipulations,
\begin{eqnarray}
\label{eq:289}
\frac{\partial}{\partial t}\tilde{\gamma}_{ij}&=&-2 N \tilde{A}_{ij}+ N^k \partial_k\tilde{\gamma}_{ij}+\tilde{\gamma}_{ik}\partial_j N^k +\tilde{\gamma}_{jk}\partial_i N^k-\frac{2}{3}\tilde{\gamma}_{ij}\partial_k N^k,\\
\label{eq:2810}
\frac{\partial}{\partial t}\phi&=&-\frac{1}{6} N K+ N^k \partial_k \phi+\frac{1}{6}\partial_k N^k,
\end{eqnarray}
\begin{eqnarray}
\label{eq:2811}
\frac{\partial}{\partial t}\tilde{A}_{ij}&=&e^{-4\phi}\bigl\{-D_i D_j N+ N R_{ij}+4\pi  N[\gamma_{ij}(S-\rho)-2 S_{ij}]\bigr\}^{\text{TF}}\\\nonumber
&&\hspace{5mm}+ N(K\tilde{A}_{ij}-2\tilde{A}_{ik}\tilde{A}^k_j)+ N^k \partial_k\tilde{A}_{ij}+\tilde{A}_{ik}\partial_j N^k +\tilde{A}_{jk}\partial_i N^k-\frac{2}{3}\tilde{A}_{ij}\partial_k N^k,\\
\label{eq:2812}
\frac{\partial}{\partial t}K&=&-D_iD^i N + N\left(\tilde{A}_{ij}\tilde{A}^{ij}+\frac{1}{3}K^2\right)+4\pi N(\rho+S),
\end{eqnarray}
where TF denotes the tracefree part of the expression inside the brackets, and indices of conformal quantities are raised and lowered using $\tilde{\gamma}_{ij}$ (e.g. $\tilde{A}^{ij}=\tilde{\gamma}^{ik}\tilde{\gamma}^{jl}\tilde{A}_{kl}=e^{4\phi}\gamma^{ik}\gamma^{jl} A_{kl}=e^{4\phi} A^{ij}$).
The Ricci tensor appearing in \eqref{eq:2811} must be expressed in terms of the evolution variables. This can be done by writing $R_{ij}$ as $R_{ij}:=\tilde R_{ij}+R^\phi_{ij}$, where $\tilde R_{ij}$ is the Ricci tensor associated with the conformal metric $\tilde\gamma_{ij}$, i.e.,
\begin{equation}
\tilde{R}_{ij}=-\frac{1}{2}\tilde\gamma^{lm}\partial_l\partial_m\tilde\gamma_{ij}+\tilde\gamma_{k(i}\partial_{j)}\tilde\Gamma^k+\tilde\Gamma^k\tilde\Gamma_{(ij)k}+\tilde\gamma^{lm}\left(2\tilde\Gamma^k_{\phantom kl(i}\tilde\Gamma_{j)km}+\tilde\Gamma^k_{\phantom kim}\tilde\Gamma_{klj}\right),
\end{equation}
and $R^\phi_{ij}$ denotes additional terms that depend on $\phi$,
\begin{equation}
R^\phi_{ij}=-2\tilde D_i \tilde D_j\phi-2\tilde\gamma_{ij}\tilde D^k \tilde D_k\phi+4 \tilde D_i\phi \tilde D_j\phi-4\tilde\gamma_{ij}\tilde D^k\phi\tilde D_k\phi,
\end{equation}
with $\tilde D$ the Levi-Civita covariant derivative associated with $\tilde\gamma_{ij}$.

To conclude, we need to provide an evolution equation for $\tilde\Gamma^k$. Taking the $t$-derivative of \eqref{eq:Gammai} and using \eqref{eq:evogamma}, we get
\begin{equation}
\label{eq:evGamma1}
\begin{split}
\partial_t\tilde\Gamma^i=&\tilde\gamma^{jk}\partial_j\partial_k N^i+\frac{1}{3}\tilde\gamma^{ij}\partial_j\partial_k N^k+ N^j\partial_j\tilde\Gamma^i-\tilde\Gamma^j\partial_j N^i\\
&\hspace{5mm}+\frac{2}{3}\tilde\Gamma^i\partial_j N^j-2\left( N\partial_j\tilde A^{ij}+\tilde A^{ij}\partial_j N\right).
\end{split}
\end{equation}

Despite this system being well-posed, it has been noted that it turns out to be numerically unstable
in practice. In order to fix this issue, we write the momentum constraint \eqref{eq:momconstr} in terms of the evolution variables of this formulation,
\begin{equation}
\partial_j\tilde A^{ij}=-\tilde\Gamma^i_{jk}\tilde A^{jk}-6\tilde A^{ij}\partial_j \phi+\frac{2}{3}\tilde\gamma^{ij}\partial_j K+8\pi \tilde j^i
\end{equation}
with $\tilde j^i=e^{4\phi}j^i$, and we use this to eliminate the divergence of $\tilde A^{ij}$ appearing in \eqref{eq:evGamma1}. We obtain
\begin{equation}
\label{eq:evGamma}
\begin{split}
\partial_t\tilde \Gamma^i=&\tilde\gamma^{jk}\partial_j \partial_k  N^i+\frac{1}{3}\tilde\gamma^{ij}\partial_j\partial_k N^k+ N^j\partial_j\tilde\Gamma^i-\tilde\Gamma^j\partial_j N^i+\frac{2}{3}\tilde\Gamma^i\partial_j N^j\\
&\hspace{5mm}-2\tilde A^{ij}\partial_j N+2 N\left(\tilde\Gamma^i_{jk}\tilde A^{jk}+6\tilde A^{ij}\partial_j\phi-\frac{2}{3}\tilde\gamma^{ij}\partial_j K-8\pi\tilde j^i\right).
\end{split}
\end{equation}

The system composed of equations \eqref{eq:289}, \eqref{eq:2810}, \eqref{eq:2811}, \eqref{eq:2812}, \eqref{eq:evGamma} is well-posed and shows better stability properties in simulations of asymptotically flat spacetimes than the ADM formulation.

\section{Generalised harmonic formalism}
\label{sec:genharform}

The BSSNOK formulation, presented in Section~\ref{sec:BSSNOK}, has proven effective to evolve asymptotically flat spacetimes, in which the Cauchy problem of general relativity is an initial value problem: the goal is to solve the evolution equations for a solution that satisfies the chosen initial conditions on a spatial hypersurface. In this section, we review a different formulation that has shown similar stability properties in simulations of asymptotically flat spacetimes \cite{Pretorius:2004jg}.
This is the so-called \emph{generalised harmonic formulation}, originally presented by \cite{Foures-Bruhat:1952grw} and employed to prove the first crucial theorems about existence and uniqueness of solutions to the Einstein equations.
This idea was later applied to obtain simulations in asymptotically AdS spacetimes \cite{Bantilan:2012vu,Bantilan:2017kok,Bantilan:2020pay,Bantilan:2020xas}, in which the Cauchy problem of general relativity is an initial-boundary value problem: the goal is to solve the evolution equations for a solution that satisfies the chosen initial conditions on a spacelike hypersurface as well as boundary conditions at the timelike boundary of AdS.
In this section, we briefly review the main theoretical aspects of the generalised harmonic formulation.

By taking the trace of the Einstein equations \eqref{eq:EFE} with respect of $g_{\alpha\beta}$, we obtain
\begin{equation}
R=-8\pi T+4\Lambda,
\end{equation}
where $T:=g^{\alpha\beta}T_{\alpha\beta}$.
Plugging this in \eqref{eq:EFE}, we obtain the Einstein equations in trace-reversed form:
\begin{equation}
\label{eq:EFEtrrev}
R_{\alpha\beta}=\Lambda g_{\alpha\beta}+8\pi\left(T_{\alpha\beta}-\frac{1}{2}T g_{\alpha\beta}\right).
\end{equation}

Let us now define \emph{generalised harmonic coordinates}, $x^\alpha$, as coordinates satisfying the scalar wave equation 
\begin{equation}
\label{eq:GHdef}
\Box x^\alpha=H^\alpha,
\end{equation}
where $H^\alpha$ are arbitrary source functions. Given any solution $g_{\alpha\beta}$ of the Einstein equations in any coordinate system, we can easily obtain the corresponding source functions using \eqref{eq:GHdef}:
\begin{equation}
\label{eq:soufunmet}
H^\alpha:=\Box x^\alpha=\frac{1}{\sqrt{-g}}\partial_\gamma \left(\sqrt{-g}g^{\gamma\delta}x^\alpha_{\phantom \alpha,\delta}\right)=\frac{1}{\sqrt{-g}}\partial_\gamma  \left(\sqrt{-g}g^{\gamma\alpha}\right)=-g^{\gamma\delta}\Gamma^\alpha_{\phantom\alpha\gamma\delta},
\end{equation}
where $\Gamma^\alpha_{\;\;\gamma\delta}$ are the Christoffel symbols.
However, in any problem of interest, the goal is to solve the Einstein equations in some set of coordinates, so we do not know $g_{\alpha\beta}$ a-priori. Consequently, we do not know a consistent choice of source functions a-priori. The generalised harmonic formalism provides a way around this issue and a well-posed system of PDEs for the unknowns $g_{\alpha\beta}$ equivalent to the Einstein equations. We proceed as follows. 
(i) We choose a coordinate system $x^\alpha$.
(ii) We use \eqref{eq:soufunmet} and its derivatives
to write the trace-reversed Einstein equations \eqref{eq:EFEtrrev} in the form
\begin{equation}
\label{eq:EFEsoufun0}
\begin{split}
-\frac{1}{2}g^{\gamma\delta}g_{\alpha\beta,\gamma\delta}-g^{\gamma\delta}_{\phantom\gamma\phantom\delta,(\alpha}g_{\beta)\gamma,\delta}-H_{(\alpha,\beta)}+&H_\gamma\Gamma^\gamma_{\phantom\gamma\alpha\beta}-\Gamma^\gamma_{\phantom\gamma\delta\alpha}\Gamma^\delta_{\phantom\delta\gamma\beta}\\
&=\Lambda g_{\alpha\beta}+8\pi\left(T_{\alpha\beta}-\frac{1}{2}T g_{\alpha\beta}\right).
\end{split}
\end{equation}
(iii) We then promote the source functions to \emph{independent quantities}, determined by four independent equations that we write schematically as 
\begin{equation}
\label{eq:eqH}
\mathcal{L}_\alpha [H_\alpha]=0 \;\;\;\;\text{(no index summation)}.
\end{equation} 
The Einstein equations are thus equivalent to the system of equations \eqref{eq:EFEsoufun0} and \eqref{eq:eqH}, provided that the \emph{generalised harmonic constraints},
\begin{equation}
\label{eq:GHcon}
C^\alpha:= H^\alpha -\Box x^\alpha =0,
\end{equation}
are satisfied for all values of $t\equiv x^0$. 
In this way we just shifted the problem to finding equations \eqref{eq:eqH} such that no coordinate singularities are formed in the resulting evolution of the metric components. This can be done in different ways depending on the type of problem we are investigating. We will study the case of asymptotically AdS spacetimes in Chapter~\ref{Chapter:NoSym}.
\noindent
(iv) We prove that solutions of \eqref{eq:EFEsoufun0} and \eqref{eq:eqH}, for which $C^\alpha|_{t=0}=\partial_tC^\alpha|_{t=0}=0$, satisfy $C^\alpha=0$ at all times. 
\emph{Sketch of proof.} We notice that equation \eqref{eq:EFEsoufun0} is equivalent to $R_{\alpha\beta}-\nabla_{(\alpha}C_{\beta)}-\Lambda g_{\alpha\beta}-8\pi\left(T_{\alpha\beta}-\frac{1}{2}T g_{\alpha\beta}\right)=0$. From this equation and the contracted Bianchi identity, $\nabla^\alpha\bigl(R_{\alpha\beta}-\frac{1}{2}Rg_{\alpha\beta}\bigr)=0$, we obtain the hyperbolic equation 
\begin{equation}
\label{eq:dalC}
\Box C_\beta=-C^\alpha\nabla_{(\alpha}C_{\beta)}-C^\alpha\left[\Lambda g_{\alpha\beta}+8\pi\left(T_{\alpha\beta}-\frac{1}{2}T g_{\alpha\beta}\right)\right].
\end{equation}
$C^\alpha=0$ for all $t$ is a trivial solution of this equation with initial data $C^\alpha|_{t=0}=\partial_tC^\alpha|_{t=0}=0$. The theory of PDEs tells us that the solution is unique, which concludes the proof. \\
(v) We notice that if the ADM Hamiltonian and momentum constraints at $t=0$, \eqref{eq:hamconstr},\eqref{eq:momconstr} are satisfied,  then also $\partial_tC^\alpha|_{t=0}=0$.

In summary, we have shown that, instead of solving the Einstein equations, we can provide initial data $g_{\alpha\beta}|_{t=0},\partial_tg_{\alpha\beta}|_{t=0},H_\alpha|_{t=0}$ satisfying the ADM constraints \eqref{eq:momconstr},\eqref{eq:hamconstr} and $C^\alpha|_{t=0}=0$, and evolve this via the well-posed system of evolution equations \eqref{eq:EFEsoufun0}, \eqref{eq:eqH}. Notice that $C^\alpha|_{t=0}=0$ is trivially satisfied by choosing $H^\alpha|_{t=0}$ as given by \eqref{eq:soufunmet} in terms of $g_{\alpha\beta}|_{t=0}$ and $\partial_tg_{\alpha\beta}|_{t=0}$.
 
However, in a numerical scheme, $C^\alpha$ will vanish at each value of $t$ only up to a numerical error and, during evolution, non-zero solutions of \eqref{eq:dalC} could grow exponentially (so the result at (iv) would no longer be valid). If the generalised harmonic constraints are not satisfied, we cannot be sure that a solution of \eqref{eq:EFEsoufun0} is also a solution of the Einstein equations. In order to drive the solution towards small values of $C^\alpha$, we add the \emph{constraint-damping terms} suggested in \cite{Gundlach:2005eh} to our evolution equations \eqref{eq:EFEsoufun0}, obtaining the \emph{modified Einstein equations (MEE)}
\begin{equation}
\label{eq:EFEsoufun}
\begin{split}
&-\frac{1}{2}g^{\gamma\delta}g_{\alpha\beta,\gamma\delta}-g^{\gamma\delta}_{\phantom\gamma\phantom\delta,(\alpha}g_{\beta)\gamma,\delta}-H_{(\alpha,\beta)}+H_\gamma\Gamma^\gamma_{\phantom\gamma\alpha\beta}-\Gamma^\gamma_{\phantom\gamma\delta\alpha}\Gamma^\delta_{\phantom\delta\gamma\beta}\\
&-\kappa\left(2n_{(\alpha}C_{\beta)}-(1+P)n^\gamma C_\gamma\right)\\
&=\Lambda g_{\alpha\beta}+8\pi\left(T_{\alpha\beta}-\frac{1}{2}T g_{\alpha\beta}\right),
\end{split}
\end{equation}
where $n_\alpha$ is the unit normal to hypersurfaces of constant $t$.
This is the evolution equation that we solve numerically in the scheme described in Chapter~\ref{Chapter:NoSym}.
In our simulations, the constants $\kappa,P$ appearing in the constraint-damping terms (the second line of \eqref{eq:EFEsoufun}) take the values $\kappa=-10$ and $P=-1$.
Notice that the highest derivative terms, $-\frac{1}{2} g^{\gamma \delta} \partial_\gamma \partial_\delta g_{\alpha\beta}$, consist of a wave operator acting on metric components. Thus, the well-posedness of the wave equation suggests that the Cauchy problem in generalized harmonic form is well-posed, if we make reasonable assumptions on the remaining components of the problem. 

In the following we will be interested in the case in which gravity is coupled to a massless real scalar field, therefore we employ the energy-momentum tensor \eqref{eq:KHmomtenscov} with $\mu=0$, which we rewrite here for completeness in terms of partial derivatives:
\begin{equation}
\label{eq:KHmomtenspar}
T_{\alpha\beta}=\partial_\alpha \varphi \partial_\beta \varphi - g_{\alpha\beta} \frac{1}{2} g^{\gamma\delta} \partial_{\gamma} \varphi \partial_{\delta} \varphi\,.
\end{equation}
The modified Einstein equations \eqref{eq:EFEsoufun} must thus be solved together with the massless Klein-Gordon equation \eqref{eq:m0KGeq}.
In terms of partial derivatives, this reads
\begin{equation}\label{eqn:eoms2cart}
g^{\alpha\beta} \partial_{\alpha} \partial_{\beta} \varphi -g^{\alpha\beta} \Gamma^\gamma_{\phantom\gamma\beta\alpha}\partial_\gamma\varphi= 0\,.
\end{equation}

\section{Finite difference methods}
\label{sec:findiffmet}

In numerical relativity, we wish to use computers to solve partial differential equations for fields defined over a continuous domain. 
For instance, in the generalised harmonic formalism, we aim to solve the evolution equations, \eqref{eq:EFEsoufun} and \eqref{eqn:eoms2cart}, and the constraint equations, \eqref{eq:hamconstr} and \eqref{eq:momconstr}.
A system of PDEs can be written in the schematic form
\begin{equation}
\label{eq:genpde}
\mathcal{L} u=0,
\end{equation}
where $u$ is the set of functions of the spacetime coordinates $(t,x^i)$ that we wish to solve for, and $\mathcal{L}$ is a differential operator. For equations \eqref{eq:EFEsoufun} and \eqref{eqn:eoms2cart}, $\mathcal{L}$ contains up to second derivatives with respect to all spacetime coordinates, while for the constraint equations \eqref{eq:hamconstr} and \eqref{eq:momconstr}, $\mathcal{L}$ contains up to second derivatives with respect to spatial coordinates.
Each of the functions in $u$ has an infinite number of degrees of freedom, one for each point in the continuous domain. Since computers can only work with a finite amount of data, it is necessary to approximate the problem with one involving only a finite number of degrees of freedom. Finite differences methods, whose main aspects are described in this section, provide a common way to perform such approximations. 
In the following, we will discuss ideas widely employed in many modern codes, and we will illustrate them with concrete examples that are relevant for our scheme.
Additional technical details that are specific to our scheme and our simulations, as well as other topics relevant for numerical evolution in asymptotically AdS spacetimes, can be found in Chapter~\ref{Chapter:NoSym} and the appendices.

\subsection{Discretizing time and space}
\label{sec:disc}

The first necessary step consists in defining a computational domain with a finite number of points.
To this end, we start by discretizing the chosen time direction, identified by the chosen time coordinate $t$, into a finite number of time levels, separated for simplicity by a constant time interval $\Delta t$. The time coordinate of each level is given by $t_n=n\Delta t$ for some integer $n$ in the range $0\leq n\leq N_t-1$, with $N_t:=t_f/\Delta t+1$ where $t_f$ is the final time at which we wish to obtain the solution.

\begin{figure*}[t!]
        \centering
        \includegraphics[width=4.0in,clip=true]{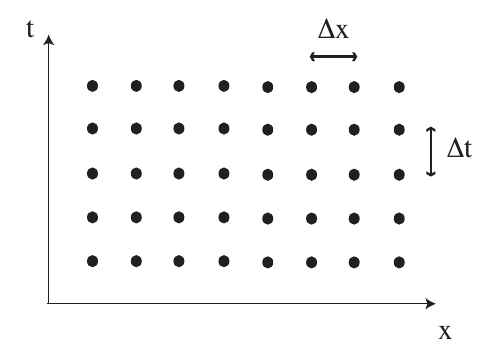}
\parbox{5.0in}{\caption{Uniform discretisation of time and space in a 1+1 case. Image from \cite{Alcubierre:1138167}.
        }\label{fig:discr}}
\end{figure*}

Each slice at fixed $t=t_n$, $\Sigma_n$, is typically an infinite hypersurface, and the boundary conditions are conditions on the behaviour of the solution at infinity. These conditions can be easily implemented if we include infinity in the computational domain. This can be done by employing coordinates $\{x,y,z\}$ that take finite values at infinity. In other words, we use coordinate that take values in a finite range: $x\in[x_{min},x_{max}],y\in[y_{min},y_{max}],z\in[z_{min},z_{max}]$.
Then, we discretize each hypersurface $\Sigma_n$ into a spatial grid (or mesh), whose points are, for simplicity, equally spaced along each spatial direction with uniform spacing $\Delta:=\Delta x=\Delta y=\Delta z$. The grid points have spatial coordinates $x_i=i\Delta+x_{min}$, $y_j=j\Delta+y_{min}$, $z_k=k\Delta+z_{min}$ for integers $i,j,k$ in the range $0\leq i\leq N_x-1$, $0\leq j\leq N_y-1$, $0\leq k\leq N_z-1$, where $N_{x}:=(x_{max}-x_{min})/\Delta+1$ and $N_y$ and $N_z$ are given by similar expressions. Therefore, each grid point can also be identified by the three integers $(i,j,k)$.
The resulting discretized spacetime, represented in Figure~\ref{fig:discr} in the case of one time and one spatial directions, has a finite number of points, identified by $(t_n,x_i,y_j,z_k)$ or, equivalently, by the integers $(n,i,j,k)$.

The goal is now to find a good approximation for the solution $u$ at the points of this discretized computational domain. We denote this approximation by $u_\Delta$ and its value at point $(n,i,j,k)$ by $(u_\Delta)^n_{ijk}:= u_\Delta(t=t_n,x=x_i,y=y_j,z=z_k)$.

\subsection{Finite difference stencils}
\label{sec:diffstenc}

After discretizing time and space, it is necessary to approximate the derivatives appearing in $\mathcal{L}$ in terms of differences between function values at nearby points of the discretized domain.

Let us now explain how these approximations, called finite difference stencils, can be obtained by discussing a few examples that are relevant for the numerical scheme presented in Chapter~~\ref{Chapter:NoSym}. We discuss this for derivatives with respect to $x$. Stencils for other spatial derivatives and for time derivatives can be obtained in similar ways. 
We start by finding a finite difference approximation for $u_{,x}:=\partial_x u$ at $t=t_n,x=x_i,y=y_j,z=z_k$.
The $x$-dependence of $u(t_n,x,y_j,z_k)$ near the point $(n,i,j,k)$ is given by the Taylor expansion
\begin{equation}
\label{eq:tayexp}
u(t_n,x,y_j,z_k)=u|_{(n,i,j,k)}+(u_{,x})|_{(n,i,j,k)}(x-x_i)+\frac{1}{2}(u_{,xx})|_{(n,i,j,k)}(x-x_i)^2+\mathcal{O}((x-x_i)^3).
\end{equation}
If we evaluate this expression at $(n,i+1,j,k)$ and we approximate $u$ with $u_\Delta$, we get the approximated expression
\begin{equation}
\label{eq:tayexpnip1jk}
(u_\Delta)^{n}_{i+1jk}=(u_\Delta)^n_{ijk}+(u_{,x})|_{(n,i,j,k)}\Delta x+\frac{1}{2}(u_{,xx})|_{(n,i,j,k)}\Delta x^2+\mathcal{O}(\Delta x^3).
\end{equation}
From this, we can easily read off an approximation for $u_{,x}$ at $(t_n,x_i,y_j,z_k)$ that can be computed from the values $(u_\Delta)^{n}_{ijk}$ and $(u_\Delta)^{n}_{i+1jk}$:
\begin{equation}
\label{eq:deruxord1for}
[(u_{,x})_\Delta]^n_{ijk}=\frac{(u_\Delta)^{n}_{i+1jk}-(u_\Delta)^n_{ijk}}{\Delta x}.
\end{equation}
From \eqref{eq:tayexpnip1jk}, we also see that the error in this approximation is $\mathcal{O}(\Delta x)$.

It is possible to obtain higher order approximations in a similar way. For instance, if we want to obtain a second order finite difference stencil, we evaluate \eqref{eq:tayexp} at $(n,i-1,j,k)$:
\begin{equation}
\label{eq:tayexpnim1jk}
(u_\Delta)^{n}_{i-1jk}=(u_\Delta)^n_{ijk}-(u_{,x})|_{(n,i,j,k)}\Delta x+\frac{1}{2}(u_{,xx})|_{(n,i,j,k)}\Delta x^2+\mathcal{O}(\Delta x^3).
\end{equation}
From \eqref{eq:tayexpnip1jk} and \eqref{eq:tayexpnim1jk} we clearly see that
\begin{equation}
\label{eq:deruxord2}
[(u_{,x})_\Delta]^n_{ijk}=\frac{(u_\Delta)^{n}_{i+1jk}-(u_\Delta)^{n}_{i-1jk}}{2\Delta x}
\end{equation}
is an approximation for $u_{,x}$ with $\mathcal{O}(\Delta x^2)$ error.
However, unlike in eq. \eqref{eq:deruxord1for}, to compute this value we need to know the value of $u_\Delta$ at $(n,i-1,j,k)$, which is to the ``left'' of the point $(n,i,j,k)$ at which we wish to know $(u_{,x})_\Delta$. 
For various reasons (see section~\ref{sec:AHfind} for one of these), this value might not be available in a simulation and we might prefer to use stencils involving only points to the ``right'' of $(n,i,j,k)$, i.e., points with $x\geq x_i$, called \emph{forward stencils}.  
\eqref{eq:deruxord1for} is an example of a first order forward stencil for $u_{,x}$.
Using the arguments presented above, it is not hard to find a forward stencil for $u_{,x}$ at $(n,i,j,k)$ with $\mathcal{O}(\Delta x^2)$ error:
 \begin{equation}
 \label{eq:deruxord2for}
[(u_{,x})_\Delta]^n_{ijk}=\frac{-3(u_\Delta)^{n}_{ijk}+4(u_\Delta)^{n}_{i+1jk}-(u_\Delta)^{n}_{i+2jk}}{2\Delta x}
\end{equation}
We can notice the following trend: to increase the accuracy of the stencil at a given point, we need to use the values of $u$ at further points.
In simulations, it might also be necessary to use stencils involving only points with $x\leq x_i$, called \emph{backward stencils}.
The first order backward stencil for $u_{,x}$ at $(n,i,j,k)$ is
\begin{equation}
\label{eq:deruxord1}
[(u_{,x})_\Delta]^n_{ijk}=\frac{(u_\Delta)^{n}_{ijk}-(u_\Delta)^n_{i-1jk}}{\Delta x}.
\end{equation}
A second order backward stencil for $u_{,x}$ at $(n,i,j,k)$ is
 \begin{equation}
 \label{eq:deruxord2back}
[(u_{,x})_\Delta]^n_{ijk}=\frac{3(u_\Delta)^{n}_{ijk}-4(u_\Delta)^{n}_{i-1jk}+(u_\Delta)^{n}_{i-2jk}}{2\Delta x}.
\end{equation}
Stencils that are neither forward nor backward, such as \eqref{eq:deruxord2}, are called \emph{centred} stencils.

So far, we have obtained approximations for first derivatives. We now consider a few examples of stencils for second derivatives.
From \eqref{eq:tayexpnip1jk} and \eqref{eq:tayexpnim1jk}, we can also obtain a second-order centred stencil for $u_{,xx}$ at $(n,i,j,k)$:
\begin{equation}
\label{eq:derderuxord2}
[(u_{,xx})_\Delta]^n_{ijk}=\frac{(u_\Delta)^{n}_{i+1jk}-2(u_\Delta)^{n}_{ijk}+(u_\Delta)^{n}_{i-1jk}}{(\Delta x)^2}.
\end{equation}
A forward stencil for $u_{,xx}$ at $(n,i,j,k)$ with $\mathcal{O}(\Delta x^2)$ error is
\begin{equation}
[(u_{,xx})_\Delta]^n_{ijk}=\frac{2(u_\Delta)^{n}_{ijk}-5(u_\Delta)^{n}_{i+1jk}+4(u_\Delta)^{n}_{i+2jk}-(u_\Delta)^{n}_{i+3jk}}{(\Delta x)^2}.
\end{equation}
A backward stencil for $u_{,xx}$ at $(n,i,j,k)$ with $\mathcal{O}(\Delta x^2)$ error is
\begin{equation}
[(u_{,xx})_\Delta]^n_{ijk}=\frac{2(u_\Delta)^{n}_{ijk}-5(u_\Delta)^{n}_{i-1jk}+4(u_\Delta)^{n}_{i-2jk}-(u_\Delta)^{n}_{i-3jk}}{(\Delta x)^2}.
\end{equation}

In the numerical scheme of Chapter~\ref{Chapter:NoSym} we approximate all derivatives using the second order stencils obtained in this section. In regions where it is necessary, we use forward or backward stencils.

\subsection{Newton-Gauss-Seidel time integration}
\label{sec:NGS}

Using the approximations discussed in sections \ref{sec:disc} and \ref{sec:diffstenc}, the system of evolution equations can be written, in the notation of \eqref{eq:genpde}, as an equation for $u_\Delta$ at each $(n,i,j,k)$ involving only finite difference stencils:
\begin{equation}
\label{eq:disceq}
(\mathcal{L}_\Delta u_\Delta)^n_{ijk}=0,
\end{equation}
where $\mathcal{L}_\Delta$ is obtained from the differential operator $\mathcal{L}$ by substituting the partial derivatives with the corresponding stencils. Similarly, conditions at the boundary of the numerical domain can also be written as finite differences involving boundary points and, possibly, interior points close to the boundary.
The goal is to find the solution $u_\Delta$ on all grid points at time level $t_{n+1}$, provided that we know $u_\Delta$ at previous time levels. In other words, at time $t_n$, we wish to perform a step of time integration to reach $t_{n+1}$.
Notice that solving \eqref{eq:disceq} exactly, i.e., inverting \eqref{eq:disceq} to obtain an explicit expression for $(u_\Delta)^{n+1}$ in terms of the solution $u_\Delta$ at previous times at each grid point, would be prohibitively expensive given the number of unknowns on a typical grid in a physical problem, therefore we look for an approximated solution.
In our scheme, time integration is carried out through a \emph{Newton-Gauss-Seidel (NGS) relaxation algorithm}, as described in this section (see also \cite{Pretorius2002} and \cite{Pretorius:2004jg}).

At each grid point $(i,j,k)$, the use of second order accurate stencils for time derivatives requires the knowledge of $(u_\Delta)^n_{ijk}$ and $(u_\Delta)^{n-1}_{ijk}$ to obtain $(u_\Delta)^{n+1}_{ijk}$ from the equations of motion.
Once the solution values at time levels $n,n-1$ are known (at the initial time level $n=0$, these values are given by the choice of initial data), the Newton-Gauss-Seidel algorithm determines $(u_\Delta)^{n+1}_{ijk}$ at each $(i,j,k)$ through the following iterative procedure.
\begin{enumerate}
\item $(u_\Delta)^{n+1}_{ijk}$ is initially set equal to $(u_\Delta)^n_{ijk}$.
\item Using $u_\Delta$ at times $t_{n+1},t_n,t_{n-1}$, we compute the left hand side of \eqref{eq:disceq} at each grid point. This quantity is called the \emph{residual} of \eqref{eq:disceq} and we denote it by $\mathcal{R}^n_{ijk}$. We also compute the diagonal elements of the Jacobian of \eqref{eq:disceq}:
\begin{equation}
\mathcal{J}^n_{ijk}=\frac{\partial[(\mathcal{L}_\Delta u_\Delta)^n_{ijk}]}{\partial [(u_\Delta)^{n+1}_{ijk}]}.
\end{equation}
\item At each internal grid point, we update $(u_\Delta)^{n+1}_{ijk}$ via
\begin{equation}
\label{eq:NGSupdate}
(u_\Delta)^{n+1}_{ijk}\to(u_\Delta)^{n+1}_{ijk}-\frac{\mathcal{R}^n_{ijk}}{\mathcal{J}^n_{ijk}}.
\end{equation}
Then, at boundary points, we set $(u_\Delta)^{n+1}_{ijk}$ equal to the value imposed by the boundary conditions.
This step is referred to as a \emph{relaxation sweep}.
\item We iterate 2. and 3. until the $L^2$-norm of the residual over the entire grid, $||\mathcal{R}^n||_2$, is below a user-specified tolerance.
\end{enumerate}
Notice that, in step 3. we are essentially solving a linearized version of \eqref{eq:disceq} for $(u_\Delta)^{n+1}_{ijk}$ at grid point $(i,j,k)$, assuming that all the other unknowns, i.e., the other values of $(u_\Delta)^{n+1}$ at the other grid points, are given.
Therefore, if the initial guess at step 1 is not too far from the solution, we expect the residual $ \mathcal{R}^n_{ijk}$ to decrease after each iteration and, thus, to get progressively closer to the exact numerical solution of \eqref{eq:disceq}. Moreover, this convergence turns out to be rather fast for the evolution equations that we are interested in\footnote{We refer here to the class of hyperbolic equations, which can be thought of as equations with solutions whose features propagate at finite speed.}, i.e., it takes only a few iterations to bring the residual below typical low tolerance values.

\begin{figure*}[t!]
        \centering
        \includegraphics[width=4.5in,clip=true]{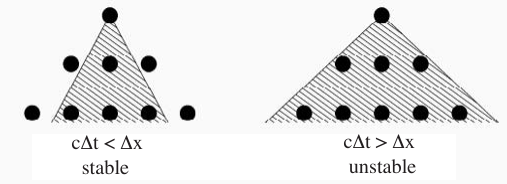}
\parbox{5.0in}{\caption{CFL condition for stability in a 1+1 case. The set of points employed to determine the solution at $(n+1,i)$ cannot be entirely contained in the interior of the causal past of $(n+1,i)$. Image from \cite{Alcubierre:1138167}.
        }\label{fig:stabcone}}
\end{figure*}

An evolution scheme in which time integration provides a finite approximated solution $u_\Delta^{n+1}$ for any $n$ in the range $0\leq n\leq N_t-1$ is said to be \emph{stable}.
Consider a point $(n+1,i,j,k)$ and its causal past. 
The two shaded regions of Figure~\ref{fig:stabcone} represent two different examples of causal past of a point in 1+1 dimensions.
Numerical stability cannot be achieved if, for some point $(n+1,i,j,k)$, the points involved in the calculation of $(u_\Delta)^{n+1}_{ijk}$ are in the interior of the causal past of $(n+1,i,j,k)$, as in the second image of Figure~\ref{fig:stabcone}. This can be understood from the fact that, if this scenario occurs, then the numerical calculation is not taking into consideration all the physical information in the causal past of $(n+1,i,j,k)$, which is what is needed to determine the exact solution $u$ at $(n+1,i,j,k)$. 
The requirement that this issue never occurs can be translated into a coordinate-dependent and scheme-dependent inequality, due to Courant, Friedrichs and Lewy (CFL), that restricts the ratio $\lambda\equiv\Delta t/\Delta$ to values smaller than some $\mathcal{O}(1)$ number, $\lambda_{\text{max}}$. 
In particular, this implies that the size of the time interval $\Delta t$ cannot be arbitrarily large, and it must get smaller as we reduce the grid spacing $\Delta$.
We did not perform a thorough stability analysis of our evolution scheme. However, the CFL condition is expected to be satisfied by setting $\lambda$ well below 1, and the fact that our simulations last for long evolution times is good evidence that our scheme is stable.

\subsection{Convergence}
\label{sec:conv}

Any reliable scheme should output a solution $u_\Delta$ of \eqref{eq:disceq} that approaches the exact solution $u$ of \eqref{eq:genpde} in the continuum limit, i.e., $\Delta \to 0$\footnote{Notice that in the continuum limit we also have $\Delta t\to 0$, since $\Delta t$ is proportional to $\Delta$ with proportionality constant given by the CFL factor $\lambda$.}.
In this section we introduce a few concepts that are useful to analyse this crucial property, called \emph{convergence}.

Let us start by defining two types of numerical error.
The \emph{truncation error} is the error introduced by approximating the differential operator $\mathcal{L}$ with its finite difference version $\mathcal{L}_\Delta$. This is defined, over the numerical domain, by
\begin{equation}
\label{eq:truerr}
\tau_\Delta:=\mathcal{L}_\Delta u-\mathcal{L}u=\mathcal{L}_\Delta u, 
\end{equation}
where in the last equality we used the fact that $\mathcal{L}u=0$. In a reliable numerical scheme, we clearly want the truncation error to go to zero as $\Delta$ goes to zero at any point. This property is called \emph{consistency}.
Let us also define the \emph{solution error} over the numerical domain as the difference between the exact solution of \eqref{eq:genpde} and the numerical solution:
\begin{equation}
\label{eq:solerr}
\epsilon_\Delta:=u-u_\Delta.
\end{equation}
Convergence is the requirement that $\epsilon_\Delta\to 0$ as $\Delta\to 0$ at any point.

Notice that consistency is a necessary but not sufficient condition for convergence. This can be understood from the fact that $\Delta t$ becomes infinitesimal as we approach the continuum limit, therefore we need to perform time integration over an infinite number of time steps to obtain the numerical solution $u_\Delta$ at some finite time. 
Even if the truncation error approaches zero in the continuum limit, the infinite sum of infinitesimal truncation error contributions may give a finite solution error, and thus violate convergence.
Nevertheless, consistency is crucial for the \emph{Lax equivalence theorem}, which provides a necessary and sufficient condition for convergence in terms of stability: given a well-posed formulation of an initial value problem and a consistent finite difference scheme, the scheme is convergent if and only if it is stable. This result was generalised to initial-boundary value problems in \cite{Serna1985}.

In addition to the theoretical tool provided by the equivalence theorem, we now present some numerical convergence tests that employ the numerical output of simulations.
We will make use of the observation, due to Richardson, that the solution error of a stable finite difference scheme can be expanded in powers of $\Delta$, so that the numerical solution can be written as the so-called \emph{Richardson expansion} \cite{doi:10.1098/rsta.1911.0009},
\begin{equation}
\label{eq:Richu}
u_\Delta=u+e_{1}\Delta+e_{2}\Delta^2+\mathcal{O}(\Delta^3), 
\end{equation}
at each point $(n,i,j,k)$, where the error coefficients $e_{1},e_{2},\dots$ depend only on the grid point $(n,i,j,k)$ under consideration but they do not depend on the grid spacing $\Delta$.
The accuracy order of the finite difference stencils is reflected on the accuracy order of the numerical solution: a scheme employing first-order finite difference stencils provides a first-oder accurate approximated solution $u_\Delta$, i.e., $e_{1}\neq 0$; a scheme employing at least second-order accurate finite difference stencils provides a second-oder accurate approximated solution $u_\Delta$, i.e., $e_{1}=0$ and $e_{2}\neq 0$; etc.
We can apply the Richardson expansion also to $\mathcal{L}_\Delta u_\Delta$. Recalling that the exact term vanishes because of \eqref{eq:genpde}, we are left with the expansion of the error term $\gamma_\Delta:=\mathcal{L} u-\mathcal{L}_\Delta u_\Delta$ in powers of $\Delta$:
\begin{equation}
\label{eq:RichL}
\mathcal{L}_\Delta u_\Delta=c_{1}\Delta+c_{2}\Delta^2+\mathcal{O}(\Delta^3).
\end{equation}

The first test aims to show that the numerical solution, $u_\Delta$, of an $m^{th}$-order finite difference scheme converges to some function (not necessarily the solution of $\mathcal{L}u=0$) in the continuum limit, i.e., that \eqref{eq:Richu} holds with some exact function $u$ and leading order error term given by $e_m\Delta^m$.
Let us suppose that we have obtained the numerical evolution of the same set of initial data on three different grids with spacing increasing by a factor $p>1$: $\Delta_1=h, \Delta_2= p h, \Delta_3=p^2 h$. From this, we can calculate the convergence factor
\begin{equation}
\label{eq:confact1}
Q^n_{ijk}=\frac{1}{\ln p}\ln\biggl(\frac{u_{p^2 h}-u_{p h}}{u_{p h}-u_{h}}\biggr)\biggr|_{(n,i,j,k)},
\end{equation}
at the common points $(n,i,j,k)$ of the three numerical domains. Let us denote the $L^2$-norm of $Q^n_{ijk}$, computed over the common grid points at each time level common to all three simulations of the triplet, by $||Q^n||_2$.
It is easy to see that $Q^n_{ijk}$, and thus also $||Q^n||_2$, approaches the order of convergence, $m$, in the limit $h\to 0$.
This result is used, in practice, to state that the scheme has good convergence properties if the time series $||Q^n||_2$ approaches the constant $m$ as more refined resolution triplets are considered.

The second test aims to show that the solution of an $m^{th}$-order finite difference scheme converges to a solution of the PDE \eqref{eq:genpde}, i.e., that \eqref{eq:RichL} holds with leading order error term given by $c_m\Delta^m$. For this test, we only need a pair of different resolutions: $\Delta_1=h, \Delta_2= p h$. Using these, we compute the corresponding approximations of left-hand side of \eqref{eq:genpde}, $\mathcal{L}_{ph} u_{p h}$ and $\mathcal{L}_{h} u_{h}$, which we use to calculate the convergence factor
\begin{equation}
\label{eq:confact2}
(Q_{\text{PDE}})^n_{ijk}=\frac{1}{\ln p}\ln\biggl(\frac{\mathcal{L}_{ph} u_{p h}}{\mathcal{L}_h u_{h}}\biggr)\biggr|_{(n,i,j,k)}
\end{equation}
at the common points $(n,i,j,k)$ of the two discretized domains. 
Similarly to the convergence factor \eqref{eq:confact1}, 
$(Q_{\text{PDE}})^n_{ijk}$, and thus also $||(Q_{\text{PDE}})^n||_2$, approaches the order of convergence, $m$, in the limit $h\to 0$.
This result is used, in practice, to state that the scheme has good convergence properties if the time series $||(Q_{\text{PDE}})^n||_2$ approaches $m$ as more refined resolution pairs are considered.

\subsection{Dissipation}
\label{sec:diss}

A finite difference scheme can only resolve modes that change substantially only over length scales much larger than the grid spacing $\Delta$. On the contrary, high frequency modes, whose wavelength is comparable with $\Delta$, will be unresolved. This high frequency noise in the numerical solution can grow in time and lead to numerical instabilities. Thus, it is convenient - and often necessary - to dissipate the noise, which is typically done by applying a low-pass filter, i.e., an operator that leaves low frequency modes unchanged and damps high frequency ones. This technique, called \emph{artificial dissipation}, is discussed in this section with particular focus on the implementation in our scheme. Our presentation follows that of \cite{Pretorius2002}.

To illustrate the problem in a simple way, let us consider a problem in one time dimension, $t$, and one spatial dimension, $x$. The numerical solution at $(n,i)$ is a a superposition of Fourier components $(\tilde{u}_\Delta)^n(k) e^{ikx_i}$ with amplitude $(\tilde{u}_\Delta)^n(k)$, depending on the  wavenumber $k$, and wavelength $\lambda=|2\pi/k|$ in the $x$ direction. An oscillation along $x$ can be detected on a grid of spacing $\Delta$ only if the wavelength is larger than the \emph{Nyquist wavelength} $\lambda_{\text{Nyq}}=2\Delta$.
Therefore, $k$ can take values in the range $[-k_{\text{Nyq}},k_{\text{Nyq}}]$, where $k_{\text{Nyq}}=\pi/\Delta$ is the maximum frequency that can be represented on a grid with spacing $\Delta$. The role of dissipation is to damp the Fourier components with $|k|$ close to $k_{\text{Nyq}}$, which are badly resolved, while leaving essentially unchanged the components with smaller $|k|$.

Dissipation is typically performed, in our scheme and many others, through the widely employed \emph{Kreiss-Oliger method}. We discuss this here for a second-order finite difference scheme, which is the relevant case for the scheme discussed in Chapter~\ref{Chapter:NoSym} (see \cite{Alcubierre:1138167} for the general case of an $m$-th order scheme).
We define the Kreiss-Oliger filter in the $x$ direction, $D_x$, as the finite difference operator whose action on $u_\Delta$ reads
\begin{equation}
\label{eq:KOterm1}
(D_x u_\Delta)^n_{i}=\frac{\epsilon_{KO}}{16}\bigl[(u_\Delta)^n_{i-2}-4(u_\Delta)^n_{i-1}+6(u_\Delta)^n_{i}-4 (u_\Delta)^n_{i+1}+(u_\Delta)^n_{i+2}\bigr],
\end{equation}
at each point $(n,i)$ such that $(n,i-2),(n,i-1),(n,i+1),(n,i+2)$ are points where the numerical solution is known. The dissipation parameter $\epsilon_{KO}$ is a positive quantity that determines the amount of dissipation, and must be less than 1 for stability (as we explain below). Notice that $(D_x u_\Delta)^n_{i}$ can be written as $\frac{\epsilon_{KO}}{16}\Delta^4 [(u_{,xxxx})_\Delta]^n_i$, where $[(u_{,xxxx})_\Delta]^n_i$ is a second-order centred stencil for the fourth derivative of $u$ with respect to $x$. Therefore, this term does not spoil the order of convergence of the scheme in the limit $\Delta\to 0$.
It should be mentioned that several schemes do not apply dissipation at $(n,i)$, i.e., $(D_x u_\Delta)^n_{i}=0$, if the solution is not available at some points in the set $S^n_i=\{(n,i-2),(n,i-1),(n,i+1),(n,i+2)\}$, which is typically the case near numerical boundaries.
On the contrary, in our scheme, $(D_x u_\Delta)^n_{i}$ is given at such points $(n,i)$ by a term proportional to $\Delta^2 [(u_{,xx})_\Delta]^n_i$, where $[(u_{,xx})_\Delta]^n_i$ is a second order stencil for $u_{,xx}$ that references only the points in $S^n_i$. This modification was suggested in \cite{Calabrese:2003vx}; the explicit expression of $(D_x u_\Delta)^n_{i}$ depends on which of the neighbouring points can be referenced. Clearly, this choice still preserves the order of convergence of the scheme.

Let us now study the properties of \eqref{eq:KOterm1}. This study can also be applied to the dissipative term near numerical boundaries proposed by \cite{Calabrese:2003vx}, and it leads to similar conclusions.
The Fourier transform of $(D_x u_\Delta)^n_{i}$ to frequency space with rescaled frequency $\xi:=\Delta k\in[-\pi,\pi]$ is
\begin{equation}
(D_\xi \tilde{u}_\Delta)^n(\xi)=\epsilon_{KO} \sin^4\biggl(\frac{\xi}{2}\biggr) (\tilde{u}_\Delta)^n(\xi).
\end{equation}
We thus see that \eqref{eq:KOterm1} is a high-pass filter, i.e., it leaves Fourier components with frequencies $k$ close to $\pm k_{\text{Nyq}}$ unchanged and it reduces components with small frequencies. Therefore, in order to obtain a low-pass filter, we must subtract \eqref{eq:KOterm1} from the solution.
Furthermore, we notice that we must have $\epsilon_{KO}\leq 1$. In fact, if $\epsilon_{KO}>1$, then $\epsilon_{KO}  \sin^4\bigl(\frac{\xi}{2}\bigr) >1$ for rescaled frequencies $\xi>2\arcsin(\epsilon_{KO}^{-4})$. Thus, for these frequencies, we would be subtracting a quantity larger than the amplitude $(\tilde{u}_\Delta)^n(\xi)$, i.e., the solution would still contain high-frequency components with amplitudes of opposite sign. If $\epsilon_{KO}$ is too large, we might even be amplifying some of these high-frequency components, thus spoiling stability properties of the scheme. 

The literature shows that the subtraction of the dissipation term can be implemented in several ways and at different stages of the simulation. In our scheme, we subtract the dissipation term for each spatial direction from $(u_\Delta)^{n-1}$ and $(u_\Delta)^{n}$, i.e., we update $(u_\Delta)^{n-1}_{ijk}$, $(u_\Delta)^{n}_{ijk}$ at each grid point $(i,j,k)$ via
\begin{equation}
\begin{split}
(u_\Delta)^{n-1}_{ijk}&\to(u_\Delta)^{n-1}_{ijk} -(D_x u_\Delta)^{n-1}_{ijk}-(D_y u_\Delta)^{n-1}_{ijk}-(D_z u_\Delta)^{n-1}_{ijk}\;,\\
(u_\Delta)^{n}_{ijk}&\to(u_\Delta)^{n}_{ijk}-(D_x u_\Delta)^{n}_{ijk}-(D_y u_\Delta)^{n}_{ijk}-(D_z u_\Delta)^{n}_{ijk}\;,
\end{split}
\end{equation}
before using these values for the NGS time integration at $t_n$ (see \eqref{eq:NGSupdate}).

\subsection{Adaptive mesh refinement}
\label{sec:AMR}

Interesting dynamics during the evolution can occur at different places of the domain and over very different length scales. When this is the case, a certain grid resolution can become insufficient to resolve dynamical features in a certain spatial region, or it can become unnecessarily refined, and thus computationally expensive, in regions where a much coarser grid would be sufficient. \emph{Adaptive mesh refinement (AMR)} is a tool meant to avoid such issues and optimise computational resources. It consists of using a certain coarse resolution to capture the relevant physics over the longest length scales involved in the problem, and add finer and finer grids in subregions of the coarse grid where smaller scale effects occur. An algorithm that performs evolution over this hierarchy of grids, called Berger and Oliger algorithm, is presented in this section. Furthermore, it is typically necessary to determine the appropriate hierarchy as the evolution proceeds. The algorithm that allows to add, remove, extend or reduce finer grids when and where necessary is called \emph{dynamical regridding}, and its main aspects are reviewed in this section, with particular focus on the implementation in our scheme. Our discussion follows that of \cite{Pretorius2002} and \cite{Pretorius:2005ua}.

Let us first specify the structure of the AMR hierarchy.
This is divided into levels containing grids with the same resolution. Each level is identified by an integer $l=1,2,\dots,l_f$ , where $l_f$ is the (user specified) maximum number of levels allowed in a simulation. For simplicity, we consider levels that double in resolution: if level $l$ has grid spacing $\Delta_l$, then level $l+1$ has grid spacing $\Delta_{l+1}=\Delta_l/2$. Grids at level $l+1$ (child grids) are fully contained in grids at level $l$ (parent grids). We also restrict to the case in which child and parent grids share the same coordinate system, the boundaries of child grids are parallel to the boundaries of parent grids, and a child grid contains all the points of its parent grid within the overlap region. As detailed below, we want to perform Newton-Gauss-Seidel time integrations (see Section~\ref{sec:NGS}) on the hierarchy levels. We denote the time interval of time integration on grids at level $l$ by $\Delta_{t_l}$. We pick $\Delta t_{l+1}=\Delta t_{l}/2$, so that $\lambda_{l}:=\Delta t_{l}/\Delta_{l}$ does not depend on $l$ and the CFL condition for stability is satisfied for all levels, once it is satisfied on the coarsest grid at $l=1$.

At any time $t_n$, the \emph{Berger and Oliger (B\&O) algorithm} has two tasks.
The first one is to obtain the solution at time $t_{n+1}=t_n+\Delta t_1$ on all levels of the hierarchy. We will achieve this through a particular sequence of time integrations.
Notice that the boundary conditions at the boundary of the coarsest grid are given by the PDE problem, but the boundary conditions for smaller grids at finer levels are not known. However, once we have integrated in time the solution on a level $l$ from some $t_n$ to $t_n+\Delta t_l$,  we can use the resulting $u_{\Delta_l}$ at $t_n+\Delta t_l$ to obtain boundary conditions for two consecutive time integrations on the child level $l+1$: from $t_n$ to $t_n+\Delta t_{l+1}$, and from $t_n+\Delta t_{l+1}$ to $t_n+2\Delta t_{l+1}=t_n+\Delta t_{l}$. For the latter, we need spatial interpolation to obtain the value at boundary points of level $l+1$, if these are not grid points of the parent level $l$. For the former, we will also need interpolation in time to extract boundary values at $t_n+\Delta t_{l+1}=t_n+\Delta t_{l}/2$ from $u_{\Delta_l}$ at $t_n$ and $t_n+\Delta_l$. After the second time integration on level $l+1$, the solution at time $t_n+\Delta t_{l}$ is known on both parent and child levels, and we say that these levels are \emph{synchronised}.
The second task is to ensure that the solution at a given point has the same value on all levels, and that this value is the one attained at the finest level that contains that point. We will achieve this by a sequence of \emph{injections}, i.e., operations that simply copy solution values from a child level to a parent synchronised level at common points.

\begin{figure*}[t!]
        \centering
        \includegraphics[width=4.5in,clip=true]{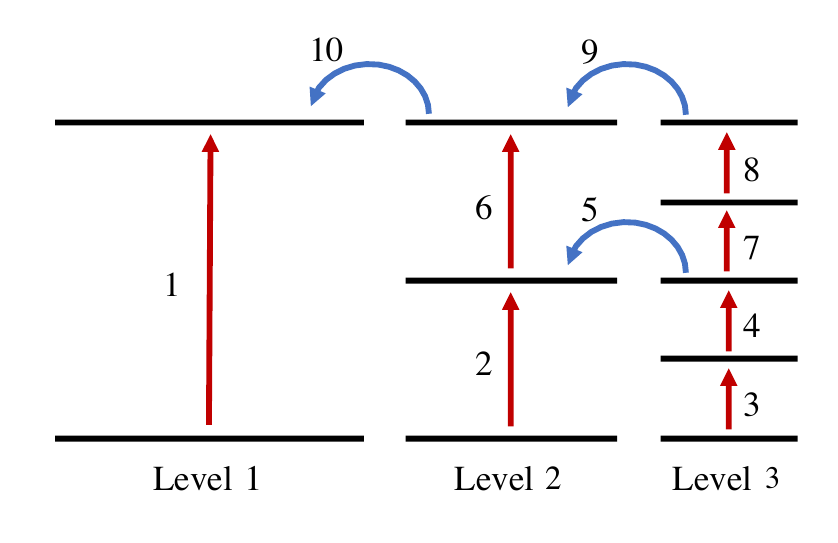}
\parbox{5.0in}{\caption{Representation of AMR time integration strategy for a hierarchy with 3 levels. Image from \cite{Kunesch:2018jeq}.
        }\label{fig:AMR}}
\end{figure*}

Both tasks can be performed by applying the following procedure in a recursive way to all levels from $l=1$ to $l=l_f-1$.
\emph{After evolving the solution on all grids at level $l$ by $\Delta t_l$, the solution on all the grids at level $l+1$ is evolved by $\Delta t_{l+1}$ twice, using the boundary conditions obtained from the evolution on level $l$. After the second evolution of level $l+1$ (i.e., when level $l+1$ and level $l$ are synchronised), we inject the solution on level $l+1$ into the solution on all coarser levels synchronised with $l+1$.}
For instance, in the case of $l_f=3$, we perform the following operations to evolve the solution from time $t_n$ to $t_{n+1}=t_n+\Delta t_1$ on all levels. Refer to Figure~\ref{fig:AMR} for a visual representation of the sequence of operations \cite{Kunesch:2018jeq}.
\begin{enumerate}
\item we integrate in time the solution on level 1 from $t_n$ to $t_{n+1}=t_n+\Delta t_1$, and extract the boundary conditions for level 2 at $t_n+\Delta t_2$ and $t_n+2\Delta t_2=t_n+\Delta t_1$;
\item we integrate in time the solution on level 2 from $t_n$ to $t_n+\Delta t_2$, and extract the boundary conditions for level 3 at $t_n+\Delta t_3$ and $t_n+2\Delta t_3=t_n+\Delta t_2$;
\item we integrate in time the solution on level 3 from $t_n$ to $t_n+\Delta t_3$;
\item we integrate in time the solution on level 3 from $t_n+\Delta t_3$ to $t_n+2\Delta t_3=t_n+\Delta t_2$;
\item we inject the solution on level 3 at $t_n+\Delta t_2$ into the solution on level 2 at $t_n+\Delta t_2$;
\item we integrate in time the solution on level 2 from $t_n+\Delta t_2$ to $t_n+2\Delta t_2=t_n+\Delta t_1$, and extract the boundary conditions for level 3 at $t_n+\Delta t_2+\Delta t_3$ and $t_n+\Delta t_2+2\Delta t_3=t_n+2\Delta t_2=t_n+\Delta t_1$;
\item we integrate in time the solution on level 3 from $t_n+\Delta t_2$ to $t_n+\Delta t_2+\Delta t_3$;
\item we integrate in time the solution on level 3 from $t_n+\Delta t_2+\Delta t_3$ to $t_n+\Delta t_2+2\Delta t_3=t_n+2\Delta t_2=t_n+\Delta t_1$;
\item we inject the solution on level 3 at $t_n+\Delta t_1$ into the solution on level 2 at $t_n+\Delta t_1$;
\item we inject the solution on level 2 at $t_n+\Delta t_1$ into the solution on level 1 at $t_n+\Delta t_1$.
\end{enumerate}

We now turn to the problem of \emph{dynamical regridding} at level $l$, i.e., identifying how the hierarchy of grids at levels higher than $l$ needs to be modified in order to have a numerical solution that tracks small dynamical features. In doing so, the fundamental quantity is an estimate of the solution error (called \emph{truncation error estimate (TE)} for historical reasons), since large solution error in a certain region can be expected to signal failure to track small features, and thus the need for finer grids.
$\epsilon_{\Delta_l}$ at time $t_n$ on a grid $g_l$ at level $l$ can be estimated by comparing the solution on $g_l$, $(u_{\Delta_l})^n$, with the solution on the parent grid at level $l-1$, $(u_{\Delta_{l-1}})^n=(u_{2\Delta_{l}})^n$, at common points. In fact, using \eqref{eq:Richu}, we see that 
\begin{equation}
(\epsilon_{\Delta_l})^n=(u_{\Delta_{l}})^n-(u_{\Delta_{l-1}})^n+\mathcal{O}(\Delta_l^m),
\end{equation}
where $m$ is the accuracy order of the finite difference scheme. Therefore, the sum over (some of) the unknowns contained in $u$ of some point-wise norm of $(u_{\Delta_{l}})^n-(u_{\Delta_{l-1}})^n$ provides a TE estimate at each point of the grid $g_l$ at level $l$:
\begin{equation}
\label{sec:TE}
(\tau_{g_l})^n:=\sum_u ||(u_{\Delta_{l}})^n-(u_{\Delta_{l-1}})^n||.
\end{equation}
We see that, when AMR is used, the AMR hierarchy should be initialised so that level 2 contains the coarsest grid that we are interested in (i.e., the grid that tracks the long wavelength dynamics), while level 1 is simply used to determine $(\tau_{g_2})^n$.
The values of the grid function $(\tau_{g_l})^n$ on all grids at levels $l,l+1,\dots, l_f$ are analysed by a clustering algorithm.
This algorithm first determines if the TE estimate is above some user-specified tolerance on the finest level covering each point. Then, it provides suitable additions and extensions of grids to cover the identified large TE regions, and suitable removals and reductions of grids on regions of small TE (see \cite{d5f0a88a11304a7686f0d59b3444c1ed} for more details on clustering algorithms; see \cite{Pretorius2002}, \cite{Pretorius:2005ua} and \cite{d5f0a88a11304a7686f0d59b3444c1ed} for more details on the clustering employed in our scheme).

\subsection{Solving constraints via multigrid algorithm}
\label{sec:MG}

In simulations of gravity theories, in addition to evolution equations, it is necessary to solve constraint equations. In general relativity, an example is provided by the equations \eqref{eq:hamconstr},\eqref{eq:momconstr} that constrain initial data on the hypersurface at $t=0$. These equations are elliptic in nature, so solving them requires techniques that differ from those employed to solve hyperbolic evolution equations. An efficient numerical tool to solve finite difference elliptic equations is the \emph{multigrid (MG) algorithm}. In our scheme, we employ a variant of the MG algorithm, called \emph{Full Approximation Storage (FAS) MG algorithm}, which is described in this section (a more extensive review of the implementation in our scheme is contained in \cite{Pretorius2002}).

Consider the finite difference version of constraint equations on a grid with spacing $\Delta$ at $t=0$, which we write again in the schematic form
\begin{equation}
\label{eq:MGPDE}
\mathcal{L}_\Delta u_\Delta=0,
\end{equation}
as well as the finite difference version of the desired boundary conditions.
Once again, solving this problem exactly is practically impossible given the number of unknowns and grid resolutions involved in a typical problem, therefore we look for approximated solutions.
Recall that, given an approximated solution $u_\Delta$ such that $\mathcal{L}_\Delta u_\Delta$ is not exactly vanishing, we refer to the left-hand-side of \eqref{eq:MGPDE} computed from this approximated $u_\Delta$ as the residual $\mathcal{R}_\Delta$.

We might think of obtaining approximated solutions of \eqref{eq:MGPDE} simply via the Newton-Gauss-Seidel method, as we do for the evolution equations (see Section~\ref{sec:NGS}), however relaxation methods have slow convergence rates for elliptic equations, i.e., it takes many iterations to reduce the residual below the desired tolerance.
On the other hand, an MG algorithm can achieve a small residual with just a few iterations by employing relaxation methods, such as NGS, over a series of grids with decreasing resolution. 
This is possible because relaxation methods efficiently smoothen the residual $\mathcal{R}_\Delta$ by damping its Fourier components with frequency of order $\pi/\Delta$. Although these methods do not directly remove much shorter frequencies, smoothening the residual on coarser grids allows to damp also shorter frequency components. After a few ``coarsening'' iterations, we reach a grid that is coarse enough for \eqref{eq:MGPDE} to be solved exactly with affordable computational resources. This provides the exact shortest frequency part of the numerical solution, which we simply use to correct the solution previously obtained on finer levels (these corrections are called \emph{coarsest grid corrections (CGC)}). The whole process is  called \emph{V-cycle}, since we are first going ``down'' towards coarser grids and then ``up'' towards finer ones.

Let us now see how these ideas are implemented in practice in the MG FAS algorithm.
We consider grids $g_0,g_1,\dots,g_N$ with spacings, respectively, $\Delta_0,\Delta_1=2\Delta_0,\Delta_2=4\Delta_0,\dots,\Delta_N=2N\Delta_0$ (notice that, unlike the notation used to describe AMR in Section~\ref{sec:AMR}, in this context larger values of the index $n$ correspond to coarser grids). 
We first find an approximated solution $u_{\Delta_0}$ of the problem \eqref{eq:MGPDE} on the finest grid $g_0$. We start from an initial guess for $u_{\Delta_0}$, and we perform NGS relaxation sweeps until the $L^2$-norm of the residual, $||\mathcal{R}_{\Delta_0}||_2$, is below some tolerance, i.e., we find $u_{\Delta_0}$ that satisfies
\begin{equation}
\label{eq:MGeqRg0}
\mathcal{L}_{\Delta_0} u_{\Delta_0}=\mathcal{R}_{\Delta_0},
\end{equation}
together with the desired boundary conditions on $u_{\Delta_0}$ in finite difference form. 

We now move the problem to the coarser grid $g_1$. Let us denote the operator that restricts grid functions on $g_0$ to grid functions on $g_1$, called \emph{restriction operator}, by $I_{\Delta_1}$. For instance, the restriction of $u_{\Delta_0}$ to $g_1$ is $I_{\Delta_1}[u_{\Delta_0}]$. We also define the truncation error on $g_1$ resulting by approximating the exact problem \eqref{eq:MGPDE} on $g_0$ with finite difference stencils on $g_1$, 
\begin{equation}
\tau_{\Delta_1}:=\mathcal{L}_{\Delta_1}[ I_{\Delta_1}[u_{\Delta_0}]]-I_{\Delta_1}[\mathcal{L}_{\Delta_0} u_{\Delta_0}]=\mathcal{L}_{\Delta_1}[ I_{\Delta_1}[u_{\Delta_0}]],
\end{equation}
where the last expression is obtained using the fact that $\mathcal{L}_{\Delta_0} u_{\Delta_0}=0$ for the exact problem on $g_0$.

As stated above, we are not interested in solving the problem on the coarser grid $g_1$, i.e., finding $u_{\Delta_1}$ such that $\mathcal{L}_{\Delta_1} u_{\Delta_1}=\mathcal{R}_{\Delta_1}$ with $||\mathcal{R}_{\Delta_1}||_2$ below some tolerance. Instead, on grid $g_1$, we want to keep smoothing the residual of the finest grid problem, so that the unknown $u_{\Delta_1}$ approaches the solution of the problem on $g_0$ at points common to $g_0$ and $g_1$. To this end, on grid $g_1$, we initialise $u_{\Delta_1}=I_{\Delta_1}[u_{\Delta_0}]$ and we perform NGS sweeps to obtain $u_{\Delta_1}$ such that
\begin{equation}
\label{eq:MGeqRg1}
\mathcal{L}_{\Delta_1} u_{\Delta_1}-I_{\Delta_1}[\mathcal{R}_{\Delta_0}]-\tau_{\Delta_1}=\mathcal{R}_{\Delta_1},
\end{equation}
with residual $||\mathcal{R}_{\Delta_1}||_2$ below some tolerance. We can see that this method serves our purpose by noticing that, if the problem on $g_0$ were solved exactly, i.e., $\mathcal{R}_{\Delta_0}=0$, then \eqref{eq:MGeqRg1} would have the (presumably unique) exact (i.e., satisfying $\mathcal{R}_{\Delta_1}=0$) solution $u_{\Delta_1}=I_{\Delta_1}[u_{\Delta_0}]$.
Once the residual on $g_1$ is sufficiently small, we move the problem to the coarser grid $g_2$ and we repeat the same procedure. When we reach the coarsest grid $g_N$, the problem $\mathcal{L}_{\Delta_N} u_{\Delta_N}-I_{\Delta_N}[\mathcal{R}_{\Delta_{N-1}}]-\tau_{\Delta_N}=0$ is solved exactly, which completes the ``down'' part of the V-cycle.

In the ``up'' part of the V-cycle, the solution on grid $g_n$ is used to correct the solution on the finer grid $g_{n-1}$, for all values of $n$ from $N$ to $1$.
The CGC correction to $u_{\Delta_{n-1}}$ is computed as
\begin{equation}
\delta u_{\Delta_{n-1}}=P_{\Delta_{n-1}}\left[u_{\Delta_{n}}-  I_{\Delta_n}[u_{\Delta_{n-1}}]\right],
\end{equation}
where $P_{\Delta_{n-1}}$, called \emph{prolongation operator}, calculates the values of a grid function on grid $g_{n-1}$ by interpolation from the values of the grid function on grid $g_n$. $u_{\Delta_{n-1}}$ is then corrected by the replacement $u_{\Delta_{n-1}}\to u_{\Delta_{n-1}}+\delta u_{\Delta_{n-1}}$. The residual corresponding to the corrected solution will have small short frequency components but the interpolation, involved in $P_{\Delta_{n-1}}$, reintroduces some high frequency noise. In order to eliminate this noise, a few iterations of the NGS method are performed on grid $g_{n-1}$, before using the resulting $u_{\Delta_{n-1}}$ to correct $u_{\Delta_{n-2}}$. This process is repeated until the solution on $g_0$ has been corrected and smoothened.

Finally, we mention that it is possible to adapt the MG algorithm to solve elliptic equations on an AMR hierarchy, and we refer the reader to \cite{Pretorius2002} and \cite{Pretorius:2005ua} for this discussion.

\subsection{Apparent horizon finder and excision}
\label{sec:AHfind}

Once the solution is obtained at a certain time $t_n$, we can search for the position of an apparent horizon (AH) at that time. The formation of an AH in a simulation typically signals the imminent formation of a spacetime singularity in the AH interior. One possible strategy to avoid dealing with singularities in the numerical domain goes under the name of excision. In this section we describe a possible apparent horizon finder algorithm, and we describe some general features of excision (see Section~\ref{sec:AH_exc} for technical details about the AH finder and the implementation of excision in our simulations).

The AH finder employs the numerical solution written in some set of compactified spatial spherical coordinates $(t,\rho,\theta,\phi)$ of the type considered in Section~\ref{subsec:apphor}, and such that the compactified radial coordinate $\rho$ takes values from 0 (at the the origin of the spherical coordinate frame) to a finite value $\rho_{\text{max}}$ (at infinity) that we can set to 1 without loss of generality. The angular coordinates take the usual values $\theta\in(0,\pi),\phi\in(0,2\pi)$, and points with $\phi$ and $\phi+2\pi$ are identified. We consider a grid over the $(\theta,\phi)$ domain. For simplicity, we use equal spacing along both angular directions, i.e., $\Delta\theta=\Delta\phi\equiv \Delta_{AH}$\footnote{The grid on which the AH finder is executed is completely independent of the specifics of the grid used for numerical evolution.}.
The AH finder algorithm starts by considering $n$ two-dimensional surfaces at constant, equally spaced, values of $\rho$ within a user-specified range included in (0,1), and selects the surface for which the $L^2$-norm of the outward null expansion, computed over the $(\theta,\phi)$ grid, is smallest.
Let $\rho_0$ be the $\rho$ coordinate on this surface.
Starting from the initial guess $R(\theta,\phi)=\rho_0$, at each of the $(\theta,\phi)$ grid points,
we find the numerical solution to the diffusion equation
\begin{equation}
\label{eq:floweq}
\frac{dR(\theta,\phi)}{d s}=-\Theta_{out}(\rho,\theta,\phi)|_{\rho=R(\theta,\phi)}\,,
\end{equation}
where $\Theta_{out}(\rho,\theta,\phi)|_{\rho=R(\theta,\phi)}$ is the outward null expansion of the two-dimensional surface given by $F(\rho,\theta,\phi)\equiv\rho-R(\theta,\phi)=0$, defined in \eqref{eq:outexp}. An approximated solution is obtained by replacing the previous guess for $R(\theta,\phi)$ by $R(\theta,\phi)-\Delta s \Theta_{out}(\rho,\theta,\phi)|_{\rho=R(\theta,\phi)}$.
Assuming that the initial guess $\rho_0$ is not too distant from the position of the AH, as we iterate this process, $R(\theta,\phi)$ is expected to progressively approach the AH, where $\Theta_{out}=0$, after each iteration. In other words, $R(\theta,\phi)$ ``flows'' towards the AH position as the fictitious time $s$ increseas. This is the reason why this method is generally referred to as \emph{flow method}. Notice that, since \eqref{eq:floweq} is a parabolic equation, the step $\Delta s$ must be at least of order $\Delta_{AH}^2$ for stability in the fictitious time $s$.
We require that this process stops when either the $L^2$-norm of $\Theta_{out}(\rho,\theta,\phi)|_{\rho=R(\theta,\phi)}$ is below some specified tolerance, i.e., $R(\theta,\phi)$ is sufficiently close to the AH, or the user-specified maximum number of iterations has been reached, i.e., either there is no AH at time $t$ or this method was not able to find it.

In order to avoid dealing with singularities in the numerical domain, at any time at which an AH is found, we do not evolve a region of the spatial grid centred at the centre of the AH and well-inside the AH.
This technique, called \emph{excision}, is effective in removing singularities if the following common assumptions are valid on the spacetimes that we consider: (i) weak cosmic censorship is not violated, i.e., geometric singularities are contained inside a black hole event horizon; (ii) the AH at any time $t$ is contained in $t$-constant slices of the event horizon; (iii) the AH at any $t$ provides a sufficiently accurate approximation for $t$-constant slices of the event horizon. 

It is important to notice that excising a region of the AH preserves stability and convergence properties only if the solution inside the AH does not affect the solution, at later times, outside the AH. In other words, when performing excision, we assume that the curves along which the PDE solution propagates, called \emph{characteristics}, flow towards the centre of the AH if they start inside the AH. This is a reasonable assumption given the definition of AH (see Section~\ref{sec:AH}), as the boundary of a region that traps light rays.
In practice, this means that there is no need to impose conditions on the solution at the excision boundary: the information needed to solve the equations of motion on and outside the excision surface at a certain time is entirely contained in the unexcised part of the domain at previous times, and we just need to retain this information by writing the equations of motion at the excision surface in terms of one-sided (i.e., either forward or backward) stencils that do not reference points inside the excised region. More precisely, in our scheme the excised surface is the same for all three time levels involved in the Newton-Gauss-Seidel time step at time level $t_n$. Therefore, we only need to use the one-sided version of the spatial stencils.

\ifpaper
\end{document}
\fi
\newif\ifpaper
\paperfalse

\ifpaper
\input{../preamble}
\begin{document}
\fi

\chapter{Cauchy evolution of asymptotically AdS spacetimes with no symmetries}
\label{Chapter:NoSym}

The techniques discussed in Chapter~\ref{chap:ovengr} opened the doors to numerical evolution of rather general initial data in asymptotically flat spacetimes. However, the Cauchy problem in asymptotically AdS spacetimes has been, until recently, solved only in settings with strong symmetry requirements; see, for example, \cite{Bantilan:2012vu,Bantilan:2017kok,Choptuik:2017cyd}.
A \emph{characteristic scheme}, i.e., a scheme based on foliating the spacetime with null slices, has been successfully employed to simulate certain dynamical asymptotically AdS spacetimes in Poincar\'{e} coordinates\footnote{Poincar\'{e} coordinates in AdS are coordinates valid between the conformal boundary and two hypersurfaces that form the so-called Poincar\'{e} horizon. The wedge-shaped region that they cover is often called Poincar\'{e} patch of AdS or simply Poincar\'{e} AdS; see Appendix~\ref{sec:poincare}.} 
in full generality, i.e., no symmetry assumptions \cite{Chesler:2013lia}.\footnote{A characteristic formulation was also used to successfully evolve single black holes in Minkowski spacetime \cite{Gomez:1998uj} and in AdS in global coordinates \cite{Chesler:2018txn,Chesler:2021ehz}.} 
The radial coordinate of this scheme is given by the affine parameter along the ingoing null geodesics that span the null slices.
This approach, however, is not suitable for certain studies, as it will fail if the ingoing null geodesics intersect within the numerical domain (we refer to this phenomenon as the formation of a \emph{caustic}): the intersection point does no longer correspond to a unique value of the radial coordinate. 
In a number of cases, this issue can be avoided by assuming the presence of an apparent horizon that includes all points where caustics are expected to form, and excising a sufficiently large region in the horizon interior, thus excluding caustic formation from the numerical domain.
However, caustics can still form outside any apparent horizon whenever a sufficiently strong and localized perturbation of the background spacetime is present.
Two examples in which this is likely to occur are provided by the dynamical formation of localized black holes in the background of the AdS soliton spacetime \cite{Bantilan:2020pay} or even a localized black hole falling through the Poincar\'{e} horizon of AdS.
Furthermore, the characteristic scheme cannot be employed when evolving initial data with no horizon, such as the data considered in the study of gravitational collapse -- the growth of curvatures that eventually leads to the formation of a singularity in spacetime -- and black hole formation in AdS.\footnote{In asymptotically flat spaces, it has not been possible thus far to simulate all stages of a black hole binary with characteristic coordinates precisely because of the formation of caustics outside the black holes \cite{Bishop:1997ik,Lehner:2001wq}.}

On the other hand, Cauchy evolution in conjunction with the generalized harmonic formulation (reviewed in Section~\ref{sec:genharform}) is well-known to successfully handle strong, highly dynamical and localized gravitational fields, such as those produced by the individual black holes in a binary. 
Whilst it is possible that many problems that have been solved using a characteristic scheme can also be solved with Cauchy evolution, the latter can be applied to situations where characteristic schemes will almost certainly fail.
Furthermore, the use of Cauchy evolution benefits from the infrastructure developed over many years to numerically solve the Cauchy problem in asymptotically flat spacetimes.

The main purpose of this chapter is to present the first proof-of-principle Cauchy evolution of asymptotically AdS spacetimes that has been achieved with no symmetry assumptions, and to describe the framework that makes Cauchy evolution in AdS possible in full generality.
In particular, the code described in the present work, previously published in \cite{Bantilan:2020xas}, has built-in AMR (described in Section~\ref{sec:AMR}) and is designed to run in large supercomputing clusters; both of these features will likely turn out to be crucial in solving certain key open problems in AdS.
We couple gravity to a massless scalar field, but the latter does not play any fundamental role in our scheme; we introduce it as a convenient mechanism to arrange for initial data whose future Cauchy development contain trapped surfaces.

A key requirement for obtaining stable evolution in AdS is a gauge choice of generalised harmonic source functions that is consistent with the conditions imposed at the AdS boundary (see, for example, \cite{Bantilan:2012vu}).
In most cases, a gauge choice leading to stable numerical evolution is typically found in spacetimes with a certain degree of symmetry.
In the present work, we detail a gauge choice in $D=4$ spacetime dimensions that leads to stable evolution in an asymptotically AdS setting with no symmetry assumptions.
This work is a direct precursor to fully general studies of gravitational collapse.
In this context, Cartesian coordinates are suitable as they are regular everywhere, do not contain coordinate singularities, and do not have the well-known limitation suffered by spherical coordinates in the form of severely shorter time steps imposed by the Courant-Friedrichs-Lewy (CFL) condition (see Section~\ref{sec:NGS}). In addition, most AMR infrastructures are designed for this type of coordinates. 
Similar coordinates were used in \cite{Bantilan:2017kok} to study the non-spherically symmetric collapse of a massless scalar field in 5-dimensional AdS in global coordinates with SO(3) symmetry. 
In anticipation of fully general studies, we choose to write our prescription in terms of global Cartesian coordinates, using second order finite difference derivative stencils to discretize the initial constraint equations and the evolution equations. 
The framework we present here straightforwardly generalizes to other settings and other discretization schemes.

The rest of this chapter is organized as follows.
In Section~\ref{sec:asyAdS}, we review theoretical notions about asymptotically AdS spacetimes that are relevant for the following discussion.
In Section~\ref{sec:pre_sta} we detail our prescription for obtaining stable Cauchy evolution with no symmetries in Cartesian coordinates.
The crucial ingredients for this prescription are reflective Dirichlet boundary conditions imposed on appropriate evolution variables, and a specific choice of generalized harmonic source functions. We also define certain boundary quantities whose evolution describes the physics at the AdS boundary.
In Section~\ref{sec:numerical_scheme} we outline the generalized harmonic scheme that we use in our simulations. 
Section~\ref{sec:results} contains preliminary results of simulations of gravitational collapse with no symmetry assumptions.
We have relegated some technical details to several appendices. In Appendix~\ref{sec:sphevvarboucon}, we follow our prescription for the interesting case of AdS in global spherical coordinates and we obtain the corresponding stable gauge. In Appendix~\ref{sec:poincare}, we do the same for the Poincar\'e patch. Appendix \ref{sec:initdata} contains a description of our construction of initial data for the class of spacetimes considered in the paper, while in Appendix \ref{sec:GCbulk} we provide the details of our complete gauge choice, including the bulk. In Appendix \ref{sec:extrapconvbdy} we explain how we carry out the extrapolation to read off the boundary quantities. Some convergence tests are presented in Appendix \ref{sec:convbulk}.

\section{Theoretical aspects of asymptotically anti-de Sitter spacetimes}
\label{sec:asyAdS}

In this section, we review the definitions of anti-de Sitter spacetime and asymptotically AdS spacetimes, as well as some of their fundamental features.

\subsection{Anti-de Sitter spacetime}
\label{subsec:pureAdS}

Let us define \emph{anti-de Sitter (AdS)} spacetime, also called \emph{pure AdS} or \emph{global AdS}. 
This is the maximally symmetric spacetime with manifold $\mathcal{M}=\mathbb{R}^4$ whose metric solves \eqref{eq:EFE} with negative cosmological constant, in vacuum.
In terms of global spherical coordinates $(t,r,\theta,\phi)\in(-\infty,+\infty)\times(0,+\infty)\times(0,\pi)\times(0,2\pi)$, the AdS metric can be expressed as
\begin{equation}
\label{eqn:ads4}
\hat{g}= -\left(1+\frac{r^2}{L^2}\right) dt^2 + \left(1+\frac{r^2}{L^2}\right)^{-1} dr^2 +r^2 d{\Omega_2}^2 \,,
\end{equation}
where $d{\Omega_2}^2=d\theta^2+\sin^2\theta d\phi^2$.
$L>0$ is a characteristic length scale, also called AdS radius, that is related to the cosmological constant by $\Lambda = - 3/L^2$. Notice that we approach the Minkowski metric in spherical coordinates \eqref{eq:Minsph} as $L\to+\infty$.
The isometries of AdS spacetime form the 10-dimensional group O(3,2), which we also refer to as the \emph{AdS isometry group}. The subgroup connected to the identity, $SO(3,2)$, is generated by 10 linearly independent Killing vector fields.

AdS is conformally compact. To see this, we define a compactified radial coordinate $\rho$ by $r=2\rho/(1-\rho^2/\ell^2)$, so that spatial infinity at $r \rightarrow +\infty$ is at a finite value of the new radial coordinate, $\rho=\ell$.\footnote{We emphasize that the arbitrary compactification scale $\ell$ is completely independent of the AdS length scale $L$.}
We hereafter set $\ell=1$ without loss of generality, so that spatial infinity is at $\rho=1$. In this way, we obtain (compactified) spherical coordinates $x^\alpha=(t,\rho,\theta,\phi)$.
Defining a convenient function $\hat{f}(\rho) = (1-\rho^2)^2+4\rho^2/L^2$, the metric of AdS in this set of coordinates reads 
\begin{equation}
\label{eqn:ads4_compact}
\hat{g} = \frac{1}{(1-\rho^2)^2} \left( -\hat{f}(\rho) dt^2 + \frac{4(1+\rho^2)^2}{\hat{f}(\rho)} d\rho^2 + 4\rho^2 d{\Omega_2}^2 \right).
\end{equation}
Notice that $\Omega=\frac{(1-\rho^2)}{2}$ is a defining function, and the conformal metric $\bar{g}=\frac{(1-\rho^2)^2}{4}\hat{g}$ can be extended to spatial infinity $\rho=1$, where it induces a metric given by $g_{(0)}=-(1/L^2)dt^2+d\theta^2+\sin^2\theta d\phi^2$. By a redefinition of the $t$ coordinate, $t\to L t$, we see that this spacetime possesses a conformal boundary $\partial \mathcal{M}$ at $\rho=1$, called \emph{AdS boundary}, with the conformal structure given by the \emph{Einstein Static Universe (ESU)}, i.e., $(\mathbb{R}\times S^2,-dt^2+d\theta^2+\sin^2\theta d\phi^2)$. A crucial feature of AdS spacetime is that the conformal boundary, regarded as a hypersurface of $(\bar{\mathcal{M}}=\mathcal{M}\cup\partial \mathcal{M},\bar{g})$, is timelike.

In our scheme we will make use of (compactified) Cartesian coordinates $x^\mu=(t,x,y,z)\in(-\infty,+\infty)\times (-1,1)\times(-1,1)\times(-1,1)$ defined by $x=\rho\cos\theta$, $y=\rho\sin\theta\cos\phi$, $z=\rho\sin\theta\sin\phi$.
As explained in Section~\ref{sec:numcauprob}, this allows us to bypass the severe CFL restriction on the time step size near $\rho=0$ on a grid in spherical coordinates.
The metric of AdS in Cartesian coordinates reads
\begin{equation}
\label{eqn:ads4_final}
\begin{split}
\hat{g} = \frac{1}{\left(1-\rho^2\right)^2 }\Big[ &-\hat{f}(\rho)dt^2+\frac{4 \left(1+\rho^2\right)^2}{\rho^2 \hat{f}(\rho)} (x dx + y dy + z dz)^2\\
&+\frac{4}{\rho^2} \Big(\left(y^2+z^2\right) dx^2 + \left(x^2+z^2\right) dy^2 + \left(x^2+y^2\right) dz^2 \\
&-2 x y\, dx dy - 2 y z\, dy dz - 2 x z\,dx dz\Big)\Big]\,,
\end{split}
\end{equation}
where $\rho=\rho(x,y,z):=\sqrt{x^2+y^2+z^2}$.
Without loss of generality, in the rest of this work we set the AdS length scale to $L=1$ ($L$ can later be reinserted by dimensional analysis). 
In particular, this implies that times, lengths, masses and temperatures are dimensionless quantities. 
With this choice, the metric \eqref{eqn:ads4_final} takes the diagonal form
\begin{equation}
\label{eq:ads4_diag}
\hat{g} =-\left(\frac{1+\rho^2}{1-\rho^2}\right)^2dt^2+\frac{4}{(1-\rho^2)^2} \left(dx^2 + dy^2+dz^2\right).
\end{equation}

\vspace{-0.5cm}
An additional, useful set of coordinates, in which the pure AdS metric with $L=1$ takes the particular simple form called Fefferman-Graham gauge, is given by $\bar{t}=t,\bar{\theta}=\theta,\bar{\phi}=\phi$ and $\bar{z}=2 (1-\rho)/(1+\rho)$.
The relation between $\bar z$ and $\rho$ can be inverted as $\rho=\frac{2-\bar{z}}{2+\bar{z}}$.
In these coordinates, the pure AdS metric reads
\begin{equation}
\label{eq:AdSFG}
\hat{g}=\frac{1}{\bar{z}^2}\left[d\bar{z}^2-\left(1+\frac{\bar z^2}{2}+\frac{\bar z^4}{16}\right) d\bar{t}^2+\left(1-\frac{\bar z^2}{2}+\frac{\bar z^4}{16}\right)d\bar\Omega_2^2\right],
\end{equation}
where $d\bar\Omega_2^2=d\bar{\theta}^2+\sin^2\bar{\theta}d\bar{\phi}^2$.

\subsubsection{Null waves in AdS}
\label{subsubsec:wavesAdS}

Here we present a simple calculation revealing that null waves in an AdS background reach the AdS boundary in finite coordinate time.
Therefore, in order to determine the subsequent evolution, boundary conditions on $\partial \mathcal{M}$ must be specified.
Since gravitational dynamics propagates along null waves, this result is the first indication that boundary conditions on $\partial \mathcal{M}$ are needed to determine the long-time evolution of dynamical scenarios in AdS. We will discuss further evidence for this claim in the next section.

According to the postulates of general relativity, null waves that move solely under the force of gravity travel along null geodesics. For simplicity, we will restrict to the case of radial null geodesics.
Consider the coordinate system $x^{\dot\alpha}=(t,r,\theta,\phi)$ in which the AdS metric reads \eqref{eqn:ads4} (recall that here we also set $L=1$). We define the time-orientation using $k=\frac{\partial}{\partial t}$, i.e., the future light cone of any point $p$ is the cone that contains the vector $k$ at $p$.
Let $x^{\dot\alpha}(\lambda)=\left(t(\lambda),r(\lambda),\theta(\lambda),\phi(\lambda)\right)$ be the value of the coordinates on a null geodesic with affine parameter $\lambda$, i.e., a curve with tangent vector $V^{\dot\alpha}=\frac{dx^{\dot\alpha}(\lambda)}{d\lambda}$ satisfying 
\begin{equation}
\label{eq:nullV}
g_{\dot\alpha\dot\beta}\left(x(\lambda)\right)V^{\dot\alpha}V^{\dot\beta}=0
\end{equation} 
and the geodesic equation \eqref{eq:geod1} with $f=0$. The requirement that the geodesic is radial means that $\frac{d\theta}{d\lambda}=0$ and $\frac{d\phi}{d\lambda}=0$.
Furthermore, we study geodesics that are future directed, i.e., $\frac{d t}{d\lambda}>0$, and outgoing, i.e., $\frac{d r}{d\lambda}>0$. This implies $\frac{d t}{dr}=\frac{d t}{d\lambda}\frac{d \lambda}{dr}>0$.

Now, \eqref{eq:nullV} gives
\begin{equation}
\label{eq:nullV2}
-(1+r^2(\lambda))\left(\frac{dt}{d\lambda}\right)^2+\frac{1}{(1+r^2(\lambda))}\left(\frac{dr}{d\lambda}\right)^2=0.
\end{equation}
Multiplying this by $\left(\frac{d\lambda}{dr}\right)^2$ and using the chain rule, we obtain
\begin{equation}
\left(\frac{dt}{dr}\right)^2=\frac{1}{(1+r^2)^2}.
\end{equation}
Since $\frac{d t}{dr}>0$, we must have $\frac{dt}{dr}=\frac{1}{1+r^2}$. This can be easily integrated between $r=0$, where we require $t=0$, and an arbitrary $r$. We obtain
\begin{equation}
\label{eq:tofr}
t(r)=\arctan (r).
\end{equation}
From this, it is evident that it takes a finite coordinate time $\Delta t=\pi/2$ for a radial null geodesic to go from $r=0$ to the AdS boundary $r\to+\infty$, as anticipated.
The time interval $\Delta t=\pi/2$ is also called \emph{light-crossing time} (in units of the AdS radius $L$).\footnote{A different convention defines the light-crossing time as $\Delta t=\pi$ (in units of the AdS radius $L$), i.e., the time that light needs to cross pure AdS from boundary to boundary.}

\subsection{Asymptotically locally anti-de Sitter spacetimes}
\label{subsec:asylocAdS}

Let us now define spacetimes that locally approach pure AdS near the conformal boundary. These are called \emph{asymptotically locally AdS spacetimes}, and are defined as spacetimes $(\mathcal{M},g)$ that admit a conformal compactification, thus allowing the definition of a conformal boundary $\partial \mathcal{M}$, and that satisfy the Einstein equations \eqref{eq:EFE} with negative cosmological constant, $\Lambda=-3$. If matter is coupled with gravity, we also require that the matter fields satisfy the corresponding equations of motion.
We emphasize that no assumption is made at this stage on the conformal structure of the boundary. We still refer to $\partial \mathcal{M}$ as the AdS boundary. In this section, we discuss a theorem, whose original formulation is due to Fefferman and Graham (FG) \cite{AST_1985__S131__95_0}, that characterises the asymptotic behaviour of asymptotically locally AdS spacetimes near the AdS boundary. We then define the energy-momentum tensor, boundary scalar field of the dual conformal field theory at $\partial \mathcal{M}$, as well as the conserved charges associated with asymptotic symmetries.

\subsubsection{Fefferman-Graham expansion}
\label{subsubsec:FGexpans}

Given an asymptotically locally AdS spacetime $(\mathcal{M},g)$, it is possible to define so-called \emph{Fefferman-Graham (FG) coordinates} $x^{\bar{\alpha}}=(\bar{t},\bar{z},\bar{\theta},\bar{\phi})$ in a bulk neighbourhood $\mathcal{U}\subset \mathcal{M}$ of $\partial \mathcal{M}$ such that $\bar{z}$ is a defining function, $x^{\bar{a}}=(\bar{t},\bar\theta,\bar\phi)$ are coordinates on $\partial \mathcal{M}$ and each hypersurface $\partial \mathcal{M}_{\bar z}$ in $\mathcal{U}$ at a fixed value of $\bar z$, and the metric reads
\begin{equation}
\label{eqn:FGmetric}
g=\frac{1}{\bar{z}^2}(d\bar{z}^2+g_{\bar{a}\bar{b}}(\bar{t},\bar{z},\bar{\theta},\bar{\phi})dx^{\bar{a}}dx^{\bar{b}}),
\end{equation}
where $g_{\bar{a}\bar{b}}(\bar{t},\bar{z},\bar{\theta},\bar{\phi})$ has a smooth limit $\bar z\to 0$, denoted by $g_{(0)\bar{a}\bar{b}}(\bar{t},\bar{\theta},\bar{\phi})$, which determines the conformal class of boundary metrics $[g_{(0)}]$.
The expression \eqref{eqn:FGmetric} is called FG gauge. 
Notice that $d\bar z$, which is a covector normal to $\partial \mathcal{M}$, is spacelike with respect to the metric $\bar{g}:=\bar z^2 g$, in fact $\left((d\bar z)_{\bar \alpha} (d\bar z)_{\bar \beta} \bar g^{\bar \alpha\bar \beta}\right)\bigr\rvert_{\partial \mathcal{M}}=\bar g^{\bar z\bar z}|_{\partial \mathcal{M}}=1>0$. Therefore, $\partial \mathcal{M}$ is a timelike hypersurface of the extended spacetime $(\mathcal{M}\cup\partial \mathcal{M},\bar g)$.
The fact that $\partial \mathcal{M}$ is timelike is a crucial observation. As we will see below, this property of the boundary implies that, given certain initial data (satisfying the constraints of general relativity) for the metric and any other field coupled with gravity on an initial spacelike slice $\Sigma$, a solution of the equations of motion in the entire future of $\Sigma$ that reduces, when evaluated on $\Sigma$, to the prescribed initial data can be obtained if we also impose boundary conditions on the near-boundary behaviour of the metric and the other fields, throughout the entire evolution.
The FG theorem, whose original version was presented in \cite{AST_1985__S131__95_0}, shows that the equations of motion restrict the allowed near-boundary behaviour of $g_{\bar{a}\bar{b}}$ and the other fields involved. In the rest of this section, we discuss the 4-dimensional version of this result.
We emphasize that the following discussion can be generalised in a straightforward way to the case of higher even spacetime dimensions, however the case of odd spacetime dimensions is more complicated and will not be discussed here. The case of spacetimes with an arbitrary number of dimensions is reviewed, for example, in \cite{Skenderis:2002wp}.

In the vacuum case, the FG theorem states that the near-boundary expansion of $g_{\bar{a}\bar{b}}$ is given by
\begin{equation}
\label{eqn:FGbdymetric}
g_{\bar{a}\bar{b}}(\bar{t},\bar{z},\bar{\theta},\bar{\phi})=g_{(0)\bar{a}\bar{b}}(\bar{t},\bar{\theta},\bar{\phi})+g_{(2)\bar{a}\bar{b}}(\bar{t},\bar{\theta},\bar{\phi})\bar{z}^2+g_{(3)\bar{a}\bar{b}}(\bar{t},\bar{\theta},\bar{\phi})\bar{z}^3+\mathcal{O}(\bar{z}^4),
\end{equation}
as a consequence of the near-boundary expansion of the vacuum Einstein equations.
Furthermore, the near-boundary expansion of the Einstein equations completely determines the coefficient $g_{(2)\bar{a}\bar{b}}$ in terms of $g_{(0)\bar{a}\bar{b}}$, as well as all the coefficients of the higher orders in terms of $g_{(0)\bar{a}\bar{b}}$ and $g_{(3)\bar{a}\bar{b}}$.
Therefore the bulk dynamics is determined by the term of order $\bar{z}^3$ in the expansion of $g_{\bar{a}\bar{b}}$.
The Einstein equations also constrain $g_{(3)\bar{a}\bar{b}}$, as can be expected from the fact that a solution of the Einstein equations must satisfy the Hamiltonian and momentum constraints at all times. The requirement is that $g_{(3)\bar{a}\bar{b}}$ is trace-free and divergence-free with respect to the boundary metric $g_{(0)\bar{a}\bar{b}}$, i.e.,
\begin{equation}
\label{eq:g3const}
g_{(0)}^{\bar{a}\bar{b}}g_{(3)\bar{a}\bar{b}}=0 \quad \text{and}\quad \mathcal{D}_{(0)}^{\bar{a}}g_{(3)\bar{a}\bar{b}}=0,
\end{equation}
where $g_{(0)}^{\bar{a}\bar{b}}$ is the inverse of $g_{(0)\bar{a}\bar{b}}$, $\mathcal{D}_{(0)}$ is the Levi-Civita covariant derivative associated with $g_{(0)\bar{a}\bar{b}}$, and indices of $\mathcal{D}_{\bar{a}}$ are raised and lowered by $g_{(0)\bar{a}\bar{b}}$.
We see that the FG expansion \eqref{eqn:FGbdymetric} provides the allowed near-boundary behaviour of $g$ in FG gauge. This, together with a specification of $g_{(0)\bar{a}\bar{b}}$ (i.e., a specification of the conformal class $[g_{(0)}]$), defines Dirichlet boundary conditions that allow for asymptotically locally AdS solutions of the vacuum Einstein equations.
As an example and a consistency check, notice that the pure AdS metric $\hat{g}$ satisfies the FG expansion, as can be easily seen from \eqref{eq:AdSFG}.

The FG theorem has been extended to the case of gravity in asymptotically locally AdS spacetimes coupled with matter fields, with particular focus on a real scalar field, in \cite{deHaro:2000vlm}. We present here the simplified version of this result that is relevant to our study.
To better understand the general case, let us start from the study of real scalar matter of mass $\mu$ in an asymptotically locally AdS spacetime background.
The general solution to the Klein-Gordon equation in the FG coordinates of the background spacetime can be written as
\begin{equation}
\varphi=\bar{z}^{\Delta_-}\chi_-(\bar t,\bar z, \bar \theta,\bar\phi)+\bar{z}^{\Delta_+}\chi_+(\bar t,\bar z, \bar \theta,\bar\phi),
\end{equation}
where $\Delta_\pm=\frac{3}{2}\pm\sqrt{\frac{9}{4}+\mu^2}$. Here, $\bar{z}^{\Delta_-}\chi_-(\bar x)$ is a slowly decaying KG solution with leading order asymptotics $\bar{z}^{\Delta_-}$ in the near-boundary limit $\bar{z}\to 0$, and $\bar{z}^{\Delta_+}\chi_+(\bar x)$ is a fastly decaying solution with leading order asymptotics $\bar{z}^{\Delta_+}$.
Notice that the fastly decaying part vanishes at $\partial \mathcal{M}$, whereas in general the slowly decaying mode does not vanish at $\partial \mathcal{M}$. When dynamically coupled with gravity, the slowly decaying mode can modify the near-boundary expansion \eqref{eqn:FGbdymetric} of the metric.\footnote{This modification introduces logarithmic terms of order $\bar{z}^3\ln \bar{z}$; see \cite{deHaro:2000vlm} for the details.}
However, if we impose the Dirichlet boundary condition $\varphi|_{\partial \mathcal{M}}=0$, which is equivalent to $\chi_-|_{\partial \mathcal{M}}=0$ at $\bar{z}=0$, then the Klein-Gordon equation implies that $\chi_-=0$ in the entire neighbourhood $\mathcal{U}$ where FG coordinates are defined. This is the case that we study in this work.
With this boundary condition, only the fastly decaying mode remains, and the solution can be written as
\begin{equation}
\label{eq:FGscal}
\varphi=\bar{z}^{\Delta_+}\left(\varphi_{(0)}(\bar t,\bar \theta,\bar\phi)+\mathcal{O}(\bar z)\right).
\end{equation}
Higher order coefficients in the expansion of $\varphi$ can be determined in terms of $\varphi_{(0)}$, using the near-boundary expansion of the Klein-Gordon equation.

We can now state the FG theorem for gravity coupled with scalar matter in asymptotically locally AdS spacetimes, restricted to the case of the Dirichlet boundary conditions that we study, i.e., the near-boundary behaviour \eqref{eqn:FGbdymetric} with a given specification of $g_{(0)\bar{a}\bar{b}}$, and the condition $\varphi|_{\partial \mathcal{M}}=0$. 
Under these conditions, bulk dynamics does not change the value of the metric and the scalar field $\varphi$ at the boundary, thus the boundary acts as a mirror that reflects gravitational and scalar null waves back into the bulk. For this reason, it is common to refer to these conditions as \emph{reflective boundary conditions.}
The theorem states that, given FG coordinates $x^{\bar{\alpha}}=(\bar t,\bar z, \bar \theta,\bar\phi)$ in a sufficiently small neighbourhood $\mathcal{U}\subset \mathcal{M}$ of $\partial \mathcal{M}$, the near-boundary expansion of a solution $(g,\varphi)$ of the Einstein and Klein-Gordon equations is given by \eqref{eqn:FGbdymetric} and \eqref{eq:FGscal}.
In other words, the presence of a scalar field satisfying $\varphi|_{\partial \mathcal{M}}=0$ does not alter the expansion \eqref{eqn:FGbdymetric}.
Moreover, higher order coefficients in the expansions of $g$ and $\varphi$ can be obtained from $g_{(0)\bar{a}\bar{b}}, g_{(3)\bar{a}\bar{b}}$ and $\varphi_{(0)}$, using the near-boundary expansion of the Einstein and Klein-Gordon equations. $g_{(3)\bar{a}\bar{b}}$ must still satisfy the constraints \eqref{eqn:FGbdymetric}.

Given our Dirichlet boundary conditions and a choice of initial data, i.e., $g_{(0)\bar{a}\bar{b}}$, $g_{(3)\bar{a}\bar{b}}$ and $\varphi_{(0)}$ and their time-derivatives on a spacelike slice $\Sigma$ in $\mathcal{U}$, the FG theorem ensures that we can solve the Cauchy problem of the Einstein-Klein-Gordon theory in the following way. 
First, the FG theorem tells us that the entire solution $(g,\varphi)$ on $\Sigma$ is determined by $g_{(0)\bar{a}\bar{b}}$, $g_{(3)\bar{a}\bar{b}}$ and $\varphi_{(0)}$ on $\Sigma$.
Second, by integrating in time the Einstein-Klein-Gordon equations for an infinitesimal time interval, we can obtain $g_{(3)\bar{a}\bar{b}}$ and $\varphi_{(0)}$ on a slice that is slightly in the future of $\Sigma$, say $\Sigma'$. From this data on $\Sigma'$, and our choice of $g_{(0)\bar{a}\bar{b}}$ at all times, the FG theorem says that we can obtain the entire solution $(g,\varphi)$ on $\Sigma'$. Iterating this procedure, we can determine the solution $(g,\varphi)$ in the entire causal future of the initial slice $\Sigma$.
In other words, the initial-boundary value problem in asymptotically locally AdS spacetimes is solvable for our choice of Dirichlet boundary conditions.
Finally, it will be convenient to use coordinates that cover a large region of the bulk, so that the solution on each spacelike slice can be extended to a region of $\mathcal{M}$ larger than $\mathcal{U}$. This is what we do in our numerical scheme.

\subsubsection{Boundary energy-momentum tensor, boundary scalar field}
\label{subsec:bdysetconscharg}

According to the dictionary of the AdS/CFT conjecture, given a choice of the conformal class of boundary metrics (through a specification of $g_{(0)}$) and the boundary condition $\varphi|_{\partial \mathcal{M}}=0$, the evolution of a set of initial data under the laws of general relativity is dual to the evolution of a system governed by the laws of a strongly coupled, quantum \emph{conformal field theory (CFT)} on the fixed spacetime $(\partial \mathcal{M},\bar{g}_{(0)})$ with no matter field sources. In particular, the metric $g$ is dual to the boundary energy-momentum tensor of the dual CFT, and $\varphi$ is dual to a scalar operator of scaling dimension $\Delta_+$\footnote{The scaling dimension $\Delta$ of a boundary operator $\mathcal{O}$ is determined by the transformation of its 1-point function $\langle \mathcal{O}\rangle$ under dilatations, given by rescalings of a global time coordinate on $\partial \mathcal{M}$: $\bar{t}\to \lambda \bar{t}$, $\lambda\in\mathbb{R}$. If, under this transformation, $\langle \mathcal{O}\rangle$ transforms as $\langle \mathcal{O}\rangle\to \lambda^{-\Delta} \langle \mathcal{O}\rangle$, then we say that $\mathcal{O}$ has scaling dimension $\Delta$.}.
In this section, we present the prescription by \cite{deHaro:2000vlm} to obtain the 1-point function of the boundary energy-momentum tensor and the 1-point function of the dual scalar operator.

Given a solution $(g,\varphi)$ of the equations of motion in FG coordinates, we can read off the 1-point function of the dual CFT operators from the FG expansion \eqref{eqn:FGbdymetric}.
For the boundary energy-momentum tensor operator in FG coordinates, the prescription of \cite{deHaro:2000vlm} gives
\begin{equation}
\label{eq:holoprebdyT}
\langle T_{\bar{a}\bar{b}}\rangle_{CFT}=\frac{3}{16\pi}g^{(3)}_{\bar{a}\bar{b}}\,.
\end{equation}
We often refer to this quantity simply as the \emph{boundary energy-momentum tensor} in FG coordinates. 
Notice that the constraints \eqref{eqn:FGbdymetric} ensure that $\langle T_{\bar{a}\bar{b}}\rangle_{CFT}$ is trace-free and conserved for a solution of the equations of motion, i.e., 
\begin{equation}
\label{eq:traconsset}
g_{(0)}^{\bar{a}\bar{b}}\langle T_{\bar{a}\bar{b}}\rangle_{CFT}=0 \quad \text{and}\quad \mathcal{D}_{(0)}^{\bar{a}}\langle T_{\bar{a}\bar{b}}\rangle_{CFT}=0,
\end{equation}
as appropriate for the energy-momentum tensor of a CFT on a spacetime with odd number of dimensions.
For the scalar operator, we have
\begin{equation}
\langle \mathcal{O}\rangle_{CFT}=(2\Delta_+ - 3) \varphi_{(0)}.
\end{equation}
We often refer to this simply as the \emph{boundary scalar field}.
In our simulations, we compute the boundary energy-momentum tensor following a different, but equivalent, prescription, presented by \cite{Balasubramanian:1999re} and applied to our study as discussed in Section~\ref{sec:bouset2}. The two prescriptions are compared in Appendix~\ref{sec:HoloRen}.
Moreover, in the simulations we compute a quantity equal to $\langle \mathcal{O}\rangle_{CFT}$ up to numerical factors. We will define this quantity precisely in Section~\ref{sec:gauge_choice}.

\subsubsection{Asymptotic symmetries and conserved charges}
\label{subsec:conscharg}

The generalisation of Noether's theorem to gravity tells us that, for each asymptotic symmetry with generator $\xi$, there is an associated charge $Q[\xi]$ that is conserved for a solution of the equations of motion (see \cite{Fischetti:2012rd} for a review).
Here, we give an expression for the charges and we show their conservation. 

Let us recall that, according to the definitions of Section~\ref{subsec:spastruc}, asymptotic symmetries are given by non-pure gauge groups of diffeomorphisms that preserve the boundary conditions, and are generated by non-pure gauge, inequivalent asymptotic Killing vector fields $\xi$ that also preserve the boundary condition on $\varphi$. 
Notice that requiring that an asymptotic Killing vector field $\xi$ preserves $\varphi|_{\partial \mathcal{M}}=0$ is trivial, since the boundary limit $\xi_{(0)}$ is tangent to $\partial \mathcal{M}$, and thus the associated boundary diffeomorphisms can only ``move around'' the vanishing values of $\varphi|_{\partial \mathcal{M}}$.
Hence, in order to characterise asymptotic symmetries associated with our Dirichlet boundary conditions, we can restrict to the gravitational sector.
It can be proved that each non-pure gauge, inequivalent asymptotic Killing vector field $\xi$ preserving the FG gauge, i.e., such that $(\mathcal{L}_\xi g)_{\bar z\bar z}=0$ and $(\mathcal{L}_\xi g)_{\bar z\bar a}=0$ in FG coordinates, asymptotes to a particular (non-vanishing) conformal Killing vector field $\xi_{(0)}$ of $g_{(0)}$ \cite{Fischetti:2012rd}, i.e., a vector field $\xi_{(0)}$ on $\partial \mathcal{M}$ that satisfies
\begin{equation}
\label{eq:confKillg0second}
\mathcal{D}_{(0)\bar a} \xi_{(0)\bar b}+\mathcal{D}_{(0)\bar b} \xi_{(0)\bar a}=\frac{2}{3}\left(\mathcal{D}_{(0) \bar c} \xi^{\bar c}_{(0)} \right)g_{(0)\bar a\bar b}.
\end{equation}
Thus, we conclude that the asymptotic symmetries associated with our Dirichlet boundary conditions are those generated by vector fields $\xi$ whose boundary limit $\xi_{(0)}$ is a (non-vanishing) conformal Killing vector field of $(\partial \mathcal{M},g_{(0)})$.
This should not be surprising since the only non-trivial structure at the boundary specified by our boundary conditions is the conformal class of boundary metrics $[g_{(0)}]$, which is invariant under conformal isometries of $(\partial \mathcal{M},g_{(0)})$.

We now define the charge $Q[\xi]$ associated with a generator $\xi$ of an asymptotic symmetry.
Let us fix an arbitrary defining function $\Omega$, which uniquely determines the boundary metric $g_{(0)}$. We will see that $Q[\xi]$ does not depend on this choice.
Given any spacelike slice of $\partial \mathcal{M}$, denoted by $\mathcal{S}$, the charge associated with a generator $\xi$ of an asymptotic symmetry is the following integral over some set of coordinates on $\mathcal{S}$:
\begin{equation}
\label{eq:charge}
Q[\xi]=\int_{\mathcal{S}} d^2x \sqrt{\sigma_{(0)}}\left( \langle T_{\bar a\bar b}\rangle_{CFT} u^{\bar a} \xi_{(0)}^{\bar b}\right),
\end{equation}
where $u^{\bar a}$ is the future pointing unit vector normal to $\mathcal{S}$, $\sigma_{(0)\bar a\bar b}=g_{(0)\bar a\bar b}+u_{\bar a} u_{\bar b}$ is the metric on $\mathcal{S}$ induced by the boundary metric $g_{(0)\bar a\bar b}$.
We can now consider a different choice of defining function $\Omega'$. We must have $\Omega'=f\Omega$ for some smooth, strictly positive function $f$.
Under $\Omega\to f\Omega$, the quantities in the right hand side of \eqref{eq:charge} are rescaled by factors of $f$.\footnote{It should be noted that this is true for $\langle T_{\bar a\bar b}\rangle_{CFT}$ only in even spacetime dimensions.}
Dimensional analysis shows that all these factors must cancel, hence the entire right hand side remains invariant. 
Therefore, $Q[\xi]$ does not depend on the choice of $\Omega$ and it can only depend, in general, on the 2-dimensional boundary slice $\mathcal{S}$.

For a solution of the equations of motion, $Q[\xi]$ does not even depend on $\mathcal{S}$. In other words, $Q[\xi]$ is a constant and we say that it is conserved. To see this, we pick $\Omega$ such that $\bar{g}=\Omega^2 g$ induces a boundary metric $g_{(0)}$ for which $\xi_{(0)}$ is a Killing vector field, i.e., \eqref{eq:confKillg0second} is satisfied with vanishing left hand side. Conservation will then hold also for all other possible choices of $\Omega$, since we argued that $Q[\xi]$ is independent of the choice of defining function.
Now, consider a different 2-dimensional spacelike slice $\mathcal{S}'$ in $\partial \mathcal{M}$, and the corresponding charge $Q'[\xi]$. Let $\partial \mathcal{U}\subset \partial \mathcal{M}$ be the region in $\partial \mathcal{M}$ between $\mathcal{S}$ and $\mathcal{S}'$. We have
\begin{equation}
\label{eq:conschar}
\begin{split}
Q'[\xi]-Q[\xi]&=\int_{\mathcal{S}'} d^2x \sqrt{\sigma_{(0)}}\left( \langle T_{\bar a\bar b}\rangle_{CFT} u^{\bar a} \xi_{(0)}^{\bar b}\right)-\int_{\mathcal{S}} d^2x \sqrt{\sigma_{(0)}}\left( \langle T_{\bar a\bar b}\rangle_{CFT} u^{\bar a} \xi_{(0)}^{\bar b}\right)\\
&=\int_{\partial \mathcal{U}} d^3x \sqrt{g_{(0)}} \mathcal{D}_{(0)}^{\bar a} \left( \langle T_{\bar a\bar b}\rangle_{CFT} \xi_{(0)}^{\bar b} \right),
\end{split}
\end{equation}
where in the second equality we used the divergence theorem to obtain an integral over $\partial \mathcal{U}$ in coordinates $x^{\bar{a}}$. Now, $\mathcal{D}_{(0)}^{\bar a} \left( \langle T_{\bar a\bar b}\rangle_{CFT} \xi_{(0)}^{\bar b}\right)=\left(\mathcal{D}_{(0)}^{\bar a} \langle T_{\bar a\bar b}\rangle_{CFT}\right) \xi_{(0)}^{\bar b}+ \langle T_{\bar a\bar b}\rangle_{CFT}\mathcal{D}_{(0)}^{\bar a}\xi_{(0)}^{\bar b}$. The first term vanishes, for a solution of the equations of motion, due to the conservation of the boundary energy-momentum tensor \eqref{eq:traconsset}. Since $\langle T_{\bar a\bar b}\rangle_{CFT}$ is symmetric under swapping of the indices, the second term can be written as $\frac{1}{2}\langle T_{\bar a\bar b}\rangle_{CFT}\left(\mathcal{D}_{(0)}^{\bar a}\xi_{(0)}^{\bar b}+\mathcal{D}_{(0)}^{\bar b}\xi_{(0)}^{\bar a}\right)$, which vanishes because $\xi_{(0)}$ is a Killing vector field of $(\partial\mathcal{M},g_{(0)})$. Therefore, the integrand in \eqref{eq:conschar} vanishes, and we obtain $Q'[\xi]=Q[\xi]$. This proves the conservation of $Q[\xi]$ for a solution of the equations of motion.

\subsection{Asymptotically (globally) anti-de Sitter spacetimes}
\label{subsec:asygloAdSsp}

An asymptotically locally AdS spacetime $(\mathcal{M},g)$ is said to be an \emph{asymptotically globally AdS spacetime} if its conformal boundary structure $(\partial \mathcal{M},[g_{(0)}])$ has manifold $\partial \mathcal{M}=\mathbb{R}\times S^2$, and conformal isometries that form the AdS isometry group $O(3,2)$. This definition was originally given in \cite{Ashtekar:1984zz}. 
It can be proved that the presence of the subgroups of time-translations, $\mathbb{R}$, and rotations, $SO(3)$, in the set of conformal isometries of $(\partial \mathcal{M},g_{(0)})$ implies that there exist global coordinates $(t,\theta,\phi)$ on $\partial \mathcal{M}$ such that a representative of the conformal class of boundary metrics is the metric of the ESU
\begin{equation}
\label{eq:ESU}
g_{(0)}=-dt^2+d\theta^2+\sin^2\theta d\phi^2.
\end{equation} 
Conversely, the conformal isometries of \eqref{eq:ESU} form the $O(3,2)$ group. Therefore, we can equivalently define asymptotically globally AdS spacetimes as asymptotically locally AdS spacetimes $(\mathcal{M},g)$ whose conformal boundary structure is the same as the conformal boundary structure of pure AdS, i.e., $(\partial \mathcal{M}=\mathbb{R}\times S^2,[g_{(0)}])$, with $g_{(0)}$ given by the metric of the ESU, \eqref{eq:ESU}.
This characterisation immediately tells us that the considerations of Section~\ref{subsec:asylocAdS} about asymptotically locally AdS spacetimes hold also for asymptotically globally AdS spacetimes with $g_{(0)}$ given by \eqref{eq:ESU}. In particular, in our Dirichlet boundary conditions, we must make the choice \eqref{eq:ESU}, in order to restrict the possible solutions of the Einstein-Klein-Gordon equations to asymptotically globally AdS spacetimes.
In the rest of this work, we consider only asymptotically globally AdS spacetimes, and we refer to them simply as asymptotically AdS spacetimes. We mention that we do not see any conceptual obstruction to the generalisation of our study to the case of asymptotically locally AdS spacetimes. 
In this section we discuss the asymptotic symmetries and conserved charges of asymptotically AdS spacetimes, we introduce two families of asymptotically AdS spacetimes containing a black hole, and we compare the definition of asymptotically AdS spacetimes in \cite{Ashtekar:1984zz} with the one in \cite{Henneaux:1985tv}.

\subsubsection{Asymptotic symmetries and conserved charges}
\label{subsubsec:asyglocha}

By definition, the asymptotic conformal isometries of an asymptotically AdS spacetime form the $O(3,2)$ group. The largest subgroup connected to the identity is $SO(3,2)$, which is generated by 10 linearly independent conformal Killing vector fields $\xi_{(0)}$ of the ESU.
According to the discussion in Section~\ref{subsec:bdysetconscharg}, the conformal Killing vector fields $\xi_{(0)}$ are in bijective correspondence with non-pure gauge, inequivalent asymptotic Killing vector fields $\xi$, which are the generators of asymptotic symmetries.
Each asymptotic symmetry corresponds to a conserved charge, therefore asymptotically AdS spacetimes have 10 conserved charges.
We will be interested, in particular, in the charges associated with a generator $k$ of asymptotic time translations and three linearly independent generators $m_i$ of asymptotic rotations. 
In order to give an explicit expression of these generators, we consider a set of coordinates $(t,x^1,\theta,\phi)$ such that, when restricted to $\partial \mathcal{M}$, $(t,\theta,\phi)$ are the global coordinates defined in Section~\ref{subsec:asygloAdSsp}.
In any such set of coordinates, the boundary limits $k_{(0)}$ and $m_{(0)i}$ are defined by $k_{(0)}=\frac{\partial}{\partial t}, m_{(0)1}=\frac{\partial}{\partial \phi}, m_{(0)2}=-\sin\phi\frac{\partial}{\partial\theta}-\cot\theta\cos\phi \frac{\partial}{\partial \phi}, m_{(0)3}=\cos\phi\frac{\partial}{\partial\theta}-\cot\theta\sin\phi \frac{\partial}{\partial \phi}$.
Notice that $k_{(0)}$ has been normalised by requiring that $k^2_{(0)}=-1$ with respect to the representative of the conformal class of boundary metrics given by the metric of ESU, \eqref{eq:ESU}, while $m_{(0)i}$ are normalised by requiring that the orbits of the corresponding $U(1)$ diffeomorphisms at the boundary have period $2\pi$.
This completely defines $k_{(0)}$ and $m_{(0)i}$, and thus it defines the asymptotic time-symmetry generator $k$ and the asymptotic rotation generators $m_{i}$, up to equivalent choices of the bulk values of $k$ and $m_{i}$. Such choices differ by the addition of a pure-gauge asymptotic Killing vector field, and are therefore considered equivalent, as per the definition in Section~\ref{subsubsec:asydiff}.
The charge associated with $k$ is called \emph{AdS energy}, $E:=Q[k]$. The charges associated with $m_i$ are the \emph{AdS angular momenta} $J_i:=-Q[m_i]$.

\subsubsection{Asymptotically AdS black hole spacetimes}
\label{subsubsec:asyblaspa}

We here introduce two families of vacuum asymptotically AdS spacetimes containing black holes. These families will be relevant for the discussions in the rest of this thesis.

The first example is the \emph{Schwarzschild-AdS spacetime}.
In the so-called \emph{Schwarzchild coordinates} $(t,r,\theta,\phi)\in(-\infty,+\infty)\times(r_+,+\infty)\times(0,\pi)\times(0,2\pi)$, the Schwarzschild-AdS metric reads
\begin{equation}
\label{eq:SchwAdS1}
g_{\text{Schw-AdS}}=-f(r)dt^2+\frac{1}{f(r)}dr^2+r^2 d{\Omega_2}^2,
\end{equation}
where $f(r)=1+r^2-\frac{2M}{r}$, and $d{\Omega_2}^2=d\theta^2+\sin^2\theta d\phi^2$ is the unit round metric on $S^2$. 
$r_+$ represents the position of the event horizon and is the real positive solution of $f(r_+)=0$. This solution exists only if $M>0$, which is what we assume in the following. If $M<0$, the spacetime displays a naked curvature singularity at $r=0$; if $M=0$, the spacetime is pure AdS.
Schwarzschild-AdS has 4 linearly independent Killing vector fields given by $K=\frac{\partial}{\partial t}$, the generator of time-translations, and $M_1=\frac{\partial}{\partial \phi}, M_2=-\sin\phi\frac{\partial}{\partial\theta}-\cot\theta\cos\phi \frac{\partial}{\partial \phi}, M_3=\cos\phi\frac{\partial}{\partial\theta}-\cot\theta\sin\phi \frac{\partial}{\partial \phi}$, generating the group of rotations $SO(3)$.
Since the orbits of the $SO(3)$ subgroup are 2-dimensional spheres, this spacetime is spherically symmetric.
Since $K$ is timelike near $\partial \mathcal{M}$ and orthogonal to hypersurfaces at constant $t$, then this spacetime is static.
Notice that the coordinates $(t,r,\theta,\phi)$ satisfy the requirement of Section~\ref{subsubsec:asyglocha}, thus we can immediately identify the generator of asymptotic time-translations as $K$, and the generators of asymptotic rotations as $M_1, M_2, M_3$.
The AdS energy of this spacetime is $E=M$, and the AdS angular momenta vanish, $J_i=0$. It can be easily seen that \eqref{eq:SchwAdS1} asymptotes to the pure AdS metric \eqref{eqn:ads4} near the AdS boundary, i.e., at large $r$.

The generalisation of Schwarzschild-AdS to a rotating black hole is the \emph{Kerr-AdS spacetime}, originally found by Carter in \cite{Carter:1968ks}. The Kerr-AdS metric can be written in \emph{Boyer-Lindquist coordinates} $(t,r,\theta,\phi)\in(-\infty,+\infty)\times(r_+,+\infty)\times(0,\pi)\times(0,2\pi)$ as
\begin{equation}
\label{eq:KerrAdS10}
g_{\text{KAdS}}=-\frac{\Delta}{\Sigma^2}\biggl(dt-\frac{a}{\Xi}\sin^2\theta d\phi\biggr)^2+
\frac{\Sigma^2}{\Delta}dr^2+\frac{\Sigma^2}{\Delta_\theta}d\theta^2+\frac{\Delta_\theta}{\Sigma^2}\sin^2\theta\biggl(a dt-\frac{r^2+a^2}{\Xi}d\phi\biggr)^2,
\end{equation}
where
\begin{eqnarray}
\label{eq:fnrule0}
&&\Delta=(r^2+a^2)(1+r^2)-2Mr, \quad\quad\Delta_\theta=1-a^2\cos^2\theta,\nonumber\\
&&\Sigma^2=r^2+a^2\cos^2\theta, \quad\quad\Xi=1-a^2,
\end{eqnarray}
and $r_+$ is the largest real solution of the equation $\Delta(r_+)=0$ and it denotes the value of the $r$-coordinate at the event horizon. In order to avoid naked singularities, the parameters $M,a$ must satisfy $M\geq M_{\text{extr}}(a)$, where the critical mass parameter $M_{\text{extr}}(a)$ is given by
\begin{equation}
\label{eq:KerrAdScondM0}
\begin{split}
M_{\text{extr}}(a)= \frac{1}{3 \sqrt{6}} \left(2 \left(a^2+1\right)+\sqrt{12a^2+\left(a^2+1\right)^2}\right)\times\\
\sqrt{-(a^2+1)+\sqrt{12a^2+\left(a^2+1\right)^2}}.
\end{split}
\end{equation}
The case $M=M_{\text{extr}}(a)$ corresponds to Kerr-AdS black holes with maximal angular velocity for given mass, called \emph{extremal} Kerr-AdS black holes. A trivial example is the case $M=0$, and thus $a=0$, which corresponds to pure AdS. We will not discuss the extremal case further. Instead, we will consider the subextremal range of parameters $M>M_{\text{extr}}(a)$.
If the rotation parameter $a$ is negative, then the redefinition $\phi\to-\phi$ leads to the metric \eqref{eq:KerrAdS10} with parameter $-a>0$. Therefore, we can choose $a\geq0$ without loss of generality. We must also have $a<1$ for \eqref{eq:KerrAdS10} to be regular. If $a=0$, \eqref{eq:KerrAdS10} reduces to the Schwarzschild metric \eqref{eq:SchwAdS1}.
It can be shown that the Kerr-AdS spacetime is stationary and axisymmetric.
$\frac{\partial}{\partial t}$ is a stationary Killing vector field, and $\frac{\partial}{\partial \phi}$ is an axial Killing vector field.
Any other Killing vector field of Kerr-AdS is a linear combination of these two.
$M$ and $a$ determine the AdS energy and AdS angular momenta as
$E=\frac{M}{\Xi^2}$, $J_1=\frac{aM}{\Xi^2}$, $J_2=J_3=0$.
Since $J_2$ and $J_3$ vanish, it is common to refer to $J_1$ as the angular momentum of a Kerr-AdS black hole and denote it by $J$.

Notice that Boyer-Lindquist coordinates $(t,r,\theta,\phi)$ are not of the type used in Section~\ref{subsubsec:asyglocha}. In fact, it is easy to verify that, in coordinates $(t,\theta,\phi)$, all metrics in the conformal class are conformally related to the metric
\begin{equation}
g_{(0)}=-dt^2+\frac{2a\sin^2\theta}{1-a^2} dtd\phi+\frac{1}{1-a^2\cos^2\theta}d\theta^2+\frac{\sin^2\theta}{1-a^2} d\phi^2,
\end{equation}
which clearly differs from \eqref{eq:ESU} for non-vanishing $a$. 
Hence, it is necessary to use other sets of coordinates to easily show that Kerr-AdS is an asymptotically AdS spacetime, and to identify the generators of asymptotic time-translations, $k$, and asymptotic rotations, $m_i$, in a simple way. 
We also mention that it is possible to define coordinates in which the Kerr-AdS metric is regular at the event horizon, thus $g_{\text{KAdS}}$, and in particular $g_{\text{Schw-AdS}}$, can be extended to the interior of the black hole. Various sets of coordinates on Kerr-AdS with the desired properties will be discussed in detail in Chapter~\ref{Chapter:KAdS}.
To conclude, we notice that the Kerr-AdS spacetime is the analog, in the case of negative cosmological constant, of the \emph{Kerr spacetime}, which describes a stationary and axisymmetric rotating black hole in a geometry that is asymptotically flat. The Kerr metric in Boyer-Lindquist coordinates $(t,r,\theta,\phi)$ can be obtained as the $r\ll1,a\ll1$ approximation of \eqref{eq:KerrAdS10}, or, equivalently, by reinserting factors of $L$ (by dimensional analysis) in \eqref{eq:KerrAdS10} and taking the limit $L\to+\infty$.

\subsubsection{Comparing definitions of asymptotically AdS spacetimes}
\label{eq:compdefasyAdS}

In this section we compare the definition of asymptotically (globally) AdS spacetimes by \cite{Ashtekar:1984zz}, presented at the beginning of Section~\ref{subsec:asygloAdSsp}, with the one given by \cite{Henneaux:1985tv}, which is also commonly used in the literature. This discussion will be useful to identify equivalent sets of boundary conditions.

The authors of \cite{Henneaux:1985tv} implicitly consider conformally compact spacetimes $(\mathcal{M},g)$ that satisfy the Einstein equations, i.e., they consider vacuum asymptotically locally AdS spacetimes.\footnote{The characterisation of \cite{Henneaux:1985tv}, and the following discussion, can be straightfowardly extended to the case of matter coupled with gravity, if we stay within the framework of asymptotically locally AdS spacetimes, i.e., we assume that the matter fields satisfy the corresponding equations of motion. Furthermore, we must require that the matter fields vanish at $\partial \mathcal{M}$, thus selecting the fastly decaying mode that does not alter the asymptotics of the metric, as explained in Section~\ref{subsubsec:FGexpans}. All the spacetimes that we study satisfy these requirements.}
Within this framework, Ref. \cite{Henneaux:1985tv} defines asymptotically AdS spacetimes as spacetimes $(\mathcal{M},g)$ satisfying the following Dirichlet boundary conditions. It is demanded that there exist coordinates $x^{\dot \alpha}=(t,r,\theta,\phi)$ near $\partial \mathcal{M}$ (with $(\theta,\phi)$ coordinates on $S^2$ and boundary $\partial \mathcal{M}$ at $r\to+\infty$), in which $g$ reads
\begin{equation}
\label{eq:HTg}
g=\hat g+h_{\dot \alpha\dot \beta}dx^{\dot \alpha} dx^{\dot \beta},
\end{equation}
where the pure AdS metric $\hat g$ has the expression \eqref{eqn:ads4} (recall that we are now setting $L=1$), and the tensor $h$, containing the deviation from the pure AdS metric, decays at large $r$ as
\begin{equation}
\label{eq:HTh}
h_{\dot m\dot n}=\mathcal{O}(r^{-1}),\quad
h_{ r\dot m}=\mathcal{O}(r^{-4}),\quad
h_{ r r}=\mathcal{O}(r^{-5}).
\end{equation}
Here, $\dot m,\dot n$ denote indices associated with the coordinates $t,\theta,\phi$.

This definition is equivalent to the one of \cite{Ashtekar:1984zz}, introduced at the beginning of Section~\ref{subsec:asygloAdSsp}, as discussed in detail in \cite{Hollands:2005wt}.
We here prove this result by showing that the metric satisfying the conditions of \cite{Henneaux:1985tv} can be brought, in a sufficiently small neighbourhood of $\partial \mathcal{M}$, to FG form \eqref{eqn:FGmetric}, satisfying the boundary conditions \eqref{eqn:FGbdymetric} with $g_{(0)}$ given by \eqref{eq:ESU}. 
We achieve this with a sequence of coordinate transformations.
Let us start by employing a compactified coordinate $\rho$ defined in terms of $r$ as in Section~\ref{subsec:pureAdS}, i.e., $r=2\rho/(1-\rho^2)$, which can be inverted as $\rho=(-1+\sqrt{1+r^2})/r$. The boundary is now at $\rho=1$.
The metric \eqref{eq:HTg} in coordinates $x^\alpha=(t,\rho,\theta,\phi)$ reads
\begin{equation}
\label{eq:HTcom}
g=\hat g+h_{\alpha\beta}dx^\alpha dx^\beta,
\end{equation}
where the pure AdS metric $\hat g$ has the expression \eqref{eqn:ads4_compact} (with $L=1$). The fall-offs of $h_{ \alpha \beta}$ can be easily obtained from \eqref{eq:HTh} using the coordinate transformation law for $\binom{0}{2}$ tensor components, i.e., the analog of \eqref{eq:tenstran12}, and the fact that $r$ is $\mathcal{O}((1-\rho)^{-1})$. We have
\begin{eqnarray}
\label{eq:sphbounconh1}
h_{\rho\alpha}&=&f_{\rho\alpha}(t,\theta,\phi)(1-\rho)^2+\mathcal{O}((1-\rho)^3), \, \textrm{ if $\alpha\neq\rho$},  \nonumber\\ 
h_{\alpha\beta}&=&f_{\alpha\beta}(t,\theta,\phi)(1-\rho)+\mathcal{O}((1-\rho)^{2}), \, \; \textrm{ otherwise},
\end{eqnarray}
for arbitrary functions $f_{\alpha\beta}(t,\theta,\phi)$. 
Then, we use a coordinate $z$ given in terms of $\rho$ by the same relation that brings the pure AdS metric \eqref{eqn:ads4} to FG form \eqref{eq:AdSFG}, i.e., $z=2(1-\rho)/(1+\rho)$ (inverted as $\rho=(2-z)/(2+z)$).\footnote{The coordinate $z$, defined in this way, is a re-definition of the radial coordinate $\rho$ and should not be confused with the third Cartesian coordinate.} Notice that $\partial\mathcal{M}$ is at $z=0$ and $z$ is $\mathcal{O}(1-\rho)$.
For small $z$, the metric in coordinates $(t,z,\theta,\phi)$ is
\begin{equation}
\label{eq:metnewcoords}
\begin{split}
g=\frac{1}{z^2}\biggl[&\quad
\left(1+f_{\rho\rho}z^3+\mathcal{O}(z^4)\right)dz^2 \\
&\hspace{0.0cm} - \left(1+z^2/2+f_{tt}z^3+\mathcal{O}(z^4)\right)dt^2  \\
&\hspace{0.0cm} + \left(1-z^2/2+f_{\theta\theta}z^3 +\mathcal{O}(z^4) \right)d\theta^2  \\
&\hspace{0.0cm} + \sin^2\theta \left(1-z^2/2+\frac{f_{\phi\phi}}{\sin^2\theta} z^3 +\mathcal{O}(z^4) \right)d\phi^2 \\ 
&\hspace{-0.0cm}+ 2 \left( f_{t\theta}z^3 +\mathcal{O}(z^4)\right)dt d\theta + 2 \left( f_{t\phi} z^3+\mathcal{O}(z^4)\right)dt d\phi  \\
&\hspace{-0.0cm}+ 2 \left(f_{\theta\phi} z^3 +\mathcal{O}(z^4)\right)d\theta d\phi - 2\left(f_{t\rho}z^4 +\mathcal{O}(z^5)\right)dtdz  \\
&\hspace{-0.0cm}-2\left(f_{\rho\theta}z^4+\mathcal{O}(z^5)\right)dzd\theta-2\left(f_{\rho\phi}z^4 +\mathcal{O}(z^5)\right)dzd\phi\biggr] ,
\end{split}
\end{equation}
where the functions $f_{\alpha\beta}(t,\theta,\phi)$ are the ones of \eqref{eq:sphbounconh1}.
Notice that \eqref{eq:metnewcoords} is not in FG form yet because the $zz$ component is not $1/z^2$, and the $tz$, $z\theta$, $z\phi$ components do not vanish up to the highest known order in $z$.
Defining 
\begin{eqnarray}
\label{eqn:FGcoords}
\bar{t}&=& t+\frac{1}{10 } \left(2 f_{t\rho}(t,\theta,\phi)+\frac{1}{3}f_{\rho\rho,t}(t,\theta,\phi)\right)z^5+\mathcal{O}(z^6),\nonumber\\
\bar{z}&=& z+\frac{1}{6}f_{\rho\rho}(t,\theta,\phi)z^4+\mathcal{O}(z^5),\nonumber\\
\bar{\theta}&=& \theta-\frac{1}{10} \left(2 f_{\rho\theta}(t,\theta,\phi)+\frac{1}{3}f_{\rho\rho,\theta}(t,\theta,\phi)\right)z^5+\mathcal{O}(z^6),\nonumber\\
\bar{\phi}&=& \phi-\frac{1}{10} \left(2 f_{\rho\phi}(t,\theta,\phi)+\frac{1}{3}f_{\rho\rho,\phi}(t,\theta,\phi)\right)z^5+\mathcal{O}(z^6),
\end{eqnarray}
which can be inverted near the boundary as
\begin{eqnarray}
\label{eqn:invertFGcoords}
t&=& \bar{t}-\frac{1}{10} \left(2 f_{t\rho}(\bar{t},\bar{\theta},\bar{\phi})+\frac{1}{3}f_{\rho\rho,\bar{t}}(\bar{t},\bar{\theta},\bar{\phi})\right)\bar{z}^5+\mathcal{O}(\bar{z}^6),\nonumber\\
z&=&\bar{z}-\frac{1}{6}f_{\rho\rho}(\bar{t},\bar{\theta},\bar{\phi})\bar{z}^4+\mathcal{O}(\bar{z}^5),\nonumber \\
\theta&=& \bar{\theta}+\frac{1}{10} \left(2 f_{\rho\theta}(\bar{t},\bar{\theta},\bar{\phi})+\frac{1}{3}f_{\rho\rho,\bar{\theta}}(\bar{t},\bar{\theta},\bar{\phi})\right)\bar{z}^5+\mathcal{O}(\bar{z}^6), \nonumber\\
\phi&=&\bar{\phi}+\frac{1}{10} \left(2 f_{\rho\phi}(\bar{t},\bar{\theta},\bar{\phi})+\frac{1}{3}f_{\rho\rho,\bar{\phi}}(\bar{t},\bar{\theta},\bar{\phi})\right)\bar{z}^5+\mathcal{O}(\bar{z}^6),
\end{eqnarray}
we finally obtain the metric in FG form:
\begin{equation}
\label{eq:asyFG}
\begin{split}
g=\frac{1}{\bar{z}^2}\biggl[ &\quad\left(1+\mathcal{O}(\bar{z}^4)\right)d\bar{z}^2  \\
&-\left(1+\frac{\bar{z}^2}{2}+\left( \textstyle{f_{tt}-\frac{1}{3}f_{\rho\rho}}\right)\bar{z}^3+\mathcal{O}(\bar{z}^4) \right)d\bar{t}^2 \\
&+ \left(1-\frac{\bar{z}^2}{2}+\left(\textstyle{f_{\theta\theta}+\frac{1}{3}f_{\rho\rho}} \right)\bar{z}^3 +\mathcal{O}(\bar{z}^4)\right)d\bar{\theta}^2\\
&\hspace{-0.0cm} +  \sin^2\bar{\theta} \left(1-\frac{\bar{z}^2}{2}+\left(\textstyle{\frac{f_{\phi\phi}}{ \sin^2\bar{\theta}}+\frac{1}{3}f_{\rho\rho}}\right)\bar{z}^3 +\mathcal{O}(\bar{z}^4)\right)d\bar{\phi}^2\\
&\hspace{-0.0cm}+ 2 \left(f_{t\theta}\bar{z}^3 +\mathcal{O}(\bar{z}^4) \right) d\bar{t} d\bar{\theta} + 2 \left( f_{t\phi}\bar{z}^3 +\mathcal{O}(\bar{z}^4)\right)d\bar{t} d\bar{\phi}  \\
&\hspace{-0.0cm}+ 2\left( f_{\theta\phi} \bar{z}^3+\mathcal{O}(\bar{z}^4)\right)d\bar{\theta} d\bar{\phi} \\
&\hspace{-0.0cm}+\mathcal{O}(\bar{z}^5)d\bar{t}d\bar{z}+\mathcal{O}(\bar{z}^5)d\bar{z}d\bar{\theta}+\mathcal{O}(\bar{z}^5)d\bar{z}d\bar{\phi}
\biggr]\,,
\end{split}
\end{equation}
where now the coefficients $f_{\alpha\beta}$ are functions of $(\bar t, \bar\theta, \bar\phi)$. Notice that $f_{\rho\rho}$ has been reabsorbed in $g_{\bar{t}\bar{t}}, g_{\bar{\theta}\bar{\theta}}, g_{\bar{\phi}\bar{\phi}}$. We see that \eqref{eq:asyFG} satisfies the FG expansion \eqref{eqn:FGbdymetric} with $g_{(0)\bar a\bar b}=-d{\bar t}^2+d{\bar\theta}^2+\sin^2{\bar\theta} d\bar\phi^2$ given by the metric of ESU, \eqref{eq:ESU}. This proves our result.
Comparing \eqref{eq:asyFG} with the pure AdS metric in FG coordinates \eqref{eq:AdSFG}, we have confirmation that the dynamics that makes $g$ differ from $\hat{g}$ arises at order $\bar{z}^3$ in the expansion of $g_{\bar{a}\bar{b}}$, as expected from the FG theorem.

We now discuss the motivations that led the authors of \cite{Henneaux:1985tv} to the boundary conditions \eqref{eq:HTh}.
Let us first mention that, given a set of coordinates $x^{\dot\alpha}=(t,r,\theta,\phi)$ in which the pure AdS metric is given by \eqref{eqn:ads4}, the asymptotic behaviour of Killing vectors $\xi$ of pure AdS spacetime in this set of coordinates is \cite{Henneaux:1985tv}
\begin{equation}
\label{eq:asyAdSKVF}
\xi^{\dot m}=\mathcal{O}(1), \quad \xi^{r}=\mathcal{O}(r),\quad
\xi^{\dot m}_{\phantom m,r}=\mathcal{O}(r^{-3}), \quad \xi^{r}_{\phantom r,r}=\mathcal{O}(r^{-1}).
\end{equation}
Ref. \cite{Henneaux:1985tv} motivates the boundary conditions \eqref{eq:HTh} by using the fact that i) they are satisfied by the Kerr-AdS metric in some set of coordinates, ii) they are preserved by diffeomorphisms generated by Killing vector fields $\xi$ of pure AdS, i.e., 
\begin{equation}
\label{eq:asyKEHT}
(\mathcal{L}_\xi h)_{\dot \alpha\dot \beta}=\mathcal{O}(h_{\dot \alpha\dot \beta})
\end{equation}
where $\xi$ satisfies \eqref{eq:asyAdSKVF};
iii) the charges $Q[\xi]$ associated with the generators of the asymptotic symmetry group of pure AdS, $SO(3,2)$, are finite.

This discussion seems to suggest that using \eqref{eq:asyAdSKVF}, plugging the ansatz $h_{\dot\alpha\dot\beta}\sim r^{p_{\dot\alpha\dot\beta}}$, with $p_{\dot\alpha\dot\beta}<0$, in the asymptotic Killing equation \eqref{eq:asyKEHT},
and solving for the exponents $p_{\dot\alpha\dot\beta}$ would lead us to the boundary conditions \eqref{eq:HTh}, i.e., $p_{\dot m\dot n}=-1, p_{r\dot m}=-4,p_{r r}=-5$.
This is exactly what is done, for an arbitrary number $D$ of spacetime dimensions, in \cite{Bantilan:2012vu}, where the authors find the $D$-dimensional generalisation of \eqref{eq:HTh}.
However, it should be noticed that this method does not lead to a unique set of exponents $p_{\dot\alpha\dot\beta}$.
For instance, it can be verified that $h'_{\dot\alpha\dot\beta}\sim r^{p'_{\dot\alpha\dot\beta}}$ with $p'_{\dot m\dot n}=-1, p'_{r\dot m}=-3,p'_{r r}=-5$ also satisfies \eqref{eq:asyKEHT}.
In fact, we can check that this method allows, in general, for fall-offs connected with each other by pure gauge diffeomorphisms.

Let us investigate this more explicitly.
Given $h_{\dot\alpha\dot\beta}$ that decays as $r^{p_{\alpha\beta}}$, i.e., $h_{\dot\alpha\dot\beta}=\mathcal{O}(r^{p_{\dot\alpha\dot\beta}})$ with $p_{\dot\alpha\dot\beta}<0$, the fact that $h'_{\dot\alpha\dot\beta}:=(h+\mathcal{L}_{\xi_{\text{pg}}} h)_{\dot\alpha\dot\beta}$, for any pure gauge vector field $\xi_{\text{pg}}$, still falls off as $r\to+\infty$ is almost trivial.
In fact, since both $h_{\dot\alpha\dot\beta}$ and $\xi_{\text{pg}}^{\dot\alpha}$ tend to 0 at $\partial \mathcal{M}$, then \eqref{eq:Lieder} tells us that $(\mathcal{L}_{\xi_{\text{pg}}} h)_{\dot \alpha\dot \beta}$ tends to 0, which implies that $h'_{\dot\alpha\dot\beta}$ tends to 0.
Let us denote by $p'_{\alpha\beta}$ the exponents of the fall-off of $h'_{\dot\alpha\dot\beta}$, i.e., $h'_{\dot\alpha\dot\beta}=\mathcal{O}(r^{p'_{\dot\alpha\dot\beta}})$ with $p'_{\dot\alpha\dot\beta}<0$. Assume now that $h$ satisfies the asymptotic Killing equation \eqref{eq:asyKEHT} for a vector field $\xi$ satisfying \eqref{eq:asyAdSKVF}. We see that, if it is possible to choose $\xi_{\text{pg}}$ such that  $\left(\mathcal{L}_\xi h'\right)_{\dot\alpha\dot\beta}=\mathcal{O}(r^{p'_{\dot\alpha\dot \beta}})$, i.e., $(\mathcal{L}_\xi \mathcal{L}_{\xi_{\text{pg}}} h)_{\dot \alpha\dot \beta}=\mathcal{O}(r^{p'_{\dot\alpha\dot \beta}-p_{\dot\alpha\dot\beta}})$, then $h'$ is another solution of the asymptotic Killing equation for $\xi$, connected with $h$ by a pure gauge vector field $\xi_{\text{pg}}$ as anticipated.

The choice of fall-offs for $h$ can be made arbitrarily within the class of fall-offs connected by pure gauge diffeomorphisms, since all the boundary conditions related by pure gauge diffeomorphisms are physically equivalent, in the sense that they allow for the same set of asymptotically AdS spacetimes, just in a different set of coordinates. In other words, it is always possible to identify a new set of coordinates $x^{\dot\alpha'}=(t',r',\theta',\phi')$ in which the asymptotics of $h'=h+\mathcal{L}_{\xi_{\text{pg}}} h$ (with $\xi_{\text{pg}}$ picked as above) are $\mathcal{O}({r'}^{p_{\alpha\beta}})$, with the same exponents $p_{\alpha\beta}$ that give the asymptotics of $h$ in coordinates $x^\alpha$. 
A practical way to prove that a given decay of $h$ in some coordinates $x^\alpha$ gives boundary conditions equivalent to \cite{Ashtekar:1984zz} and \cite{Henneaux:1985tv} is to show that it is possible to relate the coordinates $x^\alpha$ to FG coordinates $x^{\bar\alpha}$. This can be done in a way similar to the calculation outlined above, although the order of $z$ or $\bar{z}$ in the analog of relations \eqref{eqn:FGcoords} and \eqref{eqn:invertFGcoords} will in general be different.

\section{Boundary prescription}\label{sec:pre_sta}

Our goal is to combine the generalised harmonic formalism, reviewed in Section~\ref{sec:genharform}, with the theoretical notions about asymptotically AdS spacetimes, reviewed in Section~\ref{sec:asyAdS}, to develop a numerical scheme that achieves stable Cauchy evolution.
We will aim to obtain the evolution of a given set of initial data by solving the evolution equations of the generalised harmonic formalism with constraint-damping terms, i.e., the (modified) Einstein equations \eqref{eq:EFEsoufun}. For simplicity, we restrict our study to the case of gravity coupled with a massless real scalar field $\varphi$, whose evolution equation is \eqref{eqn:eoms2cart}.
However, our scheme can be generalised to massive scalar fields, as well as other types of matter fields, in a straightforward way.
As explained in Section~\ref{subsubsec:FGexpans}, a choice of boundary conditions in a given set of coordinates is a key ingredient in any evolution scheme in asymptotically AdS spacetimes.
In this section, we first state our explicit choice of boundary conditions.
We then discuss the ingredient that turns out to be crucial to achieve stability, namely an appropriate gauge choice of generalized harmonic source functions near the AdS boundary.

\subsection{Boundary conditions}
\label{subsec:bouconsphcar}

The boundary conditions that we impose are the Dirichlet boundary conditions discussed in Section~\ref{subsubsec:FGexpans}, allowing for asymptotically locally AdS solutions of the Einstein-Klein-Gordon equations. 
As mentioned above, these are often called reflective boundary conditions, since they make the AdS boundary reflect null waves.
To these, we add the choice of $g_{(0)}$ given by \eqref{eq:ESU}, which restricts the allowed solutions to the class of asymptotically AdS spacetimes, defined in Section~\ref{subsec:asygloAdSsp}. In Section~\ref{eq:compdefasyAdS}, we saw that these boundary conditions can be equivalently stated in various types of coordinates by imposing a certain fall-off of the tensor $h$, describing the deviation of the spacetime metric $g$ from the pure AdS metric $\hat g$. We now specify the coordinates picked for our simulations, and the expression of the boundary conditions in these coordinates.
Recall that a solution in the generalised harmonic formalism is given, in an arbitrary set of coordinates $x^\alpha$ in terms of the metric $g_{\alpha\beta}$, the scalar field $\varphi$, and the generalised harmonic source functions $H_\alpha$. Therefore, we will need to explicitly state the fall-offs for all these quantities.
Given the fall-offs for the metric and the scalar field, the corresponding fall-offs for the source functions can be obtained from their definition
\begin{equation}
\label{eq:defsoufunsph}
H^\alpha:= \Box x^\alpha=\frac{1}{\sqrt{-g}}\partial_\beta (\sqrt{-g}g^{\beta\gamma}x^\alpha_{\;\;,\gamma})=\frac{1}{\sqrt{-g}}\partial_\beta(\sqrt{-g}g^{\beta\alpha}).
\end{equation}

In the following, we will refer to \emph{(compactified) spherical coordinates} as all those sets of coordinates $x^\alpha=(t,\rho,\theta,\phi)$ such that the pure AdS metric is given by \eqref{eqn:ads4_compact} (we also set $L=1$), and asymptotics of $h$, descending from our choice of Dirichlet boundary conditions, is given by \eqref{eq:sphbounconh1}, which we rewrite here for completeness:
\begin{eqnarray}
\label{eq:sphbounconh}
h_{\rho\alpha}&=&f_{\rho\alpha}(t,\theta,\phi)(1-\rho)^2+\mathcal{O}((1-\rho)^3), \, \textrm{ if $\alpha\neq\rho$},\\ 
h_{\alpha\beta}& =&f_{\alpha\beta}(t,\theta,\phi)(1-\rho)+\mathcal{O}((1-\rho)^{2}), \, \; \textrm{ otherwise},
\end{eqnarray}
for arbitrary functions $f_{\alpha\beta}(t,\theta,\phi)$. 
The first few Greek indices $\alpha,\beta,\gamma,\dots$ will be used to denote spherical coordinates.\footnote{Note that this definition specifies spherical coordinates only up to some order in $(1-\rho)$ near the boundary $\rho=1$. Therefore, all sets of coordinates differing from each other at higher orders in $(1-\rho)$ belong to the class of spherical coordinates.}
Regarding the scalar sector, we recall that our choice of Dirichlet boundary conditions on $\varphi$, $\varphi|_{\partial \mathcal{M}}=0$, gives the fall-off \eqref{eq:FGscal} in FG coordinates.
In particular, we have $\Delta_+=3$ for a massless scalar field.
Hence, in spherical coordinates, obtained from FG coordinates as explained in Section~\ref{eq:compdefasyAdS} (in particular, $\bar{z}$ is $\mathcal{O}(1-\rho)$), the fall-off of $\varphi$ reads
\begin{equation}
\label{eq:sphbounconphi}
\varphi= f(t,\theta,\phi)(1-\rho)^3+\mathcal{O}((1-\rho)^4)
\end{equation}
for an arbitrary function $f(t,\theta,\phi)$.
From \eqref{eq:defsoufunsph} and \eqref{eq:sphbounconh}, we can obtain the fall-offs of the source functions in spherical coordinates.
Denoting the pure AdS values by $\hat{H}_\alpha$, we have
\begin{eqnarray}\label{eq:sphbouncondsoufunc}
H_\alpha & =& \hat{H}_\alpha+f_\alpha(t,\theta,\phi)(1-\rho)^3+\mathcal{O}((1-\rho)^4), \, \textrm{ if $\alpha\neq\rho$,}  \\
H_\rho &=&  \hat{H}_\rho+f_\rho(t,\theta,\phi)(1-\rho)^2+\mathcal{O}((1-\rho)^3),
\end{eqnarray}
for some functions $f_\alpha$, determined by the functions $f_{\alpha\beta}$ appearing in \eqref{eq:sphbounconh}.

Now, we consider another set of coordinates $x^\mu=(t,x,y,z)$, defined in terms of spherical coordinates by the relations used in Section~\ref{subsec:pureAdS} to define Cartesian coordinates in pure AdS, i.e., $x=\rho\cos\theta$, $y=\rho\sin\theta\cos\phi$, $z=\rho\sin\theta\sin\phi$.
We have already seen that the pure AdS metric in these coordinates is given by \eqref{eq:ads4_diag}. The fall-offs of $h$ and $\varphi$ can be deduced from \eqref{eq:sphbounconh} and \eqref{eq:sphbounconphi}, and are given by 
\begin{eqnarray}
\label{eq:carbouncondh}
h_{\mu\nu} &=& f_{\mu\nu}(t,x,y,z)|_{\rho=1}(1-\rho)+\mathcal{O}((1-\rho)^{2}), \\
\label{eq:carbouncondphi}
\varphi &=& f(t,x,y,z)|_{\rho=1}(1-\rho)^3+\mathcal{O}((1-\rho)^{4}), 
\end{eqnarray}
for arbitrary $f_{\mu\nu}$ and $f$, and where $\rho = \rho(x,y,z):=\sqrt{x^2+y^2+z^2}$.\footnote{Similarly to spherical coordinates, Cartesian coordinates are specified only up to some order in $(1-\rho)$ near the boundary $\rho=1$. Therefore, all sets of coordinates differing from each other at higher orders in $(1-\rho)$ belong to the class of Cartesian coordinates.}
In the following, we will refer to \emph{Cartesian coordinates} as all those sets of coordinates $x^\mu=(t,\rho,\theta,\phi)$ in which the pure AdS metric is given by \eqref{eq:ads4_diag}, and the asymptotics of $h$, descending from our choice of Dirichlet boundary conditions, is \eqref{eq:carbouncondh}.
The last few Greek indices $\mu,\nu,\rho,\dots$ will be used to denote Cartesian coordinates.
In Cartesian coordinates, denoting the pure AdS values by $\hat{H}_\mu$, \eqref{eq:defsoufunsph} and \eqref{eq:carbouncondh} imply
\begin{equation}\label{eq:carbouncondsoufun}
H_\mu=\hat{H}_\mu+f_\mu(t,x,y,z)|_{\rho=1}(1-\rho)^2+\mathcal{O}((1-\rho)^3)
\end{equation}
for some $f_\mu$, whose value at $\rho=1$ is determined by the value at $\rho=1$ of the functions $f_{\mu\nu}$ appearing in \eqref{eq:carbouncondh}.

\subsubsection{Evolution variables}
\label{subsec:cartevvarboucon}

The boundary asymptotics on asymptotically AdS spacetimes, given in Section~\ref{subsec:bouconsphcar}, can be imposed in a simple way at the AdS boundary.
This requires appropriately defining and evolving a new set of variables, from which the full solution $(g_{\mu\nu},\varphi,H_\mu)$ can be subsequently reconstructed. 
Here, we define evolution variables in the Cartesian coordinates employed by our numerical scheme.
In Appendix~\ref{sec:sphevvarboucon}, we define the corresponding evolution variables in spherical coordinates, and explicitly show how these spherical variables relate to our Cartesian evolution variables.

The Cartesian metric evolution variables, $\bar{g}_{\mu\nu}$, are defined by first considering the deviation from pure AdS in Cartesian coordinates, $h_{\mu\nu}=g_{\mu\nu}-\hat{g}_{\mu\nu}$, then stripping $h_{\mu\nu}$ of as many factors of $(1-\rho^2)$ as needed so that each component falls off linearly in $(1-\rho)$ near the AdS boundary at $\rho=1$.\footnote{Looking at the boundary conditions \eqref{eq:carbouncondh}, it seems natural to factor out $(1-\rho)$ rather than $(1-\rho^2)$. However, the latter is preferred since it preserves the even/odd character in the $\rho$ variable.}
We see from \eqref{eq:carbouncondh} that the metric evolution variables $\bar{g}_{\mu\nu}$ that satisfy these requirements are simply 
\begin{equation}\label{eq:gbarcart}
\bar{g}_{\mu\nu}=h_{\mu\nu}\,.
\end{equation}
Similarly, the Cartesian boundary condition on the scalar field \eqref{eq:carbouncondphi} suggests that we use the evolution variable
\begin{equation}
\label{eq:phibarcart}
\bar{\varphi}=\frac{\varphi }{(1-\rho^2)^2}\,.
\end{equation}
Finally, the boundary conditions \eqref{eq:carbouncondsoufun} on $H_\mu$ suggest the use of
\begin{equation}\label{eq:soufunb}
\bar{H}_\mu=\frac{H_\mu-\hat{H}_\mu}{1-\rho^2 }\,.
\end{equation}
For evolved variables defined in this way, the Dirichlet boundary conditions \eqref{eq:carbouncondh}, \eqref{eq:carbouncondphi}, \eqref{eq:carbouncondsoufun} can be easily imposed by setting the boundary values to zero:
\begin{equation}
\label{eq:dirbc}
 \bar{g}_{\mu\nu}\big|_{\rho=1}=0\,,\quad \bar{\varphi}\big|_{\rho=1}=0\,,\quad \bar{H}_\mu\big|_{\rho=1}=0\,.
 \vspace{+0.5cm}
 \end{equation}

\subsection{Gauge choice for stability}\label{sec:gauge_choice}

Coordinates over the entire spacetime are fully determined only once we choose the gauge source functions $H_\mu$. This amounts to the choice of the coordinates in which the solution for $g$ and $\varphi$ is given.
In Cartesian coordinates, as can be seen from \eqref{eq:carbouncondsoufun}, $H_\mu$ are fixed up to order $1-\rho$ by their pure AdS values $\hat{H}_\mu$ in an expansion near the AdS boundary.
We might expect that different choices of $H_\mu$ at the next order in this expansion, $(1-\rho)^2$, would amount to choosing different sets of Cartesian coordinates, and thus that any choice would allow to solve the evolution equations \eqref{eq:EFEsoufun} for arbitrarily long times. 
As we shall see, it turns out, instead, that an appropriate choice must be made if we wish to achieve stable evolution.
A specification of generalized harmonic source functions at order $(1-\rho)^2$ that provides stable Cauchy evolution can be obtained following the procedure detailed in this section.

The first step involves expanding the evolved variables, $\bar{g}_{\mu \nu}$, $\bar{H}_{\mu}$ and $\bar{\varphi}$, in a power series for small values of the defining function $q\equiv (1-\rho)$.
By construction, these evolved variables are linear in $q$ at leading order:
\begin{eqnarray}
\label{eqn:qexpg}
\bar{g}_{\mu \nu} &=& \bar{g}_{(1) \mu \nu} q + \bar{g}_{(2) \mu \nu} q^2 + \bar{g}_{(3) \mu \nu} q^3 + \mathcal{O}(q^4), \\
\label{eqn:qexpH}
\bar{H}_{\mu} &=&  \bar{H}_{(1) \mu} q + \bar{H}_{(2) \mu} q^2 + \bar{H}_{(3) \mu} q^3 + \mathcal{O}(q^4) ,\\
\label{eqn:qexpphi}
\bar{\varphi}  &=& \bar{\varphi}_{(1)} q + \bar{\varphi}_{(2)} q^2 + \bar{\varphi}_{(3)} q^3 + \mathcal{O}(q^4),
\end{eqnarray}
where all the coefficients are functions of the coordinates $(t,x,y,z)$ on the boundary $\rho(x,y,z)\equiv\sqrt{x^2+y^2+z^2}=1$ (or $q(x,y,z)=0$).
We now substitute these variables into the evolution equations \eqref{eq:EFEsoufun}, and we expand each component in powers of $q$. The three lowest orders, $q^{-2}$, $q^{-1}$, $q^0$, are fixed by the pure AdS metric $\hat{g}$ which itself is a solution of \eqref{eq:EFEsoufun}, so these terms vanish trivially. The remaining orders vanish only if $\bar{g}_{\mu \nu}$, $\bar{H}_{\mu}$, $\bar{\varphi}$ are a solution of \eqref{eq:EFEsoufun}.
Notice that $\bar\varphi_{(1)}$ is equal to the boundary scalar field $\langle \mathcal{O}\rangle_{CFT}$ up to numerical factors. For this reason, the terminology ``boundary scalar field'' will be used also for $\bar\varphi_{(1)}$, however we will keep using different notations for $\bar\varphi_{(1)}$ and $\langle \mathcal{O}\rangle_{CFT}$ to avoid confusion.

We are now interested in identifying the order of $q$ at which the second derivatives of $ \bar{g}_{(1) \mu \nu}$ with respect to $(t,x,y,z)$ appear.
For each component, we denote their combination by $\tilde{\Box}\bar{g}_{(1)\mu\nu}$, i.e., 
\begin{equation}
\label{eq:tildeboxbarg}
\tilde{\Box}\bar{g}_{(1)\mu\nu}:=\biggl(c^t_{\mu\nu}\frac{\partial^2}{\partial t^2}+c^x_{\mu\nu}\frac{\partial^2}{\partial x^2}+c^y_{\mu\nu}\frac{\partial^2}{\partial y^2}+c^z_{\mu\nu}\frac{\partial^2}{\partial z^2}\biggr)\bar{g}_{(1)\mu\nu}\,,
\end{equation}
for some functions $c^t_{\mu\nu}$, $c^x_{\mu\nu}$, $c^y_{\mu\nu}$, $c^z_{\mu\nu}$ of $(t,x,y,z)$ at $\rho(x,y,z)=1$.\footnote{None of these coefficients are tensors, despite the notation, and there is no sum over repeated indices.}
These derivative terms are included in the first piece of \eqref{eq:EFEsoufun}, namely in $-\frac{1}{2}g^{\rho \sigma} \bar{g}_{\mu \nu, \rho \sigma}$. From this, we can easily find their order of $q$ by recalling that the leading order of the inverse metric is given by its purely AdS piece, $g^{\mu\nu}=\mathcal{O}(\hat{g}^{\mu\nu})=\mathcal{O}(q^{2})$, and $\bar{g}_{(1)\mu\nu}$ is multiplied by $q$ in the near-boundary expression of $\bar{g}_{\mu\nu}$ (see eq. \eqref{eqn:qexpg}). Thus, $\tilde{\Box}\bar{g}_{(1)\mu\nu}$ must appear in the coefficient of order $q^{3}$ for every component of \eqref{eq:EFEsoufun}.\footnote{$\hat{g}^{\mu\nu}=\mathcal{O}(q^{2})$ is true in any number of dimensions but only for Cartesian coordinates. For an arbitrary set of coordinates, the leading power in $\hat{g}^{\mu\nu}$, and hence the order at which the operator \eqref{eq:tildeboxbarg} appears, depends on the specific component under consideration.
See Appendix~\ref{sec:sphevvarboucon} and \cite{Bantilan:2012vu} for examples in spherical coordinates in 4 and 5 dimensions, respectively.}
In other words, each component of the expansion of \eqref{eq:EFEsoufun} near $q=0$ can be written in the schematic form:
\begin{equation}\label{eq:efefullexp}
\begin{split}
0 
\quad=&\quad A_{(1)\mu\nu}q+A_{(2)\mu\nu}q^2+A_{(3)\mu\nu}q^3+A_{(4)\mu\nu}q^4+\mathcal{O}(q^5) \\
\quad=&\quad A_{(1)\mu\nu}q+A_{(2)\mu\nu}q^2+(\tilde{\Box}\bar{g}_{(1)\mu\nu}+B_{(3)\mu\nu})q^3+A_{(4)\mu\nu}q^4  +\mathcal{O}(q^5)
\end{split}
\end{equation}
or, rearranging the terms in order to obtain wave-like equations,
\begin{equation}
\label{eq:waveEFE}
\tilde{\Box}\bar{g}_{(1)\mu\nu}=-A_{(1)\mu\nu}\frac{1}{q^2}-A_{(2)\mu\nu}\frac{1}{q}-B_{(3)\mu\nu}-A_{(4)\mu\nu}q+\mathcal{O}(q^2).
\end{equation}
A similar argument shows that the terms involving the scalar field, with the fast fall-off that we have chosen in \eqref{eq:sphbounconphi}, appear in $A_{(4)\mu\nu}$ and higher order coefficients of \eqref{eq:efefullexp}. 
Similar results hold for matter fields with mass $\mu\geq 0$ in any number of dimensions, and in any set of coordinates $x^\alpha$ commonly used for numerical evolution: the terms involving fastly decaying matter fields appear at order higher than the order at which $\tilde{\Box}\bar{g}_{(1)\alpha\beta}$ appear in the near-boundary expansion of \eqref{eq:EFEsoufun}. This implies that the details of the matter sector, e.g., the value of $\mu$, do not affect the results of the prescription presented here, since only the lowest order coefficients in the expansion of \eqref{eq:EFEsoufun} are relevant.

We now explicitly write the lowest order terms of the modified Einstein equations in the wave-like form \eqref{eq:waveEFE}.
The near-boundary expansion is most easily obtained by first writing the Cartesian coordinates $(x,y,z)$ in terms of the boundary-adapted spherical coordinates $(q,\theta,\phi)$, and then expanding near $q=0$. We find
\begin{eqnarray}
\label{eqn:efett}
\tilde{\Box}\bar{g}_{(1)tt}&=&-(\cos \theta (3 \cos \theta \bar{g}_{(1)xx}-2 \bar{H}_{(1)x})+\sin\theta (3 \sin \theta \cos^2\phi \bar{g}_{(1) yy}\nonumber \\
&&\hspace{1.1cm}+3
  \sin \theta \sin \phi (2 \cos \phi \bar{g}_{(1) yz} +\sin\phi
  \bar{g}_{(1) zz})\nonumber \\
&&\hspace{1.1cm}-2 (\cos \phi \bar{H}_{(1) y}+\sin\phi
  \bar{H}_{(1) z}))+3 \sin 2 \theta \cos \phi \bar{g}_{(1) xy} \nonumber \\
&&\hspace{1.1cm}+3
  \sin 2 \theta \sin \phi \bar{g}_{(1) xz})q^{-2} +\mathcal{O}(q^{-1}),\\
\label{eqn:efetx}
\tilde{\Box}\bar{g}_{(1)tx}&=&-2 \cos \theta (3 \cos\theta \bar{g}_{(1) tx}+3 \sin \theta
  (\cos \phi \bar{g}_{(1) ty}+\sin \phi \bar{g}_{(1)tz})\nonumber \\
&&\hspace{1.1cm}-2
  \bar{H}_{(1) t})  q^{-2}+\mathcal{O}(q^{-1}),\\
\label{eqn:efety}
\tilde{\Box}\bar{g}_{(1)ty}&=&-2 \cos \phi \sin\theta (3 \cos\theta \bar{g}_{(1) tx}+3 \sin \theta
  (\cos \phi \bar{g}_{(1) ty}+\sin \phi \bar{g}_{(1)tz})\nonumber \\
&&\hspace{1.1cm}-2
  \bar{H}_{(1) t})  q^{-2}+\mathcal{O}(q^{-1}),\\
\label{eqn:efetz}
\tilde{\Box}\bar{g}_{(1)tz}&=&-2 \sin \theta \sin\phi (3 \cos\theta \bar{g}_{(1) tx}+3 \sin \theta
  (\cos \phi \bar{g}_{(1) ty}+\sin \phi \bar{g}_{(1)tz})\nonumber \\
&&\hspace{1.1cm}-2
  \bar{H}_{(1) t})  q^{-2}+\mathcal{O}(q^{-1}),\\
\label{eqn:efexx}
\tilde{\Box}\bar{g}_{(1)xx}&=&\frac{1}{4} (3 (-4 \cos ^2\theta (\bar{g}_{(1) tt}+2 \bar{g}_{(1)
  xx})+(\cos 2 \theta +3) (\bar{g}_{(1) yy}+\bar{g}_{(1)
zz}) \nonumber \\
&&\hspace{1.1cm}+8 \cos \theta \bar{H}_{(1) x})-8 \sin \theta \cos \phi 
  (3 \cos \theta \bar{g}_{(1)xy}+\bar{H}_{(1) y}) \nonumber \\
&&\hspace{1.1cm}-8 \sin\theta \sin\phi (3 \cos\theta \bar{g}_{(1) xz}+\bar{H}_{(1) z}) \nonumber \\
&&\hspace{1.1cm}+6 \sin^2\theta \cos 2 \phi (\bar{g}_{(1)yy}-\bar{g}_{(1)zz})+12
  \sin^2\theta \sin 2 \phi \bar{g}_{(1) yz})  q^{-2} \nonumber \\
&&\hspace{1.1cm}+\mathcal{O}(q^{-1}),\\
\label{eqn:efexy}
\tilde{\Box}\bar{g}_{(1)xy}&=&-\frac{1}{2} (2 \sin \theta \cos \phi (3 \cos \theta (\bar{g}_{(1)tt}+\bar{g}_{(1)xx}+\bar{g}_{(1)yy}-\bar{g}_{(1)zz})-4 \bar{H}_{(1) x}) \nonumber \\
&&\hspace{1.1cm}+3 \bar{g}_{(1)xy} (2 \cos 2\theta \sin ^2\phi +\cos 2 \phi +3)+6 \sin ^2\theta \sin 2 \phi 
  \bar{g}_{(1)xz} \nonumber \\
  &&\hspace{1.1cm}+6 \sin 2 \theta \sin \phi \bar{g}_{(1)yz}-8 \cos
  \theta  \bar{H}_{(1) y})    q^{-2} +\mathcal{O}(q^{-1}),\\
\label{eqn:efexz}
\tilde{\Box}\bar{g}_{(1)xz}&=&- (\sin \theta  \sin \phi  (3 \cos \theta  (\bar{g}_{(1)tt}+\bar{g}_{(1)xx}-\bar{g}_{(1)yy}+\bar{g}_{(1)zz})-4
   \bar{H}_{(1) x}) \nonumber \\
   &&\hspace{1.1cm}+3 \sin ^2\theta \sin 2 \phi  \bar{g}_{(1)xy}-3 \sin
   ^2\theta  \cos 2 \phi  \bar{g}_{(1)xz}+\frac{3}{2} (\cos 2 \theta +3)
   \bar{g}_{(1)xz} \nonumber \\
   &&\hspace{1.1cm}+3 \sin 2 \theta  \cos \phi  \bar{g}_{(1)yz}-4 \cos
   \theta  \bar{H}_{(1)z})    q^{-2} +\mathcal{O}(q^{-1}),
   \end{eqnarray}
\begin{eqnarray}
\label{eqn:efeyy}
\tilde{\Box}\bar{g}_{(1)yy}&=&-( (\sin \theta (3 \sin \theta  (2 \cos ^2\phi  \bar{g}_{(1)yy}+\sin 2 \phi  \bar{g}_{(1)yz}-\bar{g}_{(1)zz}) -6 \cos\phi \bar{H}_{(1) y}\nonumber \\
&&\hspace{1.1cm}+2 \sin \phi  \bar{H}_{(1) z})+6 \sin \theta \cos
   \theta \cos \phi  \bar{g}_{(1)xy}-6 \sin \theta  \cos \theta  \sin   \phi  \bar{g}_{(1)xz}\nonumber \\
&&\hspace{1.1cm}+2 \cos \theta  \bar{H}_{(1) x})-3 \sin ^2\theta \cos ^2\phi  \bar{g}_{(1)tt}\nonumber \\
&&\hspace{1.1cm}+\frac{3}{4} \bar{g}_{(1)xx} (2 \sin ^2\theta  \cos 2 \phi +\cos 2 \theta +3)  )  q^{-2} +\mathcal{O}(q^{-1}),\\
\label{eqn:efeyz}
\tilde{\Box}\bar{g}_{(1)yz}&=&-\frac{1}{2} \sin \theta (4 \sin \phi  (3 \cos \theta  \bar{g}_{(1)xy}-2 \bar{H}_{(1) y})+4 \cos \phi  (3 \cos \theta  \bar{g}_{(1)xz}-2 \bar{H}_{(1) z}) \nonumber \\
&&\hspace{1.1cm}+3 \sin \theta  \sin 2 \phi  (\bar{g}_{(1)tt}-\bar{g}_{(1)xx}+\bar{g}_{(1)yy}+\bar{g}_{(1)zz})+12 \sin \theta  \bar{g}_{(1)yz})  q^{-2} \nonumber \\
&&\hspace{1.1cm}+\mathcal{O}(q^{-1}),\\
\label{eqn:efezz}
\tilde{\Box}\bar{g}_{(1)zz}&=&(-2 \cos \theta  (3 \sin \theta  \sin \phi  \bar{g}_{(1)xz}+\bar{H}_{(1)x}) + \sin \theta  (3 \sin \theta  \bar{g}_{(1)yy}\nonumber \\
&&\hspace{1.1cm}-6 \sin\theta  \sin \phi  (\cos \phi  \bar{g}_{(1)yz}+\sin \phi 
   \bar{g}_{(1)zz}) -2 \cos \phi  \bar{H}_{(1) y} +6 \sin \phi  \bar{H}_{(1)z}) \nonumber \\
&&\hspace{1.1cm}-3  \sin ^2\theta  \sin ^2\phi  \bar{g}_{(1)tt}+\frac{3}{4}  \bar{g}_{(1)xx} (-2 \sin ^2\theta  \cos 2 \phi +\cos 2 \theta
   +3) \nonumber \\
&&\hspace{1.1cm}+3 \sin 2 \theta  \cos \phi  \bar{g}_{(1)xy})  q^{-2} +\mathcal{O}(q^{-1}), 
\end{eqnarray}
where the coordinates $(q,\theta,\phi)$ should be understood as functions of $(x,y,z)$. 
All that remains is to write down the generalized harmonic constraints $C_\mu := H_\mu-\Box x_\mu = 0$ at leading order in the same near-boundary expansion. We get
\begin{eqnarray}
\label{eqn:ct}
C_t&=&q^2 (-3 \cos \theta \bar{g}_{(1)tx}-3 \sin \theta \cos \phi \bar{g}_{(1)ty} \nonumber \\
&&\hspace{1.1cm}-3 \sin \theta \sin \phi \bar{g}_{(1)tz}+2
  \bar{H}_{(1) t})+\mathcal{O}(q^3),\\
\label{eqn:cx}
C_x&=&\frac{1}{2} q^2 (-3 \cos \theta  \bar{g}_{(1)tt}-3 \cos \theta  \bar{g}_{(1)xx}-6 \sin \theta  \cos \phi  \bar{g}_{(1)xy}-6 \sin
   \theta  \sin \phi  \bar{g}_{(1)xz} \nonumber \\
&&\hspace{1.1cm}+3 \cos \theta  \bar{g}_{(1)yy}+3
   \cos \theta  \bar{g}_{(1)zz}+4 \bar{H}_{(1) x})+\mathcal{O}(q^3),\\
\label{eqn:cy}
C_y&=&\frac{1}{2} q^2 (-3 \sin \theta  \cos \phi  \bar{g}_{(1)tt}+3 \sin
   \theta  \cos \phi  \bar{g}_{(1)xx}-6 \cos \theta  \bar{g}_{(1)xy} \nonumber \\
&&\hspace{1.1cm}-3
   \sin \theta  \cos \phi  \bar{g}_{(1) yy}-6 \sin \theta  \sin \phi
   \bar{g}_{(1)yz}\nonumber \\
&&\hspace{1.1cm}   +3 \sin \theta  \cos \phi  \bar{g}_{(1)zz}+4
   \bar{H}_{(1) y}) +\mathcal{O}(q^3),\\
\label{eqn:cz}
C_z&=&\frac{1}{2} q^2 (-3 \sin \theta \sin \phi  \bar{g}_{(1)tt}+3 \sin
   \theta \sin \phi \bar{g}_{(1)xx}-6 \cos \theta  \bar{g}_{(1)xz} \nonumber \\
&&\hspace{1.1cm}+3
   \sin \theta \sin \phi  \bar{g}_{(1)yy}-6 \sin \theta  \cos \phi    \bar{g}_{(1)yz}\nonumber \\
&&\hspace{1.1cm}-3 \sin \theta  \sin \phi  \bar{g}_{(1)zz}+4
   \bar{H}_{(1)z})+\mathcal{O}(q^3).
\end{eqnarray}

In the generalized harmonic formulation, choosing a gauge amounts to choosing a set of generalized harmonic source functions $\bar{H}_\mu$ for the entire evolution. Although we expect that many gauge choices are allowed,~\cite{Bantilan:2012vu} mentions a few that do not give rise to stable evolutions. We now present a procedure that provides the stable gauge in our Cartesian simulations. We believe that our prescription provides a stable gauge in a variety of settings of physical interest, such as higher spacetime dimensions, various couplings to matter fields, different types of global coordinates or Poincar\'{e} coordinates. Thus, it enables numerical Cauchy evolution in AdS in full generality, that is, with no symmetry assumptions.
The steps that lead to our stable gauge, in a form that can be easily applied to all previously mentioned cases, are the following.
\begin{enumerate}
\item Solve the leading order of the near-boundary generalized harmonic constraints for $\bar{H}_{(1)\mu}$.
For example, in the Cartesian case, the leading orders of \eqref{eqn:ct}--\eqref{eqn:cz} vanish for:
\begin{eqnarray}\label{eqn:target_gauge_txyz_step1}
\bar{H}_{(1)t}&=&\frac{3}{2\sqrt{x^2+y^2+z^2}}(x \bar{g}_{(1)tx}+y\bar{g}_{(1)ty}+z\bar{g}_{(1)tz}), \nonumber\\
\bar{H}_{(1)x}&=&\frac{3}{4\sqrt{x^2+y^2+z^2}}(2y \bar{g}_{(1)xy}+2z \bar{g}_{(1)xz}\nonumber \\
&&\hspace{1.1cm}+x(\bar{g}_{(1)tt}+ \bar{g}_{(1)xx}-\bar{g}_{(1)yy}-\bar{g}_{(1)zz})), \nonumber \\
\bar{H}_{(1)y}&=&\frac{3}{4\sqrt{x^2+y^2+z^2}}(2x \bar{g}_{(1)xy}+2z \bar{g}_{(1)yz}\nonumber \\
&&\hspace{1.1cm}+y(\bar{g}_{(1)tt}+ \bar{g}_{(1)xx}-\bar{g}_{(1)yy}-\bar{g}_{(1)zz})), \nonumber \\
\bar{H}_{(1)z}&=&\frac{3}{4\sqrt{x^2+y^2+z^2}}(2x \bar{g}_{(1)xz}+2y \bar{g}_{(1)yz}\nonumber \\
&&\hspace{1.1cm}+z(\bar{g}_{(1)tt}+ \bar{g}_{(1)xx}-\bar{g}_{(1)yy}-\bar{g}_{(1)zz})).
\end{eqnarray}

\item Let $N_{(1)}$ be the lowest order in $q$ appearing in the near-boundary expansions of all the $\tilde{\Box}\bar{g}_{(1)\mu\nu}$. Plug the source functions obtained in step 1 into the $q^{N_{(1)}}$ terms of the near-boundary expansions $\tilde{\Box}\bar{g}_{(1)\mu\nu}$.
This gives a number of independent equations that, together with their derivatives, ensure tracelessness and conservation of the boundary energy-momentum tensor (see Section~\ref{sec:bouset2}).\footnote{We show this in Appendix~\ref{sec:sphevvarboucon} using spherical coordinates, since they are adapted to the AdS boundary and make the proof less unwieldy.}
Solve these equations for an equal number of metric coefficients $\bar g_{(1)\mu\nu}$ and their derivatives.
In the Cartesian case, $N_{(1)}=-2$ and there is only one independent equation given by
\begin{equation}
\label{eq:cart_tracelessness}
\bar{g}_{(1)tt}-\bar{g}_{(1)xx}-\bar{g}_{(1)yy}-\bar{g}_{(1)zz}=0,
\end{equation}
which we can solve, for instance, in terms of $\bar g_{(1)tt}$.

\item Plug the solutions to the equations in step 2 into the gauge obtained in step 1.
In Cartesian coordinates, using \eqref{eq:cart_tracelessness} to eliminate $\bar{g}_{(1)tt}$ from \eqref{eqn:target_gauge_txyz_step1}, we have
\begin{eqnarray}\label{eqn:target_gauge_txyz}
\bar{H}_{(1)t}&=&\frac{3}{2\sqrt{x^2+y^2+z^2}}(x \bar{g}_{(1)tx}+y\bar{g}_{(1)ty}+z\bar{g}_{(1)tz}),\nonumber\\
\bar{H}_{(1)x}&=&\frac{3}{2\sqrt{x^2+y^2+z^2}}(x \bar{g}_{(1)xx}+y\bar{g}_{(1)xy}+z\bar{g}_{(1)xz}), \nonumber \\
\bar{H}_{(1)y}&=&\frac{3}{2\sqrt{x^2+y^2+z^2}}(x \bar{g}_{(1)xy}+y\bar{g}_{(1)yy}+z\bar{g}_{(1)yz}), \nonumber \\
\bar{H}_{(1)z}&=&\frac{3}{2\sqrt{x^2+y^2+z^2}}(x \bar{g}_{(1)xz}+y\bar{g}_{(1)yz}+z\bar{g}_{(1)zz}).
\end{eqnarray}
\end{enumerate}
This is the asymptotic gauge condition that we have empirically verified leads to stable 3+1 evolution of asymptotically AdS spacetimes in Cartesian coordinates.
Other choices of asymptotic source functions may enjoy similar stability properties.
The choice of $\bar{H}_\mu$ in the bulk is still completely arbitrary and the functional form that we implement in our simulations is detailed explicitly in Appendix~\ref{sec:GCbulk}.

The rationale for this procedure is as follows.
Recall that if $C_\mu=0$ and $\partial_t C_\mu=0$ are satisfied at $t=0$\footnote{This condition is satisfied by our initial data, see Appendix~\ref{sec:initdata}.}, and the boundary conditions are consistent with $C_\mu=0$ being satisfied at the boundary for all time, then, at the analytical level, the generalized harmonic constraint $C_\mu=0$ remains satisfied in the interior for all time. 
The addition of constraint damping terms to the Einstein equations, eq. \eqref{eq:EFEsoufun}, helps to ensure that deviations at the level of the discretized equations remain under control. 
Thus, in solving the expanded system of equations \eqref{eq:EFEsoufun}, we are assured that only the subset of solutions that are also solutions of the Einstein equations are being considered.
With this in mind, the near-boundary form of \eqref{eq:EFEsoufun}, given by \eqref{eq:efefullexp}, implies that our task in obtaining a solution is to satisfy $A_{(i)\mu\nu}=0$ for all $i$, and for some choice of source function variables $\bar{H}_\mu$. 
This task is significantly eased by picking a gauge, through a suitable choice of $\bar{H}_\mu$, that eliminates $A_{(1)\mu\nu}$, i.e., the lowest order of the expansion of the Einstein equations near the AdS boundary.
This is precisely what the above set of steps is designed to do, and it is why we did not stop at the gauge obtained in step 1, \eqref{eqn:target_gauge_txyz_step1}, which would have resulted in a gauge that does not explicitly set $A_{(1)\mu\nu}=0$. 

Finally, it is also important to develop an understanding of the reason why the choice of $\bar{H}_\mu$ is not completely free. 
Although identifying every cause for the instability of a simulation is usually very complicated, one practical reason is clear and can be understood with the following example in Cartesian coordinates.
Suppose we choose a gauge in which, after some time $t>t_0$, $\bar{H}_{(1)t}$ takes the value
\begin{equation} 
\bar{H}_{(1)t}(t>t_0)=\frac{3}{2\sqrt{x^2+y^2+z^2}}(x \bar{g}_{(1)tx}+y\bar{g}_{(1)ty}+z b_t),
\end{equation} 
where $b_t\in \mathbb{R}$ is a possibly vanishing constant.
According to \eqref{eqn:target_gauge_txyz_step1}, the requirement that $C_t=0$ now implies $\bar{g}_{(1)tz}=b_t$. 
Even though this condition does not violate any of the requirements above, it is an additional Dirichlet boundary condition that must be imposed for $t>t_0$ if we hope to find a solution for this example.\footnote{The Dirichlet boundary condition $\bar{g}_{tz}|_{\rho=1}=0$ clearly does not restrict $\bar{g}_{(1)tz}$.} Although imposing boundary conditions that change with time is of interest in certain studies motivated by the AdS/CFT correspondence, for simplicity we do not consider such cases in this article. It should be straightforward to generalize our prescription for time-dependent boundary conditions. 

\subsection{Boundary energy-momentum tensor, energy and angular momenta}
\label{sec:bouset2}

In the simulations we output the boundary energy-momentum tensor of the dual CFT.
Its expression in FG coordinates was given in \eqref{eq:holoprebdyT}.
In this section, we follow the prescription of \cite{Balasubramanian:1999re} to obtain the analytic expressions for the boundary energy-momentum tensor, the AdS energy and AdS angular momenta that we implemented in our scheme.
We employ spherical coordinates $x^\alpha=(t,\rho,\theta,\phi)$, as they are adapted to the AdS boundary $\partial \mathcal{M}=\mathbb{R}\times S^2$.
In Appendix~\ref{sec:HoloRen}, we compare the result obtained in this way with the prescription of Section~\ref{subsec:bdysetconscharg} by \cite{deHaro:2000vlm}.
In order to obtain the numerical values of all these quantities in spherical coordinates, we will have to convert the evolution variables in Cartesian coordinates $\bar{g}_{\mu\nu}$, provided by our numerical scheme, into their counterparts $\bar{g}_{\alpha\beta}$ in spherical coordinates. We do this in Appendix~\ref{sec:sphevvarboucon}, through the transformation \eqref{eq:cartosph}.

Let us denote by $x^a=(t,\theta,\phi)$ the coordinates on timelike hypersurfaces $\partial \mathcal{M}_q$ at fixed $\rho$ (or $q$).
To compute the holographic energy-momentum tensor of the boundary CFT, $\langle T_{ab}\rangle_{CFT}$, we first compute the \emph{quasi-local energy-momentum tensor} $^{(q)}T_{\alpha\beta}$ at $\partial \mathcal{M}_q$ as prescribed in \cite{Balasubramanian:1999re}. 
We have
\begin{equation}
\label{eq:qslocset}
^{(q)}T_{\alpha\beta}=\frac{1}{8\pi}\biggl(\;  \Theta_{\alpha\beta}-\Theta \;\omega_{\alpha\beta}-2\omega_{\alpha\beta}+ G_{\alpha\beta} \biggr),
\end{equation}
where $\Theta_{\alpha\beta}=-\omega^\gamma_{\alpha}\omega^\delta_\beta\nabla_{\gamma}S_{\delta}$ is the extrinsic curvature of $\partial \mathcal{M}_q$, $\omega_{\alpha\beta}=g_{\alpha\beta}-S_\alpha S_\beta$ is the induced metric on $\partial \mathcal{M}_q$ (in four-dimensional form), $S^\alpha$ is the spacelike, outward pointing unit vector normal to $\partial \mathcal{M}_q$ and $G_{\alpha\beta}$ is the Einstein tensor of $\partial \mathcal{M}_q$.\footnote{Notice the different sign in the last term of \eqref{eq:qslocset} with respect to \cite{Balasubramanian:1999re}. When comparing the two results, recall that in our expressions we set $L=1$.}$^{,}$\footnote{All these tensors, although defined on the tangent space of the spacetime manifold $\mathcal{M}$, are invariant under projection $\omega^\alpha_{\beta}=\delta^\alpha_{\beta}-S^\alpha S_\beta$ onto $\partial \mathcal{M}_q$. Therefore, they can be identified, under a natural (i.e., basis-independent) isomorphism, with tensors defined on the tangent space of $\partial \mathcal{M}_q$. The components of tensors on $\partial \mathcal{M}_q$ in coordinates $x^a$ is simply given by taking the components of tensors on $\mathcal{M}$ in coordinates $x^\alpha$ and disregarding every combination of indices that includes an index $\rho$. See \cite{Hawking:1973uf} for more details on this correspondence. We do not make the $\Omega$-dependence of all the quantities explicit to avoid cluttering the notation.} We will be interested in the value of $^{(q)}T_{\alpha\beta}$ for $q$ close to 0, i.e., near the AdS boundary.
Restricting to the indices corresponding to the coordinates $x^a$, we obtain the boundary energy-momentum tensor as
\begin{equation}
\label{eq:BalKrabdyset}
\langle T_{ab}\rangle_{CFT}=\lim_{q\to0}\frac{1}{q} \;^{(q)}T_{ab}\,.
\end{equation}
Notice that in this definition we have not required that $g$ solves the equations of motion.
This definition agrees with \eqref{eq:holoprebdyT} for a solution of the equations of motion, as explicitly shown in Appendix~\ref{sec:HoloRen}.

Following \cite{Balasubramanian:1999re}, we can also compute the conserved charges associated with the generators of the asymptotic symmetry group, $\xi$. At each time $t$ of evolution, we take a spacelike two-dimensional surface $\mathcal{S}_q$ in $\partial \mathcal{M}_q$, with induced metric $\sigma_{ab}=\omega_{ab}+u_a u_b$, where $u_a=-N(dt)_a$ is the future pointing unit 1-form normal to $\mathcal{S}_q$ in $\partial \mathcal{M}_q$, and $N$ is the lapse of $\mathcal{S}_q$ regarded as a surface of $\partial \mathcal{M}_q$. The charge associated with $\xi$ is given by
\begin{equation}
\label{eq:charge2}
Q[\xi]=\lim_{q\to0}\int_{\mathcal{S}_q} d\theta d\phi \sqrt{\sigma} ( ^{(q)}T_{ab} u^a \xi^b)\,.
\end{equation}
Once again, \eqref{eq:charge2} gives the same result as \eqref{eq:charge} for a solution of the equations of motion.
In particular, \eqref{eq:charge2} is conserved for a solution of the equations of motion.

The holographic energy-momentum tensor can be expressed in terms of the leading order coefficients of the near-boundary expansion of $\bar{g}_{\alpha\beta}$. We find
\begin{eqnarray}
\label{eq:set_explicit}
\langle T_{tt}\rangle_{CFT}&\hspace{-0.15cm}=&\hspace{-0.15cm}\frac{1}{16\pi} \left(2\bar{g}_{(1)\rho\rho}+3\bar{g}_{(1)\theta\theta}+3\frac{\bar{g}_{(1)\phi\phi}}{\sin^2\theta}\right), \nonumber \\
\langle T_{t\theta}\rangle_{CFT}&\hspace{-0.15cm}=&\hspace{-0.15cm}\frac{3}{16\pi}\bar{g}_{(1)t\theta}, \nonumber \\
\langle T_{t\phi}\rangle_{CFT}&\hspace{-0.15cm}=&\hspace{-0.15cm}\frac{3}{16\pi}\bar{g}_{(1)t\phi}, \nonumber \\
\langle T_{\theta\theta}\rangle_{CFT}&\hspace{-0.15cm}=&\hspace{-0.15cm}\frac{1}{16\pi} \left(3\bar{g}_{(1)tt}-2\bar{g}_{(1)\rho\rho}-3\frac{\bar{g}_{(1)\phi\phi}}{\sin^2\theta}\right), \nonumber \\
\langle T_{\theta\phi}\rangle_{CFT}&\hspace{-0.15cm}=&\hspace{-0.15cm}\frac{3}{16\pi}\bar{g}_{(1)\theta\phi}, \nonumber \\
\langle T_{\phi\phi}\rangle_{CFT}&\hspace{-0.15cm}=&\hspace{-0.15cm}\frac{\sin^2\theta}{16\pi} \left(3\bar{g}_{(1)tt}-2\bar{g}_{(1)\rho\rho}-3\bar{g}_{(1)\theta\theta}\right).
\end{eqnarray}

We can also express the conserved charges in terms of $\bar{g}_{(1)\alpha\beta}$. Here, we are interested in the AdS energy and AdS angular momenta. Recall that these are the conserved charges associated, respectively, with the generator of time translations, $k$, and the generators of $SO(3)$ rotations, $m_i$.
Since spherical coordinates are coordinates of the type considered in Section~\ref{subsubsec:asyglocha}, we can immediately identify the generator of time translations as $k=\frac{\partial}{\partial t}$, and the generators of $SO(3)$ rotations as $m_1=\frac{\partial}{\partial \phi}, m_2=-\sin\phi\frac{\partial}{\partial\theta}-\cot\theta\cos\phi \frac{\partial}{\partial \phi}, m_3=\cos\phi\frac{\partial}{\partial\theta}-\cot\theta\sin\phi \frac{\partial}{\partial \phi}$.
Plugging these individually in \eqref{eq:charge2}, we obtain
\begin{equation}
\label{eq:AdSmasscalc}
E:=Q[k]=\int_0^\pi d\theta \int_0^{2\pi}d\phi\frac{\sin\theta}{16\pi} \left(2\bar{g}_{(1)\rho\rho}+3\bar{g}_{(1)\theta\theta}+3\frac{\bar{g}_{(1)\phi\phi}}{\sin^2\theta}\right),
\end{equation}
and
\begin{eqnarray}
\label{eq:AdSangmom}
\hspace{-1cm} J_1:=-Q[m_1]&=&-\int_0^\pi d\theta \int_0^{2\pi}d\phi \frac{3\sin\theta}{16\pi}\bar {g}_{(1)t\phi}\,,\nonumber\\\
\hspace{-1cm}J_2:=-Q[m_2]&=&\int_0^\pi d\theta \int_0^{2\pi}d\phi\frac{3\sin\theta}{16\pi}\left(\sin\theta\sin\phi\bar {g}_{(1)t\theta}+\cos\theta\cos\phi \bar {g}_{(1)t\phi} \right),\nonumber\\
\hspace{-1cm}J_3:=-Q[m_3]&=&\int_0^\pi d\theta \int_0^{2\pi}d\phi\frac{3\sin\theta}{16\pi}\left(-\sin\theta\cos\phi\bar {g}_{(1)t\theta}+\cos\theta\sin\phi \bar {g}_{(1)t\phi} \right).
\end{eqnarray}

We can now use a representative of the conformal class of boundary metrics, $g_{(0)}=-dt^2+d\theta^2+\sin^2\theta d\phi^2$, to raise one index of $\langle T_{ab}\rangle_{CFT}$ and solve the eigenvalue problem $\langle {T^a}_{b}\rangle_{CFT} v^b=\Lambda_v v^a$ at each point along the AdS boundary. In this way, assuming that $\langle T_{ab}\rangle_{CFT}$ satisfies the weak energy condition, we obtain the \emph{energy density of the boundary CFT}, $\epsilon$, as minus the eigenvalue associated to the unique (up to rescaling) timelike eigenvector.\footnote{The \emph{weak energy condition} for an energy-momentum tensor $T_{ab}$ requires that $T_{ab}V^a V^b\geq 0$ for any causal vector $V^a$, at any point. If $\pm \langle T_{ab}\rangle_{CFT}$ fail to satisfy the weak energy condition, the $L^2$-norm of $\langle T_{ab}\rangle_{CFT}$, $||\langle T_{ab}\rangle_{CFT}||_2$, can have complex conjugate pairs of eigenvalues and no real timelike eigenvector, as pointed out in footnote 9 of \cite{Chesler:2013lia}.} Similarly, the \emph{boundary anisotropy} is given by $\Delta p\equiv|p_1-p_2|$, where $p_1$ and $p_2$ are the eigenvalues associated with, respectively, the remaining two spacelike eigenvectors.

In Section~\ref{subsec:bdysetconscharg}, we saw that the boundary energy-momentum tensor must be trace-free and conserved for a solution of the equations of motion.
Computing the trace of the energy-momentum tensor, $\langle \text{tr}T\rangle_{CFT}=g_{(0)}^{ab} \langle T_{ab}\rangle_{CFT}$, we obtain
\begin{equation}
\label{eq:tracecalc}
\langle \text{tr}T\rangle_{CFT}=\frac{3}{8\pi}\biggl(\bar{g}_{(1)tt}-\bar{g}_{(1)\rho\rho}-\bar{g}_{(1)\theta\theta}-\frac{\bar{g}_{(1)\phi\phi}}{\sin^2\theta}\biggr).
\end{equation}
If we convert the spherical quantities into their Cartesian counterparts 
we see that $\langle \text{tr}T\rangle_{CFT}$ depends only on the factor $\bar{g}_{(1)tt}-\bar{g}_{(1)xx}-\bar{g}_{(1)yy}-\bar{g}_{(1)zz}$. 
We saw in \eqref{eq:cart_tracelessness} that this factor vanishes.
This is an important sanity check: we see that tracelessness of the energy-momentum tensor, expected for a CFT in odd spacetime dimensions, is ensured by the lowest order in the near boundary expansion of the Einstein equations, provided that the generalized harmonic constraints are satisfied.
In other words, tracelessness of the boundary energy-momentum tensor is, in our scheme, directly tied to how close our numerical solution is to a solution of the Einstein field equations. 
We check that we are indeed converging to such a solution in Appendix~\ref{sec:convbulk}.
In practice, we monitor $\langle \text{tr}T\rangle_{CFT}$ as an estimate of the solution error \eqref{eq:solerr}.
Another important check that we performed is the conservation of $\langle T_{ab}\rangle_{CFT}$. 
The simplest way to prove this is by using the near-boundary expansion of the Einstein equations in spherical coordinates, as done in Appendix~\ref{sec:sphevvarboucon}.

\section{Numerical evolution scheme}\label{sec:numerical_scheme}

In this section we consider the core elements of the numerical scheme used in this study.
We start by discussing the numerical features on which this scheme relies for solving the initial-boundary value problem in AdS.
We then describe our apparent horizon finder and the method with which we excise trapped regions.

\subsection{Numerics of the initial-boundary value problem}
\label{sec:numcauprob}
We solve the Einstein equations in generalized harmonic form \eqref{eq:EFEsoufun} with constraint damping terms, coupled with the massless Klein-Gordon equation \eqref{eqn:eoms2cart}.
We obtain asymptotically AdS spacetimes in Cartesian coordinates $x^\mu=(t,x,y,z)$. 
The solution is determined in terms of the metric, scalar field and source function variables $(\bar{g}_{\mu\nu},\bar{\varphi},\bar{H}_\mu)$ defined in Section~\ref{subsec:cartevvarboucon}. We substitute the definitions of these variables, \eqref{eq:gbarcart}--\eqref{eq:soufunb}, in the equations of motion and analytically remove all the purely AdS terms.
The resulting PDEs are discretized with second order finite difference derivative stencils in time and space, and then integrated in time using an iterative Newton-Gauss-Seidel relaxation procedure with a three time level hierarchy, as described in Section~\ref{sec:NGS}.
The source function variables $\bar{H}_\mu$ near the AdS boundary are set as we have prescribed in \eqref{eqn:target_gauge_txyz}, whilst deep in the bulk they are set to zero. 
In between, we use smooth transition functions to interpolate between the near boundary and the bulk regions, see Appendix~\ref{sec:GCbulk} for the details of our full implementation. 

We use the PAMR/AMRD libraries \cite{PAMR} for running these simulations in parallel on Linux computing clusters.
Although these libraries have adaptive mesh refinement capabilities, numerical evolution is performed on a grid with fixed refinement.
The numerical grid is in $(t,x,y,z)$ with $t \in [0,t_{max}]$, $x \in [-1,1]$, $y \in [-1,1]$, $z \in [-1,1]$.
The typical grid resolution employed in the simulations described in this chapter uses $N_x=N_y=N_z=325$ points in each of the Cartesian directions, with equal grid spacings $\Delta x = \Delta y = \Delta z =: \Delta$. We will give the specifics of the simulation presented in Chapter~\ref{Chapter:KAdS} in that chapter.

The time step of evolution is determined by $\Delta t=\lambda \Delta$. 
Although we do not perform a detailed analysis of the stability of our finite difference scheme, the Courant-Friedrichs-Lewy (CFL) condition for stability is expected to be satisfied as long as the CFL factor $\lambda$ is set to a value well below 1. Thus, we use $\lambda=0.3$ for the simulations described in this chapter. Notice that the most remarkable advantage of using Cartesian coordinates is that the CFL condition, introduced in Section~\ref{sec:NGS}, does not severely restrict the CFL factor as it would in spherical coordinates, hence allowing simulations to reach large evolution times with modest computational resources.
In contrast, spherical coordinates $(t,\rho,\theta,\phi)$ with fixed resolution $\Delta\rho,\Delta\theta,\Delta\phi$ would necessitate $\Delta t = \lambda \min(\Delta\rho, \rho_{min} \Delta\theta, \rho_{min} \Delta\theta \Delta\phi)$.
$\rho$ takes its smallest value $\rho_{min}=\Delta\rho$ at points next to the origin of the spherical coordinate frame, i.e., $\rho=0$.
Hence, at these points, which must be evolved in studies of gravitational collapse and black hole formation, $\Delta t$ would become prohibitively small for high resolutions, i.e., for small $\Delta\rho$, $\Delta\theta$, $\Delta\phi$.

\begin{figure*}[t!]
        \centering
        \includegraphics[width=6.0in,clip=true]{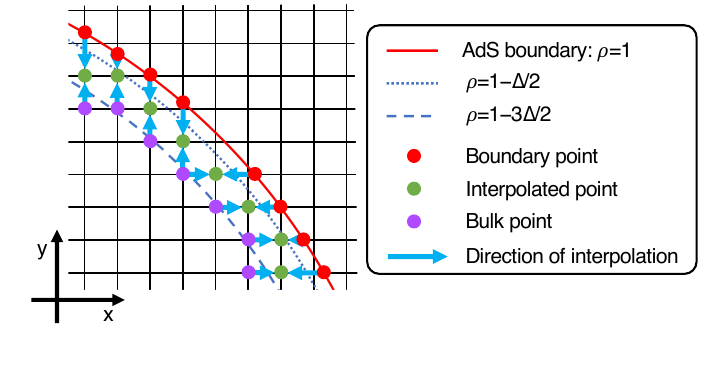}
\parbox{5.0in}{\caption{Visual description of the implementation of Dirichlet boundary conditions through first order interpolation in a portion of a surface at constant $z$ for a grid with spatial grid spacing $\Delta$.
        }\label{fig:lego_circle_dirbc}}
\end{figure*}

The following components play a fundamental role in the numerical implementation of the initial-boundary value problem.
The Dirichlet boundary conditions \eqref{eq:dirbc} are imposed at the AdS boundary $\rho=1$.
In general the AdS boundary does not lie on Cartesian grid points, so we set boundary conditions at points at most one grid point away from the boundary via interpolation. Referring to Figure~\ref{fig:lego_circle_dirbc}, for any given evolution variable, we set its value at grid points with $\rho<1-\Delta/2$ (i.e., the green dots inside the blue dotted line in this figure) by first order interpolation between the Dirichlet value at boundary points (red dots) and the value at the adjacent point further into the interior $\rho<1$ (purple dots). To identify the latter, we move along the Cartesian direction corresponding to the coordinate of the green dot with the largest absolute value. This direction is represented by light blue arrows. Notice that points with $\rho\geq1-\Delta/2$ are excised to avoid issues with quantities that would diverge 
at $\rho=1$. Finally, to obtain the values of quantities at the boundary, needed to extract the holographic observables, we use third order extrapolation from their bulk point values. 
The details of the implementation in our numerical simulations can be found in Appendix~\ref{sec:extrapconvbdy}.

Last but not least, time-symmetric initial data, sourced by a massless real scalar field, are obtained by solving the conformal decomposition of the Hamiltonian constraint \eqref{eq:hamconsfinal}. 
The solution to \eqref{eq:hamconsfinal} is computed, after second order finite discretization, through a full approximation storage (FAS) multigrid algorithm with V-cycling and Newton-Gauss-Seidel relaxation, built into the PAMR/AMRD libraries. This is the numerical technique described in Section~\ref{sec:MG}.
We ensure that initial data satisfies the generalized harmonic constraints. See Appendix~\ref{sec:initdata} for more details and the complete choice of initial data.

\subsection{Apparent horizon finder and excision}
\label{sec:AH_exc}

Once the solution is obtained at a certain time $t$, we search for the position $R(\theta,\phi)$ of an apparent horizon (AH). We use the flow method of Section~\ref{sec:AHfind} in spherical coordinates $(\rho,\theta,\phi)$, obtained in the usual way from the Cartesian coordinates of the solution.
This AH finder is based on a $(\theta,\phi)$ grid with equal grid spacings $\Delta \theta=\Delta\phi=\Delta_{AH}$.\footnote{The grid on which the AH finder is executed is completely independent of the specifics of the Cartesian evolution grid that was described in Section~\ref{sec:numcauprob}.} 
The outward null expansion at a given AH finder grid point $(\theta,\phi)$ is obtained by first order interpolation in three dimensions from the values of the expansion at Cartesian grid points that surround $(\theta,\phi)$.
These values are calculated from the definition of outward null expansion once the spacetime metric at time $t$ is known.
We observe that a $N_\theta\times N_\phi=9\times 17$ resolution is enough to find the AH in the simulations considered in Section~\ref{sec:results} in less than $10^{4}$ iterations. As noted in Section~\ref{sec:AHfind}, since \eqref{eq:floweq} is a parabolic equation, the fictitious time interval $\Delta s$ must be at least of order $\Delta_{AH}^2$ for stability. When using $n=10$ initial trial surfaces and an initial range of $\rho$ values between 0.1 and 0.5, as we do in the simulations of this chapter, we find that the AH finder works effectively if $\Delta s$ takes much smaller values. Specifically, we set $\Delta s=10^{-4}$.

When an AH is found, we excise Cartesian grid points in an ellipsoid included in the AH and centred at the centre of the AH, in order to avoid the formation of geometric singularities in the computational domain.
More specifically, the excision ellipsoid has Cartesian semi-axes, $a_x^{ex},a_y^{ex},a_z^{ex}$, determined by $a_x^{ex}=x_{AH}(1-\delta_{ex})$, where $x_{AH}$ is the $x$-coordinate value of the intersection between the AH and the $x$-axis, and similarly for $a_y^{ex}$ and $a_z^{ex}$. In the simulations described in this chapter, we set the excision buffer to $\delta_{ex}=0.4$. 
In our simulations, we assume that the characteristics of the equations of motion in the AH region flow towards the origin  of the spherical coordinate frame, $\rho=0$, although we do not compute the characteristics explicitly. 
As noted in Section~\ref{sec:AHfind}, this assumption implies that stability and convergence properties are preserved
if we simply solve the equations of motion at the excision surface by employing one-sided stencils that do not reference points inside the excised region, with no need to impose conditions at the excision boundary.
By construction, the excised surface is the same for all three time levels involved in the NGS relaxation for evolution variables at time $t$. Therefore, we only need to use the one-sided version of the spatial stencils.

It commonly occurs that the excised surface moves during evolution and previously excised points become unexcised. In this case, we initialize the value of newly unexcised points closest to the previous surface using fourth order extrapolated values from adjacent exterior points along each Cartesian direction. We do so for any variable and at all three time levels of the hierarchy.
Finally, Kreiss-Oliger dissipation~\cite{kreiss1973methods}, reviewed in Section~\ref{sec:diss}, is essential to damp unphysical high-frequency noise that arise at excision grid boundaries; we use a typical dissipation parameter of $\epsilon_{KO}=0.35$.

\section{Results}\label{sec:results}

As a proof-of-principle, we evolve initial data that undergoes gravitational collapses within one light-crossing time, and follow the subsequent ring-down to the Schwarzschild-AdS solution \eqref{eq:SchwAdS1}.
The geometry of the initial slice is sourced by a massless real scalar field with a Gaussian profile, distorted along each Cartesian direction and centred at $x=y=z=0$:
\begin{eqnarray}
\label{eq:scaGaupro}
\bar{\varphi}\big|_{t=0}&=&A e^{-(\tilde{r}(x,y,z)/\delta)^2},\\
\tilde{r}(x,y,z)&\equiv&\sqrt{ x^2(1-e_x^2)+ y^2(1-e_y^2)+ z^2(1-e_z^2)}. \nonumber
\end{eqnarray}
The amplitude of the profile is $A=0.55$ and the eccentricities are $e_x=0.3, e_y=0.2, e_z=0.25$, so that the most prominent distortion is on the $(x,y)$-plane. The width of the Gaussian is $\delta=0.2$. 
We choose the initial slice to be a moment of time symmetry, and the details of the time-symmetric initial data sourced by this matter field are collected in Appendix~\ref{sec:initdata}. 
As we see in that appendix, the momentum constraint is trivially satisfied for this type of data, so only the Hamiltonian constraint has to be solved. We evolve this initial data up to $t=31$ in units of the characteristic length scale $L=1$ (approximately 20 light-crossing times), well after the end of gravitational collapse and the resulting black hole formation. 
The initial data has zero AdS angular momenta, $J_i=0$, 
and angular momentum conservation (see Section~\ref{subsec:conscharg}) ensures that this is zero at all times. 
Therefore, we can expect the black hole to settle down to the Schwarzschild-AdS solution. However, for generic initial data with non-vanishing total angular momentum, this may not be the final state: Ref. \cite{Holzegel:2011uu} conjectured that Schwarzschild-AdS, or more generally Kerr-AdS, may suffer from a non-linear instability for generic perturbations.
We will leave this interesting problem for future work.

\subsection{Collapse and ringdown}\label{sec:rescolring}

We describe here the evolution in the bulk: this consists of an initial short phase, in which the scalar field collapses and forms a black hole, and a long ringdown stage, in which the spacetime settles down to Schwarzschild-AdS. 

\begin{figure*}[ht!]
        \centering
        \includegraphics[width=5.2in,clip=true]{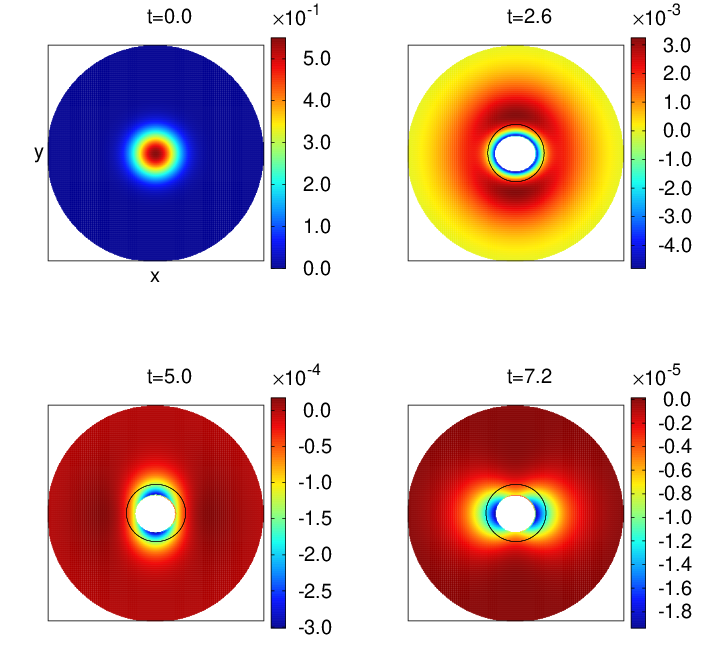}
\parbox{5.0in}{\caption{Snapshots of the scalar field profile $\bar{\varphi}$ on the $z=0$ slice in $(x,y)$ coordinates. In each plot, $x$ and $y$ are the horizontal and vertical axes, respectively, and the black square denotes the boundary of the numerical grid, i.e., $x=\pm 1$ and $y=\pm 1$. The external boundary of the coloured part is the AdS boundary. The black ellipse denotes the approximate position of the AH. This is obtained as the $z=0$ slice of the ellipsoid with Cartesian semi-axes, $x_{AH}$, $y_{AH}$, $z_{AH}$, where $x_{AH}$ is the $x$-coordinate value of the intersection between the AH and the $x$-axis, and similarly for $y_{AH}$ and $z_{AH}$.
The internal boundary of the coloured region is the excision surface: we excise points inside an ellipsoid whose semi-axes, $a_x^{ex}$, $a_y^{ex}$, $a_z^{ex}$, are given by $a_x^{ex}=x_{AH}(1-\delta_{ex})$, and similarly for $a_y^{ex}$ and $a_z^{ex}$. We use the value $\delta_{ex}=0.4$ for the excision buffer. Highest resolution: $N_x=N_y=N_z=325$.
        }\label{fig:snapshotsscalarfield}}
\end{figure*}
Figure~\ref{fig:snapshotsscalarfield} shows the profile of the scalar field variable, $\bar{\varphi}$, at four representative times on the equatorial plane $z=0$ for the highest resolution grid, with $N_x=N_y=N_z=325$ grid points along each Cartesian direction. Notice that in all of these snapshots $\bar{\varphi}=0$ at the AdS boundary, as required by the Dirichlet boundary conditions. 
At $t=0$, the asymmetry of the initial Gaussian profile is too small to be visible. 
At the beginning of evolution, we see that the scalar field lump starts propagating away from the origin $\rho=0$, and a portion of it soon forms an AH.
This occurs at $t=0.331$ in the highest resolution simulation. The rest of the scalar field remains outside the black hole, where it keeps bouncing back and forth the AdS boundary and is gradually absorbed.
The asymmetry on the $(x,y)$-plane is clearly visible at $t=2.6$, where the scalar field is stretched along the $x$-direction and squeezed along the $y$-direction. The elongation changes its direction multiple times during the evolution, as shown in the next two plots: it is along the $y$-axis at $t=5.0$ and again along the $x$-axis at $t=7.2$. 
At later times, $t\simeq 9$, the value of the scalar field becomes consistent with zero up to solution error\footnote{We estimate the solution error by comparing $\bar{\varphi}$ at different resolutions.} and the spacetime settles down to a Schwarzschild-AdS black hole spacetime with AdS energy $E=0.403$.

\begin{figure*}[t!]
        \centering
        \includegraphics[width=4.92in,clip=true]{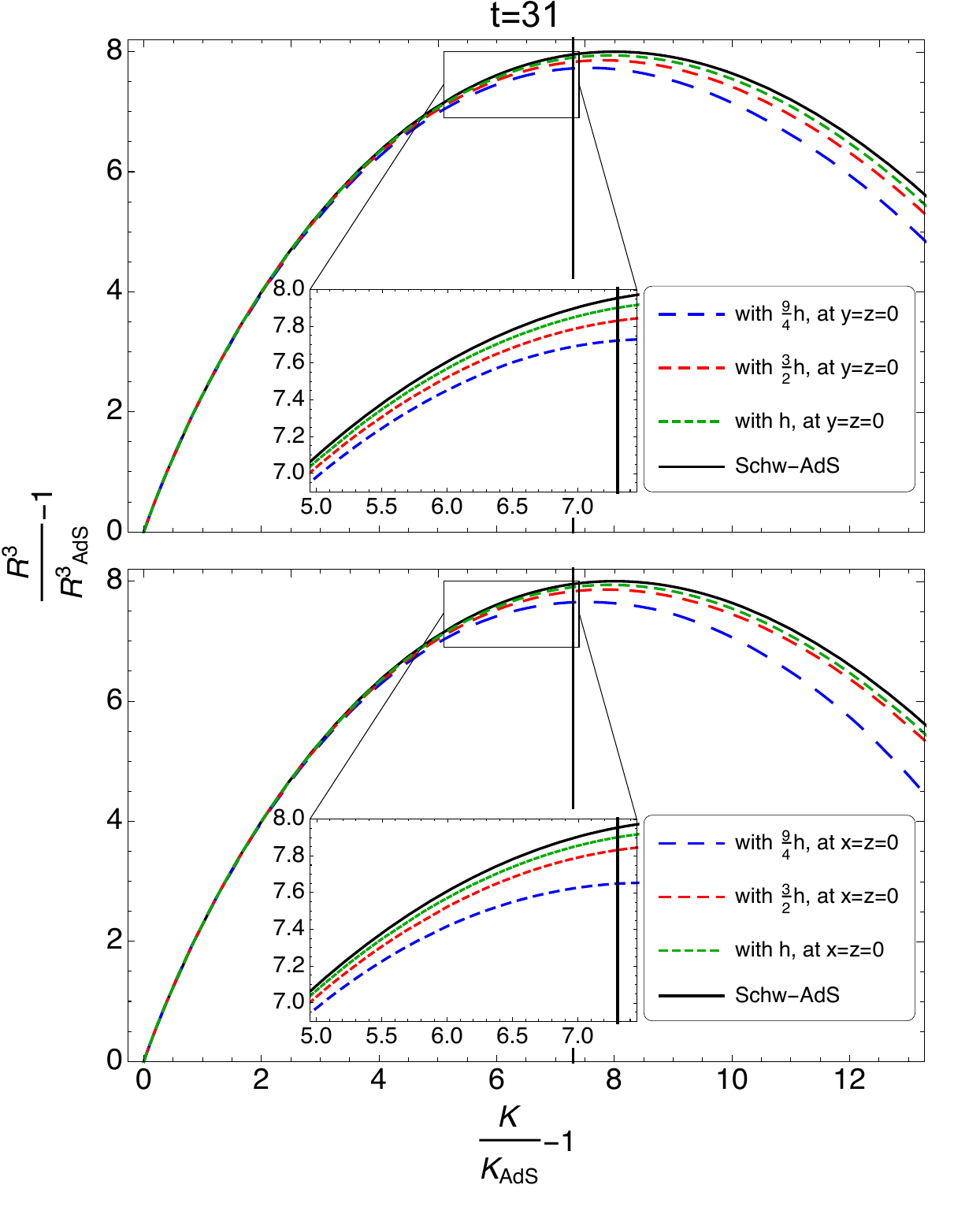}\vspace{-0.5cm}
\parbox{5.0in}{\caption{
Riemann cube scalar relative to AdS, $(R^3/R^3_{\text{AdS}})-1$, as a function of Kretschmann scalar relative to AdS, $(K/K_{\text{AdS}})-1$.
In each panel, the black curve denotes the result for a slice of Schwarzschild-AdS with AdS energy given by $E_h=0.403$ (in units of the characteristic length scale $L=1$), i.e., the value of $E$ (see eq. \eqref{eq:AdSmasscalc}) for the highest resolution run with grid spacing $h$. The black vertical line denotes the value of $(K/K_{\text{AdS}})-1$ at the horizon of the Schwarzschild-AdS black hole. The relative Kretschmann increases as we move closer to the origin of the spherical coordinate frame, $\rho=0$.
Top panel: the coloured lines denote the Riemann-Kretschmann dependence obtained from grid points on the $x$-axis (i.e., $y=z=0$) of the numerical solution at $t=31$. 
Bottom panel: the coloured lines denote the Riemann-Kretschmann dependence obtained from grid points on the $y$-axis (i.e., $x=z=0$) of the numerical solution at $t=31$.
        }\label{fig:relRiemanncube-relKretschmann-comparison-SchwAdS}}
\end{figure*}
The late-time solution is close to Schwarzschild-AdS, which can be seen explicitly in Figure~\ref{fig:relRiemanncube-relKretschmann-comparison-SchwAdS}.
Here, we compare the numerical solution at the last time slice, i.e., $t=31$, to a slice of the Schwarzschild-AdS metric with conserved AdS energy obtained from our highest resolution run ($E=0.403$).
This comparison is achieved with the following procedure.
First, we compute the \emph{Riemann cube scalar} $R^3=R_{\mu\nu\rho\sigma}R^{\rho\sigma\gamma\delta}{R_{\gamma\delta}}^{\mu\nu}$, and the \emph{Kretschmann scalar} $K=R_{\mu\nu\rho\sigma}R^{\mu\nu\rho\sigma}$.
Second, we compute the corresponding values, $R^3_{\text{AdS}}$ and $K_{\text{AdS}}$, for pure AdS.
We then use all four quantities to represent the relative Riemann scalar $(R^3/R^3_{\text{AdS}})-1$ as a function of the relative Kretschmann scalar $(K/K_{\text{AdS}})-1$ for the Schwarzschild-AdS black hole with $M=0.403$.
The same Riemann-Kretschmann dependence is estimated for our numerical solution at different resolutions from the values of $(R^3/R^3_{\text{AdS}})-1$ and $(K/K_{\text{AdS}})-1$ at each grid point along the $x$-axis ($y=z=0$ coloured lines of top panel) and the $y$-axis ($x=z=0$ coloured lines of bottom panel).

The black vertical lines in Figure~\ref{fig:relRiemanncube-relKretschmann-comparison-SchwAdS} denotes the value of $\frac{K}{K_{\text{AdS}}}-1$ at the horizon of the Schwarzschild-AdS black hole. 
Notice that $\frac{K}{K_{\text{AdS}}}-1=0$ at the AdS boundary by construction, so going to larger values of $\frac{K}{K_{\text{AdS}}}-1$ is equivalent to moving towards the centre of the grid, and closer to the singularity.
Therefore the black vertical lines give an indication of the position of the AH relative to the AdS boundary.
The two panels of Figure \ref{fig:relRiemanncube-relKretschmann-comparison-SchwAdS} indicate that, sufficiently close to the AdS boundary, the curvature invariants of the numerical solution are almost identical to Schwarzschild-AdS. For clarity, this is shown using only values of the Riemann cube and Kretschmann scalars along the $x$ and $y$ axes, but we verified this for values from the entire grid. 
At any given resolution, the numerical curvature invariants start to differ from their Schwarzschild-AdS values as we get closer to the AH. This is expected since the gradients become larger as we approach the centre of the grid.
However, these differences converge away as resolution is increased.
Finally, although there is an asymmetry at any given resolution between the $x$ and $y$ axes even at this last time slice, this late-time asymmetry also converges away as resolution is increased.

\subsection{Boundary scalar field and energy-momentum tensor}
\label{sec:resbouset}

\begin{figure*}[t!]
        \centering
        \includegraphics[width=5.0in,clip=true]{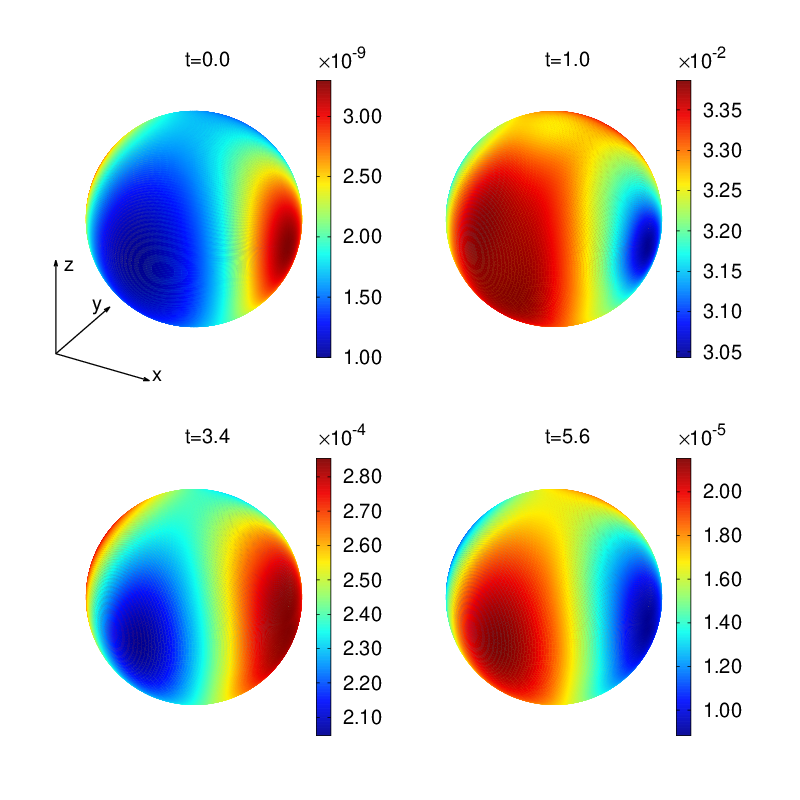}
\parbox{5.0in}{\caption{
Snapshots of the 1-point function of the dual scalar field operator $\bar{\varphi}_{(1)}$. 
The first snapshot is obtained analytically from the initial scalar field profile. The remaining three are obtained by third order extrapolation and subsequent smoothening via a low-pass filter;
see Appendix~\ref{sec:extrapconvbdy}. Highest resolution: $N_x=N_y=N_z=325$.
        }\label{fig:snapshotsbdyphi}}
\end{figure*}

In this section we consider the evolution of the holographic quantities at the AdS boundary in spherical coordinates. 
These quantities are obtained, from the expressions of  Section~\ref{sec:bouset2}, via third order extrapolation from points in the interior, with the only exception of the $t=0$ plot of Figure~\ref{fig:snapshotsbdyphi}, which is computed analytically from the initial distorted Gaussian profile \eqref{eq:scaGaupro}. 
See Appendix~\ref{sec:extrapconvbdy} for a detailed explanation of the extrapolation scheme.

We start by noting that the numerical values for the AdS energy $E$ in AdS, obtained from equation \eqref{eq:AdSmasscalc}, are approximately constant during the evolution, as expected by the AdS energy conservation discussed in Section~\ref{subsec:conscharg}. More precisely, a small drift of the AdS energy is observed numerically, however this becomes smaller as we increase the resolution and it is consistent with zero within our error estimate for boundary quantities that we will discuss shortly.

\begin{figure*}[t!]
        \centering
        \includegraphics[width=5.0in,clip=true]{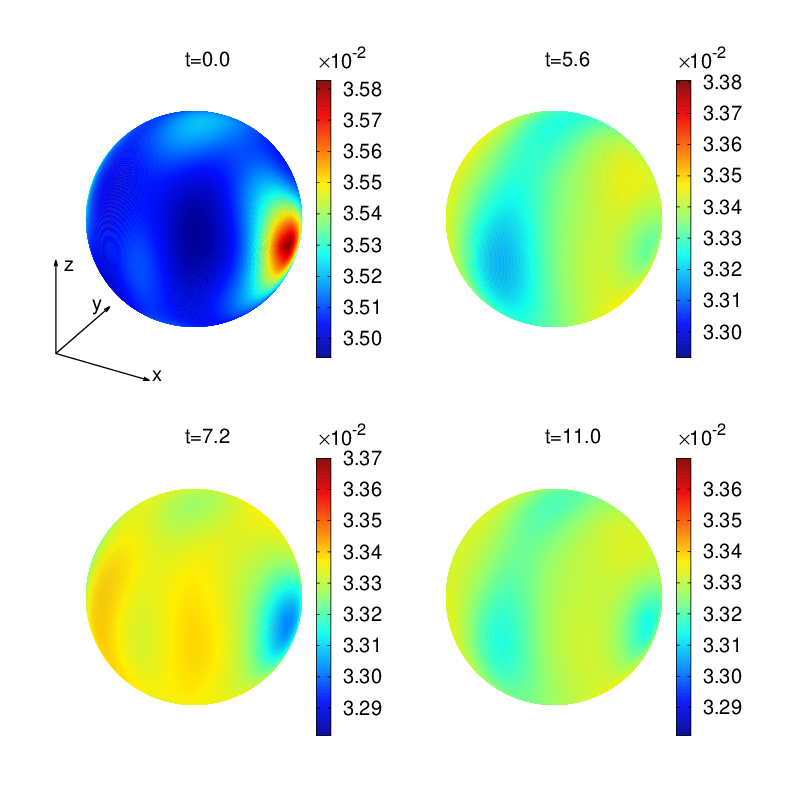}
\parbox{5.0in}{\caption{Snapshots of energy density $\epsilon$ of the dual boundary CFT, obtained by third order extrapolation and smoothened via a low-pass filter;
see Appendix~\ref{sec:extrapconvbdy}. The scale of each snapshot has fixed interval length centred at the mean value of $\epsilon$ at the corresponding evolution time to make the approach to a uniform configuration more visible. Highest resolution: $N_x=N_y=N_z=325$.
        }\label{fig:snapshotsenergydensity}}
\end{figure*}
Figure~\ref{fig:snapshotsbdyphi} shows four snapshots of the 1-point function of the boundary scalar field operator $\bar{\varphi}_{(1)}$, obtained from the near-boundary expansion of the bulk scalar field in \eqref{eqn:qexpphi}.
Unlike the $z=0$ slice snapshots of Figure~\ref{fig:snapshotsscalarfield}, these plots of the boundary $S^2$ encode the asymmetry in all three Cartesian directions in the bulk, as they appear on the boundary at $\rho = \sqrt{x^2+y^2+z^2}=1$. 
In particular, the asymmetry of the initial data is visible in the first plot, as expected given the different values of eccentricities along the three Cartesian direction (largest along $x$ and smallest along $y$) in the initial scalar field profile. At $t=0$, the boundary scalar field is overall very small, which is also expected since the initial $\bar{\varphi}$, given by \eqref{eq:scaGaupro}, is localized near $\rho=0$.
Notice from Figure \ref{fig:snapshotsbdyphi} that the asymmetry changes axes during evolution, but interestingly it is always strongest along $x$ and weakest along $y$ or vice-versa. Furthermore, a direct comparison with Figure~\ref{fig:snapshotsscalarfield} shows that the features present at a certain $t$ at the boundary take approximately $\pi/2\simeq1.6$ to reach the interior of the bulk, i.e., about a light-crossing time, as expected. At later times, mirroring the evolution in the bulk, $\bar{\varphi}_{(1)}$ decays exponentially in time as the bulk spacetime settles down to Schwarzschild-AdS.

\begin{figure*}[ht!]
        \centering
        \includegraphics[width=5.0in,clip=true]{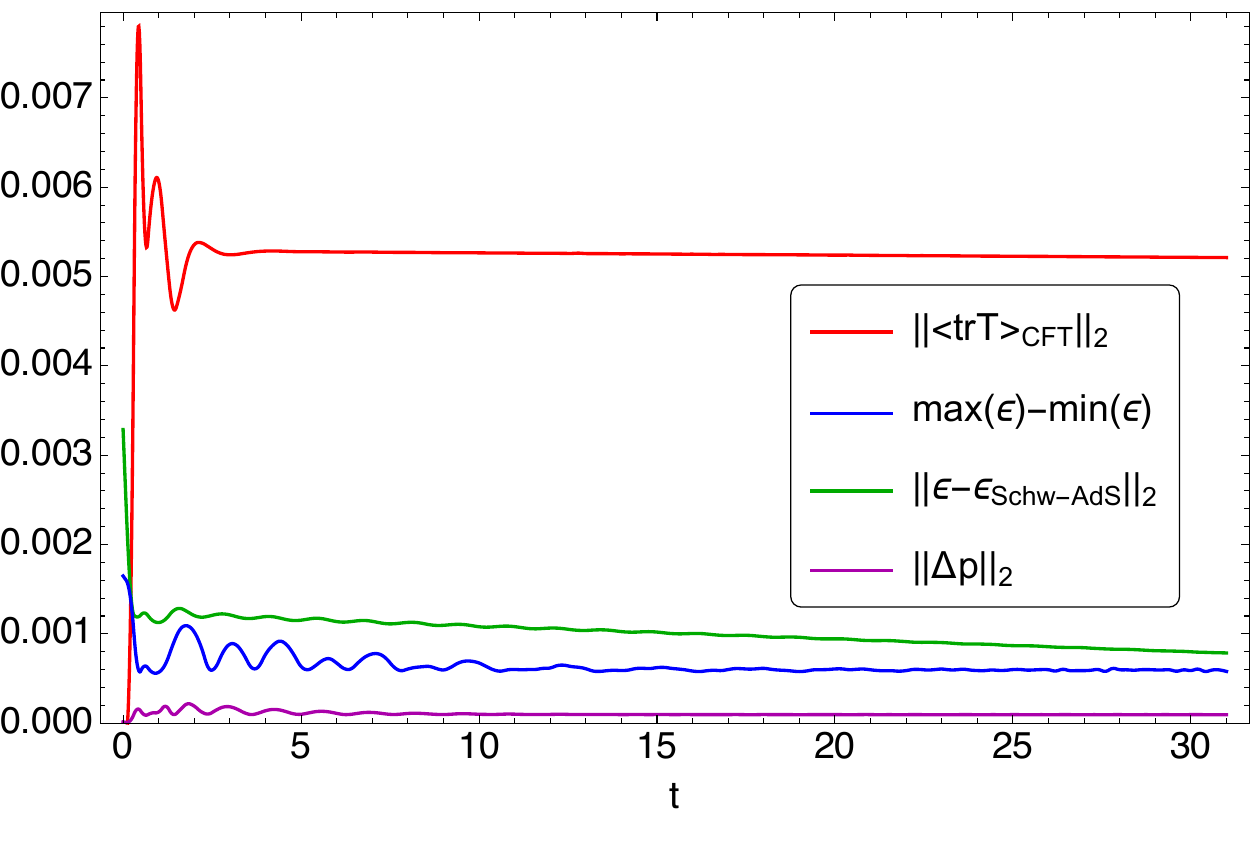}\vspace{-0.2cm}
\parbox{5.0in}{\caption{Comparison of boundary quantities with error estimate given by the deviation of the $L^2$-norm of $\langle \text{tr}T\rangle_{CFT}$ from its predicted zero value for the 2+1 CFT (red line). The following boundary quantities are displayed: difference between maximum and minimum of boundary energy density $\epsilon$ (blue line), $L^2$-norm of difference between $\epsilon$ and the Schwarzschild-AdS value $\epsilon_{\text{\text{Schw-AdS}}}=\frac{E}{4\pi}$ (green line), with Schwarzschild AdS energy $E=E_h=0.403$ (i.e., the value of $E$ for the resolution with grid spacing $h$), $L^2$-norm of boundary anistropy $\Delta p$ (magenta line). This plot is obtained from the data of the highest resolution run ($N_x=N_y=N_z=325$), but at any resolution these quantities exhibit the same hierarchy, although at different scales. Boundary quantities are computed by third order extrapolation.
        }\label{fig:fullplotfillregttraceanisotropyenergydensityminusschwmaxminusminbdyenergydensity.pdf}}
\end{figure*}
Figure \ref{fig:snapshotsenergydensity} displays the energy density $\epsilon$ of the boundary CFT.
At $t=0$ this is strongly asymmetric along the $x$-direction, as expected from the shape of the initial scalar field profile \eqref{eq:scaGaupro}. 
After that, $\epsilon$ undergoes a phase of strong evolution with several changes of elongation axes, sampled at $t=5.6$ and terminating at approximately $t=7.2$. From that time onwards, $\epsilon$ settles down to a uniform configuration, as appropriate for the Schwarzschild-AdS black hole. Approach to uniformity is emphasized by using colour scales with fixed interval length, centred at the mean value of $\epsilon$ at the corresponding evolution time.

More information about the energy density of the boundary field theory can be deduced from Figure~ \ref{fig:fullplotfillregttraceanisotropyenergydensityminusschwmaxminusminbdyenergydensity.pdf}. 
The trace $\langle \text{tr}T\rangle_{CFT}$ vanishes for a CFT on a spacetime with odd number of dimensions, such as our $\mathbb{R} \times S^2$ boundary.
In Section~\ref{sec:bouset2}, we had spelled out how this trace, in our scheme, is tied to how well we are solving the Einstein field equations.
We thus use the $L^2$-norm of the numerical values of $\langle \text{tr}T\rangle_{CFT}$ (red line) as an error estimate for boundary quantities. We compare this error with the difference between maximum and minimum of $\epsilon$ (blue line), the $L^2$-norm of the difference between $\epsilon$ and its Schwarzschild-AdS value $\epsilon_{\text{\text{Schw-AdS}}}=\frac{E}{4\pi}$ (green line), with $E=E_h=0.403$, i.e., the highest resolution value of $E$, and the $L^2$-norm of $\Delta p$ (magenta line). We compute these quantities from the data of the highest resolution simulation, but at any resolution the hierarchy is the same, although it appears at different scales.
If we exclude very early times, we see that $\max(\epsilon)-\min(\epsilon)$ is consistent with zero, which confirms that the energy density becomes uniform in time.
We also see that $||\epsilon-\epsilon_{\text{\text{Schw-AdS}}}||_2$ is consistent with zero and decreasing in time, which shows that the energy density settles down to $\epsilon_{\text{\text{Schw-AdS}}}$, as expected.
Finally, $||\Delta p||_2$ is consistent with zero, as appropriate for the boundary anistropy of the Schwarzschild-AdS black hole.

\ifpaper
\end{document}
\fi
\newif\ifpaper
\paperfalse

\ifpaper
\input{../preamble}
\begin{document}
\fi

\chapter{Towards Cauchy evolution of the superradiant instability of Kerr-AdS}
\label{Chapter:KAdS}

According to the dictionary of AdS/CFT, stationary black hole states in AdS with Hawking temperature $T_H$ are dual to CFT states in thermal equilibrium at temperature $T_H$. Therefore, perturbing a black hole in AdS and observing its return to equilibrium provides a controlled way to study the corresponding thermalisation process in the dual CFT.
The study of perturbations of the rotating Kerr-AdS black hole (introduced in Section~\ref{subsubsec:asyblaspa}) has attracted particular attention, mainly due to the existence of an effect that forces the black hole to depart from equilibrium.
This is the superradiance effect, namely the amplification of waves scattering off a rotating object.
In asymptotically AdS spacetimes with reflective boundary conditions, waves are forced to remain in the bulk and keep interacting with the hole. Thus, waves can be superradiated multiple times and become large enough to modify the properties of the initial Kerr-AdS spacetime. This is the so-called superradiant instability, whose end-state is unknown.
In particular, it was conjectured by \cite{Niehoff:2015oga} that dynamical oscillations on ever smaller length scales might eventually lead to a violation of the weak cosmic censorship conjecture (WCCC). 
The interpretation of this process from the purely quantum viewpoint of the dual CFT is currently an open question of great interest.

Great effort went into investigating the problem of superradiance for rotating black holes, both in an asymptotically flat and asymptotically AdS setting.
In this chapter, we discuss some of the most important results in the literature, with particular focus on the Kerr-AdS case.
We then present a setup, obtained as a simple modification of the numerical scheme described in Chapter~\ref{Chapter:NoSym}, that can lead to the simulation of the superradiant instability of Kerr-AdS.
In fact, this setup has made possible to simulate oscillating perturbations of Kerr-AdS for longe evolution times. We show preliminary, unpublished results from this simulation.

The chapter is structured as follows.
In Section~\ref{sec:KAdSsp}, we discuss fundamental properties of the Kerr-AdS spacetime. 
In Section~\ref{subsec:superrad}, we characterise the phenomenon of superradiance and review some of the fundamental advancements in this field, with particular focus on the problem of superradiance in Kerr-AdS.
In Section~\ref{subsec:superrad}, we present the adaption of our numerical scheme to the evolution of perturbed Kerr-AdS, and argue that we can employ this scheme to simulate superradiance in Kerr-AdS.
Initial data for the simulation of Section~\ref{subsec:superrad} are chosen in terms of scalar, vector and tensor spherical harmonics, which are defined, according to our conventions, in Appendix~\ref{sec:svtsphharm}.

\section{Kerr-AdS spacetime}
\label{sec:KAdSsp}

Kerr-AdS spacetime has been presented in Section~\ref{subsubsec:asyblaspa} in Boyer-Lindquist coordinates.
In this section we review a few properties of the (sub-extremal) Kerr-AdS spacetime. 
To avoid a cumbersome notation, only in this section, we will use the notation $g$ to denote the Kerr-AdS metric, which we previously denoted by $g_{KAdS}$.

In Boyer-Lindquist (BL) coordinates $(t,r,\theta,\phi)\in(-\infty,+\infty)\times(r_+,+\infty)\times(0,\pi)\times(0,2\pi)$, the Kerr-AdS metric is given by \eqref{eq:KerrAdS10}.
We rewrite it here for completeness:
\begin{equation}
\label{eq:KerrAdS1}
g=-\frac{\Delta}{\Sigma^2}\biggl(dt-\frac{a}{\Xi}\sin^2\theta d\phi\biggr)^2+
\frac{\Sigma^2}{\Delta}dr^2+\frac{\Sigma^2}{\Delta_\theta}d\theta^2+\frac{\Delta_\theta}{\Sigma^2}\sin^2\theta\biggl(a dt-\frac{r^2+a^2}{\Xi}d\phi\biggr)^2,
\end{equation}
where
\begin{eqnarray}
\label{eq:fnrule}
&&\Delta=(r^2+a^2)(1+r^2)-2Mr, \quad\quad\Delta_\theta=1-a^2\cos^2\theta\nonumber\\
&&\Sigma^2=r^2+a^2\cos^2\theta, \quad\quad\Xi=1-a^2,
\end{eqnarray}
Again for completeness, in the following we repeat a few facts that we stated in Section~\ref{subsubsec:asyblaspa}.
$r_+$ is the largest real solution of the equation $\Delta(r_+)=0$, i.e., $(r_+^2+a^2)(1+r_+^2)=2Mr_+$, and it denotes the value of the $r$-coordinate at the event horizon.
In order to avoid naked singularities, we assume that the parameters $M,a$ satisfy $M> M_{\text{extr}}(a)$, where the critical mass parameter $M_{\text{extr}}(a)$ is given by
\begin{equation}
\label{eq:KerrAdScondM}
\begin{split}
M_{\text{extr}}(a)= \frac{1}{3 \sqrt{6}} \left(2 \left(a^2+1\right)+\sqrt{12a^2+\left(a^2+1\right)^2}\right)\times\\
\sqrt{-(a^2+1)+\sqrt{12a^2+\left(a^2+1\right)^2}}.
\end{split}
\end{equation}
We must also have $a^2<1$ for \eqref{eq:KerrAdS1} to be regular.
Without loss of generality, we can restrict to $a\geq 0$. The case $a=0$ reduces to Schwarzschild-AdS \eqref{eq:SchwAdS1}, so we restrict to $0<a<1$.
Kerr-AdS spacetime is stationary and axisymmetric. $K=\frac{\partial}{\partial t}$ is a stationary vector field and $M=\frac{\partial}{\partial\phi}$ is an axial vector field.
As noted in Section~\ref{subsubsec:asyblaspa}, there is no representative of the conformal class of boundary metrics in these coordinates that has the expression \eqref{eq:ESU}, therefore this frame is not well-suited to prove that Kerr-AdS is asymptotically AdS, nor to identify the asymptotic time-symmetry generator $k$ and the asymptotic rotation generators $m_i$. In particular $t$ is not a time coordinate and $\phi$ is not an azimuthal coordinate in the sense of the definition of Section~\ref{subsubsec:asyphysnot}. Therefore, in the BL frame, the angular velocity of the spacetime is not given by the formula $\Omega=-g_{t\phi}/g_{\phi\phi}$.
However, a certain notion of angular velocity $\Omega'$ can be obtained from $\Omega'=-g_{t\phi}/g_{\phi\phi}$ even in this frame.
At the horizon, this gives the result
\begin{equation}
\Omega'_H=\frac{a(1-a^2)}{r^2_+ +a^2},
\end{equation}
whereas at the boundary, $r\to+\infty$, we obtain
\begin{equation}
\Omega'_{\infty}=-a.
\end{equation}
We see that, in this way, we are measuring the angular velocity of the spacetime with respect to an observer at $\partial \mathcal{M}$ that, in turn, is also rotating with angular velocity $\Omega'_{\infty}=-a$ with respect to a stationary observer at $\partial \mathcal{M}$.\footnote{Notice that this ambiguity does not arise for the analogous discussion in asymptotically flat Kerr spacetime (whose metric in BL coordinates can be obtained as explained at the end of Section~\ref{subsubsec:asyblaspa}). 
One can verify that BL coordinates on asymptotically flat Kerr are coordinates of the type that we use for the formula $\Omega=-g_{t\phi}/g_{\phi\phi}$, i.e., $t$ is a time coordinate, $\phi$ is an azimuthal coordinate, and $g_{tr}=g_{t\theta}=g_{r\phi}=g_{\theta\phi}=0$.
It is therefore expected that the formula \eqref{eq:angveldef} in BL coordinates gives the notion of angular velocity with respect to a stationary observer at $\partial \mathcal{M}$. In particular, we find $\Omega'_\infty=0$ at $r\to+\infty$.}
This already indicates that the physical notion of angular velocity is given by $\Omega'-\Omega'_{\infty}$, as originally pointed out by \cite{Caldarelli:1999xj}. In particular, the physical angular velocity of the horizon is $\Omega'_H-\Omega'_{\infty}=\frac{a(1+r_+^2)}{r^2_+ +a^2}$.
The Killing vector $\xi=\frac{\partial}{\partial t}+\Omega'_H\frac{\partial}{\partial \phi}$ is normal to the event horizon $r=r_+$, therefore the event horizon is a Killing horizon.
Its surface gravity $\kappa$, defined by \eqref{eq:surgrav}, is given by
\begin{equation}
\label{eq;surgra}
\kappa=\frac{r_+(1+r_+^2)}{r_+^2+a^2}-\frac{1-r_+^2}{2r_+}.
\end{equation}
The area of the event horizon, i.e., the area of a slice of the event horizon at fixed $t$, is
\begin{equation}
\label{eq;areahor}
A=\frac{4\pi (r_+^2+a^2)}{(1-a^2)}.
\end{equation}
The metric \eqref{eq:KerrAdS1} is the generalisation, to the case of negative cosmological constant, of the asymptotically flat Kerr metric, which describes a stationary and axisymmetric rotating black hole in a geometry that is flat near infinity. The Kerr metric in Boyer-Lindquist coordinates $(t,r,\theta,\phi)$ can be obtained as the $r\ll1,a\ll1$ approximation of \eqref{eq:KerrAdS1}, or, equivalently, by reinserting factors of $L$ (by dimensional analysis) in \eqref{eq:KerrAdS1} and taking the limit $L\to+\infty$.

We now discuss a coordinate system $x^{\dot\alpha}=(T,R,\Theta,\Phi)\in(-\infty,+\infty)\times(R_+(\theta),+\infty)\times(0,\pi)\times(0,2\pi)$ with the desired property that the Kerr-AdS metric at the boundary is manifestly conformally related to the metric of ESU in the form \eqref{eq:ESU}.
These coordinates were proposed by Hawking, Hunter and Taylor-Robinson (HHT) in \cite{Hawking:1998kw}, and are related to BL coordinates through
\vspace{-0.3cm}
\begin{eqnarray}
\label{eq:nonrotnonhorpen}
T&=&t,\nonumber\\
R^2\cos^2\Theta&=&r^2\cos^2\theta,\nonumber\\
\Xi R^2\sin^2\Theta&=&(r^2+a^2)\sin^2\theta,\nonumber\\
\Phi&=&\phi+at.
\end{eqnarray}
Explicitly, \emph{HHT coordinates}, chosen so that they coincide with the BL coordinates for $a=0$, are given by the following (smooth) relations.
\vspace{-0.3cm}
\begin{eqnarray}
\label{eq:nonrotnonhorpenexpl}
T&=&t,\nonumber\\
R&=&\sqrt{r^2+\frac{a^2 \left(r^2+1\right) \sin ^2\theta }{1-a^2}},\nonumber\\
\Theta&=&
\begin{cases}
 \arcsin\left(\sin \theta  \sqrt{\frac{a^2+r^2}{a^2 \left(r^2+1\right) \sin
   ^2\theta +\left(1-a^2\right) r^2}}\right) &  0 <\theta\leq \frac{\pi}{2} \\
 \pi -\arcsin\left(\sin \theta  \sqrt{\frac{a^2+r^2}{a^2 \left(r^2+1\right) \sin
   ^2\theta +\left(1-a^2\right) r^2}}\right) &  \frac{\pi}{2} <\theta<\pi
\end{cases}, \nonumber \\
\Phi&=&\phi+at.
\end{eqnarray}
These can be (smoothly) inverted as
\vspace{-0.3cm}
\begin{eqnarray}
\label{eq:nonrotnonhorpenexplinv}
\hspace{-0.5cm}&&t=T,\nonumber \\
\hspace{-0.5cm}&&r=\frac{1}{\sqrt{2}}\sqrt{R^2 (1-a^2 \sin ^2\Theta) -a^2+\sqrt{a^4 R^4 \sin ^4\Theta +2 a^2 R^2 \left(a^2-R^2-2\right) \sin ^2\Theta
   +\left(a^2+R^2\right)^2}},\nonumber \\
\hspace{-0.5cm}&&\theta=
\begin{cases}
 \arcsin\left(\frac{\sqrt{R^2 (1-a^2 \sin ^2\Theta) +a^2-\sqrt{a^4 R^4 \sin ^4\Theta +2 a^2 R^2 \left(a^2-R^2-2\right) \sin ^2\Theta
   +\left(a^2+R^2\right)^2}}}{\sqrt{2} a}\right) &  0 <\Theta\leq \frac{\pi}{2} \nonumber \\
 \pi - \arcsin\left(\frac{\sqrt{R^2 (1-a^2 \sin ^2\Theta) +a^2-\sqrt{a^4 R^4 \sin ^4\Theta +2 a^2 R^2 \left(a^2-R^2-2\right) \sin ^2\Theta
   +\left(a^2+R^2\right)^2}}}{\sqrt{2} a}\right) &  \frac{\pi}{2} <\Theta<\pi
\end{cases},  \nonumber \\
\hspace{-0.5cm}&&\phi=\Phi-aT.
\end{eqnarray}
$R_+(\theta)$ identifies the position of the event horizon and is given by \\
$R_+(\theta)=\sqrt{r_+^2+\frac{a^2 \left(r_+^2+1\right) \sin ^2\theta }{1-a^2}}$, or in terms of $\Theta$ by 
$R_+(\Theta)=r_+ \sqrt{\frac{2(a^2+r_+^2)}{a^2 \left(r_+^2+1\right)
   \cos 2 \Theta -\left(a^2-2\right)r_+^2+a^2}}$.
   
The Kerr-AdS metric in these coordinates, which we do not write explicitly for brevity, asymptotes to the pure AdS metric in the form \eqref{eqn:ads4}.\footnote{Of course, we refer to the expression \eqref{eqn:ads4} in which we also set $L=1$ and we relabel the coordinates with the corresponding HHT coordinate symbol, i.e., $t\to T, r\to R,\theta\to\Theta,\phi\to \Phi$.} The non-vanishing components of the deviation from pure AdS, $h$, go to zero as we approach the boundary $R\to+\infty$ with fall-offs consistent with those suggested by \cite{Henneaux:1985tv}, given by \eqref{eq:HTh}.
More precisely, $h_{TR}=h_{T\Theta}=h_{R\Phi}=h_{\Theta\Phi}=0$ and
\begin{eqnarray}
\label{eq:HHth}
&&h_{TT}=\mathcal{O}(R^{-1}), \quad h_{T\Phi}=\mathcal{O}(R^{-1}),\quad  h_{ RR}=\mathcal{O}(R^{-5}),\nonumber\\
&& h_{ R\Theta}=\mathcal{O}(R^{-4}), \quad h_{ \Theta\Theta}=\mathcal{O}(R^{-3}), \quad h_{ \Phi\Phi}=\mathcal{O}(R^{-1}).
\end{eqnarray}
Therefore, these coordinates make the asymptotically AdS nature of Kerr-AdS manifest. Furthermore, we can identify the asymptotic time-symmetry generator as $k=\frac{\partial}{\partial T}$ and the asymptotic rotation generators as $m_1=\frac{\partial}{\partial \Phi}, m_2=-\sin\Phi\frac{\partial}{\partial\Theta}-\cot\Theta\cos\Phi \frac{\partial}{\partial \Phi}, m_3=\cos\Phi\frac{\partial}{\partial\Theta}-\cot\Theta\sin\Phi \frac{\partial}{\partial \Phi}$.
The AdS energy, $E:=Q[k]$, and AdS angular momenta, $J_1:=-Q[m_1]$, $J_2:=-Q[m_2]$, $J_3:=-Q[m_3]$, are the conserved charges associated with these asymptotic Killing vector fields.
Their expressions in terms of the parameters $M,a$ were given in Section~\ref{subsubsec:asyblaspa}, and are rewritten here for completeness:
\begin{equation}
\label{eq:kerradsconscharge}
E=\frac{M}{\Xi^2}, \quad \quad J_1=\frac{aM}{\Xi^2},  \quad \quad J_2=J_3=0.
\end{equation}
Since $J_2$ and $J_3$ vanish, it is common to refer to $J_1$ as the angular momentum of a Kerr-AdS black hole and denote it by $J$. We often also denote $m_1$ by $m$.
In terms of BL coordinates, we have $k=\frac{\partial}{\partial t}-a\frac{\partial}{\partial \phi}$ and $m=\frac{\partial}{\partial \phi}$.
The angular velocity of the spacetime is given in HHT frame by the formula $\Omega=-g_{T\Phi}/g_{\Phi\Phi}$.
At the horizon, we have
\begin{equation}
\label{eq:HHTangvelhor}
\Omega_H=\frac{a(1+r_+^2)}{r^2_+ +a^2}.
\end{equation}
The angular velocity at the boundary vanishes, i.e., $\Omega_{\infty}=0$.
The co-rotating Killing vector field is given in HHT coordinates by $\xi=\frac{\partial}{\partial T}+\Omega_H\frac{\partial}{\partial \Phi}$.
It was noted in \cite{Caldarelli:1999xj,Gibbons:2004ai} that the physical notions of energy, angular momentum and angular velocity in Kerr-AdS are precisely those naturally identified by working in HHT frame.
In particular, Ref. \cite{Gibbons:2004ai} points out that, under a small variation of the parameters $M,a$ of the Kerr-AdS spacetime, the corresponding variations, $\delta E,\delta A,\delta  J$, satisfy the first law of black hole mechanics, $\delta E=\frac{\kappa}{8\pi}\delta A+\Omega_H\delta J$, only if $\Omega_H$ is the angular velocity of the horizon given by \eqref{eq:HHTangvelhor}, and $E,J$ are the AdS energy and AdS angular momentum given by \eqref{eq:kerradsconscharge}.

Given HHT coordinates, we can define a compactified radial coordinate $\rho$ by the usual relation $R=2\rho/(1-\rho^2)$. The resulting compactified HHT coordinates $(T,\rho,\Theta,\Phi)$ are spherical coordinates for Kerr-AdS, according to the definition of Section~\ref{subsec:bouconsphcar}. We have verified that the Kerr-AdS metric components in these coordinates satisfy \eqref{eqn:target_gauge_trhothetaphi}.
This shows that the gauge choice for spherical coordinates, obtained through our boundary prescription for numerical stability, allows for a Kerr-AdS solution of the evolution equations. This is an important sanity check of the boundary prescription that we introduced.

\section{Superradiance}
\label{subsec:superrad}

It was noted by Penrose \cite{Penrose:1969pc,Penrose:1971uk} that the scattering of a massive particle off a rotating black hole can be employed to extract energy and angular momentum from the hole. 
An analogous phenomenon, known as \emph{superradiance}, occurs when a wave scatters off a sufficiently rotating object. This was originally presented for rotating horizonless objects by Zeldovich \cite{Zeldovich1971GenerationOW,Zeldovich1972AmplificationOC}, and for rotating black holes by Misner \cite{PhysRevLett.28.994}, and later by Starobinsky \cite{Starobinsky:1973aij}. See \cite{Brito:2015oca} for an extensive review of these topics.
In this section, we first present simple arguments that describe these phenomena and their fundamental aspects in general settings. 
We then review more advanced studies of the potential consequences of superradiant effects.

\subsubsection{Penrose process}

Let us start with a description of the \emph{Penrose process}.
This denomination typically refers to the scattering of a massive particle off a black hole in such a way that the hole loses energy and angular momentum.
In a stationary and axisymmetric black hole spacetime, such a process can occur if there exists an \emph{ergoregion}, that is, a spacetime region just outside of the horizon where the stationary vector field $K$, used to define a notion of time, becomes spacelike (recall that $K$ is timelike near $\partial \mathcal{M}$ by definition). In this region stationary observers do not exist and every observer must rotate, with respect to a stationary observer at infinity, in the same direction as the black hole.
Crucially, the measure at $\partial \mathcal{M}$ of the energy per unit mass of a particle, which is given by $E:=-g_{\alpha\beta}u^\alpha K^\alpha$, is not necessarily positive in the ergoregion. Therefore, it is possible to imagine a setup in which a particle with initial energy $E_{in}>0$ approaches a black hole with energy $E$, arrives in the ergoregion, and splits into a particle with negative energy, $E_{1}<0$, and a particle with positive energy, $E_2>0$. 
It can be shown, at least in the asymptotically flat Kerr spacetime, that we can arrange for a scenario in which the negative energy particle enters the black hole, whereas the positive energy particle returns to its initial spatial position. 
This is the so-called Penrose process.
At the end of this process, the energy of the black hole is $E+E_1<E$, i.e., the black hole has lost energy. From energy conservation $E_{in}=E_1+E_2$, the energy lost by the black hole is carried by the escaped particle, whose energy is now $E_2=E_{in}-E_1>E_{in}$. 
In summary, this is a method to extract energy from the black hole and, in principle, carry it somewhere else where it can be used for the most diverse purposes.

Moreover, it can be shown that a negative energy particle must have negative angular momentum, hence this scattering process also extracts angular momentum from the black hole.
To see this, we use the future-directed normal $\xi$ to the event horizon $\mathcal{H}^+$ (which we assume is a Killing horizon), given by \eqref{eq:corotfie}. Since $\xi$ and the 4-velocity $u$ of a particle crossing the horizon are both future-directed causal vectors, then they are contained in the same light cone of any point $p$ at the horizon, hence $g_{\alpha\beta}u^\alpha \xi^\beta\leq0$ at $\mathcal{H}^+$. Now, $g_{\alpha\beta}u^\alpha \xi^\beta=g_{\alpha\beta}u^\alpha K^\beta+\Omega_Hg_{\alpha\beta}u^\alpha M^\beta=-E+\Omega_HL$, where $L=g_{\alpha\beta}u^\alpha M^\beta$ is the angular momentum per unit mass of the particle (as measured at $\partial \mathcal{M}$). Therefore, $L\leq E/\Omega_H$ (recall that, by convention, we choose the azimuthal coordinate in such a way that $\Omega_H$ is positive). Hence, if $E<0$, $L$ must be negative.
Notice that the event horizon does not play any crucial role, other than not letting the negative energy particle escape, 
therefore the Penrose process can also occur for horizonless rotating objects, such as stars, with an ergoregion, as long as the gravitational field of the object confines negative energy particles.

\subsubsection{Superradiance from the laws of black hole mechanics}
\label{subsubsec:supblaholmech}

We now turn to the study of \emph{superradiance}, i.e., the amplification of a wave that scatters off a rotating object. 
An instructive argument, outlined in the following and originally presented in \cite{Bekenstein:1973mi}, employs physical insights and the laws of black hole mechanics to identify the condition under which superradiance occurs. 

Consider a stationary and axisymmetric black hole with energy $E$, angular momentum $J$, angular velocity at the event horizon $\Omega_H$, surface gravity $\kappa$ and horizon area $A$.
Let $K$ be the stationary vector field normalised so that its boundary limit is the generator of the time symmetry, and let $M$ be the axial vector field normalised so that its boundary limit is the generator of a rotation symmetry.
Let $t$ and $\varphi$ be a notion of time and azimuthal angle, as defined in Section~\ref{subsubsec:asyphysnot}, and let us work with a frame $x^\alpha=(t,r,x^2,\phi)$, where $r$ is a coordinate describing some notion of distance from the event horizon at $r=r_+$, and that goes to infinity as we approach $\partial \mathcal{M}$. 
These could be, for instance, BL coordinates for Kerr or HHT coordinates for Kerr-AdS.
For simplicity, we study the simple case of a wave (also called a \emph{mode}) of scalar massless matter, with amplitude small enough so that  we can neglect the effect of the perturbation on the spacetime, and regard the wave as propagating in a fixed background spacetime.
Given the isometries of the spacetime, the Klein-Gordon equation admits a mode solution of the form $\varphi=\operatorname{Re}(\varphi_0(r,x^2)e^{-i\omega t+im\phi})$, which oscillates in $t$ with real frequency $\omega>0$ and in $\phi$ with real azimuthal number $m$ (not to be confused with the asymptotic rotation symmetry generator).\footnote{Note that, for solutions of this form, $m$ is an integer satisfying $-l\leq m\leq l$. Here $l$ is a non-negative integer, $l\geq 0$.}
The functional dependence of $\varphi_0(r,x^2)$ on the coordinates can then be obtained by solving the Klein-Gordon equation on the given spacetime with boundary conditions, chosen on physical grounds, that only allow for ingoing modes at the event horizon. 
In a scattering process with a rotating black hole, a wave naturally splits into a portion that enters the horizon, called transmitted wave, and a portion that escapes to infinity, called reflected wave (unlike the Penrose process in which the splitting of the scattering particle must be arranged).
The first step of the calculation consists of obtaining the energy and angular momentum of matter that crosses the event horizon, i.e., the energy and angular momentum flux.

In a region where $K=\frac{\partial}{\partial t}$ is timelike, consider a timelike hypersurface at constant $r>r_+$.
Let $\Delta t$ be the time interval during which energy and angular momentum cross points of this hypersurface, carried either by the incident wave or the transmitted wave.
We focus on the portion of the hypersurface within the time interval $\Delta t$, and denote this region by $\Sigma_r$.
Each 2-dimensional spacelike slice $S$ of $\Sigma_r$ at constant $t$ is a surface surrounding the cross-section of the event horizon at $t$. If we view one of these slices $S$ as a surface in space that remains at constant values of the spatial coordinates in the time interval $\Delta t$, the \emph{net energy flux} on $S$, i.e., the ingoing energy minus the outgoing energy, over the interval $\Delta t$ is given by the energy on $\Sigma_r$. The latter has the expression
\begin{equation}
\label{eq:enmatt}
\mathcal{E}=\int_{\Sigma_r}{dtdx^2 d\phi \sqrt{\gamma} J^\alpha n_\alpha },
\end{equation}
where $\gamma$ is the determinant of the metric on $\Sigma_r$ induced by the spacetime metric $g$, $n^\alpha=Ng^{\alpha\beta}(dr)_\beta$ is the outward-pointing unit normal to $\Sigma_r$, and $N=1/\sqrt{g^{\alpha\beta}(dr)_\alpha (dr)_\beta}$ is the normalisation factor.
Here, $J^\alpha$ is given by $J^\alpha:=-T^\alpha_\beta K^\beta$ where $T^\alpha_\beta$ is the energy-momentum tensor of the matter field. $J^\alpha$ is the enegy-momentum current measured by a stationary observer, i.e., one that travels along integral curves of $K$. 
Similarly, the \emph{net angular momentum flux} on a surface surrounding a cross-section of the event horizon at times within the time interval $\Delta t$ is given by the angular momentum on $\Sigma_r$, i.e.,
\begin{equation}
\label{eq:angmommatt}
\mathcal{J}=\int_{\Sigma_r}{dtdx^2 d\phi \sqrt{\gamma} L^\alpha n_\alpha },
\end{equation}
where $L^\alpha$, called \emph{angular momentum current}, is given by $L^\alpha:=T^\alpha_\beta M^\beta$.
Now, $J^\alpha n_\alpha=-T_{\alpha\beta}n^\alpha  K^\beta=-N T^r_t$. From the expression \eqref{eq:KHmomtenspar} for the energy-momentum tensor we obtain $J^\alpha n_\alpha=-N(g^{r\alpha} \partial_\alpha\varphi) \partial_t\varphi=\omega\varphi Ng^{r\alpha} \partial_\alpha\varphi$. 
We also have $L^\alpha n_\alpha=T_{\alpha\beta}n^\alpha  M^\beta=N T^r_\phi$. Using again \eqref{eq:KHmomtenspar}, we obtain $L^\alpha n_\alpha=N (g^{r\alpha} \partial_\alpha\varphi) \partial_\phi\varphi=m \varphi N  g^{r\alpha} \partial_\alpha\varphi$. 
Hence, the ratio between the angular momentum flux and the energy flux on a surface surrounding the event horizon in the time interval $\Delta t$ is
\begin{equation}
\label{eq:EJmatt}
\frac{\mathcal{J}}{\mathcal{E}}=\frac{m}{\omega}.
\end{equation}
By energy and angular momentum conservation, $\mathcal{J}$ and $\mathcal{E}$ must also be the fluxes at the event horizon, carried by the transmitted wave, at some later time. One can prove this explicitly by using the fact that $J^\alpha$ and $L^\alpha$ are conserved currents, i.e., $\nabla_\alpha J^\alpha=\nabla_\alpha L^\alpha=0$, which is a direct consequence of the fact that $T_{\alpha\beta}$ is conserved and $K,M$ satisfy the Killing equation (see, e.g., Chapter 12 of \cite{wald:1984}).
Then, using again energy and angular momentum conservation, the absorption of energy $\mathcal{E}$ and angular momentum $\mathcal{J}$ by the black hole must cause a variation of $E$ and $J$ given, respectively, by $\delta E=\mathcal{E}$ and $\delta J=\mathcal{J}$. 
Furthermore, $\delta E$ and $\delta J$ must be related to the change in horizon area $\delta A$ by the first law of black hole mechanics (in its ``physical process'' version discussed in Section~\ref{subsec:blaholmech}):
\begin{equation}
\label{eq:firstlaw1p5}
\delta E=\frac{\kappa}{8\pi}\delta A+\Omega_H \delta J.
\end{equation}
Using $\frac{\delta J}{\delta E}=\frac{\mathcal{J}}{\mathcal{E}}=\frac{m}{\omega}$, as given by \eqref{eq:EJmatt}, we obtain
\begin{equation}
\label{eq:firstlaw2}
\delta E\left(1-\Omega_H\frac{m}{\omega}\right)=\frac{\kappa}{8\pi}\delta A.
\end{equation}
From the second law of black hole mechanics, we have $\delta A\geq 0$. Moreover, $\kappa$ is non-negative. This implies that $\delta E\left(1-\Omega_H\frac{m}{\omega}\right)\geq 0$. 
We see that, if the perturbation satisfies the condition
\begin{equation}
\label{eq:supcond}
0<\omega<m\Omega_H,
\end{equation}
then we must have $\delta E<0$, i.e., the black hole loses energy (the case $\delta E=0$ is trivial since it would imply $\mathcal{E}=0$, i.e., the incident wave is completely reflected and there is no transmitted wave). From $\frac{\delta J}{\delta E}=\frac{m}{\omega}$, we see that, if $m>0$, the black hole also loses angular momentum, $\delta J<0$.
The lost energy and angular momentum must be carried away by the reflected wave, which is therefore amplified.
In summary, we saw that, if there exists a solution of the matter equations of motion on the given background that satisfies $0<\omega<m\Omega_H$, then that solution undergoes superradiance.
We refer to \eqref{eq:supcond} as the \emph{superradiance condition}.

The advantage of this argument is that it also applies to the case of a horizonless object with energy $E$, angular momentum $J$, angular velocity $\Omega$, temperature $T$ and entropy $S$. In fact, loosely speaking, the laws of black hole mechanics, which we applied above, turn into the laws of thermodynamics, satisfied by any thermodynamical object, under the replacements $A\to 4S$, $\kappa\to2\pi T$, and $\Omega_H\to \Omega$.
In the next section, we will consider a shorter, but less versatile, argument.

\subsubsection{Alternative derivation of black hole superradiance}
\label{subsubsec:aldersup}

Since energy and angular momentum conservation requires that the fluxes at $\Sigma_r$ with $r>r_+$ are the same as the fluxes at the horizon, it should be possible to derive the superradiance condition \eqref{eq:supcond} for black holes by a straightforward evaluation of \eqref{eq:enmatt},\eqref{eq:angmommatt} at the horizon. 
This calculation is outlined in Chapter 12 of \cite{wald:1984}, and proceeds as follows.\footnote{Strictly speaking, this calculation would require the use of coordinates in which the metric is regular at the horizon, which is not the case of BL coordinates in Kerr and HHT coordinates in Kerr-AdS. This is just a technicality which does not affect the conclusions that we can draw from this argument.}
At $\mathcal{H}^+$, the energy flux is given by \eqref{eq:enmatt} where $n^\alpha$ is the past-directed unit normal $n^\alpha=-\xi^\alpha$, with $\xi=K+\Omega_H M$ the co-rotating Killing vector field, and $M=\frac{\partial}{\partial\phi}$ the axial vector field normalised to generate an asymptotic rotation symmetry. 
Then, the integrand of \eqref{eq:enmatt} is given by $n_\alpha J^\alpha=T_{\alpha\beta}\xi^\alpha  K^\beta=(\xi^\alpha\partial_\alpha\varphi)(K^\beta\partial_\beta\varphi)-(1/2)g_{\alpha\beta}\xi^\alpha K^\beta g^{\gamma\delta}(\partial_\gamma\varphi )(\partial_\delta\varphi)$. 
Note that $K$ must be tangent to $\mathcal{H}^+$ because an isometry must preserve the geometry of the event horizon. 
Since $\xi$ is normal to $\mathcal{H}^+$, we have $g_{\alpha\beta}\xi^\alpha K^\beta=0$ at the horizon. 
Moreover, $K^\beta\partial_\beta\varphi=\partial_t\varphi=-\omega\varphi$ and $\xi^\alpha\partial_\alpha\varphi=\partial_t\varphi+\Omega_H\partial_\phi\varphi=(-\omega+\Omega_Hm)\varphi$. Hence, $n_\alpha J^\alpha=\omega(\omega-m\Omega_H)\varphi^2$. Writing $\varphi$ in the form $\varphi=|\varphi_0(r,x^2)|\cos\left(-\omega t+m\phi+\alpha(r,x^2)\right)$, where $\alpha(r,x^2)$ is the phase of the complex amplitude $\varphi_0(r,x^2)$, we have $n_\alpha J^\alpha=\omega(\omega-m\Omega_H)|\varphi_0|^2\cos^2\left(-\omega t+m\phi+\alpha\right)$.
Hence, the integrand of \eqref{eq:enmatt} is positive if $\omega>m\Omega_H$, but is negative if $0<\omega<m\Omega_H$. Since $\mathcal{E}$ must equal the variation of energy $\delta E$ of the black hole, we have again found that the black hole loses energy if the condition \eqref{eq:supcond} is satisfied.
Similarly, we can show that the integrand of \eqref{eq:angmommatt} at $r=r_+$ is given by
$n_\alpha L^\alpha=-T_{\alpha\beta}\xi^\alpha  M^\beta=(\xi^\alpha\partial_\alpha\varphi)(M^\beta\partial_\beta\varphi)=m(\omega-m\Omega_H)|\varphi_0|^2\cos^2\left(-\omega t+m\phi+\alpha\right)$. As above, we see that the black hole loses angular momentum, $\delta J=\mathcal{J}<0$, if the superradiance condition \eqref{eq:supcond} is satisfied and $m>0$.

\subsubsection{Necessity of the ergoregion}
\label{subsubsec:necerg}

The argument of Section~\ref{subsubsec:aldersup} shows explicitly that the sign of $n_\alpha J^\alpha=-\xi_\alpha J^\alpha$, or equivalently the sign of the energy flux $\mathcal{E}$, at $\mathcal{H}^+$ determines whether black hole superradiance can occur.
Using this, it is possible to obtain conditions that rule out superradiance.
In particular, under the assumption that the Dominant Energy Condition is satisfied (recall that this is the case for ``physical'' matter or vacuum; see Section~\ref{subsec:EinEq}), it is found that superradiance in the presence of an event horizon cannot occur if the stationary vector field $K$ defining time at the boundary does not have an ergoregion.
Let us prove this result. If there is no ergoregion, i.e., $K$ is timelike everywhere outside the horizon, then $K$ can be used to define a notion of time and the time-orientation in the exterior of $\mathcal{H}^+$. Thus, $K$ is timelike and future-directed in this region.
Now, the Dominant Energy Condition implies that $J^\alpha=-T^\alpha_\beta K^\beta$ is causal and future-directed everywhere outside $\mathcal{H}^+$ (or it vanishes), and, by continuity, also at $\mathcal{H}^+$.
Therefore, $\xi$ and $J$ are both causal and future-directed at $\mathcal{H}^+$ (or $J=0$), hence they are contained in the same light cone at any point $p\in\mathcal{H}^+$. This implies that $\xi_\alpha J^\alpha\leq 0$, i.e., $n_\alpha J^\alpha\geq 0$, at $\mathcal{H}^+$. Therefore, the energy flux $\mathcal{E}$ at the horizon is non-negative and superradiance cannot occur.

The previous paragraph tells us that, similarly to the discussion about the Penrose process, the existence of an ergoregion is a necessary condition for superradiance in the presence of an event horizon (at least when the Dominant Energy Condition is satisfied).
Nevertheless, the existence of an ergoregion is not a sufficient condition for superradiance, regardless of the presence of an event horizon: even if an ergoregion is present, the matter equations of motion on the given background may not admit a mode solution that grows with time.
Moreover, when there is no horizon, the presence of an ergoregion is not a necessary condition for superradiance.
In fact, superradiance can occur for any sufficiently fastly rotating body that is able to absorb part of an incident wave.
This is explicitly discussed in \cite{Richartz:2013unq} within the framework of general relativity (see also \cite{Vicente:2018mxl}). 
Outside of this framework, an example is provided by the original paper about the topic \cite{Zeldovich1972AmplificationOC}, which investigates superradiance for a classical cylinder rotating in Minkowski spacetime. 
We also mention that fermionic matter fields, i.e., matter fields with half-integer spin, do not undergo superradiance \cite{Unruh:1973bda,wald:1984}.
In the following, we will focus on studies of black hole superradiance of bosonic, i.e., integer spin, perturbations.

\subsubsection{Mode analysis}

The arguments of the previous two sections suggest that a wave is amplified if the superradiance condition is satisfied. However, the amplitude $\varphi_0$ of the perturbation that we considered does not depend on $t$. As a consequence, in our simplified treatment, we cannot track the growth of the perturbation amplitude due to superradiance. In a more thorough discussion, one could allow for time dependence in the amplitude of the perturbation, and then verify that $\varphi_0$ increases when the superradiance condition \eqref{eq:supcond} is satisfied. Going further, stationarity of the background spacetime implies that the $t$-dependence of the amplitude can be factorised and written as the factor $e^{\lambda t}$, for some real constant $\lambda$.
We can then look for the values of $\lambda$ for which the perturbation solves the matter equations of motion.
If $\lambda>0$, the perturbation grows unboundedly in time, in which case we say that the mode is \emph{unstable}.
If $\lambda=0$, then the perturbation amplitude is independent of time, and we refer to the mode as \emph{normal}.
 If $\lambda<0$, the perturbation decreases in time and we refer to it as a \emph{quasi-normal mode} of the matter field. In the latter case,  the terminology comes from the fact that the value of $\lambda$ is typically very close to $\lambda=0$.
Clearly, including this exponential factor is equivalent to considering a perturbative mode of the form $\operatorname{Re}(\varphi_0(r,x^2)e^{-i\omega' t+im\phi})$, where $\varphi_0(r,x^2)$ does not depend on $t$ and $\omega'$ is a complex number with real part given by the oscillation frequency $\omega$ and imaginary part given by the \emph{growth rate} $\lambda$, i.e., $\omega'=\omega+i\lambda$. The superradiance condition in terms of the complex frequency
$\omega'$ becomes $0<\operatorname{Re}(\omega')<m\Omega_H$. Whenever this is satisfied, we expect to find that $\operatorname{Im}(\omega')>0$, i.e., that the wave is amplified.

Furthermore, if our goal is to study perturbations, we can restrict ourselves to the linearised version of the equations of motion, i.e., we only need to consider the terms that are linear in the perturbations, since higher order terms are subleading. For the resulting linear equations, the superposition principle holds: a linear combination of solutions is also a solution.
Now, since any real perturbation $\varphi=\operatorname{Re}(\varphi_c)$ can be obtained from the complex perturbation $\varphi_c$ as the linear combination $\frac{1}{2}\left(\varphi_c+\varphi_c^\ast\right)$ (where $\ast$ denotes the complex conjugate operation), we conclude that the study of real modes can be performed by studying complex modes of the form $\varphi_c=\varphi_0(r,x^2)\exp(-i\omega' t+im\phi)$, with $\omega'\in\mathbb{C}$. 
This is the strategy followed in \emph{mode analysis}, i.e., the study of perturbative modes and their growth rates $\operatorname{Im}(\omega')$ (for example, see the historical works \cite{Damour:1976kh,Zouros:1979iw,Detweiler:1980uk}).
Once this kind of studies has been carried out, the next step consists of (at least) numerically solving the full problem of matter coupled with gravity, in which we can also track the evolving, dynamical features of the spacetime.

\subsubsection{Superradiance in Kerr}
\label{subsubsec:superkerr}

Superradiance is a prediction of general relativity that has been investigated in asymptotically flat Kerr spacetime for a long time.
Therefore, it seems appropriate to briefly review, as we do in this section, some of the many works that shed light on this phenomenon.
Moreover, certain notions that one learns in this context carry over to the case of asymptotically AdS spacetimes that we consider in the following.

In asymptotically flat spacetimes, the consequences of the amplification of massless fields are limited. A reflected massless wave carries away energy and angular momentum to infinity, and the black hole settles down to a black hole with slightly smaller energy and angular momentum. On the other hand, it was noted \cite{Bardeen:1972fi,Press:1972zz} that the radial behaviour of a field with mass $\mu> 0$ is governed by an effective potential which displays a potential barrier. The barrier partially confines the wave in a cavity of size $\mu$, thus forcing multiple interactions between the wave and the black hole.
It is thus conceivable for an initially small perturbation to be repeatedly amplified, through multiple superradiant scatterings, and enter the regime in which its effects on the background spacetime are significant. 
For this reason, if the matter equations of motion on a certain background admit a mode solution that undergoes superradiance, we say that the given spacetime background is \emph{superradiantly unstable}. We typically also say that the mode itself is superradiantly unstable.
The existence of superradiantly unstable modes, and the corresponding growth rates, for various types of matter fields on the asymptotically flat Kerr spacetime background were obtained in \cite{Damour:1976kh,Zouros:1979iw,Detweiler:1980uk} under some assumptions on the range of the Kerr black hole parameters.
For massive scalar perturbations on a Kerr background, \cite{Shlapentokh-Rothman:2013ysa} proved that superradiantly unstable modes can be found in the full range of parameters, i.e., the Kerr black hole is in general superradiantly unstable under massive scalar perturbations.

The next step consists of investigating the effects of a growing unstable perturbation on the spacetime, which can only be found by solving the non-linear Einstein equations coupled with the equations of motion of matter.
It was suggested in \cite{Press:1972zz} that the consequences of a perturbation entering the non-linear regime can be dramatic. The authors even suggest that, in the future, it might be possible to place a mirror around a black hole in order to force a perturbation to grow even faster. Then, one could harness the black hole energy, e.g., through cuts in the mirror surface, or release all the energy contained in the mirror at once to obtain a \emph{black hole bomb}.
The major obstacle to observing the non-linear effects of superradiance in numerical simulations is the long evolution time needed for the perturbation to become sizeable and begin to modify the features of the initial spacetime. 
A rough estimate for the time scale of the superradiant instability is given by $\tau\sim\frac{1}{\operatorname{Im}(\omega)}$, where $\omega$ is the complex frequency of the fastest growing mode.
Although the value of $\operatorname{Im}(\omega)$ depends on all the parameters of the problem, it is generally true that the instability time scale decreases by several orders of magnitude if we increase the spin of the perturbation. 
For instance, the instability time scale for a vector field is (spin 1), roughly speaking, three orders of magnitude shorter than the instability time scale for a scalar field (spin 0). 
Nevertheless, $\operatorname{Im}(\omega)$ is in general very small. For example, the shortest time scale for a scalar field propagating on a Kerr spacetime with mass parameter $M$ is roughly $\tau\sim 6.7\times 10^6 M$, which is achieved for a scalar field of mass $\mu\sim 0.42 M^{-1}$ \cite{Brito:2015oca}.
Furthermore, the fully general evolution with no symmetry requirements on the solution requires even longer computing times.

As a consequence, to this date, only few numerical simulations have reached the final state of the superradiant instability in asymptotically flat spacetimes \cite{Sanchis-Gual:2015lje,Baake:2016oku,East:2017ovw,East:2018glu}.
Among these, only the simulation of \cite{East:2018glu} does not impose symmetries on the solution.
In this study, the author considers the case of a Kerr black hole with mass parameter $M$ and rotation parameter $a=0.99$ that interacts with a vector field mode with mass $\mu=0.4 M^{-1}$.
It was found that the perturbation grows and, by time $t\approx10^3 M$, forms a cloud with increasing angular velocity around the black hole. The transfer of energy and angular momentum from the black hole to the cloud starts to slow down at $t\approx 4\times 10^3 M$, and it stops almost completely at $t\approx 8\times 10^3 M$. We say that the instability has saturated. This occurs when the cloud rotates with almost the same angular velocity as the black hole.
Soon after saturation, the cloud starts radiating a large amount of gravitational waves that carry energy and angular momentum to infinity. The final state of the instability is reached at $t\approx10^5 M$, when the cloud has completely dissipated, leaving behind a Kerr black hole with energy and angular momentum smaller than those of the initial black hole.
From this discussion, it is clear that it should be possible to extract energy and angular momentum from the black hole for longer times by adding a mode with smaller growth rate, and thus longer saturation time scale, to the initial data. Due to the long evolution time that this problem requires, the details of this multiple-mode instability and its final state remain unexplored.

\section{Superradiance in Kerr-AdS}
\label{sec:HRres}

As can be seen from the calculation of Section~\ref{subsubsec:wavesAdS}, the conformal boundary of AdS is reached by massless waves in finite coordinate time. If we impose the reflective boundary conditions discussed in Chapter~\ref{Chapter:NoSym}, energy and angular momentum cannot flow across the AdS boundary, which then acts as a potential barrier that completely reflects any incident wave.
In this context, the presence of the AdS boundary implies that even massless perturbations, such as gravitational ones, can undergo multiple superradiant amplifications and give rise to an instability. 
Thus, Kerr-AdS provides a suitable environment to investigate strong gravitational effects in the black hole bomb scenario suggested by \cite{Press:1972zz}, without the need to model a mirror in our general relativity framework.
Nevertheless, the main physical reason to study superradiantly unstable perturbations of Kerr-AdS is the insights on the dual CFT physics that this study can provide. 
In this section, we discuss important advancements in the field of superradiance in Kerr-AdS.

It turns out that black holes that rotate with an angular velocity $\Omega_H<1$, with respect to a stationary observer at $\partial \mathcal{M}$, do not admit perturbations that grow unboundedly in time, and in particular superradiantly unstable modes, as proved by Hawking and Reall \cite{Hawking:1999dp} under the assumption that matter satisfies the Dominant Energy Condition (this includes the vacuum case). The condition $\Omega_H<1$, or equivalently $r_+^2> a$, is referred to as the \emph{Hawking-Reall bound}.
Let us review this result. Consider Kerr-AdS in HHT coordinates (defined in Section~\ref{sec:KAdSsp}) and let $k_{(0)}=\frac{\partial}{\partial T}$ be the generator of the time symmetry at the boundary. In \cite{Hawking:1999dp} it is noted that, if $\Omega_H<1$\footnote{The condition of \cite{Hawking:1999dp} reads $|\Omega_H|<1$, since the authors consider also the case of negative rotation parameter $a$. As argued in Section~\ref{subsubsec:asyblaspa}, we can restrict to positive values of $a$ without loss of generality.} (or, equivalently, $r_+^2> a$), then the co-rotating Killing vector field $\xi=\frac{\partial}{\partial T}+\Omega_H\frac{\partial}{\partial \Phi}$ is timelike everywhere outside the horizon $\mathcal{H}^+$.\footnote{Notice that this does not occur in the asymptotically flat case, i.e., $\xi$ becomes spacelike sufficiently far from $\mathcal{H}^+$.
In other words, observers that rotate with the same angular velocity of the horizon, i.e., those moving along integral curves of $\xi$, can only exist sufficiently close to the horizon.}
In this case, either $K=\frac{\partial}{\partial T}$ or $\xi$ can be normalised to asymptote to $k_{(0)}$, and consequently used to define a notion of time.
The two choices give, in general, different notions of time in the bulk. However, the relevant notion of time for measurements, i.e, the notion of time at the boundary, is the same for both choices.
If we choose $\xi$, there is no ergoregion. Then, the argument of Section~\ref{subsubsec:necerg} tells us that superradiance cannot occur (when repeating that argument, recall that the stationary vector field defining time at the boundary, whose role is here played by the co-rotating vector field $\xi$, was denoted by $K$ in Section~\ref{subsubsec:necerg}).

If $\Omega_H>1$ (or, equivalently, $r_+^2< a$), then $\xi$ is not timelike near $\partial \mathcal{M}$,\footnote{Notice that in this case the event horizon rotates faster than the speed of light for an observer near the boundary moving along integral curves of $K$, i.e., a stationary observer. At the same time, the boundary rotates faster than the speed of light for an observer near the event horizon moving along integral curves of $\xi$, i.e., co-rotating with the horizon.} so $K=\frac{\partial}{\partial T}$ must be used as our choice of stationary vector field. 
However, $K$ has an ergoregion (this is true in any Kerr-AdS spacetime), therefore the arguments of Section~\ref{subsubsec:supblaholmech} and Section~\ref{subsubsec:aldersup} tell us that it may be possible to find oscillating solutions of the matter equations of motion on the Kerr-AdS background that satisfy \eqref{eq:supcond}, and thus are superradiantly unstable.
The superradiant instability under massive scalar field perturbations of all the Kerr-AdS black holes that violate the Hawking-Reall bound was proved in \cite{Dold:2015cqa}. Namely, it was proved that, in the full range $r_+^2<a$, it is possible to identify a range of the scalar mass for which the Klein-Gordon equation admits a mode solution with positive growth rate. The properties of these growing solutions strongly suggest that they satisfy $0<\operatorname{Re}(\omega)<m\Omega_H$ , i.e., they grow due to superradiance. 
Previously, Ref.~\cite{Cardoso:2004hs} had proved the superradiant instability of Kerr-AdS in the regime $r_+\ll1,a\ll 1,a\ll r_+$. In the range $r_+ \lesssim 0.16$, Ref.~\cite{Uchikata:2009zz} had numerically found that there exist values of $a$ for which Kerr-AdS is superradiantly unstable to scalar perturbations.
The case of gravitational perturbations of the Kerr-AdS metric, which is relevant for our study, will be discussed in the next section.

\subsection{Superradiantly unstable gravitational perturbations of Kerr-AdS}
\label{subsec:suungraKAdS}

We now focus on the study of the behaviour of gravitational perturbations (i.e., spin 2 perturbations of the metric) of Kerr-AdS.
In this case the perturbations must satisfy the Einstein equations with a given set of boundary conditions.
Once again, as most studies do, we focus on the case of reflective boundary conditions.
Furthermore, on physical grounds, one imposes conditions at the horizon that only allow for ingoing modes.
Since the objective is to study small perturbations, we can consider only the terms of the Einstein equations that are linear in the deviations from the background, and neglect higher order terms. The resulting equations are the so-called \emph{linearised Einstein equations} (see, for example, Chapter 4 of \cite{wald:1984} for a review).
If there exists a finite energy solution to these equations that grows unboundedly in time, we say that the background spacetime is \emph{linearly unstable}.
This is clearly the case if there exists an unstable mode, i.e., a mode with \emph{complex} frequency $\omega$ such that $\operatorname{Im}(\omega)>0$.\footnote{On the other hand, rigorously speaking, the non-existence of unstable modes, referred to as \emph{mode stability} of the background, does not imply linear stability, due to the following obstructions. First, there may be solutions to the linearised Einstein equations that cannot be written as a superposition of modes. 
Second, a solution given by the superposition of infinitely many modes may grow unboundedly in time even if all its single mode components do not.}
In particular, Kerr-AdS spacetimes that admit unstable modes are linearly unstable.
In physical terms, the frequencies of these modes are all the frequencies at which a perturbed Kerr-AdS black hole can oscillate, at least at early times, with the given boundary conditions.
The linearised Einstein equations for the deviations can be manipulated to give an equation, due to Teukolsky \cite{Teukolsky:1972my,Teukolsky:1973ha}, for a scalar quantity $\delta \Psi_4$ from which the deviations of the metric tensor can be computed. A description of the so-called Newman-Penrose formalism that leads to the Teukolsky equation is beyond the scopes of this section; a review can be found, for instance, in \cite{Chandrasekhar:1985kt}. 
It is important to mention that the Teukolsky equation identifies all the possible perturbations of Kerr-AdS that solve the linearised Einstein equations with the given boundary conditions, except for those that can only change the energy $E$ and the angular momentum $J$ \cite{Kodama:2003jz,Dias:2013hn}.

This framework was used to show that small, i.e., $r_+\ll1$, Kerr-AdS black holes admit unstable gravitational modes, and thus are linearly unstable under gravitational perturbations (satisfying reflective boundary conditions)~\cite{Cardoso:2006wa} (see also Ref. \cite{Kunduri:2006qa} for a higher dimensional study that employs numerical tools outside of the $r_+\ll1$ regime).
Later, Ref. \cite{Cardoso:2013pza} employed numerical methods to perform a detailed study of gravitational quasi-normal modes for the entire allowed range of parameters of the Kerr-AdS spacetime, as we now describe.
The authors work in the Chambers-Moss frame $(t,r,\chi,\phi)$ in which the Teukolsky equation is naturally written \cite{Chambers:1994ap}. 
Although this frame is not adapted to the asymptotic symmetries, and observers at infinity along integral curves of $\frac{\partial}{\partial t}$ rotate with respect to stationary observers, the final results can be straightforwardly translated into a frame $(T,R,\Theta,\Phi)$ adapted to the asymptotic symmetries, by using the defining relations of one frame in terms of the other. 
Given the isometries of the Kerr-AdS background, solutions of the linearised equations can be assumed to oscillate as $e^{-i\tilde\omega t}$ in $t$ with complex frequency $\tilde\omega$, and as $e^{im \phi}$ in $\phi$ with azimuthal number $m$. Moreover, the $r,\chi$-dependence of the amplitude of these oscillations can be separated, i.e., it can be written as $(r-i\chi)^{-2}R(r)S(\chi)$ where $R(r)$ solves an ordinary differential equation (ODE), called radial ODE, and $S(\chi)$ solves a different ODE, called angular ODE. The two ODEs are coupled through $\tilde\omega$ and a parameter $\lambda$, and thus they must be solved simultaneously.
The objective then translates into finding solutions for this ODE system with reflective conditions at $\partial \mathcal{M}$ and conditions at the horizon that only allow for ingoing modes.
The explicit expressions of these conditions, in a form suitable for this formulation of the problem, were previously obtained in \cite{Dias:2013sdc}.
There are two sets of boundary conditions, and each identifies a class of solutions labelled by $s=1,2$. We refer to modes in the first class as \emph{longitudinal}, and to modes in the second class as \emph{transverse}.\footnote{Ref. \cite{Cardoso:2013pza} refers to $s=1$ modes as scalar and $s=2$ modes as vector, but we make a different choice of nomenclature in order to make contact with works in the literature that we review in the next section.}
The following discussion holds for each class separately.
Regular solutions to the ODE system exist if and only if the azimuthal number $m$ is an integer satisfying $-l\leq m\leq l$, where $l$ is an integer larger than or equal to the spin of the perturbation, i.e., $l\geq 2$. $l$ and $m$ determine certain features of the corresponding solutions of the angular ODE, similarly to the more familiar case of spherical harmonics.
Given any value of the Kerr-AdS parameters $(M,a)$, for each pair $(l,m)$, the ODE system admits infinitely many solutions $R(r),S(\chi)$, each with a different number $p=0,1,2,\dots$ of zeros of the radial function $R(r)$. The number $p$ is called \emph{radial overtone}.
Hence, each solution is completely identified (up to an overall multiplication constant) by a triplet $(p,l,m)$. 
In practice, one fixes the values of $(l,m)$, assumes that $R(r)$ has a certain number of zeros $p$, and numerically solves the ODE system for the unique solution $R(r),S(\chi)$ and the corresponding unique values $\tilde\omega,\lambda$.
It is also possible to use the frequency $\tilde\omega$ of a solution to replace one of the labels. It is common to replace $p$ in this way. 
Hence, each solution (with given $s$) can be labelled by $(\tilde\omega,l,m)$. Given a solution $(\tilde\omega,l,m)$, the corresponding $\lambda$ is uniquely determined in terms of $\tilde\omega,l,m$ (or $p,l,m$).
If we include the number $s$ that specifies whether the mode is longitudinal or transverse, then a mode is uniquely identified by the numbers $s,\tilde\omega,l,m$ (or $s,p,l,m$).

The results obtained in this way can then be stated in terms of quantities defined with respect to the frame $(T,R,\Theta,\Phi)$, adapted to the asymptotic symmetries. These are the ``physical'' quantities that an observer at infinity would measure by employing the notion of time and azimuthal coordinate that we defined in Section~\ref{subsubsec:asyphysnot}.
Let $\omega$ denote the physical frequency and $\Omega_H$ denote the physical angular velocity at the horizon.
The linear mode solutions are labelled by the quadruplets $(s,\omega,l,m)$ or $(s,p,l,m)$.
If we fix the radial overtone $p$, modes can be identified by triplets $(s,l,m)$.
The authors of \cite{Cardoso:2013pza} focus on $p=0$ modes, since they are the first ones to become unstable, and thus interesting for the study of superradiance, as the angular velocity $\Omega_H$ of the background is increased from the instability threshold $\Omega_H=1$. 
The numerical values of the mode frequencies provide strong evidence\footnote{Of course, conclusions drawn from numerical studies cannot be considered rigorous proofs.} for the following statements.
\begin{enumerate}
\item Black holes with $\Omega_H<1$ admit no mode with $\operatorname{Im}(\omega)>0$, as expected from the linear stability result by Hawking and Reall discussed in Section~\ref{sec:HRres}. 
\item Black holes with $\Omega_H>1$ admit at least one mode with $\operatorname{Im}(\omega)>0$, thus these black holes are unstable under gravitational perturbations.
\item A mode is unstable, i.e., $\operatorname{Im}(\omega)>0$, if and only if the superradiance condition $0<\operatorname{Re}(\omega)<m\Omega_H$ is satisfied.
\end{enumerate}
In particular, the second and third point imply that, if a Kerr-AdS black hole violates the Hawking-Reall bound, then it is linearly unstable due to superradiance (a similar prediction in a higher dimensional context was obtained in \cite{Kunduri:2006qa}).
This result was later rigorously proved as a corollary of a theorem by \cite{Green:2015kur}, which can be stated as follows.
Given an asymptotically AdS black hole spacetime $(\mathcal{M},g)$ whose event horizon is a Killing horizon with corresponding Killing vector field $\xi$, if there exists a region outside of the horizon where $\xi$ is spacelike, then there exist solutions to the linearised Einstein equations on the background $(\mathcal{M},g)$ that undergo superradiance, i.e., the spacetime is linearly superradiantly unstable.
Notice that this theorem clearly applies to Kerr-AdS black holes that violate the Hawking-Reall bound, but it also applies to more general spacetimes. In particular, it is not required that the spacetime has other Killing vector fields, except for the one that is normal at the event horizon.
Finally, the authors of \cite{Cardoso:2013pza} analyse $l=m$ modes in detail, and find out the following facts that are relevant for our study of superradiance.
\begin{enumerate}
\setcounter{enumi}{3}
\item The longitudinal mode with largest $\operatorname{Im}(\omega)$ over the entire parameter space has $l=m=2$. The transverse mode with largest $\operatorname{Im}(\omega)$ over the entire parameter space has $l=m=2$. The growth of the latter mode is twice faster than the growth of the former mode.
\item
If a longitudinal mode with $\bar l=\bar m$ is unstable, then some longitudinal modes with $l=m\gg\bar m$ are also unstable. In other words, the large longitudinal $l=m$ modes are the first ones to become unstable as we increase $\Omega_H$ from the instability threshold $\Omega_H=1$. 
This statement can be strengthened for transverse modes: if a transverse mode $\bar l=\bar m$ is unstable, then all transverse modes with $l=m>\bar m$ are unstable.
\end{enumerate}

\subsection{Black resonators}
\label{subsec:blares}

As mentioned at various points above, once a spacetime is known to be unstable under certain perturbations, the next question that one would like to answer is the one regarding the non-linear evolution of the instability. 
If we impose reflective boundary conditions, the energy and angular momentum lost by a Kerr-AdS black hole during the evolution of the superradiant instability must return in the bulk in the form of energy and angular momentum of the fields that surround the black hole. As a consequence, the surrounding fields will grow in amplitude and their back-reaction on the black hole will become significant. Hence, an unstable Kerr-AdS black hole must eventually transition to a different spacetime.
This has led to predict the existence of novel black hole solutions that have no asymptotically flat analog.
One example is provided by rotating black holes surrounded by a cloud of massless scalar field, also called \emph{black holes with scalar hair} or \emph{hairy black holes}. Spacetimes of this type were numerically constructed in \cite{Dias:2011at} as solutions to the Einstein equations coupled to two complex massless Klein-Gordon fields in 5 spacetime dimensions.
Reasoning along the same lines that lead to hairy black holes, it was conjectured that the superradiant instability of Kerr-AdS under gravitational perturbations would display a transition to a class of spacetimes called \emph{black resonators} \cite{Cardoso:2013pza}.
In this section, we describe fundamental properties of black resonators, and discuss their connection with the superradiant linear instability of Kerr-AdS. 

Let us first review how the existence of these solutions can be anticipated.
The argument was originally proposed in \cite{Kunduri:2006qa} in a higher dimensional context.
Point 3 of the previous section strongly suggests that a mode $(\omega,l,m)$ is normal, i.e., $\operatorname{Im}(\omega)=0$, if and only if it satisfies $\operatorname{Re}(\omega)=m\Omega_H$. In \cite{Cardoso:2013pza}, this was was shown numerically up to some small error.
First, notice that this implies that a Kerr-AdS spacetime can have at most one longitudinal normal mode and one transverse normal mode with a fixed pair $(l,m)$, and the two would have the same frequency.
This implies that a normal mode is invariant under the diffeomorphisms generated by $\xi=\partial_T+\Omega_H\partial_\Phi$, i.e., $\mathcal{L}_\xi\left(e^{-i\omega T+im\Phi}\right)=0$, although the mode is not invariant under the diffeomorphisms generated by $\partial_T$ and $\partial_\Phi$ individually. 
Although we have obtained these modes by assuming that their back-reaction on the Kerr-AdS background is negligible, and thus solving the linearised Einstein equations, nothing seems to prevent us from considering an oscillating deviation from Kerr-AdS of the type $e^{-i\omega T+i m\Phi}$, with $\operatorname{Im}(\omega)=0$ and $\operatorname{Re}(\omega)=m\Omega_H$, with amplitude $h_0(r,\chi)$ satisfying the non-linear Einstein equations.
If a deviation of this kind exists, it would neither die off in time nor grow (since $\operatorname{Im}(\omega)=0$). Therefore, it would constitute a new long-lived vacuum spacetime solution, describing a rotating black hole surrounded by a ``lump'' of gravitational field.
This new spacetime would have the isometries preserved by the generating normal mode, that is, the diffeomorphisms generated by $\xi$. 
For small $h_0(r,\chi)$, we must recover the results of the case with no back-reaction. In particular, this implies that $\xi$ would be spacelike near the boundary and timelike near the horizon. Hence, these solutions would be neither stationary nor axisymmetric. 
The existence of such a spacetime is not ruled out by the rigidity theorems, since these apply to stationary spacetimes.

Non-linear numerical solutions of the Einstein equations with all the above-mentioned features were explicitly constructed in 4 dimensions in \cite{Dias:2015rxy} by obtaining the non-linear generalisation of the longitudinal $l=m=2$ normal mode of several Kerr-AdS black holes. 
Their numerical construction in 5 dimensions was later presented in \cite{Ishii:2018oms}.
These spacetimes were called \emph{black resonators}, and have the following properties.
The event horizon is still a Killing horizon with respect to the Killing vector field $\xi$, which is said to be a \emph{helical vector field} since its integral curves draw helices in space.
Let $k_{(0)}$ be the boundary limit of asymptotic time symmetries generators and $m_{(0)}$ the boundary limit of asymptotic rotation symmetry generators.
We use these to define a notion of time and azimuthal angle, as explained in Section~\ref{subsubsec:asyphysnot}.
Consider a frame $x^\alpha=(t,x^1,x^2,\phi)$ where $t$ is a time coordinate and $\phi$ is an azimuthal coordinate.
For a spacetime that is neither stationary nor axisymmetric, the angular velocity cannot be defined as we did in Section~\ref{subsubsec:asyphysnot} (for instance, there is no natural notion of locally non-rotating observers). 
However, we can still define the angular velocity, measured at $\partial \mathcal{M}$, of the null curves at the horizon with tangent $\xi$ (see again Section~\ref{subsubsec:asyphysnot}), which is given by $\frac{d\phi}{dt}=\xi^\phi/\xi^t$ in the frame $x^\alpha$.
It is customary to refer to this quantity as the angular velocity of the black resonator and denote it by $\Omega$.
Black resonators have $\Omega>1$, as expected since the co-rotating Killing vector field is spacelike near the boundary where $\Omega$ is measured. 
Diffeomorphisms that send a point with coordinates $(t,r,\theta,\phi)$ to the point with coordinates $(t+\frac{2\pi}{\Omega},r,\theta,\phi)$ are (discrete) isometries of the spacetime.
For this reason, we say that a black resonator is a \emph{time-periodic} spacetime with period $\frac{2\pi}{\Omega}$ and frequency $\Omega$.
Typically, we also normalise $\xi$ in such a way that its boundary limit $\xi_{(0)}$ is given by $k_{(0)}+\Omega m_{(0)}$. 
If $\xi$ is normalised in this way, then it can be written as $\xi=K+\Omega M$ for some \emph{asymptotic} Killing vector fields $K$ and $M$, which asymptote to $k_{(0)}$ and $m_{(0)}$, respectively.

It is natural to speculate that a Kerr-AdS black hole transitions to a black resonator at some stage of the superradiant instability.
The second law of black hole mechanics (or the second law of thermodynamics) allows this transition.
In fact, for given values of energy and angular momentum, black resonators have larger area $A$, and thus entropy $S=A/4$, than Kerr-AdS black holes \cite{Cardoso:2013pza,Dias:2015rxy}.
However, these new spacetimes cannot be the end-point of the instability, since they are also superradiantly unstable according to the theorem by \cite{Green:2015kur}, stated in Section~\ref{subsec:suungraKAdS}.
For a special class of 5-dimensional black resonators, this was numerically confirmed in \cite{Ishii:2020muv} in an Einstein-Klein-Gordon theory.
For this class of solutions, Ref. \cite{Ishii:2020muv} also finds that any perturbation that does not grow in time (we refer to this as a normal perturbation) is a superposition of multiple normal modes, each oscillating with a different (purely real) frequency given by an integer multiple of $\Omega$, and still invariant under the diffeomorphisms generated by the helical Killing vector field.
Under the assumption that this result holds for general 4-dimensional black resonators, then one could repeat the reasoning above and anticipate the existence of solutions to the Einstein equations that oscillate at several frequencies.
In turn, these are likely to have other multi-oscillating normal perturbations. Therefore, one would end up with a tower of multi-frequency solutions, that were called \emph{multi-black resonators}.
A special class of 5-dimensional multi-black resonators generated from black resonators by scalar field perturbations, called \emph{hairy black resonators}, was obtained in \cite{Ishii:2021xmn}. 
The thermodynamical properties and relations with other known 5-dimensional black hole solutions were also discussed.
The evolution of the superradiant instability of Kerr-AdS can be expected to go through one or more multi-black resonator phases before reaching its end-point.
However, these solutions are also likely to have their own instabilities, hence the question regarding the end-point remains.
One of two scenarios were conjectured to occur \cite{Niehoff:2015oga}: either the oscillating horizon pinches off in finite time, exposing a naked singularity, or there is no stationary end-point.
The former is a clear violation of WCCC, similar to the one proposed by Gregory and Laflamme \cite{Gregory:1993vy} for black strings and black branes in higher dimensions (see \cite{Gregory:2011kh} for a review).
In the latter scenario, spacetime oscillations would create ripples at smaller and smaller scales, until reaching the so-called Planck scale where a quantum theory is needed to describe the dynamics. 
If sub-Planck scales become relevant in finite time, this scenario is also a violation of the spirit of the WCCC, since initial data that have a consistent classical description reaches a state to which the classical theory no longer applies.
As anticipated in Section~\ref{subsec:WCCC}, any such WCCC violation in AdS/CFT would provide an opportunity to study the dynamical appearance of a naked singularity from the fully quantum viewpoint of the boundary CFT.
A thorough description of the details of the superradiant instability of Kerr-AdS, and its end-point, is still elusive.

\subsection{The first simulations of the superradiant instability}
\label{subsec:simsupin}

The first long-time simulations of the superradiant instability of Kerr-AdS were presented in \cite{Chesler:2018txn} and \cite{Chesler:2021ehz}.
The numerical scheme employed in these works, which is reviewed in \cite{Chesler:2013lia}, is based on the characteristic formulation that we mentioned in the introduction of Chapter~\ref{Chapter:NoSym}.
This formulation differs from the Cauchy one at a fundamental level, as we now explain.
One of the main differences can be identified in the fact that the characteristic scheme employs coordinates $(t,u,\theta,\phi)$ where $(t,\theta,\phi)$ are coordinates adapted to the boundary $\mathbb{R}\times S^2$ of the type defined in Section~\ref{subsec:asygloAdSsp}, and $u$ is the affine parameter along \emph{ingoing null} geodesics, i.e., geodesics that start from the boundary (at $u=0)$ and enter the bulk.
Hence points at fixed $t$ define a null hypersurface.
The main issue of this approach is the fact that spacetime points at which the null geodesics intersect cannot be described by these coordinates, since they would not correspond to a unique value of the coordinate $u$.
We say that such points constitute a \emph{coordinate singularity}. Clearly, this type of singularity has no physical relevance, and differs from the physical notion of  curvature singularity (defined in Section~\ref{subsec:WCCC}). It simply means that the coordinate system $(t,u,\theta,\phi)$ cannot cover the entire spacetime.
In practice, the components of the metric and other physically relevant tensors in coordinates $(t,u,\theta,\phi)$ would diverge at the intersection points, making it impossible for the scheme to handle these quantities.
For this reason, the characteristic scheme of \cite{Chesler:2013lia} is only suitable to studies in which an apparent horizon is present. In fact, in these scenarios, one can excise a region inside the horizon in order to eliminate all possible coordinate singularities from the numerical domain.
The simulation can then proceed as long as geodesics do not intersect outside of the excised region.

Let us now discuss the specific application of the characteristic formulation that leads to the simulations in \cite{Chesler:2018txn} and \cite{Chesler:2021ehz}.
The usual reflective boundary conditions are imposed, i.e., the boundary metric is required to be in the conformal class of the metric of the ESU, which reads \eqref{eq:ESU} in coordinates $x^a=(t,\theta,\phi)$.
In this scheme, a choice of initial data on a null hypersurface at $t=0$, $\mathcal{N}_0$, consists of a choice for the $m,n$ components of the metric $\tilde\gamma$ on $\mathcal{N}_0$, rescaled in order to fix its determinant to a desired function, and a choice of the boundary energy-momentum tensor $\langle T_{ta}\rangle_{CFT}|_{t=0}$ in coordinates $x^a$.
$\tilde\gamma$ is chosen to be equal to its pure Kerr-AdS value. 
The Kerr-AdS parameters are slightly different for the two numerical studies.
The deviation of the boundary energy-momentum tensor from the pure Kerr-AdS value, $\langle \Delta T_{ta}\rangle_{CFT}|_{t=0}$, is chosen to seed the superradiant instability.
At $t=0$, the components of the boundary energy-momentum tensor are functions of the coordinates $x^i=(\theta,\phi)$ on $S^2$. Hence they can be expanded in either scalar, vector or tensor spherical harmonics (defined, for our choice of conventions, in Appendix~\ref{sec:svtsphharm}), according to how they transform under a change of coordinates on $S^2$. For instance, $\langle  \Delta T_{tt}\rangle_{CFT}|_{t=0}$ is a scalar with respect to indices associated with $x^i$, hence it can be expanded in scalar spherical harmonics. Similarly, $ti$ components are vectors with respect to $x^i$, hence $\langle  \Delta T_{ti}\rangle_{CFT}|_{t=0}$ can be expanded in terms of vector harmonics.
The fact that one uses indices $s$ and $l,m$ to identify both spherical harmonics and the Teukolsky perturbations, discussed in Section~\ref{subsec:suungraKAdS}, is not an accident.
A choice of a given pair $(l,m)$ for the scalar spherical harmonics in the expansion of $\langle \Delta T_{tt}\rangle_{CFT}|_{t=0}$ determines initial data that correspond to a $t=0$ slice of Kerr-AdS perturbed with a combination of longitudinal, i.e., $s=1$, Teukolsky modes with the same $(l,m)$ (but, in general, different radial overtone).
Similarly, a choice of a given $s$ and a given pair $(l,m)$ for the vector spherical harmonics in the expansion of $\langle \Delta T_{ti}\rangle_{CFT}|_{t=0}$ determines initial data that correspond to a $t=0$ slice of Kerr-AdS perturbed with a combination of Teukolsky modes with the same $s$ and $(l,m)$ (but, in general, different radial overtone).\footnote{The perturbations given by spherical harmonics with $l=0,1$ can only change the energy and angular momentum of the black hole, and are those that are not included in the set of solutions to the Teukolsky equation.}'\;\footnote{In the Schwarzschild case $a=0$, Ref. \cite{Dias:2013sdc} provides the explicit map between perturbations in the Kodama-Ishibashi formalism \cite{Kodama:2003jz}, which are given by combinations of scalar, vector and tensor spherical harmonics with $s=1,2$ times a time and radius-dependent part, and Teukolsky perturbations. Perturbations given by  tensor spherical harmonics with $s=3$ (with the labelling of Appendix~\ref{sec:svtsphharm}) would violate the condition that the boundary metric is in the conformal class of the metric of the ESU, thus they are not allowed for asymptotically globally AdS spacetimes.}
Then, for any choice of Kerr-AdS parameters, the work by \cite{Cardoso:2013pza}, described in the previous section, provides the frequency of those $(s,l,m)$ modes. 
As can be expected from point 4 of Section~\ref{subsec:suungraKAdS}, for the backgrounds of \cite{Chesler:2018txn} and \cite{Chesler:2021ehz}, the fastest growing $s=1$ and $s=2$ modes have $l=m=2$. Hence, in order to minimise the time scale of the instability, both references perturb Kerr-AdS with the $s=1,l=m=2$ and $s=2,l=m=2$ harmonics.
Moreover, since black resonators are expected to be unstable to modes with higher $m$ (this expectation descends from point 5 of Section~\ref{subsec:suungraKAdS}), Ref. \cite{Chesler:2018txn} includes harmonics with $l=m=3$, while Ref. \cite{Chesler:2021ehz} includes harmonics with $l=m=3$ and $l=m=4$
(both scalar and vector with $s=1,2$, added to the corresponding components of $\langle \Delta T_{ta}\rangle_{CFT}|_{t=0}$).
For instance, the choice of \cite{Chesler:2018txn} is
\begin{eqnarray}
\label{eq:indatchlo}
\langle \Delta T_{tt}\rangle_{CFT}|_{t=0}&=&-\operatorname{Re}\left(\frac{4}{30} Y^{22}+\frac{1}{30}Y^{33}\right),\nonumber\\
\langle \Delta T_{ti}\rangle_{CFT}|_{t=0}&=&-\operatorname{Re}\left(\frac{1}{10}\left(\mathcal{V}_i^{122}+\mathcal{V}_i^{222}\right)+\frac{1}{40}\left(\mathcal{V}^{133}+\mathcal{V}^{233}\right)\right).
\end{eqnarray}
Notice that these perturbations do not change the $t=0$ value of the energy, $E|_{t=0}=\int d\theta d\phi \sin\theta \langle T_{tt}\rangle_{CFT}|_{t=0}$, and angular momentum, $J|_{t=0}=-\int d\theta d\phi \sin\theta \langle T_{t\phi}\rangle_{CFT}|_{t=0}$, since the integral over the unit round sphere of any spherical harmonic vanishes.
The fastest growing mode in this initial data is the one with $s=2,l=m=2$ and zero radial overtone.\footnote{To be precise, the authors do not explicitly state that this is the frequency of the mode with zero radial overtone. It is a guess that we make based on the fact that, for the reason explained in Section~\ref{subsec:suungraKAdS}, one typically focuses on zero radial overtone modes.}
This has frequency $\omega\approx 3.38+ 0.0225i$ for the chosen Kerr-AdS background with $E=0.2783,J=0.0635$.
Notice that all the spherical harmonics have amplitudes of the same order.

Let us describe the salient features of the observed superradiant evolution.
We first focus on the simulation described in \cite{Chesler:2018txn}, which follows the instability for approximately 900 units of time.
\begin{figure*}[t!]
        \centering
        \includegraphics[width=6.0in,clip=true]{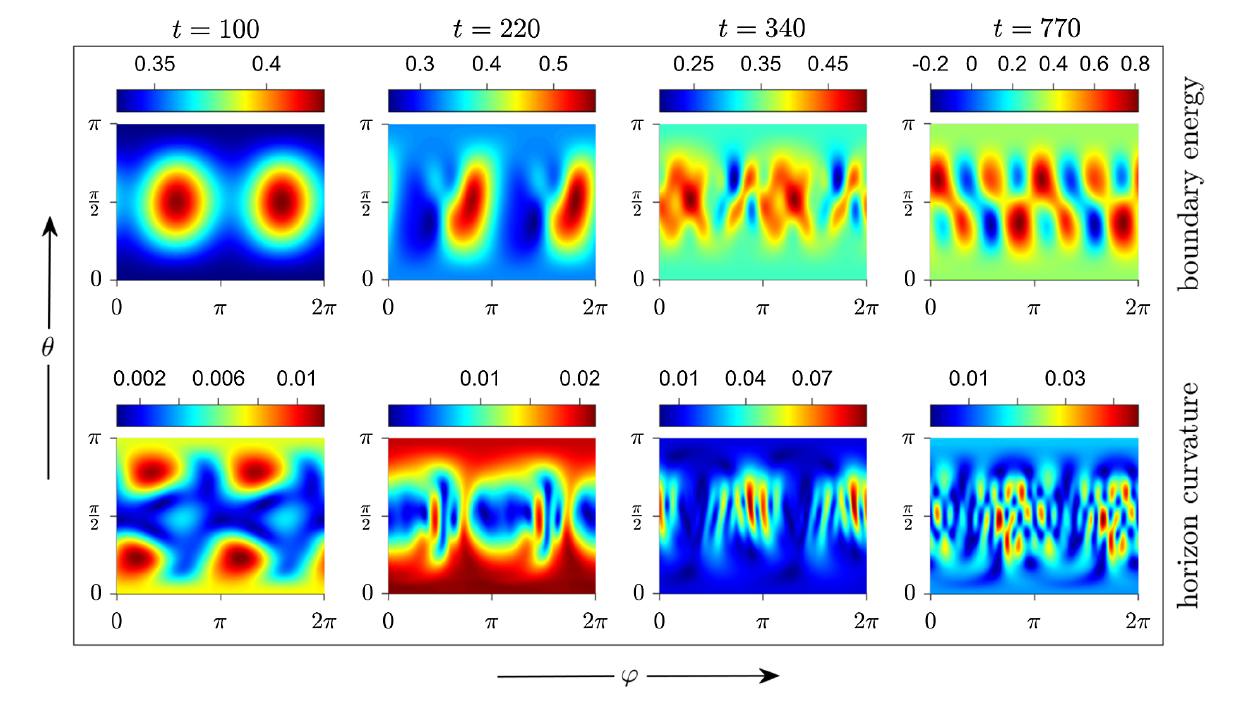}
\parbox{5.0in}{\caption{Snapshots from the simulation of the superradiant instability of a perturbed Kerr-AdS spacetime presented in \cite{Chesler:2018txn}. 
Top: bondary energy density.
Bottom: Extrinsic curvature at the event horizon.
        }\label{fig:chloev}}
\end{figure*}
The top line of Figure~\ref{fig:chloev} shows the boundary energy density at various times during the evolution of \cite{Chesler:2018txn}.
At $t=100$, the two peaks separated by $\Delta\phi\approx \pi$ indicate that the dominant mode has $m=2$.\footnote{
This is simply deduced by comparing the energy density profile at $t=100$ with the one obtained by adding a scalar spherical harmonic $Y^{lm}$ with $m=2$ (and any $l$) to the pure Kerr-AdS value of the boundary energy density. It would be visually evident that the two profiles are very similar.}
Soon after, the amplitude of this mode becomes large enough to enter the non-linear regime, and leads the transition to a black resonator phase, which is the dominant phase between $150\lesssim t \lesssim 400$. 
During this phase, the boundary energy density rotates with angular velocity $\Omega\approx 1.7$.
This is the same value that we get from $\operatorname{Re}(\omega)/2$ where $\operatorname{Re}(\omega)\approx 3.38$ is the real part of the frequency of fastest growing mode with $s=2,l=m=2$.
This is an indication that we reached the black resonator solution that we would obtain, as in \cite{Dias:2015rxy}, by computing the non-linear generalisation of a transverse (i.e., $s=2$) $l=m=2$ mode on a Kerr-AdS background for which that mode is normal, i.e., one with $\Omega_H=\operatorname{Re}(\omega)/m\approx1.7$.
The authors also verify that $\partial_t+\Omega\partial_\phi$ is an approximate Killing vector field, by showing that all the amplitudes of the excited modes rotate at the same $\Omega\approx 1.7$ in the interval $100\lesssim t \lesssim 400$.
A spherical harmonic decomposition of the boundary energy-momentum tensor reveals that the amplitude of the $s=2,l=m=2$ mode reaches a plateau during the black resonator phase.
After this epoch, a so-called secondary instability takes place, led by modes with higher $m$, as expected for the unstable modes of the black resonator. 
Modes with even $m=4,6$, that were not originally contained in the initial data, grow and start competing with the $l=m=2$ mode, as can be seen from the complicated patterns in the boundary energy density.
At a later stage, each of these perturbations rotates with a different angular velocity, thus the spacetime appears to lose the helical symmetry.
The bottom line of Figure~\ref{fig:chloev} displays the values of the extrinsic curvature at the event horizon, and it  shows that structure forms also at the horizon.
Notice that deformations form on small scales, and thus require fine grids to be resolved.
The simulation of \cite{Chesler:2018txn} stops during the secondary instability phase.

The recent work by \cite{Chesler:2021ehz} extends these results by following the evolution for 3000 time units.
They perform two simulations which differ due to slightly different choices of the amplitudes of the spherical harmonics in the initial data for $\langle \Delta T_{ta}\rangle_{CFT}|_{t=0}$. The phenomena observed in these simulations are similar. 
\begin{figure*}[t!]
        \centering
        \includegraphics[width=6.0in,clip=true]{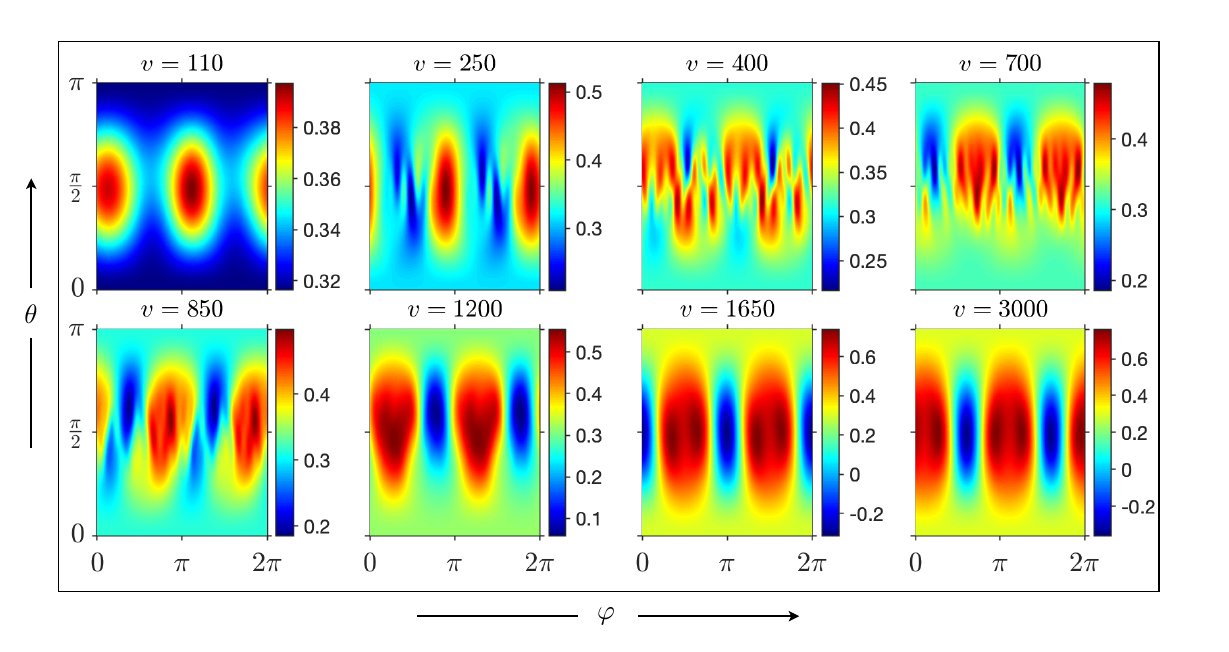}
\parbox{5.0in}{\caption{Snapshots of the boundary energy density from one of the simulations presented in \cite{Chesler:2021ehz}. $v$ is the time coordinate that we denote by $t$.
        }\label{fig:chev}}
\end{figure*}
The primary instability appears to be driven by the fastest growing mode $s=2,l=m=2$ with frequency $\omega=3.41+0.02i$, as can be noticed from the boundary energy density profile in Figure~\ref{fig:chev}. 
However, other modes with even $m$ acquire non-negligible energy during this phase. Thus, unlike in \cite{Chesler:2018txn}, the first transition does not seem to lead to a pure black resonator.
During this epoch, the transverse modes are larger than the longitudinal ones.
Following this phase, a second epoch of growth is identified, in which certain longitudinal modes with even $m$ grow, while the transverse mode amplitudes decay.
The system reaches a state in which there are only a handful of modes with non-negligible amplitude. Among these, the mode with $(s=1,l=2,m=2)$ is 30 times larger than the next largest mode, and the transverse modes are approximately $10^{-3}$ times smaller than any longitudinal mode.
These amplitudes remain essentially constant between $t\approx 1650$ and $t\approx 3000$, and their constant values are the same for both sets of initial data.
The fact that the longitudinal modes are dominant at late times might have a fundamental explanation, which is not known. 
In summary, this simulation ends with a long-lived phase described by a black hole oscillating with multiple fundamental frequencies, reminiscent of the multi black resonators described above but with no Killing vector field. This state is only roughly approximated by a black resonator generated by the $(s=1,l=2,m=2)$ mode. 
The dual CFT is in an exotic state with negative energy density.
Whether this is the final state of the instability, and whether the dual CFT state is in thermal equilibrium, remains an open question.

These simulations are crucial steps towards a complete understanding of the superradiant instability.
Remarkably, they were obtained with very limited computational resources, despite the fact that no symmetry is imposed on the solution. In particular, the simulations of \cite{Chesler:2021ehz} only needed a handful of weeks to run on a single computing unit of a laptop computer.
This is a consequence of the fact that the evolution equations of the underlying scheme are not solved for the unknowns by approximating the derivatives with finite difference stencils, as we do in our scheme. Instead, the numerical evolution strategy consists of writing, at each grid point, each unknown as a linear combination of the special functions called Chebyschev polinomials. The evolution equations provide relations between the coefficients of this combination. From these relations, the coefficients can be obtained with basic operations once initial data and boundary conditions are specified. Numerical methods that employ an expansion in terms of special functions go under the name of \emph{spectral methods}.
This time integration strategy converges faster than the typical time integration for finite difference schemes. As a consequence, it is possible to employ coarser grids and a time step that is less stringently constrained by the CFL condition.
Nevertheless, the use of finer grids would be necessary to resolve small scale dynamics, and its consequences for the superradiant instability.
This would require AMR techniques, which in turn require to run simulations in parallel over many computing units.
Our numerical scheme has these capabilities, in addition to allowing for fully general simulations in AdS with no symmetries.
Moreover, unlike the scheme of \cite{Chesler:2018txn,Chesler:2021ehz}, our scheme solves the Cauchy problem in general relativity, i.e., it foliates the spacetime by spacelike slices and evolves initial data on one of such slices.
Therefore, studying the superradiant instability with our techniques provides an opportunity to confirm previous results obtained with significantly different numerical techniques, and improve on such results to achieve a better understanding of the features of the instability.
Thus, we set out to simulate the superradiant instability of Kerr-AdS, as explained in the rest of this chapter.

\section{Cauchy evolution of perturbed Kerr-AdS}
\label{sec:cauevKAdS}

We now describe how the numerical scheme presented in Chapter~\ref{Chapter:NoSym} can be adapted to achieve long time evolution of perturbed Kerr-AdS with no symmetry assumptions, and thus employed for the study of the superradiant instability of Kerr-AdS.
This section contains original, unpublished work.

\subsection{Kerr-AdS in suitable coordinates}
\label{sec:horpenbackgr}

Let us first discuss our implementation of the Kerr-AdS metric $g_{\text{KAdS}}$ in the scheme.
We have discussed two sets of coordinates on Kerr-AdS in Section~\ref{sec:KAdSsp}.
As explained in Section~\ref{sec:AHfind}, retaining the information contained at points external to the event horizon at fixed evolution time is essential to preserve stability and convergence in a simulation. This implies that we can only excise a region inside the event horizon. Therefore, we must employ coordinates in which $g_{\text{KAdS}}$ is regular at the horizon, also called \emph{horizon-penetrating} coordinates.
Coordinates $x^{\alpha'}=(T,r,\theta,\Phi)\in(-\infty,+\infty)\times(0,+\infty)\times(0,\pi)\times(0,2\pi)$ with this property, called Kerr-Schild coordinates, are provided in \cite{Gibbons:2004uw}. They are related to Boyer-Lindquist coordinates $(t,r,\theta,\phi)$ by 
\begin{equation}
\label{eq:BLtoKS}
dT=dt+\frac{2M}{(1+r^2)(V-2M)}dr, \quad d\Phi=d\phi-adt+\frac{2aM}{(r^2+a^2)(V-2M)}dr,
\end{equation}
where $
V=\frac{1}{r}(1+r^2)(r^2+a^2)$.
The Kerr-AdS metric in Kerr-Schild coordinates reads
\begin{equation}
\label{eq:KSKAdS}
g_{\text{KAdS}}=\hat{g}+\frac{2M}{U}\left(\lambda_{\alpha'} dx^{\alpha'} \right)^2,
\end{equation}
where $U=\frac{1}{r}(r^2+a^2\cos^2\theta)$ , $\hat g$ is the pure AdS metric, which in these coordinates reads
\begin{equation}
\hat g=-\frac{(1+r^2)\Delta_\theta}{\Xi}dT^2+\frac{\Sigma^2}{(1+r^2)(r^2+a^2)}dr^2+\frac{\Sigma^2}{\Delta_\theta}d\theta^2+\frac{(r^2+a^2)}{\Xi}\sin^2\theta d\Phi^2,
\end{equation}
and $\lambda=\lambda_{\alpha'} dx^{\alpha'}$ is a null covector given by
\begin{equation}
\lambda_{\alpha'} dx^{\alpha'}=\frac{\Delta_\theta}{\Xi}dT+\frac{\Sigma^2}{(1+r^2)(r^2+a^2)}dr-\frac{a}{\Xi}\sin^2\theta d\Phi.
\end{equation}
The functions $\Delta_\theta,\Sigma,\Xi$ are given in \eqref{eq:fnrule}.
Notice that the coordinate transformation \eqref{eq:BLtoKS} depends on the parameters of the Kerr-AdS spacetime under consideration. 
As a consequence, the $M=0$ case of the Kerr-AdS metric in Kerr-Schild coordinates, i.e., the pure AdS metric, depends on $a$. In other words, Kerr-Schild coordinates cannot be considered a generalisation of the spherical coordinates on pure AdS, employed in Section~\ref{subsec:pureAdS}, to the case of Kerr-AdS. For this reason, we refer to Kerr-Schild coordinates as \emph{spheroidal coordinates}. 
This fact also implies that there is no representative of the boundary metric in spheroidal Kerr-Schild coordinates which is conformal to the metric of ESU in the form \eqref{eq:ESU}, i.e., Kerr-Schild coordinates are not adapted to the asymptotic symmetries (i.e., they are not coordinates of the type defined in Section~\ref{subsec:asygloAdSsp}).

Now we recall that the boundary prescription for numerical stability, presented in Section~\ref{sec:pre_sta}, assumes the use of coordinates adapted to the asymptotic symmetries.
Therefore, in order to stably evolve pure Kerr-AdS initial data in our scheme with our prescription, we must find coordinates on Kerr-AdS that are adapted to the asymptotic symmetries, while also being horizon-penetrating.
Coordinates that satisfy these requirements, which we call \emph{Kerr-Schild spherical coordinates} and we denote by $(T,R,\Theta,\Phi)\in(-\infty,+\infty)\times(0,+\infty)\times(0,\pi)\times(0,2\pi)$ can be obtained from Kerr-Schild spheroidal coordinates $(T,r,\theta,\Phi)$ by defining new $R,\Theta$ coordinates, related to $r,\theta$ by the same relation that brings the $r,\theta$ BL coordinates to the $R,\Theta$ HHT coordinates, i.e.,
\begin{eqnarray}
R^2\cos^2\Theta&=&r^2\cos^2\theta,\nonumber\\
\Xi R^2\sin^2\Theta&=&(r^2+a^2)\sin^2\theta.
\end{eqnarray}
The explicit functions $R(r,\theta)$ and $\Theta(r,\theta)$ are given by the second and third equation in \eqref{eq:nonrotnonhorpenexpl}. Their inverse functions are given by the second and third equation in \eqref{eq:nonrotnonhorpenexplinv}. The expressions below \eqref{eq:nonrotnonhorpenexplinv} also provide the $R$ value of the event horizon, i.e., $R_+(\theta)$ or, equivalently, $R_+(\Theta)$.
We emphasize that Kerr-Schild spherical coordinates are different from HHT coordinates. In fact, the $T,\Phi$ Kerr-Schild coordinates do not correspond to the $T,\Phi$ HHT coordinates, as is evident from the fact that they are related to BL coordinates through different relations. In particular, spherical Kerr-Schild coordinates are horizon-penetrating but HHT coordinates are not. To the best of our knowledge, spherical Kerr-Schild coordinates have not been previously introduced in the literature.
It is straightforward to verify that, in the near-boundary limit, i.e., at large $R$, the Kerr-AdS metric asymptotes to the pure AdS metric in its spherical form \eqref{eqn:ads4}, as is the case in HHT coordinates.
This implies that, at the boundary, $(T,\Theta,\Phi)$ are coordinates of the type defined in Section~\ref{subsec:asygloAdSsp}.
Hence, in order to make contact with previous notation, we denote $(T,\Theta,\Phi)$ by $x^{\dot m}=(t,\theta,\phi)$. One should not confuse these with BL coordinates, which we will not use in the rest of this thesis.
Then, the set of spherical Kerr-Schild coordinates on Kerr-AdS is denoted by $x^{\dot\alpha}=(t,R,\theta,\phi)$.
We do not write the Kerr-AdS metric in spherical Kerr-Schild coordinates explicitly because it would be too cumbersome.
However, it is straightforward to show that, in spherical Kerr-Schild coordinates $x^{\dot\alpha}$, none of the components of $g_{\text{KAdS}}$ vanish and the asymptotics of $h_{\text{KAdS}}:=g_{\text{KAdS}}-\hat g$ is not consistent with the fall-offs \eqref{eq:HTh} proposed by \cite{Henneaux:1985tv}. In fact, the latter is given by
\begin{eqnarray}
\label{eq:sphKSfo}
&&h^{\text{KAdS}}_{tt}=\mathcal{O}(R^{-1}), \quad h^{\text{KAdS}}_{t R}=\mathcal{O}(R^{-3}), \quad h^{\text{KAdS}}_{t \theta}=\mathcal{O}(R^{-2}),\quad h^{\text{KAdS}}_{t \phi}=\mathcal{O}(R^{-1}), \nonumber\\
&&h^{\text{KAdS}}_{RR}=\mathcal{O}(R^{-5}), \quad h^{\text{KAdS}}_{R\theta}=\mathcal{O}(R^{-4}), \quad h^{\text{KAdS}}_{R\phi}=\mathcal{O}(R^{-3}), \nonumber\\
 &&h^{\text{KAdS}}_{\theta\theta}=\mathcal{O}(R^{-3}), \quad h^{\text{KAdS}}_{\theta\phi}=\mathcal{O}(R^{-2}), \nonumber\\
&& h^{\text{KAdS}}_{\phi\phi}=\mathcal{O}(R^{-1}).
\end{eqnarray}
As explained in Section~\ref{eq:compdefasyAdS}, this is not worrisome: it is always possible to find a change of coordinates that relates the asymptotics of $h_{\text{KAdS}}$ in spherical Kerr-Schild coordinates with the asymptotics \eqref{eq:HTh}. One way to see this is to revert from spherical Kerr-Schild coordinates back to BL coordinates, and then go to HHT coordinates, for which the fall-offs \eqref{eq:HTh} are satisfied.
In summary, we have obtained horizon-penetrating coordinates in which the boundary limit of the Kerr-AdS metric is conformal to the metric of ESU in the form \eqref{eq:ESU}.
Now, in order to obtain Kerr-AdS in a form that can be implemented in our scheme, we need to convert Kerr-Schild spherical coordinates into a coordinate set within the Cartesian class that our scheme employs, i.e., coordinates in which the pure AdS metric reads \eqref{eq:ads4_diag} and the deviation tensor $h_{\text{KAdS}}$ has the asymptotics given by \eqref{eq:carbouncondh}.
It is clear that, starting from Kerr-Schild spherical coordinates, this can be achieved by using the expressions that we used to go from $(t,r,\theta,\phi)$ to $(t,x,y,z)$ coordinates in Section~\ref{subsec:pureAdS}.

The first step consists of defining a compactified radial coordinate $R=2\rho/(1-\rho^2)$, so that $\partial \mathcal{M}$ is at $\rho=1$. We thus obtain Kerr-Schild compactified spherical coordinates $x^\alpha=(t,\rho,\theta,\phi)$. As per previous notation, we denote the boundary coordinates by $x^a=(t,\theta,\phi)$. We also denote the coordinates on the boundary $S^2$ at fixed $t$ by $x^i=(\theta,\phi)$.
Although the pure AdS metric in these coordinates reads as \eqref{eqn:ads4_compact} (where we also set $L=1$), the asymptotics of $h_{\text{KAdS}}$ are different from those in \eqref{eq:sphbounconh}.
In fact, the first non-vanishing term of $h^{\text{KAdS}}_{\alpha\beta}$ appears at order $(1-\rho)$ in every component. We will call coordinates in which the deviation tensor $h$ has these fall-offs \emph{quasi-spherical (Kerr-Schild) coordinates}.
A consequence of the different asymptotics is that, if we wanted to write a code in quasi-spherical coordinates, for instance, in order to evolve perturbations of Kerr-AdS in these coordinates, we would have to define evolution variables by expressions that are different from the ones in Appendix~\ref{subsec:evvarbouconsphcoo}.
In fact, since the asymptotics in these coordinates reads as the asymptotics in Cartesian coordinates, the correct definition of evolution variables for quasi-spherical coordinates is given by the expression used to define the Cartesian variables \eqref{eq:gbarcart}, i.e., $\bar{g}_{\alpha\beta}:=h_{\alpha\beta}$. 
In particular, this implies that we can obtain evolution variables in quasi-spherical coordinates from the Cartesian coordinate variables simply from the tensor transformation law, i.e., from \eqref{eq:cartosph} without the $1/(1-\rho^2)$ factor. 
One can verify that the boundary quantities in terms of evolution variables for quasi-spherical coordinates are still given by the expressions of Section~\ref{sec:bouset2}. This is expected from the fact that quasi-spherical and spherical coordinates coincide on $\partial \mathcal{M}$, and the boundary quantities cannot depend on the radial coordinate that we use to approach the boundary.
For reference, the exact Kerr-AdS values of the boundary quantities are
 \begin{eqnarray}
\label{eq:set_kads_explicit}
\langle T_{tt}\rangle^{\text{KAdS}}_{CFT}&\hspace{-0.0cm}=&\hspace{-0.0cm}\frac{M \left(4+a^2-a^2 \cos 2 \theta \right)}{2 \sqrt{2} \pi  \left(2-a^2+a^2 \cos 2
   \theta\right)^{5/2}}, \nonumber \\
\langle T_{t\theta}\rangle^{\text{KAdS}}_{CFT}&\hspace{-0.0cm}=&\hspace{-0.0cm}0, \nonumber \\
\langle T_{t\phi}\rangle^{ \text{KAdS}}_{CFT}&\hspace{-0.0cm}=&\hspace{-0.0cm}-\frac{3 a M \sin ^2\theta }{\sqrt{2} \pi  \left(2-a^2+a^2 \cos 2 \theta
   \right)^{5/2}}, \nonumber \\
\langle T_{\theta\theta}\rangle^{\text{KAdS}}_{CFT}&\hspace{-0.0cm}=&\hspace{-0.0cm}\frac{M}{2 \sqrt{2} \pi  \left(2-a^2+a^2 \cos 2 \theta \right)^{3/2}}, \nonumber \\
\langle T_{\theta\phi}\rangle^{\text{KAdS}}_{CFT}&\hspace{-0.0cm}=&\hspace{-0.0cm}0, \nonumber \\
\langle T_{\phi\phi}\rangle^{\text{KAdS}}_{CFT}&\hspace{-0.0cm}=&\hspace{-0.0cm}\sin ^2\theta  \frac{M \left(1+a^2-a^2 \cos 2 \theta \right)}{\sqrt{2} \pi 
   \left(2-a^2+a^2 \cos 2 \theta \right)^{5/2}}.
\end{eqnarray}
It is also straightforward to verify that $\langle T_{ab}\rangle^{\text{KAdS}}_{CFT}$ is traceless with respect to $g_{(0)}=-dt^2+d\theta^2+\sin^2\theta d\phi^2$ and conserved, i.e., $\mathcal{D}_{(0)a} \langle T^{ab}\rangle^{\text{KAdS}}_{CFT}=0$, where $\mathcal{D}_{(0)}$ is the Levi-Civita covariant derivative associated with $g_{(0)}$.

Finally, we obtain Kerr-Schild Cartesian coordinates $x^\mu=(t,x,y,z)\in(-\infty,+\infty)\times (-1,1)\times(-1,1)\times(-1,1)$ as $x=\rho\cos\theta$, $y=\rho\sin\theta\cos\phi$, $z=\rho\sin\theta\sin\phi$.
We have verified that the first non-vanishing term of $h^{\text{KAdS}}_{\mu\nu}$ appears at order $(1-\rho(x,y,z))$ (recall $\rho(x,y,z)=\sqrt{x^2+y^2+z^2}$) in every component, which is precisely the fall-offs of \eqref{eq:carbouncondh}. Hence, these $x^\mu$ coordinates on Kerr-AdS are in the class of coordinates that we called Cartesian coordinates in Section~\ref{sec:pre_sta}.
Therefore, we expect the generalised source functions of Kerr-AdS in these coordinates to satisfy the relation \eqref{eqn:target_gauge_txyz}.
We verified that this is indeed the case.
We have also verified that our scheme successfully evolves initial data given by a $t=0$ slice of the exact Kerr-AdS metric in Kerr-Schild Cartesian coordinates. More precisely, the initial data for this test, $\bar g_{\mu\nu}|_{t=0},\partial_t\bar g_{\mu\nu}|_{t=0}$, was obtained from \eqref{eq:gbarcart} with $h_{\mu\nu}$ given by the Kerr-AdS values $h^{\text{KAdS}}_{\mu\nu}:=g^{\text{KAdS}}_{\mu\nu}-\hat{g}_{\mu\nu}$. In particular, we have $\partial_t\bar g_{\mu\nu}|_{t=0}=0$.
Such test runs give the opportunity to verify that our boundary extrapolation, described in Appendix~\ref{sec:extrapconvbdy}, reproduces the expected exact Kerr-AdS values of all the boundary quantities listed in Section~\ref{sec:bouset2} to a good degree of accuracy.
This suggests that our scheme would allow to add perturbations to such initial data, and obtain stable evolution for a perturbed Kerr-AdS black hole, regarded as a deviation from the pure AdS metric $\hat g$.

However, a different route seems more natural to follow, as we now explain.
Although the initial data in the test runs described above solves the evolution equations exactly, the finite difference version of these equations will still have some non-vanishing truncation error $\tau_\Delta$, defined in \eqref{eq:truerr}, even at $t=0$. The residual that one has in the first iteration of the NGS time integration method at time level $t=0$ is precisely equal to $\tau_\Delta$, and non-vanishing.
With just a few relaxation sweeps, NGS brings the residual below the desired tolerance and provides a solution at time $t=\Delta t$ that is close, but not identical, to the Kerr-AdS solution at that time. In other words, in these test runs, we recover the entire Kerr-AdS spacetime with the addition of small solution error $\epsilon_\Delta$.
Although $\epsilon_\Delta$ is expected to remain small during the evolution, it would still pollute the numerical solution.
This can clearly be avoided by using $g_{\text{KAdS}}$ as a background metric, instead of pure AdS $\hat g$. In other words, denoting the full metric solution in Kerr-Schild Cartesian coordinates by $g_{\mu\nu}$, we define the evolution variables $\bar{g}_{\mu\nu}$ for our scheme by \eqref{eq:gbarcart} where $h_{\mu\nu}:=g_{\mu\nu}-g^{\text{KAdS}}_{\mu\nu}$.
Clearly, if we do so, the initial data for pure Kerr-AdS becomes trivial, i.e., $\bar{g}_{\mu\nu}|_{t=0}=\partial_t\bar{g}_{\mu\nu}|_{t=0}=0$, thus the truncation error at $t=0$ vanishes to machine precision, and the solution error at later times will also vanish to machine precision.
This is the strategy that we follow to obtain the simulations discussed in Section~\ref{sec:preresK}.
Crucially, since the near-boundary asymptotics of deviation of Kerr-AdS from pure AdS in Kerr-Schild Cartesian coordinates is consistent with the Cartesian asymptotics \eqref{eq:carbouncondh}, the results of Section~\ref{sec:gauge_choice} are still valid. In particular, we can employ the gauge choice of source functions given by \eqref{eqn:target_gauge_txyz}, and we find further confirmation that this choice leads to stable evolution.

\subsection{Initial data}
\label{sec:pert}

We now describe the choice of initial data for our first long-time simulation of perturbed Kerr-AdS.
The parameters of the Kerr-AdS background are set to
\begin{equation}
M=0.2783,\quad a=0.208723.
\end{equation}
These give
\begin{equation}
\label{eq:propback}
r_+=0.5566,\quad \Omega_H=1.30667,\quad E=0.30423,\quad J=0.0635.
\end{equation}
We wish to perturb this background with $s=1,l=m=2$ and $s=2,l=m=2$ modes.
With this in mind, and following \cite{Chesler:2018txn}, we wish to prescribe our initial data, $\bar g_{\mu\nu}|_{t=0},\partial_t \bar g_{\mu\nu}|_{t=0}$, in such a way that 
\begin{eqnarray}
\label{eq:Tpert}
\langle T_{tt}\rangle_{CFT}|_{t=0}&=&\langle T_{tt}\rangle^{\text{KAdS}}_{CFT}|_{t=0}+\epsilon_Y \operatorname{Re}\left(Y^{22}\right),\nonumber\\
\langle T_{ti}\rangle_{CFT}|_{t=0}&=&\langle T_{ti}\rangle^{\text{KAdS}}_{CFT}|_{t=0}+\epsilon_{\mathcal{V}} \operatorname{Re}\left(\mathcal{V}_i^{122}+\mathcal{V}_i^{222}\right),\nonumber\\
\langle T_{ij}\rangle_{CFT}|_{t=0}&=&\langle T_{ij}\rangle^{\text{KAdS}}_{CFT}|_{t=0},
\end{eqnarray}
where $\langle T_{ab}\rangle^{\text{KAdS}}_{CFT}$ are given by \eqref{eq:set_kads_explicit}, and $\epsilon_Y,\epsilon_{\mathcal{V}}$ are user-specified values that determine the amplitudes of the perturbations.
This is achieved by the following choice of initial data in quasi-spherical Kerr-Schild coordinates $\bar{g}_{\alpha\beta}|_{t=0},\partial_t \bar{g}_{\alpha\beta}|_{t=0}$ (recall that $\bar{g}_{\alpha\beta}:=h_{\alpha\beta}$ where now $h_{\alpha\beta}$ is the deviation from Kerr-AdS in quasi-spherical coordinates, i.e., $h_{\alpha\beta}:=g_{\alpha\beta}-g^{\text{KAdS}}_{\alpha\beta}$).
\begin{eqnarray}
\label{eq:indatkads}
&&\bar{g}_{tt}|_{t=0}=0,\quad \bar{g}_{t\rho}|_{t=0}=0, \quad
\bar{g}_{t\theta}|_{t=0}= \frac{16\pi}{3}\epsilon_{\mathcal{V}}(1-\rho)\operatorname{Re}\left(\mathcal{V}_\theta^{122}+\mathcal{V}_\theta^{222}\right) ,\nonumber\\
&& \bar{g}_{t\phi}|_{t=0}=\frac{16\pi}{3}\epsilon_{\mathcal{V}}(1-\rho)\operatorname{Re}\left(\mathcal{V}_\phi^{122}+\mathcal{V}_\phi^{222}\right), \nonumber\\
&& \bar{g}_{\rho\rho}|_{t=0}=\frac{8\pi}{3}\epsilon_{Y}(1-\rho)\operatorname{Re}\left(Y^{22}\right),\quad
\bar{g}_{\rho\theta}|_{t=0}=0,\quad \bar{g}_{\rho\phi}|_{t=0}=0, \nonumber\\
&& \bar{g}_{\theta\theta}|_{t=0}=\frac{16\pi}{9}\epsilon_{Y}(1-\rho)\operatorname{Re}\left(Y^{22}\right),\quad
  \bar{g}_{\theta\phi}|_{t=0}=0, \quad
\bar{g}_{\phi\phi}|_{t=0}=\frac{16\pi}{9}\epsilon_{Y}(1-\rho)\operatorname{Re}\left(Y^{22}\right),\nonumber\\
&&\partial_t \bar{g}_{\alpha\beta}|_{t=0}=0.
\end{eqnarray}
Using \eqref{eq:set_explicit} and recalling the definition $\bar{g}_{(1)\alpha\beta}:=\frac{\partial \bar g_{\alpha\beta}}{\partial (1-\rho)}\bigr|_{\rho=1}$, it is straightforward to verify that \eqref{eq:indatkads} implies \eqref{eq:Tpert}.
One can then obtain the corresponding initial data in Cartesian coordinates $\bar{g}_{\mu\nu}|_{t=0},\partial_t \bar{g}_{\mu\nu}|_{t=0}$ from the transformation law $\bar{g}_{\mu\nu}=\frac{\partial x^\alpha}{\partial x^\mu}\frac{\partial x^\beta}{\partial x^\nu}\bar{g}_{\alpha\beta}$ (as noted above, no factor of $(1-\rho)$ is needed for the transformation between quasi-spherical and Cartesian coordinates).
However, nothing ensures that the choice of initial data specified in this way satisfies the Hamiltonian and momentum constraints.
In fact, one can verify that the boundary energy-momentum tensor that corresponds to the initial data \eqref{eq:indatkads}, given at $t=0$ by \eqref{eq:Tpert}, is neither traceless nor conserved for non-vanishing values of the perturbation amplitudes $\epsilon_Y,\epsilon_{\mathcal{V}}$.
Therefore, if we wish to evolve this initial data, we need a mechanism that efficiently damps constraint violations of the solution, thus ensuring that we quickly recover a solution of the Einstein equations within a small number of time steps.
The constraint-damping terms that we add to our evolution equation \eqref{eq:EFEsoufun} act precisely in this sense.
Thus, if $\epsilon_Y,\epsilon_{\mathcal{V}}$ are sufficiently small, our numerical solution is a solution to the Einstein equations after a few time steps.
A drawback of this strategy is the fact that, in general, we can expect the parameters of the solution obtained in this way to differ from the corresponding parameters of the chosen Kerr-AdS background. If the differences in the parameters are large, the resulting spacetime might be stable to the chosen perturbative modes, which would imply that the superradiant instability is not triggered.\footnote{As noted in Section~\ref{subsec:simsupin}, recall that the initial data that we choose cannot change the background energy $E$ and angular momentum $J$ of the background at $t=0$. This is because $E|_{t=0}=\int d\theta d\phi \sin\theta \langle T_{tt}\rangle_{CFT}|_{t=0}$ and $J|_{t=0}=-\int d\theta d\phi \sin\theta \langle T_{t\phi}\rangle_{CFT}|_{t=0}$, and the integral over the unit round sphere of any spherical harmonic vanishes. Such a change could only occur at later times if the constraint-damping terms lead the constraint-violating numerical solution to a constraint-preserving solution of the Einstein equations with different $E$ and $J$.}
In Section~\ref{sec:preresK}, we show evidence that this issue does not occur for the set of initial data that we consider. Namely, we monitor the energy and the angular momentum of the solution, and we show that these charges, which in turn determine the value of all the parameters, remain very close to those of the chosen background.
However, for other types of initial data, this issue might occur. In that case, one would have to work harder in order to ensure that the solution recovered via constraint-damping is superradiantly unstable to the chosen perturbations.
Of course, it would be ideal to implement a method that provides initial data, containing the chosen perturbative modes, as a solution of the Hamiltonian and momentum constraints. We leave this for future work.

\subsection{Specifics of the simulation}
\label{sec:detnumev}

The simulation employs the scheme described in Section~\ref{sec:numerical_scheme}.
We use a grid with $N_x=N_y=N_z=257$ points in each of the Cartesian directions, with equal grid spacings $\Delta x = \Delta y = \Delta z\equiv \Delta=2/(N_x-1)$.
We employ a CFL factor of $\lambda=0.4$.
Throughout the entire simulation, we excise a fixed region inside the event horizon of the Kerr-AdS background at $t=0$. The excised region is determined as follows. 
Consider the values of the $\rho$ coordinate at the event horizon. 
Consider the values of the (uncompactified) Kerr-Schild radial coordinate $R$ at the event horizon.
These are given by the function
\begin{equation}
R_+(\theta)=r_+ \sqrt{\frac{2(a^2+r_+^2)}{a^2 \left(r_+^2+1\right)
   \cos 2 \theta -\left(a^2-2\right)r_+^2+a^2}}.
\end{equation}
From this, we obtain the values of the $\rho$ coordinate at the event horizon, $\rho_+(\theta)$ by inverting $R_+(\theta)=2\rho_+(\theta)/(1-\rho_+^2(\theta))$, i.e., $\rho_+(\theta)=\frac{-1+\sqrt{1+R_+^2(\theta)}}{R_+(\theta)}$.
For our choice of background, the maximum and minimum values of this function are, respectively, $\rho_{+}^{ max}= 0.208525$ and $\rho_+^{min}=0.179897$.
We excise the region with $\rho<(1-\delta_{ex})\rho_+^{min}$.
The excision buffer is set to $\delta_{ex}=0.2$.
We have verified that, with this setup, constraint violations are efficiently damped if we choose amplitudes for the perturbations given by $\epsilon_Y=0.01,\epsilon_{\mathcal{V}}=0.01$.
To do so, we check that the independent residual, defined in Appendix~\ref{sec:convbulk}, takes values that vanish up to the expected solution error of order $\Delta^2$ at all times $t>\delta t$, with $\delta t\lesssim 64 \Delta t$ (recall that $\Delta t:=\lambda \Delta$ is the evolution time step).

Notice that we employ amplitudes that are one order of magnitude smaller than the perturbation amplitudes of \eqref{eq:indatchlo}. We have seen that, if we choose amplitudes of order $10^{-1}$ or larger, the numerical solution rapidly develops infinities near the excision surface, which we believe being due to the fact that large perturbations build up and make the numerical solution depart from a solution of the Einstein equations. When this departure is too fast, the constraint-damping terms are not able to recover a solution of the Einstein equations. 
One possibility that allows for slightly larger amplitudes is the use of a smaller excision buffer, which corresponds to a larger excised region. However, if we excise a region whose boundary is too close to the horizon at $t=0$, then spacetime oscillations could make the horizon intersect the excision boundary at late times. As mentioned in Section~\ref{sec:AHfind}, this would cause a loss of convergence.
If this occurs, then one needs to consider a previous time where the solution was convergent, determine the apparent horizon at that time, and excise a region inside the apparent horizon, before restarting the evolution from that time.
In order to avoid a similar scenario, or at least postpone it for as long as possible, we decide to avoid using an excision buffer smaller than $\delta_{ex}=0.2$.
It is worth mentioning a different strategy that could allow for perturbations of the boundary energy-momentum tensor with larger amplitudes. Namely, one could employ amplitudes $\epsilon_Y,\epsilon_{\mathcal{V}}$ that vanish near the horizon and smoothly increase until reaching the desired value near the boundary.
In this setup, only a limited region of spacetime is perturbed, thus we can hope that, by the time the perturbation reaches the horizon, it has not built up to the point that constraint-damping becomes ineffective.
At the time of writing this thesis, we have not tested this potential improvement.
Further investigation is needed to evolve larger perturbation amplitudes.

\subsection{Preliminary results}
\label{sec:preresK}

We here present preliminary results of the simulation of perturbed Kerr-AdS, in view of an investigation of the superradiant instability.

\begin{figure*}[t!]
        \centering
        \includegraphics[width=5.0in,clip=true]{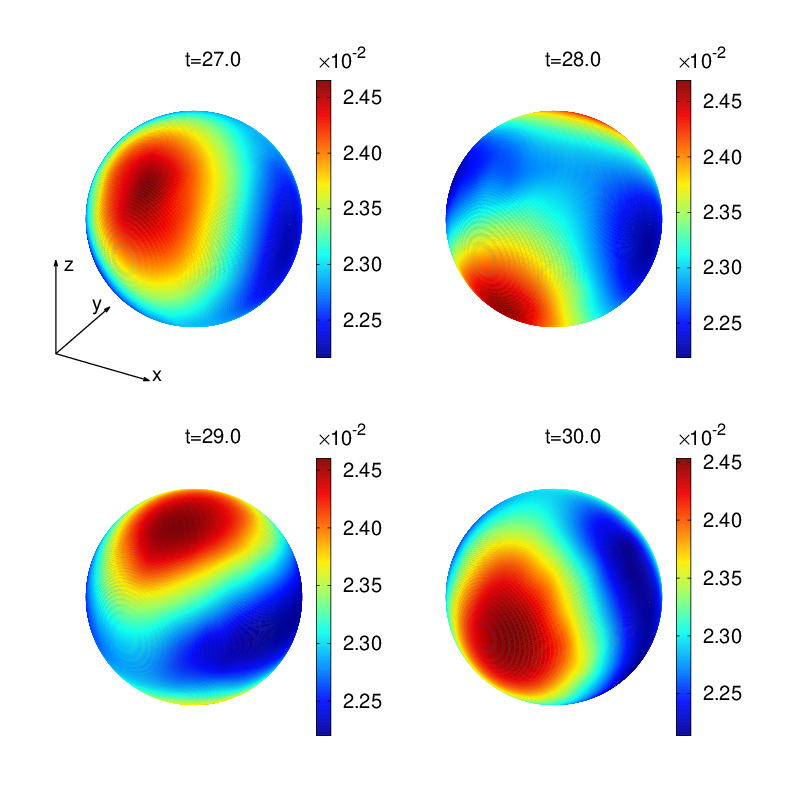}
\parbox{5.0in}{\caption{Snapshots of the boundary energy density in our simulation. The pattern of an $l=m=2$ mode is evident. The profile rotates in $\phi$ with the angular velocity of the Kerr-AdS background $\Omega_H$.
        }\label{fig:bdyendensperKAdS}}
\end{figure*}
\begin{figure*}[t!]
        \centering
        \includegraphics[width=3.5in,clip=true]{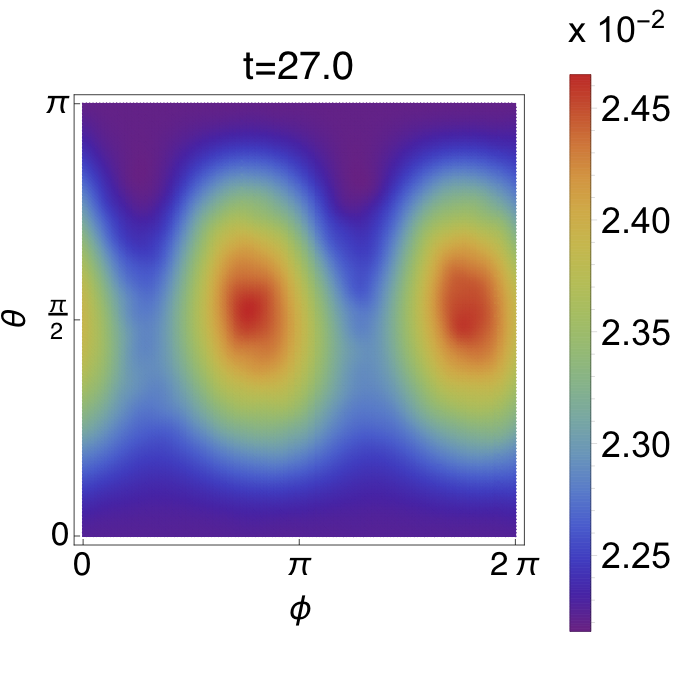}
\parbox{5.0in}{\caption{A planar plot of the boundary energy density at $t=27$ in our simulation. The profile clearly resembles the early boundary energy density profile in the simulations of \cite{Chesler:2018txn} and \cite{Chesler:2021ehz}.
        }\label{fig:bdyendensperKAdS2}}
\end{figure*}
\begin{figure*}[t!]
        \centering
       \hspace*{-0.3cm} 
           \includegraphics[width=5.0in,clip=true]{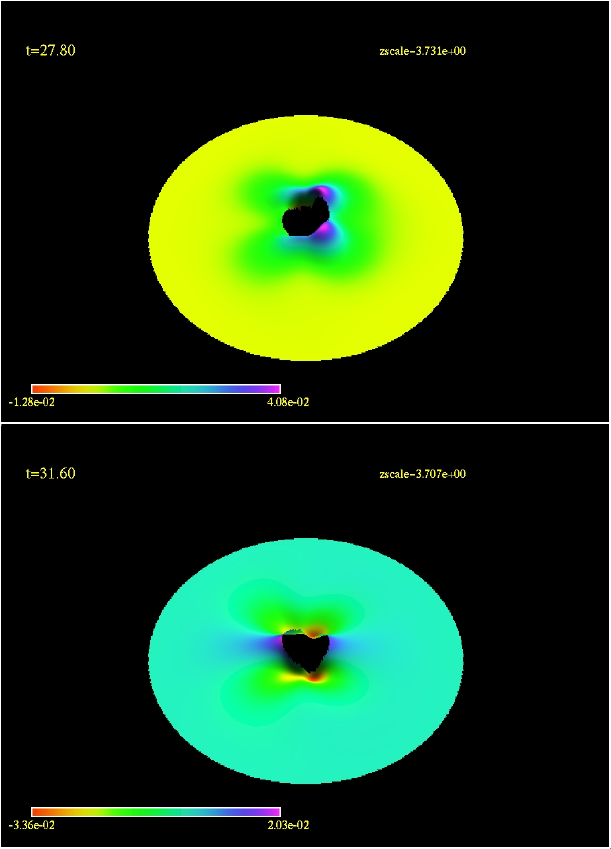}
\parbox{5.0in}{\caption{Snapshots of the evolution variable $\bar{g}_{xx}$ on the $z=0$ bulk slice, showing non-trivial dynamics. 
The height of the plot surface is determined as the magnitude of the function times the constant $\text{zscale}$ displayed in the top right corner of each plot.
        }\label{fig:bulkgbxx}}
\end{figure*}

We have evolved the initial data discussed in Section~\ref{sec:pert} until $t=110$ (in units of the AdS radius $L=1$).
Figure~\ref{fig:bdyendensperKAdS} shows four snapshots of the boundary energy density separated by a small time interval.
We obtain this quantity as explained in Section~\ref{sec:bouset2}, after extrapolating the values of the boundary energy-momentum tensor by means of the boundary extrapolation scheme described in Appendix~\ref{sec:extrapconvbdy}.
The following observations can be made.
\begin{itemize}
\item Two peaks are separated by an angle $\Delta \phi\approx \pi$. This is the typical pattern of an $l=m=2$ perturbation that can be also seen in Figure~\ref{fig:chloev} and Figure~\ref{fig:chev}. To allow for a more straightforward comparison, in Figure~\ref{fig:bdyendensperKAdS2} we also show the energy density at $t=27$ on a planar plot.
\item The amplitude of the boundary energy density is one order of magnitude smaller than the amplitude of Figure~\ref{fig:chloev}, as expected from the fact that our perturbation amplitude $\epsilon_Y,\epsilon_{\mathcal{V}}$ are one of magnitude smaller than the perturbation amplitudes of \eqref{eq:indatchlo}.
\item The boundary energy density rotates in the $\phi$ direction. A peak returns to its position after a time $\Delta t\approx 4.8$, hence the angular velocity of the profile is $2\pi/\Delta t\approx1.308$. This is consistent with the fact that the chosen Kerr-AdS background rotates in $\phi$ with angular velocity $\Omega_H=1.30667$.
\end{itemize}
From the fact that this profile remains essentially unchanged throughout the evolution, we can infer that the simulation has not yet reached the onset of a superradiant instability, and that longer evolution times are necessary.
In Figure~\ref{fig:bulkgbxx}, four snapshots of the $\bar{g}_{xx}$ evolution variable on a $z=0$ slice are displayed as examples of the non-trivial dynamics that takes place in the bulk.
As in the simulations of Section~\ref{sec:results}, the numerical solution has no symmetries.

\begin{figure*}[t!]
        \centering
        \includegraphics[width=5.0in,clip=true]{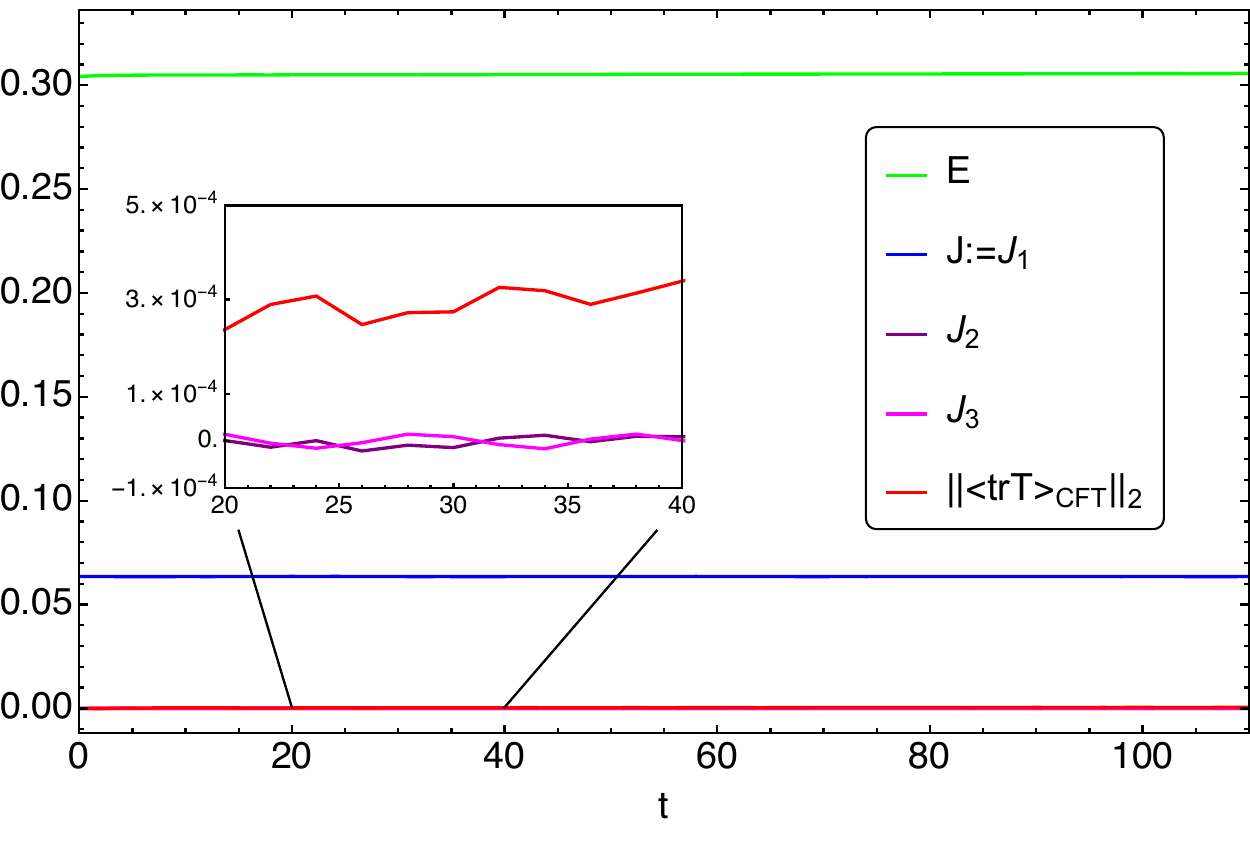}
\parbox{5.0in}{\caption{Evolution of energy $E$, angular momentum $J$, conserved charges $J_2,J_3$ and $L^2$-norm of $\langle \text{tr} T\rangle_{CFT}$. All these quantities are conserved, as expected. Moreover $E$ and $J$ are very close to the background Kerr-AdS values, despite the fact that the initial data constitutes a small violation of the constraints.
        }\label{fig:EJJ2J3trace}}
\end{figure*}

Finally, Figure~\ref{fig:EJJ2J3trace} shows the evolution of the AdS energy $E$ and AdS angular momentum $J$ of the spacetime. We also include the conserved charges $J_2,J_3$ associated with the other generators of the asymptotic rotation symmetry, i.e., $m_2$ and $m_3$, respectively, as well as the $L^2$-norm of the trace of the boundary energy-momentum tensor. 
Let us start by noting that the values of $J_2,J_3$ and $||\langle \text{tr}T\rangle_{CFT}||_2$ are very close to 0, as they should. In fact, the lines of these three quantities cannot be distinguished by eye if we use a scale of order $10^{-1}$, thus we show them over a smaller scale of order $10^{-4}$ for a restricted time interval.
Moreover, as expected, $E$ and $J$ are constant throughout the entire simulation, and given by $E\approx0.305$, $J\approx0.063$. These are almost the exact values of energy and angular momentum of the Kerr-AdS background.
As expected, $E$ and $J$ are constant throughout the entire simulation. In particular, they take the values $E\approx0.305$, $J\approx0.063$, which are very close to the values of the chosen Kerr-AdS background \eqref{eq:propback}.
This is an important sign of the fact that, although the initial data violates the constraint, the constraint-damping terms recover a solution with almost the same energy and angular momentum as the ideal, constraint-preserving solution that one would want to evolve. This implies that the user has good control on the energy and angular momentum of the solution, since these are determined, to a very good approximation, by the user-specified parameters $M,a$. Whether this is the case also for larger perturbation amplitudes remains to be seen.

In summary, the early stage of the simulation provides data that match the expectations. Moreover, the evolution appears to proceed without any sign of numerical instability. Thus, with our tools, we expect to be able to simulate, and investigate, the onset of the superradiant instability in the foreseeable future. The strong stability and convergence properties of our scheme let us hope that it can be used to eventually reach the potentially weak-cosmic-censorship-violating end-point of the instability.

\ifpaper
\end{document}
\fi
\newif\ifpaper
\paperfalse

\ifpaper
\input{../preamble}
\begin{document}
\fi

\chapter{Conclusions and Outlook}
\label{Chapter:outconc}

We have presented the first proof-of-principle Cauchy evolution scheme with no symmetry assumptions that solves the Einstein-Klein-Gordon equations for asymptotically AdS spacetimes with reflective boundary conditions.
The scheme is based on the generalised harmonic formulation.
Numerical stability is achieved through a gauge choice of generalised harmonic source functions near the AdS boundary, obtained by following the prescription of Section~\ref{sec:pre_sta}. In particular, the gauge choice in Cartesian coordinates is given by \eqref{eqn:target_gauge_txyz}. 
In this work we limited ourselves to $D=4$ spacetime dimensions, but the calculation outlined in Section~\ref{sec:gauge_choice} would be almost identical if we were to study Cartesian evolution of asymptotically AdS spacetimes in any $D\geq4$ dimensions. In particular, the stable gauge found with this method would be the same up to a numerical factor. 
Interestingly, a comparison between \eqref{eqn:target_gauge_txyz} and the corresponding result in \cite{Bantilan:2017kok} (see eq. (S10) in that previous work) clearly suggests a trend for the expression of the stable gauge as we relax symmetries, and thus increase the number of spatial coordinates on which the solution depends.
If this trend were confirmed, repeating the calculations of Section~\ref{sec:pre_sta} would not be necessary when increasing the number of spatial degrees of freedom.
See \cite{Bantilan:2020pay} for another successful application of our prescription in a higher dimensional case.
Furthermore, the scheme presented here can be applied to cases with different types of matter fields, different types of global coordinates, and to coordinates on the Poincar\'{e} patch.
For instance, in Appendix~\ref{sec:sphevvarboucon} we followed the prescription of Section~\ref{sec:pre_sta} to obtain the stable gauge also in spherical coordinates.
In Appendix~\ref{sec:poincare}, the same procedure leads to a gauge that stabilizes evolution on a Poincar\'e patch of AdS$_4$.
In other words, this framework makes numerical Cauchy evolution in asymptotically AdS spacetimes possible in full generality, with no need to impose symmetries on the solution.

We have used this scheme to evolve stationary initial data constructed from completely asymmetric Gaussian initial profiles of a massless scalar field.
We observe the collapse of the scalar field into a black hole and the subsequent ringdown to a Schwarzschild-AdS black hole spacetime, in both bulk and boundary quantities.
Deviations from Schwarzschild-AdS at late times are consistent with zero within estimates of the numerical error. 
At very late times, the spatial profiles of these small deviations appear to cascade towards higher harmonics.
Even though these deviations are consistent with our error estimates, they may nevertheless trigger a non-linear instability that can only be revealed by evolving for longer times and with higher spatial resolutions (see below for a brief discussion about the problem of the non-linear instability of black holes in AdS).

We have also described an adaptation of the scheme that allows to study the superradiant instability of Kerr-AdS. 
Specifically, we have discussed numerical output from the evolution of slightly constraint-violating initial perturbations on a Kerr-AdS background for approximately 70 light-crossing times. We have shown that the initial violation of the constraints is efficiently damped, thus the numerical solution quickly becomes a solution of the Einstein equations. 
Moreover, we have argued that, at least for sufficiently small perturbations, the solution that is approached has energy and angular momentum close to those of the chosen Kerr-AdS background, which allows to use the results of \cite{Cardoso:2013pza} to determine modes that are superradiantly unstable and will grow during the evolution.
In the future, we aim to observe and analyse the early stages of the superradiant instability of Kerr-AdS.
The stability of our scheme, even in strongly dynamical scenarios, suggests that we might be able to reach the potentially weak-cosmic-censorship-violating end-point of the instability, and investigate its properties from both the bulk and boundary perspective.

In addition to the study of superradiance, we expect to be able to tackle many other problems in asymptotically AdS spacetimes through the numerical scheme presented in this thesis.
We want to highlight two of the most important ones here.
The first is the study of the \emph{non-linear instability of AdS}, and the closely related topic of gravitational collapse in AdS (see \cite{Martinon:2017ppj} for a review). 
Pure AdS is well-known to be linearly stable \cite{Abbott:1981ff}, in the sense that no solution to the linearised Einstein equations on a pure AdS background can grow unboundedly in time.
In fact, all AdS modes are normal, i.e., with purely real frequencies, and thus not growing in time.
It should be mentioned that the non-linear generalisation of a single gravitational AdS mode has been shown to lead to new solutions of the Einstein equations, called \emph{geons}~\cite{Dias:2011ss,Horowitz:2014hja,Martinon:2017uyo,Fodor:2017spc,Dias:2017tjg}.
These solutions can be regarded as the zero-horizon radius limit of the black resonators introduced in Section~\ref{subsec:blares}; they have an helical symmetry, and appear to be linearly stable to the great majority of perturbations \cite{Ishii:2020muv}.
In \cite{DafHolz:1981ff}, the authors suggest that AdS might suffer from a non-linear instability, which arises when two or more modes interact with each other through the non-linear terms of the Einstein equations.
This was confirmed in the seminal work by Bizo\'n and Rostorowski \cite{Bizon:2011gg}, which presented a numerical study of gravitational collapse $D=4$ spacetime dimensions (later generalised to higher dimensions \cite{Jalmuzna:2011qw}) in spherical symmetry. 
Here, it was shown that a class of small perturbations of amplitude $\epsilon$ undergoes gravitational collapse and forms a black hole on a time-scale $\mathcal{O}(\epsilon^{-2})$, due to a turbulent cascade of energies from large to small distances until a horizon forms.
Subsequently, \cite{Bantilan:2017kok} considered the same massless scalar field model in 5-dimensional AdS in a 2+1 setting, and it was observed that for a certain class of initial data, the subsequent evolution resulted in collapse that happens faster away from spherical symmetry.
On the other hand, the authors in \cite{Choptuik:2017cyd} used a particular metric ansatz in a 1+1 setting to consider the inclusion of angular momentum, and observed delayed collapse.
A promising direction is provided in \cite{Moschidis:2018ruk}, \cite{Moschidis:2018kcf} with a proof of the instability of AdS in spherical symmetry for the Einstein-massless Vlasov system.
Studies of the gravitational collapse with no symmetry assumptions and with angular momentum can shed light on the intricate phenomenology of non-linear interactions in AdS.
The scheme described in this thesis makes it possible for numerical investigations to incorporate all the relevant physics needed to study gravitational collapse in AdS in full generality, and thus help to settle the question regarding the non-linear instability of AdS.

The second important problem we wish to highlight is the study of the \emph{non-linear instability of Kerr-AdS} and, in particular, Schwarzschild-AdS.
Schwarzschild-AdS has been shown to be non-linearly stable under spherically symmetric perturbations in the context of the Einstein-Klein-Gordon theory \cite{Holzegel:2011rk}.
However, Refs.~\cite{Holzegel:2011uu,Holzegel:2013kna} have found that fields on a Kerr-AdS background decay slowly in time (namely with the logarithm of a time coordinate). This suggests that they might not decay at all if gravitational interactions are taken into account, thus signalling an instability at the non-linear level away from spherical symmetry. Heuristically, this can be explained by the fact that waves, coming from the boundary and travelling in the Schwarzschild-AdS or Kerr-AdS geometry, encounter a potential barrier. Waves that are not energetic enough to overcome the barrier will remain trapped in the region between the barrier and the AdS boundary (assuming that reflective conditions are imposed).
They could then grow by repeated interactions, until entering the non-linear regime.
Since Kerr-AdS is also affected by the linear superradiant instability, one needs to ``isolate'' the non-linear instability problem. This can be done by evolving a perturbed Kerr-AdS spacetime below the Hawking-Reall threshold for the superradiant instability. 
The discussion of Chapter~\ref{Chapter:KAdS} strongly suggests that our scheme can perform this type of evolution.
As a first step, we could, for instance, corroborate the heuristic explanation of the instability by showing that the scalar field profile is non-vanishing near the boundary of a Kerr-AdS background for a very long time.
We could then go further and consider the full non-linear evolution of perturbed Kerr-AdS to probe the conjectured instability, and potentially reach the unknown end-point. 

In conclusion, the Cauchy formulation of the initial-boundary value problem in full generality, implemented by the robust and stable numerical techniques described in this thesis, is well-suited for investigations of a variety of fundamental problems about dynamics in AdS, such as the problem of the linear superradiant instability of Kerr-AdS, the non-linear instability of pure AdS, and the non-linear instability of Schwarzschild-AdS and Kerr-AdS. 
This is perhaps the main advantage of our scheme: it can be used for a multiple of purposes with minor modifications.

\ifpaper
\end{document}
\fi

\begin{appendix}
\newif\ifpaper
\paperfalse

\ifpaper
\input{../preamble}
\begin{document}
\fi
\chapter{Boundary prescription for spherical coordinates}
\label{sec:sphevvarboucon}

Although spherical coordinates $x^\alpha=(t,\rho,\theta,\phi)$ are not suitable for numerically evolving points near the origin (see discussion in Section~\ref{sec:numcauprob}), they are convenient to extract the physics of the CFT at the AdS boundary, since they are adapted to the boundary topology $\mathbb{R}\times S^2$.
In this section we apply the prescription outlined in Section~\ref{sec:pre_sta} to the case of asymptotically AdS spacetimes in $D=4$ spacetime dimensions in spherical coordinates. Similarly to the Cartesian case, we first define the spherical coordinate version of the evolution variables $(\bar{g}_{\alpha\beta},\bar{\varphi},\bar{H}_\alpha)$. We also write down the transformations between these variables and their Cartesian version, \eqref{eq:gbarcart}--\eqref{eq:soufunb}. Then, we obtain the stable gauge in spherical coordinates by following the steps introduced in Section~\ref{sec:gauge_choice}. We compare this with a different potentially stable gauge that can be inferred from the one used in~\cite{Bantilan:2012vu}. Finally, we show that tracelessness and conservation of the boundary stress-energy tensor $\langle T_{ab}\rangle_{CFT}$, whose expressions in spherical coordinates are given in Section~\ref{sec:bouset2}, is a consequence of the lowest order of the Einstein equations in the near boundary expansion, provided that the leading order of the generalized harmonic constraints is satisfied.\footnote{In fact, tracelessness was already proved in Section~\ref{sec:bouset2} by converting Cartesian variables into spherical ones. We prove it again in this section employing only spherical coordinates.}

\section{Evolution variables and boundary conditions}
\label{subsec:evvarbouconsphcoo}

We remind the reader that new evolution variables are defined in order to impose the Dirichlet boundary conditions of Section~\ref{subsec:bouconsphcar} in a simple way at the AdS boundary $\rho=1$. In the same way as in the Cartesian coordinate case, for which we defined metric evolution variables $\bar{g}_{\mu\nu}$ in \eqref{eq:gbarcart}, the metric evolution variables in spherical coordinates $\bar{g}_{\alpha\beta}$ are defined by (i) considering the deviation from pure AdS tensor $h_{\alpha\beta}=g_{\alpha\beta}-\hat{g}_{\alpha\beta}$ in spherical coordinates, and (ii) stripping $h_{\alpha\beta}$ of as many factors of $(1-\rho^2)$ as needed so that they fall off linearly in $(1-\rho)$ near the AdS boundary.

The boundary conditions on $h_{\alpha\beta}$ \eqref{eq:sphbounconh} tell us that
\begin{eqnarray}
\label{eq:gbarsph}
\bar{g}_{\rho\alpha}&=&\frac{h_{\rho\alpha} }{1-\rho^2}\,,\qquad \textrm{ if $\alpha\neq\rho$}\,, \nonumber\\
\bar{g}_{\alpha\beta}&=&\quad h_{\alpha\beta}\,, \qquad\;\;\;\;\, \textrm{ otherwise}.
\end{eqnarray}
Despite the notation, we emphasize that $\bar{g}_{\alpha\beta}$ and $\bar{g}_{\mu\nu}$ are not in general components of the same tensor (as it should be clear from their definition), therefore the usual transformation between tensor components in different sets of coordinates cannot be applied. The correct transformation can be easily deduced from \eqref{eq:gbarsph} and \eqref{eq:gbarcart}, remembering that $h$ is indeed a tensor: 
\begin{eqnarray}\label{eq:cartosph}
\bar{g}_{\rho\alpha}&=&\frac{1}{(1-\rho^2)}\frac{\partial x^\mu}{\partial \rho}\frac{\partial x^\nu}{\partial x^\alpha}\bar{g}_{\mu\nu}\,,\quad \textrm{ if $\alpha\neq\rho$}\,, \nonumber\\
\bar{g}_{\alpha\beta}&=&\frac{\partial x^\mu}{\partial x^\alpha}\frac{\partial x^\nu}{\partial x^\beta}\bar{g}_{\mu\nu}\,,\qquad\quad \;\;\;\;\;\; \textrm{ otherwise}.
\end{eqnarray}
Similarly, the boundary conditions on the scalar field \eqref{eq:sphbounconphi} suggest that we use the evolution variable
\begin{equation}
\bar{\varphi}=\frac{\varphi }{(1-\rho^2)^2}\,,
\end{equation}
which the same as the one in Cartesian coordinates, as expected for a scalar field.
Finally, the boundary conditions \eqref{eq:sphbouncondsoufunc} on $H_\alpha$ suggest the use of the evolution variables
\begin{eqnarray}
 \bar{H}_\alpha&=&\frac{H_\alpha-\hat{H}_\alpha}{(1-\rho^2)^2 }\,, \qquad \textrm{ if $\alpha\neq\rho$\,,} \\ \nonumber
 \bar{H}_\rho&=&\frac{H_\rho-\hat{H}_\rho}{1-\rho^2 }\,,
 \end{eqnarray}
in spherical coordinates.
Neither $H_\alpha,\hat{H}_\alpha,\bar{H}_\alpha$ nor $H_\mu,\hat{H}_\mu,\bar{H}_\mu$ are components of the same tensor, so there is no simple transformation from one set to the other. The two triplets of quantities can only be obtained from the definition of source functions in terms of the full metric $g$ in the appropriate set of coordinates, e.g., equation \eqref{eq:sphbouncondsoufunc} in spherical coordinates.

In a numerical scheme in spherical coordinates employing the framework presented in this article, reflective Dirichlet boundary conditions can be easily imposed as
\begin{equation}
\label{eq:dirbc_sphcoords}
\bar{g}_{\alpha\beta}\big|_{\rho=1}=0\,,\quad \bar{\varphi}\big|_{\rho=1}=0\,,\quad \bar{H}_\alpha\big|_{\rho=1}=0\,.
 \end{equation}

\section{Gauge choice for stability}\label{sec:gau_choice_sphcoords}

Since the evolution variables in spherical coordinates, $(\bar{g}_{\alpha\beta},\bar{\varphi},\bar{H}_\alpha)$, are linear in $q=1-\rho$ by construction, we can borrow the near-boundary expansions \eqref{eqn:qexpg}--\eqref{eqn:qexpphi}. 
We now substitute these into the evolution equations \eqref{eq:EFEsoufun}, and we expand each component in powers of $q$. Rewriting the resulting equations in the wave-like form \eqref{eq:waveEFE}, we obtain
\begin{eqnarray}\label{eqn:efett_3p1}
\hspace{-10mm}\tilde{\Box}\bar{g}_{(1)tt}&=&q^{-2} (2 \bar{H}_{(1) \rho }-3 \bar{g}_{(1) \rho \rho })+O(q^{-1}),\\
\label{eqn:efetrho_3p1}
\hspace{-10mm}\tilde{\Box}\bar{g}_{(1)t\rho}&=&\frac{1}{2}q^{-2} (-\bar{g}_{(1)\theta \theta,t}+\bar{g}_{(1) \rho \rho ,t}-\bar{g}_{(1)
  \text{$tt$},t}+\csc ^2\theta (2 \bar{g}_{(1) \text{$t$$\phi $},\phi }-\bar{g}_{(1)
  \phi \phi ,t}) \nonumber\\
   &&\hspace{0.5cm}+2 \bar{g}_{(1) \text{$t$$\theta $},\theta }-2 \bar{H}_{(1) \rho ,t}-3
   \cot \theta  \bar{g}_{(1) \text{$t$$\theta $}}-40 \bar{g}_{(1) \text{$t$$\rho $}}+20
   \bar{H}_{(1) t})+\mathcal{O}(q^{-1}),\\
\label{eqn:efettheta_3p1}
\hspace{-10mm}\tilde{\Box}\bar{g}_{(1)t\theta}&=&\mathcal{O}(q^{-1}),\\
\label{eqn:efetphi_3p1}
\hspace{-10mm}\tilde{\Box}\bar{g}_{(1)t\phi}&=&\mathcal{O}(q^{-1}),
\end{eqnarray}
\begin{eqnarray}
\label{eqn:eferhorho_3p1}
\hspace{-4mm}\tilde{\Box}\bar{g}_{(1)\rho\rho}&=&3q^{-2} (\csc ^2\theta \bar{g}_{(1) \phi \phi }+\bar{g}_{(1)\theta \theta}-2 \bar{g}_{(1)
   \rho \rho }-\bar{g}_{(1) \text{$tt$}}+2 \bar{H}_{(1) \rho })+\mathcal{O}(q^{-1}),\\
\label{eqn:eferhotheta_3p1}
\hspace{-4mm}\tilde{\Box}\bar{g}_{(1)\rho\theta}&=&\frac{1}{2} q^{-2}(\bar{g}_{(1)\theta \theta,\theta }+\csc ^2\theta (-\bar{g}_{(1)
   \phi \phi ,\theta }+2 \bar{g}_{(1)\theta \phi,\phi }+5 \cot \theta  \bar{g}_{(1) \phi
   \phi }) \nonumber\\
   &&\hspace{0.5cm}+\bar{g}_{(1) \rho \rho ,\theta }-2 \bar{g}_{(1) \text{$t$$\theta
   $},t}+\bar{g}_{(1) \text{$tt$},\theta }-2 \bar{H}_{(1) \rho ,\theta } \nonumber\\
   &&\hspace{0.5cm}-3 \cot \theta 
   \bar{g}_{(1)\theta \theta}-40 \bar{g}_{(1) \rho \theta }+20 \bar{H}_{(1)\theta})+\mathcal{O}(q^{-1}),\\
\label{eqn:eferhophi_3p1}
\hspace{-4mm}\tilde{\Box}\bar{g}_{(1)\rho\phi}&=&\frac{1}{2} q^{-2}(2 \bar{g}_{(1)\theta \phi,\theta }+\csc ^2\theta \bar{g}_{(1) \phi
   \phi ,\phi }-\bar{g}_{(1)\theta \theta,\phi }+\bar{g}_{(1) \rho \rho ,\phi }-2
   \bar{g}_{(1) \text{$t$$\phi $},t}\nonumber\\
   &&\hspace{0.5cm}+\bar{g}_{(1) \text{$tt$},\phi }-2 \bar{H}_{(1) \rho ,\phi }-13
   \cot \theta  \bar{g}_{(1)\theta \phi}-40 \bar{g}_{(1) \rho \phi }+20 \bar{H}_{(1) \phi
   }) \nonumber\\
   &&\hspace{0.5cm}+\mathcal{O}(q^{-1}),\\
\label{eqn:efethetatheta_3p1}
\hspace{-4mm}\tilde{\Box}\bar{g}_{(1)\theta\theta}&=&q^{-2}  (3 \bar{g}_{(1) \rho \rho }-2 \bar{H}_{(1) \rho })+\mathcal{O}(q^{-1}),\\
\label{eqn:efethetaphi_3p1}
\hspace{-4mm}\tilde{\Box}\bar{g}_{(1)\theta\phi}&=&\mathcal{O}(q^{-1}),\\
\label{eqn:efephiphi_3p1}
\hspace{-4mm}\tilde{\Box}\bar{g}_{(1)\phi\phi}&=&q^{-2}  \sin^2\theta(3 \bar{g}_{(1) \rho \rho }-2 \bar{H}_{(1) \rho })+\mathcal{O}(q^{-1}).
\end{eqnarray}

Doing the same for the generalized harmonic constraints $0=C_\alpha\equiv H_\alpha-\Box x_\alpha$, we have
\begin{eqnarray}
\label{eqn:ct_3p1}
C_t&=&\frac{1}{2} q^3 (-\bar{g}_{(1)\theta \theta,t}-\bar{g}_{(1) \rho \rho ,t}-\bar{g}_{(1)
   \text{$tt$},t}+\csc ^2\theta (2 \bar{g}_{(1) \text{$t$$\phi $},\phi }-\bar{g}_{(1)
   \phi \phi ,t})\nonumber\\
   &&\hspace{0.5cm}+2 \bar{g}_{(1) \text{$t$$\theta $},\theta }-16 \bar{g}_{(1)
   \text{$t$$\rho $}}+8 \bar{H}_{(1) t})+\mathcal{O}(q^4),\\
\label{eqn:crho_3p1}
C_\rho&=&q^2 (2 \bar{H}_{(1) \rho }-\frac{3}{2} (\csc^2\theta (-\bar{g}_{(1) \phi
   \phi })-\bar{g}_{(1)\theta \theta}+\bar{g}_{(1) \rho \rho }+\bar{g}_{(1)
   \text{$tt$}}))+\mathcal{O}(q^3),\\
\label{eqn:ctheta_3p1}
C_\theta&=&\frac{1}{2} q^3 (\bar{g}_{(1)\theta \theta,\theta }+\csc ^2\theta
   (-\bar{g}_{(1) \phi \phi ,\theta }+2 \bar{g}_{(1)\theta \phi,\phi }+2 \cot \theta
    \bar{g}_{(1) \phi \phi })\nonumber\\
   &&\hspace{0.5cm}-\bar{g}_{(1) \rho \rho ,\theta }-2 \bar{g}_{(1)
   \text{$t$$\theta $},t}+\bar{g}_{(1) \text{$tt$},\theta }-16 \bar{g}_{(1) \rho \theta }+8
   \bar{H}_{(1)\theta})+\mathcal{O}(q^4),\\
\label{eqn:cphi_3p1}
C_\phi&=&\frac{1}{2} q^3 (\bar{g}_{(1)\theta \theta,\theta }+\csc ^2\theta
   (-\bar{g}_{(1) \phi \phi ,\theta }+2 \bar{g}_{(1)\theta \phi,\phi }+2 \cot \theta
    \bar{g}_{(1) \phi \phi })\nonumber\\
   &&\hspace{0.5cm}-\bar{g}_{(1) \rho \rho ,\theta }-2 \bar{g}_{(1)
   \text{$t$$\theta $},t}+\bar{g}_{(1) \text{$tt$},\theta }-16 \bar{g}_{(1) \rho \theta }+8
   \bar{H}_{(1)\theta})+\mathcal{O}(q^4).
\end{eqnarray}

We now follow the three steps of Section~\ref{sec:gauge_choice} to obtain a stable gauge choice.
\begin{enumerate}
\item Solve the leading order of the near-boundary generalized harmonic constraints, \eqref{eqn:ct_3p1}--\eqref{eqn:cphi_3p1}, for $\bar{H}_{(1)\alpha}$. We obtain
\begin{eqnarray}\label{eqn:target_gauge_trhothetaphi_step1}
\bar{H}_{(1)t}&=&\frac{1}{8} (\bar{g}_{(1)\theta \theta,t} +\csc ^2\theta (\bar{g}_{(1) \phi \phi ,t}-2 \bar{g}_{(1) \text{$t$$\phi$},\phi })\nonumber \\
   &&\hspace{0.5cm}+\bar{g}_{(1) \rho \rho ,t}+\bar{g}_{(1) \text{$tt$},t}-2 \bar{g}_{(1) \text{$t$$\theta $},\theta }+16 \bar{g}_{(1) \text{$t$$\rho
   $}})\,,\nonumber \\
\bar{H}_{(1)\rho}&=&\frac{3}{4} (-\csc ^2\theta \bar{g}_{(1) \phi \phi }-\bar{g}_{(1)\theta \theta}+\bar{g}_{(1) \rho \rho }+\bar{g}_{(1) \text{$tt$}}),\nonumber\\
\bar{H}_{(1)\theta}&=&\frac{1}{8} (-\bar{g}_{(1)\theta \theta,\theta } +\csc ^2\theta (\bar{g}_{(1)
   \phi \phi ,\theta }-2 (\bar{g}_{(1)\theta \phi,\phi }+\cot \theta  \bar{g}_{(1)
   \phi \phi }))\nonumber\\
   &&\hspace{0.5cm}+\bar{g}_{(1) \rho \rho ,\theta }+2 \bar{g}_{(1) \text{$t$$\theta
   $},t}-\bar{g}_{(1) \text{$tt$},\theta }+16 \bar{g}_{(1) \rho \theta })\,,\nonumber\\
\bar{H}_{(1)\phi}&=&\frac{1}{8} (-2 \bar{g}_{(1)\theta \phi,\theta } +\csc ^2\theta (-\bar{g}_{(1)
   \phi \phi ,\phi })\nonumber \\
   &&\hspace{0.5cm}+\bar{g}_{(1)\theta \theta,\phi }+\bar{g}_{(1) \rho \rho ,\phi
   }+2 \bar{g}_{(1) \text{$t$$\phi $},t}-\bar{g}_{(1) \text{$tt$},\phi }\nonumber\\
   &&\hspace{0.5cm}+4 \cot \theta 
   \bar{g}_{(1)\theta \phi}+16 \bar{g}_{(1) \rho \phi }).
\end{eqnarray}
\item Plug \eqref{eqn:target_gauge_trhothetaphi_step1} into the $q^{-2}$ terms of \eqref{eqn:efett_3p1}--\eqref{eqn:efephiphi_3p1}. This gives the following independent equations.
\begin{eqnarray}
\label{eq:indeq1_3p1}
&&\hspace{-0.4cm}\bar{g}_{(1) \text{$tt$}}-\csc ^2\theta\bar{g}_{(1) \phi \phi }-\bar{g}_{(1)\theta \theta }-\bar{g}_{(1) \rho \rho }=0, \\
\label{eq:indeq2_3p1}
&&\hspace{-0.4cm} \csc ^2\theta (-\bar{g}_{(1) \phi \phi ,\theta }+\bar{g}_{(1)\theta \phi,\phi}+\cot \theta  \bar{g}_{(1) \phi \phi })\nonumber \\
   &&\hspace{2.0cm} -\frac{2}{3} \bar{g}_{(1) \rho \rho,\theta } +\bar{g}_{(1) \text{$tt$},\theta }+\cot \theta  \bar{g}_{(1)\theta \theta}=0,\\
   \label{eq:indeq3_3p1}
&&\hspace{-0.4cm}\csc ^2\theta 
   (\bar{g}_{(1) \phi \phi ,t}-\bar{g}_{(1) \text{$t$$\phi $},\phi })+\bar{g}_{(1)\theta \theta,t}+\frac{2}{3} \bar{g}_{(1) \rho \rho ,t}-\cot \theta\bar{g}_{(1) \text{$t$$\theta $}}=0,\\
      \label{eq:indeq4_3p1}
&&\hspace{-0.4cm}\bar{g}_{(1)\theta \phi,\theta }-\bar{g}_{(1)\theta \theta,\phi }-\frac{2}{3} \bar{g}_{(1)\rho \rho ,\phi }+\bar{g}_{(1) \text{$tt$},\phi }+\cot \theta \bar{g}_{(1)\theta \phi}=0.
\end{eqnarray}
We prove below that these equations ensure tracelessness and conservation of the boundary stress-energy tensor $\langle T_{ab}\rangle_{CFT}$, defined in Section~\ref{sec:bouset2}.
\item Use \eqref{eq:indeq1_3p1}--\eqref{eq:indeq4_3p1} to eliminate $\bar{g}_{(1)tt}$, $\bar{g}_{(1)t\theta,t}$, $\bar{g}_{(1)t\theta,\theta}$, $\bar{g}_{(1)t\phi,t}$ from \eqref{eqn:target_gauge_trhothetaphi_step1}.
In this way we obtain a stable gauge in spherical coordinates
\begin{eqnarray}\label{eqn:target_gauge_trhothetaphi}
\bar{H}_{(1)t}&=&\frac{1}{12} (\bar{g}_{(1) \rho \rho ,t}+3 \cot \theta  \bar{g}_{(1) \text{$t$$\theta
   $}}+24 \bar{g}_{(1) \text{$t$$\rho $}}), \nonumber\\
\bar{H}_{(1)\rho}&=&\frac{3}{2} \bar{g}_{(1) \rho \rho }\,,\nonumber\\
\bar{H}_{(1)\theta}&=&\frac{1}{12} (3 \cot \theta  (\bar{g}_{(1)\theta\theta}-\csc ^2\theta \bar{g}_{(1) \phi \phi })+\bar{g}_{(1) \rho \rho ,\theta }+24 \bar{g}_{(1) \rho \theta }),\nonumber\\
\bar{H}_{(1)\phi}&=&\frac{1}{12}(\bar{g}_{(1) \rho \rho ,\phi }+9 \cot \theta  \bar{g}_{(1)\theta \phi}+24 \bar{g}_{(1) \rho \phi }).
\end{eqnarray}
\end{enumerate}

By looking at the gauge choice employed in~\cite{Bantilan:2012vu} (see eq. (74)) to obtain stability in simulations of 5-dimensional asymptotically AdS spacetimes with an SO(3) symmetry, and choosing numerical factors consistent with \eqref{eqn:target_gauge_trhothetaphi}, we can infer the following potentially stable gauge for the 4-dimensional case with no symmetry assumptions:
\begin{eqnarray}
\label{eq:hbold_3p1}
\bar{H}_{(1)t}&=&2 \bar{g}_{(1)\text{$t$$\rho $}}\,, \nonumber\\
\bar{H}_{(1)\rho}&=&\frac{3}{2} \bar{g}_{(1) \rho \rho }\,,\nonumber\\
\bar{H}_{(1)\theta}&=&2 \bar{g}_{(1) \rho \theta }\,,\nonumber\\
\bar{H}_{(1)\phi}&=&2 \bar{g}_{(1) \rho \phi }\,.
\end{eqnarray}
Notice that by setting certain terms in \eqref{eqn:target_gauge_trhothetaphi} to zero, one recovers \eqref{eq:hbold_3p1}.
As mentioned at the end of Section~\ref{sec:KAdSsp}, we verified that \eqref{eqn:target_gauge_trhothetaphi} is satisfied by Kerr-AdS in HHT (spherical) coordinates.
On the contrary, \eqref{eq:hbold_3p1} is not satisfied by Kerr-AdS in HHT coordinates.
It will be interesting to confirm numerical stability of \eqref{eqn:target_gauge_trhothetaphi} and \eqref{eq:hbold_3p1} with empirical studies.

\section{Tracelessness and conservation of boundary energy-momentum tensor}

We conclude this section by showing that tracelessness and conservation of $\langle T_{ab}\rangle_{CFT}$ follow from \eqref{eq:indeq1_3p1}--\eqref{eq:indeq4_3p1}, i.e., from the lowest order of the Einstein equations, provided that the leading order of the generalized harmonic constraints are satisfied. 

With the notation of Section~\ref{sec:bouset2}, let $x^a=(t,\theta,\phi)$ be the coordinates along the AdS boundary, $g_{(0)ab}dx^a dx^b=-dt^2+d\theta^2+\sin^2\theta d\phi^2$ be a representative of the conformal class of metrics at the AdS boundary, and $\mathcal{D}_{(0)}$ be the Levi-Civita connection of $g_{(0)ab}$, i.e., $\mathcal{D}_{(0)}$ is torsion-free and $\mathcal{D}_{(0)a}g_{(0)bc}=0$. Then, $\langle \text{tr}T\rangle_{CFT}=g_{(0)}^{ab}\langle T_{ab}\rangle_{CFT}$ is the trace of the boundary-energy-momentum tensor and $\mathcal{D}_{(0)}^a \langle T_{ab}\rangle_{CFT}=g_{(0)}^{ac}\mathcal{D}_{(0)c} \langle T_{ab}\rangle_{CFT}$ is its divergence. 
We want to prove that $\langle \text{tr}T\rangle_{CFT}=0$ and $\mathcal{D}_{(0)}^a \langle T_{ab}\rangle_{CFT}=0$. 
The expression of $\langle \text{tr}T\rangle_{CFT}$ in terms of the leading order of the metric variables in spherical coordinates was already written in \eqref{eq:tracecalc}. We repeat it here for completeness:
\begin{equation}
\label{eq:tracecalc2}
\langle \text{tr}T\rangle_{CFT}=\frac{3}{8\pi}\biggl(\bar{g}_{(1)tt}-\bar{g}_{(1)\rho\rho}-\bar{g}_{(1)\theta\theta}-\csc ^2\theta \bar{g}_{(1)\phi\phi}\biggr).
\end{equation}
The divergence of the boundary energy-momentum tensor is given by
\begin{eqnarray}
\label{eq:divergence_t}
\mathcal{D}_{(0)}^a \langle T_{at}\rangle_{CFT}&=&\frac{1}{16\pi}(-3 \csc ^2\theta  \bar{g}_{(1) \phi \phi ,t}-3 \bar{g}_{(1)\theta \theta ,t}\nonumber\\
&&\hspace{-1.0cm}-2 \bar{g}_{(1) \rho \rho ,t}+3 \bar{g}_{(1) \text{$t$$\theta $},\theta }+3 \csc^2\theta \bar{g}_{(1) \text{$t$$\phi $},\phi }+3 \cot \theta  \bar{g}_{(1)\text{$t$$\theta $}})\,, \\
\label{eq:divergence_theta}
\mathcal{D}_{(0)}^a \langle T_{a\theta}\rangle_{CFT}&=&\frac{1}{16\pi}(3 \csc ^2\theta  \bar{g}_{(1)\theta \phi ,\phi }-2 \bar{g}_{(1) \rho \rho ,\theta }  -3 \csc^2\theta \bar{g}_{(1) \phi \phi ,\theta }\nonumber \\
&&\hspace{-1.0cm}-3 \bar{g}_{(1) \text{$t$$\theta $},t}+3\bar{g}_{(1) \text{$tt$},\theta } +3 \cot \theta  \bar{g}_{(1)\theta \theta }+3 \cot   \theta \csc ^2\theta  \bar{g}_{(1) \phi \phi })\,,\\
\label{eq:divergence_phi}
\mathcal{D}_{(0)}^a \langle T_{a\phi}\rangle_{CFT}&=&\frac{1}{16\pi}(3 \bar{g}_{(1)\theta \phi ,\theta }-3 \bar{g}_{(1)\theta \theta ,\phi }-2 \bar{g}_{(1) \rho\rho ,\phi }\nonumber\\
   &&\hspace{-1.0cm}-3 \bar{g}_{(1) \text{$t$$\phi $},t}+3 \bar{g}_{(1) \text{$tt$},\phi }+3 \cot\theta \bar{g}_{(1)\theta \phi })\,.
\end{eqnarray}
We immediately see that $\langle \text{tr}T\rangle_{CFT}=0$ as a consequence of \eqref{eq:indeq1_3p1}.
Moreover, by solving the system of 6 equations given by the first derivatives of \eqref{eq:indeq1_3p1} with respect to $t,\theta,\phi$ and \eqref{eq:indeq2_3p1}, \eqref{eq:indeq3_3p1}, \eqref{eq:indeq4_3p1} for $\bar{g}_{(1) \text{$tt$},t},\bar{g}_{(1) \text{$tt$},\theta},\bar{g}_{(1) \text{$tt$},\phi },\bar{g}_{(1) \text{$t$$\theta $},t}$,$\bar{g}_{(1)\text{$t$$\theta $},\theta },\bar{g}_{(1) \text{$t$$\phi $},t}$, and substituting the solution into the right hand side of \eqref{eq:divergence_t}--\eqref{eq:divergence_phi}, we see that $\mathcal{D}_{(0)}^a \langle T_{ab}\rangle_{CFT}=0$.

\section{Boundary energy-momentum tensor from holographic renormalization}
\label{sec:HoloRen}

In Section~\ref{eq:compdefasyAdS} we related spherical coordinates to FG coordinates. We can use that result, and in particular \eqref{eq:asyFG} to straightforwardly read off the the boundary energy-momentum tensor, through the holographic renormalization prescription of \cite{deHaro:2000vlm} reviewed in Section~\ref{subsec:bdysetconscharg}.
We repeat the expression for the boundary energy-momentum tensor for completeness:
\begin{equation}
\label{eq:bdysetFGform}
\langle T_{\bar{a}\bar{b}}\rangle_{CFT}=\frac{3}{16\pi}g^{(3)}_{\bar{a}\bar{b}}\,,
\end{equation}
where $g^{(3)}_{\bar{a}\bar{b}}$ are the $z^3$ terms of the metric components in FG form, \eqref{eq:asyFG}.
The explicit components of the energy-momentum tensor in \eqref{eq:bdysetFGform} are given by
\begin{eqnarray}
\label{eq:set_explicit_2}
\langle T_{\bar{t}\bar{t}}\rangle_{CFT}&=&\frac{1}{16\pi}(3f_{tt}-f_{\rho\rho})\,, \nonumber \\
\langle T_{\bar{t}\bar{\theta}}\rangle_{CFT}&=&\frac{3}{16\pi}f_{t\theta}\,, \nonumber \\
\langle T_{\bar{t}\bar{\phi}}\rangle_{CFT}&=&\frac{3}{16\pi}f_{t\phi}\,, \nonumber \\
\langle T_{\bar{\theta}\bar{\theta}}\rangle_{CFT}&=&\frac{1}{16\pi} (3f_{\theta\theta}+f_{\rho\rho})\,, \nonumber \\
\langle T_{\bar{\theta}\bar{\phi}}\rangle_{CFT}&=&\frac{3}{16\pi}f_{\theta\phi}\,, \nonumber \\
\langle T_{\bar{\phi}\bar{\phi}}\rangle_{CFT}&=&\frac{\sin^2\bar{\theta}}{16\pi} \biggl(\frac{f_{\phi\phi}}{ \sin^2\bar{\theta}}+\frac{1}{3}f_{\rho\rho}\biggr).
\end{eqnarray}

On the other hand, in Section~\ref{sec:bouset2} we compute the boundary stress-tensor starting from the metric in global spherical coordinates and then using the prescription of \cite{Balasubramanian:1999re}. Of course, the expressions \eqref{eq:set_explicit} and \eqref{eq:set_explicit_2} are equivalent, as we now explain. To obtain \eqref{eq:set_explicit}, we have not imposed that the metric components satisfy the Einstein equations. On the other hand, \eqref{eq:bdysetFGform} gives the correct boundary stress-energy tensor if the bulk metric solves the Einstein equations, in agreement with the assumptions of the FG theorem. It is thus expected that \eqref{eq:set_explicit} and \eqref{eq:set_explicit_2} agree if we assume the validity of the lowest order of the Einstein equations in the form that takes into account the generalized harmonic constraints, i.e., \eqref{eq:indeq1_3p1}--\eqref{eq:indeq4_3p1}.
In fact, we only need \eqref{eq:indeq1_3p1}. For example, starting from \eqref{eq:set_explicit}, imposing \eqref{eq:indeq1_3p1} and using the fact that $\bar{t}=t,\bar{\theta}=\theta,\bar{\phi}=\phi$ at the boundary $\rho=1$ together with $\bar{g}_{(1)\alpha\beta}=f_{\alpha\beta}$,\footnote{Note that $\bar{t}=t,\bar{\theta}=\theta,\bar{\phi}=\phi$ at $\rho=1$ (i.e., $\bar{z}=0$) from \eqref{eqn:invertFGcoords}, while $\bar{g}_{(1)\alpha\beta}\equiv\frac{\partial \bar{g}_{\alpha\beta}}{\partial q}\bigr|_{q=0}=f_{\alpha\beta}$, where the second equality is obtained by comparing \eqref{eq:sphbounconh} with \eqref{eq:gbarsph} to write $\bar{g}_{\alpha\beta}$ in terms of $f_{\alpha\beta}$ and the corresponding $\rho$-dependent factors.} we find precisely the expressions \eqref{eq:set_explicit_2}.

\ifpaper
\end{document}
\fi
\newif\ifpaper
\paperfalse

\ifpaper
\input{../preamble}
\begin{document}
\fi
\chapter{Boundary prescription for the Poincar\'e patch}
\label{sec:poincare}

Here we follow the prescription of Section~\ref{sec:pre_sta} in the case of asymptotically AdS spacetimes in Poincar\'e coordinates and display a choice of generalized harmonic source functions that stabilizes the evolution in this case.

The metric of pure AdS in \emph{Poincar\'e coordinates} $x^\mu=(t,z,x_1,x_2)\in\mathbb{R}^4$, with AdS radius set to $L=1$, can be written as
\begin{equation}
\label{eq:AdSpoincare}
\hat{g} = \frac{1}{z^2} \left( -dt^2 + dz^2 + dx_1{}^2 + dx_2{}^2  \right).
\end{equation}
These coordinates only cover a wedge-shaped part of the entire pure AdS spacetime, called \emph{Poincar\'e patch of AdS} or \emph{Poincar\'e AdS}, bounded by the AdS boundary, $z=0$, and the so-called \emph{Poincar\'e horizon}, $z\to+\infty$.
To include the Poincar\'e horizon in our computational domain, we compactify the bulk coordinate $z=(1-\rho^2)/\rho^2$ to have the Poincar\'e horizon at $\rho=0$ and the AdS boundary at $\rho=1$. This gives the following form for the metric of AdS:
\begin{equation}
\hat{g} = \frac{\rho^4}{(1-\rho^2)^2} \left( -dt^2 + (4/\rho^6)d\rho^2 + dx_1^2 + dx_2^2  \right).
\end{equation}

Let us now consider asymptotically AdS spacetimes.
Since \eqref{eq:AdSpoincare} is in the form given by the leading order of the FG expansion, \eqref{eqn:FGmetric}--\eqref{eqn:FGbdymetric}, we see that $(t,z,x_1,x_2)$ are FG coordinates.
We can thus read off the fall-offs of the tensor $h$, describing the deviation from pure AdS, from the rest of the FG expansion. As in the other sets of coordinates, this provides boundary conditions allowing for asymptotically locally AdS solutions of the evolutions equations.
We also impose that the boundary metric is conformally flat, i.e., that a representative that the conformal class of boundary metrics is $g_{(0)}=-dt^2 + dx_1^2 + dx_2^2 $. This restricts the class of solutions to asymptotically (globally) AdS spacetimes.
In presence of a scalar field, we impose the usual Dirichlet boundary condition, i.e., $\varphi|_{\partial \mathcal{M}}=0$, which selects the fastly-decaying mode, as explained in Section~\ref{subsubsec:FGexpans}.
We refer to all sets of coordinates in which $g$ and $\varphi$ satisfy these boundary conditions as Poincar\'e coordinates.
In this appendix, we use the last few Greek indices, $\mu,\nu,\rho,\dots$, to denote Poincar\'e coordinates.

The evolved fields in the generalised harmonic formalism consist of the spacetime metric $g_{\mu\nu}$, possibly a scalar field $\varphi$, and the generalized harmonic source functions $H_\mu$.
The fall-offs of the metric components $g_{\mu\nu}$ read the same as \eqref{eq:carbouncondh}, with $f_{\mu\nu}(t,x_1,x_2)$ coefficients.
The scalar field fall-off that preserves the metric asymptotics is given by \eqref{eq:carbouncondphi}, with $c(t,x_1,x_2)$ coefficient.
The fall-offs of the source functions can be inferred from the metric fall-offs, which are given by \eqref{eq:carbouncondsoufun}, with $f_{\mu}(t,x_1,x_2)$ coefficients.
As a result, the corresponding evolution variables in this Poincar\'e setting are given exactly by the same expressions as we had written in \eqref{eq:gbarcart}--\eqref{eq:soufunb}.

Using the same steps as in Section~\ref{sec:gauge_choice}, we obtain the following gauge:
\begin{eqnarray}
\label{eq:hbold_poincare}
\bar{H}_{(1)t}&=&\frac{3}{2} \bar{g}_{(1)\text{$t$$\rho $}}\,, \nonumber\\
\bar{H}_{(1)\rho}&=&\frac{3}{2} \bar{g}_{(1) \rho \rho }\,,\nonumber\\
\bar{H}_{(1)x_1}&=&\frac{3}{2} \bar{g}_{(1) \rho x_1 }\,,\nonumber\\
\bar{H}_{(1)x_2}&=&\frac{3}{2} \bar{g}_{(1) \rho x_2 }\,.
\end{eqnarray}
We have verified that this gauge leads to stable evolution in asymptotically AdS spacetimes in Poincar\'e coordinates. 
We close by noting that \cite{Bantilan:2020pay} obtained a similar stable gauge to evolve dynamical black holes in the background of the AdS soliton.

\ifpaper
\end{document}
\fi
section\newif\ifpaper
\paperfalse

\ifpaper
\input{../preamble}
\begin{document}
\fi
\chapter{Initial data}
\label{sec:initdata}

The Cauchy problem in general relativity requires the prescription of initial data on a spacelike hypersurface $\Sigma$ and a choice of gauge throughout the entire evolution. In an asymptotically AdS spacetime, in addition, we have to specify boundary conditions at the boundary of AdS; we have dealt with boundary conditions in Section \ref{subsec:bouconsphcar}. We pick Cartesian coordinates $x^\mu=(t,x,y,z)$ such that $t=0$ on $\Sigma$. The spatial Cartesian coordinates on $\Sigma$ are denoted by $x^i=(x,y,z)$, and the corresponding indices by $i,j,k,\dots$. 
With this notation, the data needed for the Cauchy evolution in the generalized harmonic scheme is composed of the initial data $\bar{\varphi}|_{t=0}$, $\bar{g}_{ij}|_{t=0}$, $\partial_t\bar{\varphi}|_{t=0} $, $\partial_t\bar{g}_{ij}|_{t=0}$ and the source functions $\bar{H}_\mu$ at all times. The gauge used in our numerical scheme at $t>0$ is discussed in Appendix~\ref{sec:GCbulk}. With regard to the gauge at $t=0$, we do not set $\bar{H}_\mu|_{t=0}$ explicitly, but we make an equivalent choice for $\bar{g}_{t\mu}|_{t=0}$, and $\partial_t\bar{g}_{t\mu}|_{t=0}$, and then compute $\bar{H}_\mu|_{t=0}$ from \eqref{eq:defsoufunsph}.
In summary, the complete set of initial data that we prescribe is $\bar{\varphi}|_{t=0}$, $\bar{g}_{\mu\nu}|_{t=0}$, $\partial_t\bar{\varphi}|_{t=0}$ and $\partial_t\bar{g}_{\mu\nu}|_{t=0}$. In this section we explain how this is done in our simulations, taking into account two crucial facts. Firstly, initial data cannot be chosen in a completely arbitrary way, but it must satisfy the Hamiltonian and momentum constraints of GR, \eqref{eq:hamconstr}--\eqref{eq:momconstr}. Secondly, the choice of the initial degrees of freedom must be consistent with the desired gauge \eqref{eqn:target_gauge_txyz} near the AdS boundary. 

\section{Constraints}
\label{sec:constr}

Here we explain how the Hamiltonian and momentum constraints of GR on the initial spacelike hypersurface $\Sigma$, \eqref{eq:hamconstr}--\eqref{eq:momconstr}, are solved for massless real scalar matter, whose energy-momentum tensor is \eqref{eq:KHmomtenspar}, in the simplified case of \emph{time-symmetric data}. 
We will use the notation of Section~\ref{sec:3p1splitt}.
Time symmetry in the scalar sector,
\begin{equation}
\partial_t \bar{\varphi}\big|_{t=0}=0,
\end{equation}
implies $j^i=0$. Time symmetry in the gravitational sector,
\begin{equation}
\label{eq:gij0}
\partial_t \bar{g}_{ij}\big|_{t=0}=0,
\end{equation}
together with the initial gauge choice
\begin{equation}
\label{eq:gti0}
\bar{g}_{ti}\big|_{t=0}=0,
\end{equation}
implies $K_{ij}=0$. This can be proved by recognising that \eqref{eq:gij0} implies $\partial_t\gamma_{ij}=0$, \eqref{eq:gti0} implies that the shift vanishes, $ N^i=0$, and then using \eqref{eq:evogamma}.
Thus, we see that the momentum constraint is trivially satisfied.
The Hamiltonian constraint, instead, reduces to
\begin{equation}
\label{eq:redhamconstr}
^{(3)}R-2\Lambda=16\pi\rho.
\end{equation}
This can be solved through the conformal approach, initiated in \cite{Lichnerowicz:1994}, which assumes that the spatial metric $\gamma_{ij}$ is conformal to the spatial metric $\hat{\gamma}_{ij}$ of the $t=0$ slice of pure AdS in Cartesian coordinates:
\begin{equation}
\label{eq:confdec}
\gamma_{ij}=\zeta^4 \hat{\gamma}_{ij},
\end{equation}
where $\zeta$ is a smooth positive function on $\Sigma$, satisfying the AdS boundary condition $\zeta|_{\rho=1}=1$. Let $\hat{D}$ be the Levi-Civita connection of $\hat{\gamma}_{ij}$ and $^{(3)}\hat{R}$ the corresponding Ricci scalar. Using \eqref{eq:confdec} and its inverse, $\gamma^{ij}=\zeta^{-4} \hat{\gamma}^{ij}$, we obtain
\begin{equation}
\label{eq:confRicci}
^{(3)}R=\frac{1}{\zeta^4}\left(^{(3)}\hat{R}-\frac{8}{\zeta}\hat{\gamma}^{ij}\hat{D}_i \hat{D}_j \zeta\right).
\end{equation}
Plugging \eqref{eq:confRicci} into \eqref{eq:redhamconstr} gives
\begin{equation}
\label{eq:rehamconstr2}
^{(3)}\hat{R}\zeta-8\hat{\gamma}^{ij}\hat{D}_i \hat{D}_j \zeta-2\Lambda \zeta^5=16\pi\rho \zeta^5.
\end{equation}
$^{(3)}\hat{R}$ can be computed from the spatial part of the pure AdS metric \eqref{eqn:ads4_final}: $^{(3)}\hat{R}=-6/L^2=2\Lambda$. Thus, equation \eqref{eq:rehamconstr2} can be written as
\begin{equation}
\label{eq:rehamconstr3}
\hat{\gamma}^{ij}\hat{D}_i \hat{D}_j \zeta-\frac{1}{4}\Lambda\zeta+\frac{1}{4}(\Lambda+8\pi\rho)\zeta^5=0.
\end{equation}
Finally, the version of the Hamiltonian constraint that we are going to solve is obtained by writing the matter energy density $\rho$ in terms of $\zeta$. The time-symmetry requirement $\partial_t \varphi\big|_{t=0}=0$ gives
\begin{equation}
\label{eq:mattendens}
\rho=\frac{1}{2\zeta^4}\hat{\gamma}^{ij}\partial_i\varphi\partial_j\varphi,
\end{equation}
so the Hamiltonian constraint reads
\begin{equation}
\label{eq:hamconsfinal}
\hat{\gamma}^{ij}\hat{D}_i \hat{D}_j \zeta-\frac{1}{4}\Lambda\zeta+\frac{1}{4}(\Lambda\zeta^5+4\pi\zeta\hat{\gamma}^{ij}\partial_i\varphi\partial_j\varphi)=0.
\end{equation}
For any given choice of scalar field $\varphi$ on $\Sigma$, \eqref{eq:hamconsfinal} is an elliptic equation that can be solved for $\zeta$ with boundary condition $\zeta|_{\rho=1}=1$. In our simulations we pick the initial scalar field profile $\varphi|_{t=0}=\bar{\varphi}|_{t=0}(1-\rho^2)^2$ with $\bar{\varphi}|_{t=0}$ specified by \eqref{eq:scaGaupro}, and we solve \eqref{eq:hamconsfinal} with the multigrid algorithm of Section~\ref{sec:MG}, built into the PAMR/AMRD libraries.
The initial metric variables $\bar{g}_{ij}|_{t=0}$ are then easily reconstructed from \eqref{eq:confdec} and $\gamma_{ij}=g_{ij}|_{t=0}=\hat{g}_{ij}|_{t=0}+\bar{g}_{ij}|_{t=0}$, i.e.,
\begin{equation}
\bar{g}_{ij}\big|_{t=0}=\zeta^4\hat{\gamma}_{ij}-\hat{g}_{ij}\big|_{t=0}.
\end{equation}

\section{Consistency at the boundary}
\label{sec:consistbound}

In the previous section we explained how some components of the initial data for our simulations are obtained: (i) we impose time-symmetry, namely $\partial_t\bar{\varphi}|_{t=0}=0$ and $\partial_t\bar{g}_{ij}|_{t=0}=0$; (ii) we make the initial gauge choice $\bar{g}_{ti}|_{t=0}=0$; (ii) we choose the massless real scalar field profile $\bar{\varphi}|_{t=0}$ given by \eqref{eq:scaGaupro}; (iii) we determine $\bar{g}_{ij}|_{t=0}$ through the conformal decomposition of the Hamiltonian constraint. In this section we determine the remaining necessary components for Cauchy evolution based on the generalized harmonic scheme: $\bar{g}_{tt}|_{t=0}$ and $\partial_t\bar{g}_{t \mu}|_{t=0}$.

In doing so, the only restriction to consider is the one already obtained in step 2 of our gauge prescription in Section~\ref{sec:gauge_choice}: the Einstein equations in a gauge that satisfies the generalized harmonic constraints impose the condition $\bar{g}_{(1)tt}=\bar{g}_{(1)xx}+\bar{g}_{(1)yy}+\bar{g}_{(1)zz}$ near the boundary. This will hold at all times of the evolution and it must be imposed on initial data. Given that there is no requirement on the value of $\bar{g}_{tt}$ in the bulk, we make the simplest choice and set that to zero. In order to smoothly transition from the bulk value of $\bar{g}_{tt}$ to its required boundary value, we use the smooth transition function
  \begin{equation}
  \label{eq:transfunc}
    f(\rho) =
    \begin{cases*}
      1\,, & if $\rho\geq \rho_{b}$, \\
      1-R^3(\rho)\left(6 R^2(\rho)-15 R(\rho)+10\right)\,, & if $\rho_{b} > \rho\geq\rho_{a}$, \\
      0\,,        & otherwise,
    \end{cases*}
  \end{equation}
where $R(\rho)=(\rho_{b}-\rho)/(\rho_{b}-\rho_{a})$ and $\rho_{a},\rho_{b}$ are the values between which the transition takes place, set to $\rho_{a}=0.5,\rho_{b}=0.9$ in the simulations of Chapter~\ref{Chapter:NoSym}, and $\rho_{a}=0.3,\rho_{b}=0.6$ in the simulations of Chapter~\ref{Chapter:KAdS}.
Thus, our choice of $\bar{g}_{tt}|_{t=0}$ is
\begin{equation}
\bar{g}_{tt}\big|_{t=0}=f(\bar{g}_{xx}|_{t=0}+\bar{g}_{yy}|_{t=0}+\bar{g}_{zz}|_{t=0}).
\end{equation}

To conclude, the remaining initial variables can be chosen in a completely arbitrary way so we make the simplest choice everywhere on the grid:
\begin{equation}
\partial_t\bar{g}_{t \mu}|_{t=0}=0.
\end{equation}

\ifpaper
\end{document}
\fi
\newif\ifpaper
\paperfalse

\ifpaper
\input{../preamble}
\begin{document}
\fi
\chapter{Complete gauge choice}
\label{sec:GCbulk}

In Section~\ref{sec:gauge_choice} we discussed the gauge choice of source functions that we impose near the boundary in order to obtain stable evolutions. Furthermore, the gauge at $t=0$, $\bar{H}_{\mu}|_{t=0}$, is determined from the initial data, detailed in Appendix~\ref{sec:initdata}, through the definition of source functions \eqref{eq:defsoufunsph} at $t=0$. All that remains is to make a gauge choice of $\bar{H}_\mu$ in the bulk, and smoothly join this with the target boundary values \eqref{eqn:target_gauge_txyz} on each spatial slice and with the initial values $\bar{H}_{\mu}|_{t=0}$ during evolution. In this section we describe how all this is implemented in our numerical scheme.

We start by choosing a zero value for $\bar{H}_\mu$ in the bulk, as this is the simplest choice. Therefore, the values of the source functions on each spatial slice, after the time transition from $t=0$, are given by
\begin{eqnarray}
\label{eqn:extend_gauge_txyz}
F_t&\equiv&\frac{3f_1}{2\sqrt{x^2+y^2+z^2}}(x \bar{g}_{tx}+y\bar{g}_{ty}+z\bar{g}_{tz})\,, \nonumber \\
F_x&\equiv&\frac{3f_1}{2\sqrt{x^2+y^2+z^2}}(x \bar{g}_{xx}+y\bar{g}_{xy}+z\bar{g}_{xz})\,, \nonumber \\
F_y&\equiv&\frac{3f_1}{2\sqrt{x^2+y^2+z^2}}(x \bar{g}_{xy}+y\bar{g}_{yy}+z\bar{g}_{yz})\,, \nonumber \\
F_z&\equiv&\frac{3f_1}{2\sqrt{x^2+y^2+z^2}}(x \bar{g}_{xz}+y\bar{g}_{yz}+z\bar{g}_{zz})\,,
\end{eqnarray}
where the spatial transition function $f_1(\rho)$ is defined as in \eqref{eq:transfunc} with transition occurring between $\rho_{1a}=0.05$ and $\rho_{1b}=0.95$.

Then, we define the time-transition function
\begin{equation}
g(t,\rho)=\left(\frac{t}{\xi_2 f_0(\rho)+\xi_1(1-f_0(\rho))}\right)^4\,,
\end{equation}
where $f_0(\rho)$ is defined as in \eqref{eq:transfunc} with transition interval between $\rho_{0a}=0.0$ and $\rho_{0b}=0.95$. Notice that $g(0,\rho)=0$, $g(t,\rho)\gg 1$ for $t\gg\xi_1,\xi_2$ and, in particular, $g(t,\rho)$ takes large values with characteristic time $\xi_1$ in the interior region $\rho\leq\rho_{0a}$ (i.e., where $f_0=0$) and characteristic time $\xi_2$ in the near-boundary region $\rho\geq\rho_{0b}$ (i.e., where $f_0=1$).

With these ingredients, we can finally write the complete gauge choice made in our simulations
\begin{equation}
\bar{H}_\mu=\bar{H}_{\mu}\big|_{t=0}\exp(-g)+F_\mu[1- \exp(-g)]\,.
\end{equation}
From the properties of $g(t,\rho)$, we see that $\bar{H}_\mu=\bar{H}_{\mu}|_{t=0}$ at $t=0$ and $\bar{H}_\mu=F_\mu$ for $t\gg\xi_1$ in the interior and $t\gg\xi_2$ near the boundary. Since the target gauge is crucial for stability and needs to be reached quickly, $\xi_2$ is typically set to a small value. On the other hand, it is not necessary, and perhaps even troublesome, to deal with a fast transition in the bulk, therefore $\xi_1$ takes a larger value. In the simulations of Chapter~\ref{Chapter:NoSym} we set $\xi_1=0.1,\xi_2=0.0025$; in the simulations of Chapter~\ref{Chapter:KAdS} we set $\xi_1=0.1,\xi_2=0$.

\ifpaper
\end{document}
\fi
\newif\ifpaper
\paperfalse

\ifpaper
\input{../preamble}
\begin{document}
\fi
\chapter{Boundary extrapolation}\label{sec:extrapconvbdy}

As explained in Section~\ref{sec:bouset2}, since the AdS boundary generally does not lie on points of the Cartesian grid, we can only obtain the approximated value of any boundary quantity $f$ through extrapolation from the numerical values of $f$ on grid points near the boundary. In this section we describe how extrapolation is implemented in our scheme. 

For simplicity, we consider first order extrapolation, i.e., extrapolation from two grid points. The following can be generalized to higher extrapolation orders in a straightforward way. In particular, third order extrapolation is used for the plots in Section~\ref{sec:resbouset} and Section~\ref{sec:preresK}, since this improves the accuracy of the extrapolated numerical values.\footnote{This fact was tested by comparing values obtained with increasing extrapolation order and exact values, in cases where the latter are known, e.g. boundary scalar field values at $t=0$.}
A few minor details of the extrapolation technique differ between the simulations of Chapter~\ref{Chapter:NoSym} and the simulation of Chapter~\ref{Chapter:KAdS}. We first discuss the details of extrapolation for the simulations of Chapter~\ref{Chapter:NoSym}, and we then point out the differences in the extrapolation used in Chapter~\ref{Chapter:KAdS}.

\section{Extrapolation in Chapter~\ref{Chapter:NoSym}}
\label{sec:extcha3}

\begin{figure*}[t!]
        \centering
        \includegraphics[width=6.1in,clip=true]{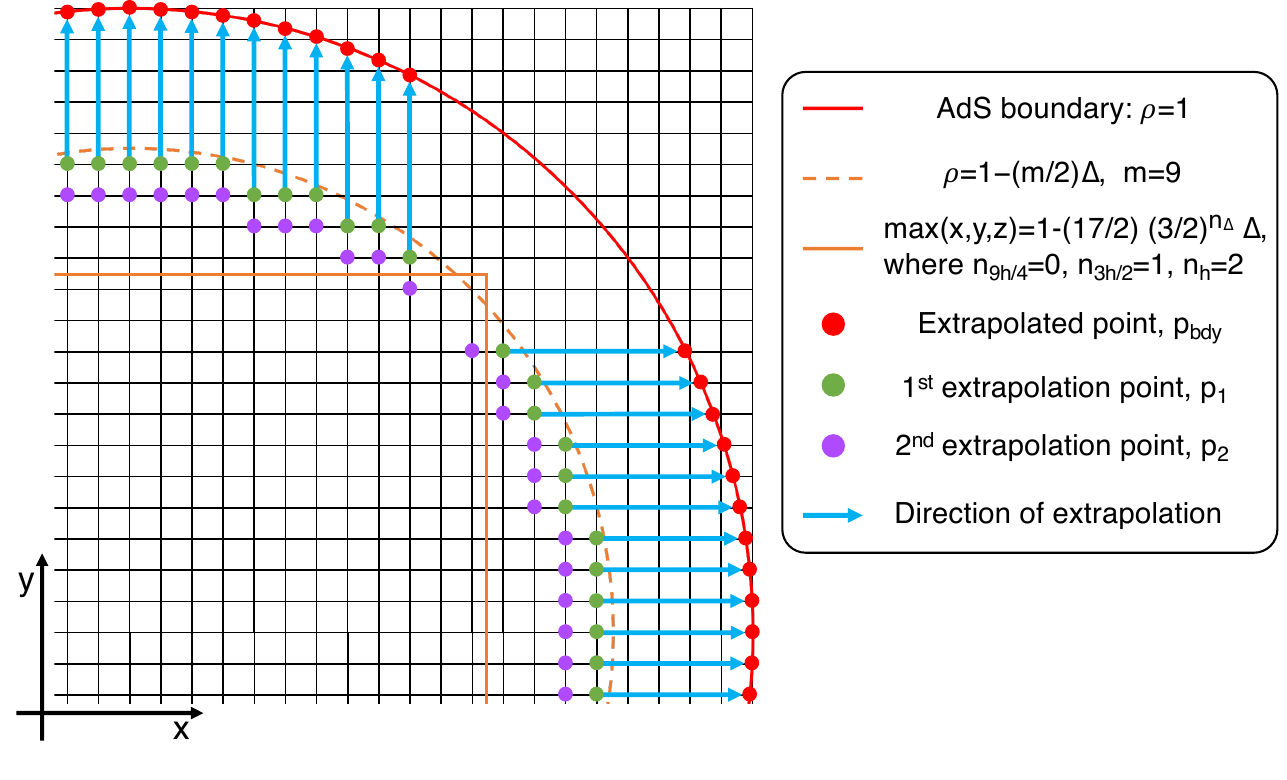}
\parbox{5.0in}{\caption{Visual description of first order extrapolation technique in the first quadrant of a surface at constant $z$ for a grid with spatial refinement $\Delta$.
        }\label{fig:lego_circle}}
\end{figure*}

Given a Cartesian grid with spacing $\Delta$, let $f_\Delta$ denote the values of $f$ at bulk grid points and $f^{bdy}_{\Delta}$ denote the extrapolated values of $f$ at boundary points. 
We extrapolate the values $f^{bdy}_{\Delta}$ through the following procedure.
 \begin{enumerate}
 \item Restrict to the points with Cartesian coordinates $(x,y,z)$ satisfying $\rho(x,y,z)<1-m\Delta/2$ for an integer $m$ (inside the orange dashed line of Figure~\ref{fig:lego_circle}), and $\max(x,y,z)>1-\frac{17}{2}\left(\frac{3}{2}\right)^{n_\Delta}\Delta$ (outside the continuous orange line of Figure~\ref{fig:lego_circle}), where $n_\Delta$ denotes the degree of the three resolutions used for convergence, $n_{9h/4}=0,n_{3h/2}=1,n_{h}=2$ (notice that $\left(\frac{3}{2}\right)^{n_\Delta}\Delta$ is a constant for all three resolutions). We pick $m=9$, since we have empirically found that considering points closer to the boundary leads to unphysical or non-converging values, already at early times. See Section~\ref{subsec:testextrconv} for a possible explanation.
 \item For any point in the range defined at step 1, identify the coordinate with the largest absolute value, e.g., $x$, and its sign, say $x>0$. If two coordinates have the same absolute value, then we pick $x$ over $y$ and $z$, and $y$ over $z$. Each direction identified in this way is represented by a light blue arrow. Among all the points along the identified direction ($x$ in our example) and within the range of step 1, pick the closest point to the boundary. We denote this point by $p_1$ and its coordinates by $(x_1,y_1,z_1)$. For each direction identified as above, the corresponding $p_1$ point is represented as a green dot in Figure~\ref{fig:lego_circle}.
 \item Consider the nearest point to $p_1$ along the identified axis in the direction of the bulk (decreasing $x$ in the example). We denote this point by $p_2$ and its coordinates by $(x_2,y_2,z_2)$. For each $p_1$ point, the corresponding $p_2$ is represented as a purple dot in Figure~\ref{fig:lego_circle}. In our example $x_2=x_1-\Delta,y_2=y_1,z_2=z_1$.
 \item Use first order extrapolation on $f_\Delta(p_1),f_\Delta(p_{2})$ to determine the value of $f^{bdy}_{\Delta}(p_{bdy})$ where $p_{bdy}$ is the boundary point along the identified axis in the direction of the boundary. 
For each pair $p_1,p_2$, the corresponding $p_{bdy}$ is represented by a red dot in Figure~\ref{fig:lego_circle} and the AdS boundary is represented by a red line.
In our example, $p_{bdy}$ is the point with coordinates $(x_{bdy},y_{bdy},z_{bdy})=(\sqrt{1-y_1^2-z_1^2},y_1,z_1)$ and
 \begin{equation}
 \label{eq:firstordextrap}
 f^{bdy}_{\Delta}(p_{bdy})=\frac{x_{bdy}-x_2}{x_1-x_2}f_\Delta(p_1)+\frac{x_{bdy}-x_1}{x_2-x_1}f_\Delta(p_2).
 \end{equation}
 \item In order to avoid issues arising from singularities in the definition of spherical coordinates in terms of Cartesian coordinates, we do not extrapolate boundary points with $y_{bdy}=z_{bdy}=0$. Furthermore, extrapolation will in general miss some of the points at the boundary of the $(\theta,\phi)$ domain, i.e., $\theta=0,\pi$ or $\phi=0,2\pi$.
We fill each of these points by copying the mean value of the closest boundary extrapolated points. This ensures continuity at the semi-circle $z_{bdy}=0, y_{bdy}\geq 0$, i.e., points with $\phi=0\sim 2\pi$.
 \end{enumerate}

Figure~\ref{fig:lego_circle} shows that the extrapolated values are not uniformly distributed on the boundary. 
We aim to improve this in the future by extrapolating the values at points on a uniform $(\theta,\phi)$ grid with given resolution on the $S^2$ at the boundary. For now, we fill the empty regions by linearly interpolating boundary values. The data obtained in this way displays high-frequency noise that does not allow for a clear visualisation of physical features. To eliminate these, we first create a $(\theta,\phi)$ grid with $N_\theta\times N_\phi$ grid points and uniform spacing, then we compute the value of $f$ at each point of this grid by interpolating the known values at boundary points. Finally, we apply a low-pass filter, with user-specified frequency threshold $\omega$, to the resulting grid function $f$.
More precisely, we apply the filter on three copies of the boundary sphere joined along the semi-circle $z_{bdy}=0$, $y_{bdy}\geq 0$ and then we plot the smooth data of the central copy. After re-enforcing continuity at the semi-circle, which consists of repeating step 5 above, this strategy provides regular smooth data at the semi-circle if the original raw data is approximately periodic in $\phi$ with period $2\pi$, which is expected for data on a sphere.

The plots of Section~\ref{sec:resbouset} have been obtained by third order Cartesian extrapolation, which straightforwardly generalises the technique described above. We smoothen the raw data on a grid with $N_\theta= N_\phi=55$ points. The plots of Figure~\ref{fig:snapshotsbdyphi} have been smoothened with a frequency threshold of $\omega=0.5$. We have verified that this threshold accurately reproduces the exact known values for the boundary scalar field $\bar{\varphi}_{(1)}$ at $t=0$.
The plots of Figure~\ref{fig:snapshotsenergydensity} have been smoothened with $\omega=0.25$.

\subsection{Testing boundary convergence}
\label{subsec:testextrconv}

Notice that, as \eqref{eq:firstordextrap} shows, second order convergence of boundary values $f^{bdy}_{\Delta}$ is a direct consequence of second order bulk convergence of $f_\Delta$, which is confirmed by Figure~\ref{fig:L2norm_iresallconvergence-crop} for the simulations of Chapter~\ref{Chapter:NoSym}. Despite this fact, some modifications must be made to our extrapolation scheme if we wish to perform explicit convergence tests on our boundary data. We now explain the reason for this and the necessary modifications. We assume the validity of the Richardson expansion \eqref{eq:Richu} for $f_{\Delta}$ at any grid point $p$,
\begin{equation}
\label{eq:Richexp}
f_\Delta(p)=f(p)+e(p)\Delta^2+\mathcal{O}(\Delta^3),
\end{equation}
where $f(p)$ is the true value of $f$ at $p$ and the rest of the right hand side is the solution error of $f_\Delta(p)$. The validity of this expansion is confirmed by bulk convergence of $f_\Delta$ to $f$.
Then, from \eqref{eq:firstordextrap}, we obtain the Richardson expansion for $f^{bdy}_{\Delta}$ at any extrapolated boundary point $p_{bdy}$:
 \begin{eqnarray}
 \label{eq:bdyRichexp}
 f^{bdy}_{\Delta}(p_{bdy})
&=&f(p_{bdy})+e_{extr}(p_{bdy},p_1,p_2)\nonumber \\
&&\hspace{0.8cm}+e_\Delta(p_1,p_2)\Delta^2 +\mathcal{O}(\Delta^3)\,,
 \end{eqnarray}
where the $f(p_{bdy})$ is the true value of $f$ at $p_{bdy}$, $e_{extr}(p_{bdy},p_1,p_2)$ is the error due to the extrapolation approximation. The remaining error terms come from the solution error in $f_\Delta$. The typical convergence test involves the computation of the convergence factor \eqref{eq:confact1},
\begin{equation}\label{eq:qconv}
Q(p_{bdy})=\frac{1}{\ln(3/2)}\ln\left( \frac{f_{9h/4}(p_{bdy})-f_{3h/2}(p_{bdy})}{f_{3h/2}(p_{bdy})-f_{h}(p_{bdy})} \right)
\end{equation}
at each boundary point $p_{bdy}$. We clearly see that $Q(p_{bdy})$ can be expected to asymptote to 2 as $\Delta\rightarrow0$, thus confirming second order convergence in the continuum limit, only if the points $p_1,p_2$ are the same for all 3 resolutions involved. Therefore, our extrapolation scheme must be modified to select pair of bulk points, $p_1$ and $p_2$, for extrapolation that are present in all three grids involved in the convergence test. 
In practice, we saw that boundary convergence follows the trend of bulk convergence only if, in addition to this modification, we restrict to $p_1$ points in the range mentioned in step 1 above. 
This fact might occur because the transition in the initial data, discussed in Section~\ref{sec:consistbound}, affects the results of extrapolation at early times, if the latter employs points in the transition region. This hypothesis is supported by the fact that in Chapter~\ref{Chapter:KAdS}, where the transition occurs far from the boundary (between $\rho=0.3$ and $\rho=0.6$), extrapolation works well in a different range, simply identified to be well outside the transition region. 
(see Section~\ref{sec:extcha4} below for the details of this range). However, the reason for this should be investigated further. 

Finally, \eqref{eq:bdyRichexp} shows that this type of test does not prove convergence to the true value $f(p_{bdy})$, but rather to its approximation $f(p_{bdy})+e_{extr}(p_{bdy},p_1,p_2)$. For this reason, the convergence test \eqref{eq:qires} cannot be performed at the boundary for functions with vanishing true value (such as $\langle trT \rangle_{CFT}$), because their extrapolated value is not just the term linear in $\Delta^2$ but it also includes the extrapolation error $c_{extr}$. A more detailed analysis must be made to examine the explicit form $e_{extr}(p_{bdy},p_1,p_2)$ and be able to find the rate of convergence to $f(p_{bdy})$. In our study, we simply make the natural assumption that $e_{extr}(p_{bdy},p_1,p_2)$ decreases as we increase resolution, so $f^{bdy}_{\Delta}(p_{bdy})$ is a sufficiently accurate approximation of $f(p_{bdy})$ for sufficiently high resolution (i.e., sufficiently small $\Delta$).

\section{Changes in the extrapolation in Chapter~\ref{Chapter:KAdS}}
\label{sec:extcha4}

The boundary extrapolation for the simulation presented in Chapter~\ref{Chapter:KAdS} proceeds exactly as explained in Section~\ref{sec:extcha3}, except for few minor details that we now point out.
We recall here that in Chapter~\ref{Chapter:KAdS} we consider only one simulation with number of grid points in the bulk given by $N_x=N_y=N_z=257$.

We set the value of $m$, appearing in step 1 above, to $m=5$, thus extrapolation employs points $p_1$ that are closer to the boundary with respect to those in Chapter~\ref{Chapter:NoSym}. Moreover, instead of the ad-hoc restriction $\max(x,y,z)>1-\frac{17}{2}\left(\frac{3}{2}\right)^{n_\Delta}\Delta$, we require that the points $p_1$ also satisfy the more ``natural'' condition $\rho>1-6\Delta$, which has the simple role of not allowing for points that are too far from the AdS boundary, and might thus lead to inaccurate extrapolated values.
In step 5 above, instead of filling the value at each missing boundary point by copying the mean value of the closest boundary extrapolated points, we fill that value by copying the value of the closest boundary extrapolated point. The same is done after the low-pass filter, when we repeat step 5.
We have found that the two methods are essentially equivalent, therefore we employ the cheapest one in terms of computing time.
For the plots of Figure~\ref{fig:bdyendensperKAdS} and Figure~\ref{fig:bdyendensperKAdS2}, we applied a low-pass filter with frequency threshold of $\omega=0.5$ on a grid with $N_\theta= N_\phi=65$ points. 
We previously verified that these specifics reproduce exact known values for all the boundary quantities defined in Section~\ref{sec:bouset2} with great accuracy.

For the simulation of Chapter~\ref{Chapter:KAdS}, we have compared the results of Cartesian extrapolation with those of a radial extrapolation method. The latter consists of identifying, for each boundary point $p_{bdy}$, two points (or more, for higher order extrapolation) $p_1,p_2$ along the line connecting $p_{bdy}$ with the origin $\rho=0$, and then i) extrapolate the value of $f_\Delta$ at $p_1,p_2$ by using the closest Cartesian grid points, ii) using $f_\Delta(p_1), f_\Delta(p_2)$, and the $\rho$-value of $p_1$ and $p_2$, denoted by $\rho_1, \rho_2$, to obtain  $f^{bdy}_{\Delta}(p_{bdy})$ from
 \begin{equation}
 f^{bdy}_{\Delta}(p_{bdy})=\frac{\rho_{bdy}-\rho_2}{\rho_1-\rho_2}f_\Delta(p_1)+\frac{\rho_{bdy}-\rho_1}{\rho_2-\rho_1}f_\Delta(p_2),
 \end{equation}
 where $\rho_{bdy}=1$ is the $\rho$-value of $p_{bdy}$.
We have seen that second order radial extrapolation gives results approximately as accurate as third order Cartesian extrapolation. We have thus decided to employ Cartesian extrapolation because it is computationally cheaper  for simulations in Cartesian coordinates, as only one extrapolation per boundary point is needed.

\ifpaper
\end{document}
\fi
\newif\ifpaper
\paperfalse

\ifpaper
\input{../preamble}
\begin{document}
\fi
\chapter{Convergence of the independent residual}\label{sec:convbulk}

To show that the solution is converging to a solution of the Einstein equations, we compute the \emph{independent residual} that is obtained by taking the numerical solution, and substituting it back into the discretized version of
$S_{\mu\nu}:=R_{\mu\nu}-\frac{1}{2}R g_{\mu\nu}+\Lambda g_{\mu\nu}-8\pi T_{\mu\nu}$.
At each grid point, we then take the maximum value over all components of $S_{\mu\nu}$, which we denote by $\Phi_\Delta$. 
Since the exact value of the independent residual is zero, we can test convergence, as explained in Section~\ref{sec:conv}, by computing the convergence factor \eqref{eq:confact2}, which in this case reads
\begin{equation}\label{eq:qires}
Q_{EFE}(t,x,y,z)=\frac{1}{\ln(3/2)}\ln\left( \frac{\Phi_{3h/2}(t,x,y,z)}{\Phi_{h}(t,x,y,z)} \right).
\end{equation}
Again, with second-order accurate finite difference stencils and with a factor of 3/2 between successive resolutions, we expect $Q$ to approach $Q=2$ as $\Delta\rightarrow0$.
 \begin{figure*}[t!]
        \centering
        \includegraphics[width=5.0in,clip=true]{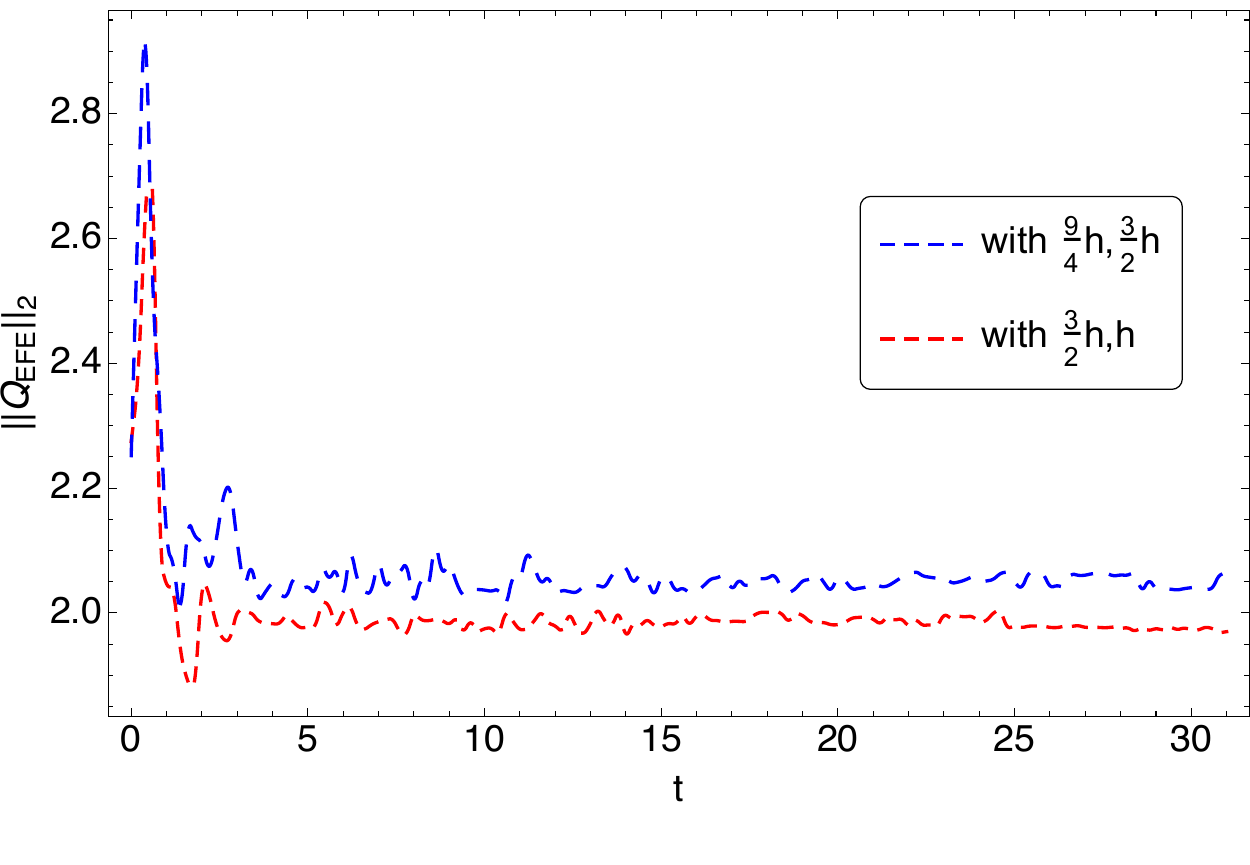}
\parbox{5.0in}{\caption{Time evolution for $L^2$-norm of convergence factor for independent residual of Einstein equations at different resolutions on the $z=0$ slice. This data is obtained from the simulations discussed in Chapter~\ref{Chapter:NoSym}.
        }\label{fig:L2norm_iresallconvergence-crop}}
\end{figure*}
Figure~\ref{fig:L2norm_iresallconvergence-crop} displays the $L^2$-norm of the convergence factor \eqref{eq:qires} for two pairs of resolutions on the $z=0$ slice, computed from the data of the simulations described in  Chapter~\ref{Chapter:NoSym}. It clearly shows second order convergence to a solution of the Einstein equations, after an initial transition phase.

\ifpaper
\end{document}
\fi
\newif\ifpaper
\paperfalse

\ifpaper
\input{../preamble}
\begin{document}
\fi
\chapter{Scalar, vector, tensor spherical harmonics}
\label{sec:svtsphharm}

In this appendix, we provide formulae for the scalar, vector, and tensor spherical harmonics according to the conventions used in Chapter~\ref{Chapter:KAdS} and in \cite{Chesler:2018txn,Chesler:2021ehz}.
Let $x^i=(\theta,\phi)$ be coordinates on the unit round sphere $(S^2,d\Omega^2)$, in which the metric reads $d\Omega^2=d\theta^2+\sin^2\theta d\phi^2$.
We use indices $i,j,k,p$ to denote indices associated with the coordinates $x^i$.
Let $D$ be the Levi-Civita covariant derivative associated with $d\Omega^2$.
Let $l,m$ be integer numbers satisfying $l\geq 0$ and $-l\leq m\leq l$.
For each pair $(l,m)$ we define \emph{scalar spherical harmonics} by
\begin{equation}
Y^{lm}=\left[\frac{2l+1}{4\pi}\frac{(l-m)!}{(l+m)!}\right]^{1/2}P^{lm}(\cos\theta)e^{im\phi},
\end{equation}
where $P_{lm}(u)$ are the associated Legendre polynomials
\begin{equation}
P_{lm}(u)=\frac{(-1)^m}{2^l l!}(1-u^2)^{m/2}\frac{d^{l+m}}{du^{l+m}}(u^2-1)^l.
\end{equation}
There are two classes of vector spherical harmonics. \emph{Even vector spherical harmonics} are denoted by $s=1$ and given by
\begin{equation}
\mathcal{V}^{1lm}_i=\frac{1}{\sqrt{l(l+1)}}D_iY^{lm}.
\end{equation}
\emph{Odd vector spherical harmonics} are denoted by $s=2$ and given by
\begin{equation}
\mathcal{V}^{2lm}_i=\frac{1}{\sqrt{l(l+1)}}\epsilon_i^{\phantom i j} D_j Y^{lm},
\end{equation}
where the only non-vanishing components of $\epsilon_i^{\phantom i j}$ are $\epsilon_\theta^{\phantom\theta\phi}=\csc\theta$ and $\epsilon_\phi^{\phantom\phi \theta}=-\sin\theta$.
The \emph{tensor spherical harmonics} are labelled by $s=1,2,3$ and given by
\begin{equation}
\label{eq:tensharm}
\begin{split}
\mathcal{T}^{1lm}_{ij}&=\frac{1}{\sqrt{l(l+1)(l(l+1)/2-1)}}\left(D_i D_j+\frac{l(l+1)}{2}h_{ij}\right) Y^{lm},\\
\mathcal{T}^{2lm}_{ij}&=\frac{1}{\sqrt{l(l+1)(l(l+1)/2-1)}}\left(\epsilon_{(i}^{\phantom{j)} k} D_{j)} D_k Y^{lm}\right),\\
\mathcal{T}^{3lm}_{ij}&=\frac{1}{\sqrt{2}}h_{ij}Y^{lm},\\
\end{split}
\end{equation}
where $h_{ij}$ are the components of $d\Omega^2$ in coordinates $x^i$.
Notice that a different labelling is common in the literature, i.e., the one obtained from our labelling by exchanging $s=3$ and $s=1$. We choose labels as in \eqref{eq:tensharm} since in this way the correspondence between Teukolski modes with $s=1,2$ and spherical harmonics with $s=1,2$, discussed in Section~\ref{subsec:blares}, is more straightforward to make.
The normalisation of the spherical harmonics is chosen in order to satisfy the following orthonormality relations:
\begin{equation}
\begin{split}
\int_{S^2} d\Omega Y^{lm} \left(Y^{l'm'}\right)^\ast&=\delta_{ll'}\delta_{mm'},\\
\int_{S^2} d\Omega h^{ij}\mathcal{V}_i^{slm} \left( \mathcal{V}_j^{s'l'm'}\right)^\ast&=\delta_{ss'}\delta_{ll'}\delta_{mm'},\\
\int_{S^2} d\Omega h^{ik}h^{jp}\mathcal{T}_{ij}^{slm} \left( \mathcal{T}_{kp}^{s'l'm'}\right)^\ast&=\delta_{ss'}\delta_{ll'}\delta_{mm'},
\end{split}
\end{equation}
where $\ast$ denotes complex conjugation and $d\Omega=\sin\theta d\theta d\phi$ is the area element of the unit round sphere.
Finally, all the spherical harmonics vanish when integrated over $S^2$, i.e., $\int_{S^2} d\Omega Y^{lm}=\int_{S^2} d\Omega \mathcal{V}_i^{slm}=\int_{S^2} d\Omega \mathcal{T}_{ij}^{slm}=0$.
We refer the reader to standard textbooks for an exhaustive list of other properties of spherical harmonics.

\ifpaper
\end{document}
\fi
\end{appendix}

\clearpage
\bibliographystyle{JHEP}
\markboth{References}{References}
\bibliography{biblio}

\end{document}